\documentclass[letterpaper,11pt]{article}
\pdfoutput=1 

\usepackage{jheppub} 

\usepackage[T1]{fontenc} 
\usepackage{lmodern}
\usepackage[mathscr]{euscript}
\usepackage{graphicx}
\usepackage{ytableau}
\usepackage{amsmath}
\usepackage{amssymb}
\usepackage{hyperref}
\usepackage{xcolor}
\usepackage{color}
\usepackage{dsfont}
\usepackage{graphicx}
\usepackage{tikz}
\usetikzlibrary{calc,decorations.markings}
\usepackage{textcomp}
\usetikzlibrary{shapes.geometric}
\newcommand{\dotr}[1]{#1^{\bullet}} 
\usepackage{extarrows} 
\definecolor{navyblue}{rgb}{0.0, 0.0, 0.5}
\definecolor{firebrick}{rgb}{0.7, 0.13, 0.13}
\usetikzlibrary{shapes.misc}

\DeclareFontFamily{U}{mathx}{\hyphenchar\font45}
\DeclareFontShape{U}{mathx}{m}{n}{
      <5> <6> <7> <8> <9> <10>
      <10.95> <12> <14.4> <17.28> <20.74> <24.88>
      mathx10
      }{}
\DeclareSymbolFont{mathx}{U}{mathx}{m}{n}
\DeclareFontSubstitution{U}{mathx}{m}{n}
\DeclareMathSymbol{\bigplus}        {1}{mathx}{"90}
\DeclareMathSymbol{\bigtimes}       {1}{mathx}{"91}

\DeclareFontFamily{OT1}{pzc}{}
\DeclareFontShape{OT1}{pzc}{m}{it}{<-> s * [1.10] pzcmi7t}{}
\DeclareMathAlphabet{\mathpzc}{OT1}{pzc}{m}{it}
\newenvironment{rcases}
  {\left.\begin{aligned}}
  {\end{aligned}\right\rbrace}

\def\beq{\begin{equation}}
\def\eeq{\end{equation}}
\def\Det{{\mathrm{Det}}}
\def\Tr{{\mathrm{Tr}}}

\newcommand{\fm}{\mathfrak{m}}

\newcommand{\qe}{\mathfrak{q}}

\newcommand{\BA}{{\mathbb{A}}}
\newcommand{\BB}{{\mathbb{B}}}
\newcommand{\BC}{{\mathbb{C}}}

\newcommand{\BR}{{\mathbb{R}}}

\newcommand{\ve}{{\varepsilon}}
\newcommand{\ii}{{\mathrm{i}}}

\newcommand{\BD}{{\mathbb{D}}}

\newcommand{\uCalC}{\underline{\EuScript{C}}}
\newcommand{\CalD}{\mathcal{D}}

\newcommand{\CalS}{{\mathcal S}}

\newcommand{\CalP}{{\mathcal P}}

\title{\boldmath
{Opers, surface defects, and Yang-Yang functional}}

\author[a]{Saebyeok Jeong}
\author[a, b]{and Nikita Nekrasov}
\preprint{YITP-SB-18-17}

\affiliation[a]{C.N. Yang Institute for Theoretical Physics, Stony Brook University,\\Stony Brook, NY 11794-3840, USA}
\affiliation[b]{Simons Center for Geometry and Physics, Stony Brook University,\\ Stony Brook, NY 11794-3636, USA}

\emailAdd{saebyeok.jeong@gmail.com, nnekrasov@scgp.stonybrook.edu}

\abstract{We explore the non-perturbative Dyson-Schwinger equations obeyed by the partition functions of the $\Omega$-deformed $\EuScript{N}=2, d=4$ supersymmetric  linear quiver gauge theories in the presence of surface defects. We demonstrate that the partition functions of different types of defects (orbifold or vortex strings) are related by analytic continuation. We introduce Darboux coordinates on a patch of the moduli space of flat $SL(N)$-connections on a sphere with special punctures, which generalize the NRS coordinates defined in the $SL(2)$ case. Finally, we compare the generating function of the Lagrangian variety of opers in these Darboux coordinates with the effective twisted superpotential of the linear quiver theory in the two-dimensional $\Omega$-background, thereby proving the NRS conjecture and its generalization to the $SL(3)$ case.}

\begin{document} 
\maketitle
\flushbottom

\section{Introduction} \label{sec:intro}
The dynamics of supersymmetric gauge theories is a rewarding research subject. 
The exact low-energy description of the four-dimensional gauge theories with $\EuScript{N}=2$ supersymmetry was proposed in \cite{sw1, sw2} for the $SU(2)$ theories with various matter multiplets. The proposal has been generalized in the subsequent papers, allowing for different gauge groups and matter representations. In many cases the Coulomb branch of the moduli space of vacua is a family of algebraic curves (called the Seiberg-Witten curves) equipped with meromorphic differential. The periods of the differential compute the central charges of the supersymmetry algebra determining the masses of the BPS particles at this vacuum. The microscopic study of these theories using direct quantum field theory methods and supersymmetric localization was initiated in \cite{nek1}, leading to the exact computation of the partition functions of a deformed version of the theory, the realization they coincide with the partition functions of some two dimensional chiral theory, and connecting that theory to the $M$- and string theory fivebranes \cite{nek1, N1, N1bis, gai1, agt}. 

{}The method of \cite{nek1} reduces the computation of the path integral to a problem of counting fixed points under the action of the global symmetry group on a finite dimensional BPS field configurations. More specifically, the partition function can be written as a product of analytic functions, 
\begin{align} \label{partfunc}
\EuScript{Z}(\mathbf{a}, \mathbf{m}, \boldsymbol{\varepsilon}, \mathfrak{q}) = \EuScript{Z}^{\text{classical}} (\mathbf{a}, \boldsymbol{\varepsilon}, \mathfrak{q})  \; \EuScript{Z}^{\text{1-loop}} (\mathbf{a}, \mathbf{m}, \boldsymbol{\varepsilon})  \; \EuScript{Z}^{\text{inst}} (\mathbf{a}, \mathbf{m}, \boldsymbol{\varepsilon}, \mathfrak{q}).
\end{align}
Here $\mathfrak{q}$ schematically denotes the gauge couplings of the theory, while $\mathbf{a}$, $\mathbf{m}$, and $\boldsymbol{\varepsilon} = (\varepsilon_1 ,\varepsilon_2)$ denote the equivariant parameters for the group of global gauge symmetry, the group of flavor symmetry, and the group of Lorentz symmetry, respectively. $\varepsilon_{1,2}$ are also called $\Omega$-deformation parameters (See appendix \ref{appA} for a more detailed review of the $\EuScript{N}=2$ partition functions). The effective prepotential is then obtained by taking the limit (while keeping $\mathbf{a},\mathbf{m}, \mathfrak{q}$ generic)
\begin{align}
\EuScript{F} (\mathbf{a}, \mathbf{m}, \mathfrak{q}) = \lim_{\varepsilon_{1}, \varepsilon_2 \to 0} \varepsilon_1 \varepsilon_2 \log{\EuScript{Z}(\mathbf{a}, \mathbf{m}, \boldsymbol{\varepsilon}, \mathfrak{q})},
\end{align}
which provides the direct microscopic derivation of the results in \cite{sw1, sw2} (either using the limit shape approach \cite{nekokoun}, or the blowup equations \cite{ny}). 

Meanwhile, it was observed in \cite{swint1,swint2,swint3} that the Coulomb branch of vacua of a $\EuScript{N}=2$ theory canonically has a structure of a base $\EuScript{B}$ of an algebraic integrable system. The full structure is revealed when the theory is compactified \cite{sw3} on a circle $S^{1}_R$ \cite{nekpes}. The moduli space of the effective $\EuScript{N}=4, d=3$ theory is a hyper-K\"{a}hler manifold which metrically collapses to the Coulomb moduli space $\EuScript{B}$ of the four-dimensional theory in the limit $R \to \infty$ \cite{sw3}. In this limit, one of the complex structures, say, $I$ is singled out, with respect to which we have a holomorphic symplectic form $\Omega_I$. For finite $R$, the moduli space is a $\Omega_I$-Lagrangian fibration over $\EuScript{B}$ by abelian varieties. More specifically,  the Coulomb branch $\EuScript{B}$ is parametrized by the expectation values $u_{k} = \langle \EuScript{O}_k \rangle$ of chiral observables (these are local operators anticommuting with the four nilpotent supercharges of one Lorentz chirality). 

These observables carry over to the theory with finite $R$. We define the Hamiltonians to be the $I$-holomorphic functions on  
the moduli space of the compactified theory by
\begin{equation}
H_{k} = \langle \EuScript{O}_k \rangle \ , \quad k = 1, \cdots, \dim \EuScript{B},
\end{equation}
and it is not difficult to show that these functions Poisson-commute with respect to the
${\Omega}_{I}^{-1}$. 

\subsection{Quantization via gauge theory: \\ Effective twisted superpotential as Yang-Yang functional}
The remarkable correspondence between the gauge theory and integrable system was promoted to the quantum level in \cite{neksha3}, by placing the gauge theory into the realm of Bethe/gauge correspondence \cite{neksha1, neksha2}. We consider the theory in the $\Omega$-background affecting two out of four dimensions of spacetime. Equivalently, we take \textit{the NS limit} $(\varepsilon_1 = {\hbar} \neq 0, \varepsilon_2 \to 0)$ of the general $\Omega$-background, so that the theory  retains the two-dimensional $\EuScript{N}=(2,2)$ supersymmetry. The effective action includes the twisted $F$-term given by the effective twisted superpotential,
\begin{align} \label{efftw}
\widetilde{\EuScript{W}} (\mathbf{a}, \mathbf{m} ,\hbar, \mathfrak{q}) = \lim_{\varepsilon_2 \to 0} \varepsilon_2  \log{\EuScript{Z}(\mathbf{a}, \mathbf{m}, \varepsilon_1 = \hbar, \varepsilon_2, \mathfrak{q})}.
\end{align}
Typically, the theories with four supercharges have isolated vacua. In this way the $\Omega$-deformation of the four dimensional theory lifts the continuous moduli of vacua. The discrete set of vacua is in one-to-one correspondence with the solutions to the following equation,
\begin{align} \label{vaceq}
\exp{ \frac{ \partial \widetilde{\EuScript{W}} (\mathbf{a}, \mathbf{m} ,\hbar, \mathfrak{q}) }{\partial a_{\alpha}}} =1, \quad \alpha = 1, \cdots, \dim{\EuScript{B}}
\end{align}
In the context of Bethe/gauge correspondence, this equation is identified with the Bethe equation which determines the set of joint eigenvalues of the mutually commuting Hamiltonians. The Coulomb moduli $\mathbf{a}$ in \eqref{vaceq} map to the quasi-momenta, or Bethe roots, of the integrable system. The spectrum of the Hamiltonians for a given solution $\mathbf{a}_{*} $ of \eqref{vaceq} is computed as
\begin{align} \label{spec}
u_k (\mathbf{a}_{*}, \mathbf{m}, \hbar, \mathfrak{q}) = \langle \EuScript{O}_k \rangle^{{\ve}_{1} = \hbar, {\ve}_{2} = 0 ; \, \mathbf{m}, \mathfrak{q}}_{\mathbf{a} = \mathbf{a}_{*} },
\end{align}
The $\Omega$-deformation parameter $\hbar$ plays the role of the Planck constant of the quantum integrable system. The potential $\widetilde{\EuScript{W}}$ of the Eqs. \eqref{vaceq} determining the Bethe roots is identified with \textit{the Yang-Yang functional} \cite{yy} in the context of the integrable system. 
The effective twisted superpotential, or the Yang-Yang functional, can be written in the following form according to the decomposition of \eqref{partfunc},
\begin{align}
\widetilde{\EuScript{W}} (\mathbf{a}, \mathbf{m} ,\hbar, \mathfrak{q}) = \widetilde{\EuScript{W}}^{\text{classical}} (\mathbf{a}, \mathbf{m} ,\hbar) \log{\mathfrak{q}} + \widetilde{\EuScript{W}}^{\text{1-loop}} (\mathbf{a}, \mathbf{m} ,\hbar) + \widetilde{\EuScript{W}}^{\text{inst}} (\mathbf{a}, \mathbf{m} ,\hbar, \mathfrak{q}).
\end{align}
The 1-loop part depends on the regularization scheme but is independent of the gauge coupling $\mathfrak{q}$, while the instanton part is expanded as a series in $\mathfrak{q}$. The series can be exactly computed by taking the NS limit of the Young diagram expansion of the instanton partition function. See appendix \ref{appA} for more background on the localization computation of the effective twisted superpotential.

\subsection{Hitchin systems, flat connections, and opers}
In this paper, we study a specific subclass of the four-dimensional $\EuScript{N}=2$ theories, which is called the class-$\EuScript{S}$ theories \cite{gai1}. The class-$\EuScript{S}$ theory $\mathcal{T}[\mathfrak{g}, \EuScript{C}]$ $(\mathfrak{g} = ADE)$ is the four-dimensional $\EuScript{N}=2$ superconformal theory engineered by compactifying the 6-dimensional $(2,0)$ superconformal theory of type $\mathfrak{g}$ on the Riemann surface $\EuScript{C}$, with a partial topological twist. As we discussed earlier, the further compactification of $\mathcal{T}[\mathfrak{g}, \EuScript{C}]$ on a circle $S^1$ yields a three-dimensional $\EuScript{N}=4$ gauge theory whose Coulomb moduli space is the phase space of the Seiberg-Witten integrable system. By changing the order of compactification on $\EuScript{C} \times S^1$\cite{gmn1}, it can be verified that the moduli space is equivalent to the moduli space $\EuScript{M}_H (G, \EuScript{C})$ of the Hitchin pairs $({\CalP}, \varphi)$, that is, the locus of the Hitchin equations on $\EuScript{C}$ \cite{hit1},
\begin{align}\label{hiteq}
\begin{split}
&F_A + [ \varphi, \bar{\varphi}] =0  \\
&\bar{\partial}_A \varphi = 0, \;\; \partial_A \bar{\varphi} =0.
\end{split}
\end{align}
modulo the $G$-gauge transformations. Here, $G$ is the simple Lie group corresponding to $\mathfrak{g}$, $A$ is a $G$-connection on the principal $G$-bundle ${\CalP} \to \EuScript{C}$, and $\varphi \in {\Gamma} (\EuScript{C}, K_{\EuScript{C}} \otimes \text{ad}_{\CalP})$ is the $\mathfrak{g}_{\BC}$-valued (1,0)-form called the Higgs field. Note that $\EuScript{C}$ may have punctures, and the Higgs field is prescribed  to have specific singular behaviors at those punctures. Therefore, the Seiberg-Witten integrable system for the class-$\EuScript{S}$ theory $\mathcal{T}[\mathfrak{g}, \EuScript{C}]$ is the Hitchin integrable system with the phase space $\EuScript{M}_H (G, \EuScript{C})$.

As discussed in \cite{kapwit}, we can view the Hitchin moduli space $\EuScript{M}_H$ as a hyper-K\"{a}hler quotient of the affine space $\mathcal{W}$ of all the field configurations of $(A, \varphi)$. $\mathcal{W}$ is hyper-K\"{a}hler with a natural $\mathbb{P}^1$-family of complex structures,
\begin{align}
\mathcal{I} = a I + b J + c K, \quad \mathcal{I}^2 = -1, \quad \text{for}\quad a^2+b^2+c^2=1,
\end{align}
where we may choose the convention that $I$, $J$, and $K$ are the complex structures with the holomorphic coordinates $(A_{\bar{z}}, \varphi_z)$, $(\mathcal{A}_z \equiv A_z + i \varphi_z, \mathcal{A}_{\bar{z}} \equiv A_{\bar{z}} + i \varphi_{\bar{z}})$, and $(A_z+\varphi_z , A_{\bar{z}} - \varphi_{\bar{z}})$, respectively. The corresponding K\"{a}hler forms are
\begin{align}
\begin{split}
&\omega_I = -\frac{1}{4\pi} \int_{\EuScript{C}} \text{Tr} \; (\delta A \wedge \delta A - \delta \varphi \wedge \delta \varphi), \\
&\omega_J = \frac{1}{2\pi}\int_{\EuScript{C}} \vert d^2 z \vert \; \text{Tr} \; (\delta \varphi_{\bar{z}} \wedge \delta A_z + \delta \varphi_z \wedge \delta A_{\bar{z}}),\\
&\omega_K = \frac{1}{2\pi} \int_{\EuScript{C}} \text{Tr} \; (\delta A \wedge \delta \varphi).
\end{split}
\end{align}
Then the Hitchin equations \eqref{hiteq} are just the moment map equations for these K\"{a}hler forms. Therefore $\EuScript{M}_H (G, \EuScript{C})$ is also hyper-K\"{a}hler with the same complex structures and K\"{a}hler forms. We also define $\Omega_I = \omega_J + i \omega_k$ and its cyclic permutations,
\begin{align}
\begin{split} \label{holsymp}
&\Omega_I = \frac{1}{\pi} \int_{\EuScript{C}} \vert d^2 z \vert \; \text{Tr} \; (\delta \varphi_z \wedge \delta A_{\bar{z}}), \\
&\Omega_J = -\frac{i}{4\pi} \int_{\EuScript{C}} \text{Tr } (\delta \mathcal{A} \wedge \delta \mathcal{A}), \\
&\Omega_K = -\frac{i}{2\pi} \int_{\EuScript{C}} \vert d^2 z \vert \; \text{Tr} (\delta A_{\bar{z}}  \wedge  \delta A_z - \delta \varphi_{\bar{z}} \wedge \delta \varphi_z - \delta \varphi_{\bar{z}} \wedge \delta A_z - \delta \varphi_z \wedge \delta A_{\bar{z}} ),
\end{split}
\end{align}
each of which is a holomorphic symplectic (2,0)-form with respect to the complex structure $I$, $J$, and $K$, respectively.

The complete integrability of $\EuScript{M}_H (G, \EuScript{C})$ is manifest when we work in the complex structure $I$. We restrict our attention to the case $\mathfrak{g} = A_{N-1}$ from now on. Let us define the Hitchin fibration by the map,
\begin{align}
\begin{split}
\pi : \EuScript{M}_H (A_{N-1}, \EuScript{C}) &\longrightarrow \EuScript{B} \equiv \bigoplus_{k=2} ^N H^0 (\EuScript{C}, K_\EuScript{C} ^k), \\
(\CalP, \varphi) &\longmapsto \left( \text{Tr}\varphi^k \right)_{k=2} ^N.
\end{split}
\end{align}
It is possible to show that under the partial topological twist, the vacuum expectation values of the chiral observables of $U(1)$ $R$-charge $k$ exactly span $H^0 (\EuScript{C}, K_{\EuScript{C}} ^k)$. Therefore, we observe that the base $\EuScript{B}$ of the Hitchin fibration is precisely the Coulomb moduli space of $\mathcal{T}[A_{N-1}, \EuScript{C}]$. It is clear from the expression for $\Omega_I$ in \eqref{holsymp} that all the base elements are mutually Poisson-commuting under $\Omega_I$. A dimension counting also shows that $\dim{\EuScript{B}} = \frac{1}{2}\dim{\EuScript{M}_H (A_{N-1}, \EuScript{C})}$. Finally, the preimage of $u = \left( u_k (z) \right)_{k=2} ^N \in \EuScript{B}$ can be shown to be an abelian variety, the Jacobian $\text{Jac} (\Sigma_u)$ of the spectral curve
\begin{align} \label{swc}
\Sigma_u =\{(z,x) \in T^\ast \EuScript{C} \; \vert \; x^N + \sum_{k=2} ^N u_k (z) x^{N-k}=0 \} \subset T^\ast \EuScript{C},
\end{align}
establishing the algebraic integrable structure of $\EuScript{M}_H (A_{N-1}, \EuScript{C})$. The spectral curve $\Sigma_u$ is identified with the Seiberg-Witten curve of the theory $\mathcal{T}[A_{N-1}, \EuScript{C}]$.

On the other hand, we can alternatively view $\EuScript{M}_H (A_{N-1}, \EuScript{C})$ through the complex structure $J$. Up to some stability issue that we do not discuss here, the hyper-K\"{a}hler quotient can be equivalently performed by imposing only the moment map equation for $\Omega_J$, 
\begin{align}
\mathcal{F} \equiv d \mathcal{A} + \mathcal{A} \wedge \mathcal{A}=0,
\end{align}
and moding out the $G_{\mathbb{C}} (= SL(N))$-gauge transformations. Thus, the Hitchin moduli space $\EuScript{M}_H (A_{N-1}, \EuScript{C})$ is identified with the moduli space of flat $SL(N)$-connections on $\EuScript{C}$, $\EuScript{M}_{\text{flat}} (SL(N), \EuScript{C})$. It is convenient to use the holonomy map to express $\EuScript{M}_{\text{flat}} (SL(N), \EuScript{C})$ as the character variety, i.e., the representations of the fundamental group of $\EuScript{C}$,
\begin{align} \label{flatmoduli}
\EuScript{M}_{\text{flat}} (SL(N), \EuScript{C}) = \left\{ \rho \in \text{Hom} (\pi_1 (\EuScript{C}), SL(N)) \; \vert \; \left[ \rho(\gamma_i) \right] \text{ fixed} \right\} /SL(N),
\end{align}
where $\{i\}$ enumerates all the punctures in $\EuScript{C}$, $\gamma_i$ is the loop encircling the $i$-th puncture only, and the bracket $[\cdots]$ denotes the conjugacy class. The Poisson structure induced by $\Omega_J$ on $\EuScript{M}_{\text{flat}} (SL(N), \EuScript{C})$ can be explicitly written as the skein-relations on the Wilson loops \cite{gold, tur}.

To see the quantization at work, the class-$\EuScript{S}$ theory $\mathcal{T}[A_{N-1}, \EuScript{C}]$ is subject to  the $\Omega$-deformation in the NS limit. This is most effectively implemented by deforming the underlying geometry into the product of a cylinder and a cigar-like geometry, $X^4 = \mathbb{R} \times S^1 \times {\CalD}^2$ \cite{nekwit}. The following metric on ${\CalD}^2$ is taken,
\begin{align} \label{cigar}
\begin{split}
&ds^2 = dr^2 + f(r) d\theta^2, \quad r \in {\bf I}  = [ 0, \infty] , \quad  \theta \in [0, 2\pi), \\
&\text{with } f(r) \sim r^2 \quad \quad \:\: \text{for } r \sim 0, \\
& \quad\quad \; f(r) \sim \text{const} \quad \text{for sufficiently large } r
\end{split}
\end{align}
Note that this metric asymptotes to $X^4 \sim \mathbb{R} \times S^1 \times {\bf I} \times \widetilde{S^1}$.  One recalls that the $\Omega$-deformation with respect to the isometries of the two-torus can be undone by a redefinition of the fields of the theory \cite{nekwit}. In the limit where both circles $S^1$ and ${\widetilde{S^1}}$ are small we can approximate the theory by its reduction. The dependence  of the theory on the radii of the  circles $S^1$ and ${\widetilde{S^1}}$ is $Q$-exact, where $Q$ is the supercharge preserved by the $\Omega$-deformation. The dimensional reduction along the two-torus $S^1 \times \widetilde{S^1} $ results in a two-dimensional $\EuScript{N}=(4,4)$ sigma model, with the worldsheet  $\mathbb{R} \times {\bf I}$ and the target space $\EuScript{M}_H (A_{N-1}, \EuScript{C})$. The quantization of the Hitchin integrable system arises by correctly specifying the boundary conditions at $0, \infty \in {\bf I}$ \cite{nekwit}. The boundary condition at $\infty \in {\bf I}$ determines the space of states in the integrable system, implemented by a $\omega_K$-Lagrangian brane. It is also argued in \cite{nekwit} that the effect of the $\Omega$-deformation is correctly accounted by the boundary condition at $0\in {\bf I}$ corresponding to the canonical coisotropic brane of $\EuScript{M}_H (A_{N-1}, \EuScript{C})$ \cite{kap1}. Surprisingly, this brane could be T-dualized along the fibers of the Hitchin fibration to produce a brane supported on a distinguished $J_{\hbar}$-holomorphic $\Omega_{J_{\hbar}}$-Lagrangian submanifold of $\EuScript{M}_H (A_{N-1}, \EuScript{C})$: conjecturally, \textit{the variety of opers} \cite{bedrin}. Here, $J_{\hbar}$ differs from $I, -I$, and is determined by the $\Omega$-deformation parameter $\hbar$. In the absence of punctures on $\EuScript{C}$ all complex structures different from $I,-I$ are diffeomorphic. When punctures are present the diffeomorphism rotating $J_{\hbar}$ to $J$ changes the masses of the matter hypermultiplets, and, accordingly, the eigenvalues of the monodromy around the punctures. With this subtlety understood, we shall skip the subscript $\hbar$ in the notation for the complex structure $J$ in what follows. 

The variety $\mathcal{O}_N [\EuScript{C}] = \{ \widehat{\mathfrak{D}} \}$ of opers can be locally represented as a set of $N$-th order meromorphic differential operators
\begin{align} \label{oper}
\widehat{\mathfrak{D}} = {\partial}_{z}^{N} + t_2 (z)  {\partial}_{z}^{N-2} + \cdots + t_{N} (z),
\end{align}
acting on $\left( -\frac{N-1}{2} \right)$-differentials $K_\EuScript{C} ^{ -\frac{N-1}{2}}$. Here we view $\widehat{\mathfrak{D}}$ as an element of $\EuScript{M}_{\text{flat}} (SL(N),  \EuScript{C})$ by associating it to the representation
\begin{align}
\begin{split}
\rho_{\widehat{\mathfrak{D}}} : \pi_1 (\EuScript{C}) &\longrightarrow SL(N)\\
\gamma &\longmapsto M_{\gamma} (\widehat{\mathfrak{D}}),
\end{split}
\end{align}
where $M_{\gamma} (\widehat{\mathfrak{D}})$ is the $SL(N)$-valued monodromy of the solutions of $\widehat{\mathfrak{D}}$ along the loop $\gamma$. More specifically, the conjugacy class of the monodromy around each puncture is fixed, so that
\begin{align}
\mathcal{O}_N [\EuScript{C}] = \left\{ \; \widehat{\mathfrak{D}}  \;\; \Big\vert \; \left[ M_{\gamma_i} (\widehat{\mathfrak{D}}) \right] \text{ fixed} \right\},
\end{align}
leaving only $\dim{\mathcal{O}_N [ \EuScript{C}]} = \dim{\EuScript{B}}$ degrees of freedom for the meromorphic functions $\left( t_k (z) \right)_{k=2} ^N$ which is equal to the half of the dimension of the full moduli space $\EuScript{M}_{\text{flat}} (SL(N), \EuScript{C})$. In fact, as an oper \eqref{oper} can be regarded as a quantization of the Seiberg-Witten curve \eqref{swc}, the variety of opers $\mathcal{O}_N [\EuScript{C}]$ provides a quantization of the Coulomb moduli space $\EuScript{B}$, and the holomorphic functions on $\mathcal{O}_N [\EuScript{C}]$ precisely correspond to the \textit{off-shell} spectra of the mutually commuting quantum Hitchin Hamiltonians \cite{bedrin2}. 

The $\omega_K$-Lagrangian brane at infinity $\infty \in {\bf I}$ is T-dualized to another $\omega_K$-Lagrangian brane $L$. The ground states of open strings with two ends on $\mathcal{O}_N [\EuScript{C}]$ and $L$, respectively, define the space of morphisms in Fukaya category
\begin{align}
\mathcal{H} = \text{Hom}(\mathcal{O}_N [\EuScript{C}], L).
\end{align}
The space of morphisms between two Lagrangian branes in Fukaya category is the symplectic Floer homology $\dotr{HF} _{\text{symp}} (\mathcal{O}_N [\EuScript{C}], L)$, which can be obtained as a cohomology of a complex spanned by the intersection points with the differential obtained by studying pseudo-holomorphic disks with boundaries on $\mathcal{O}_N [\EuScript{C}]$ and $L$. For hyper-K\"{a}hler manifolds, such as the Hitchin space in our case, there is no contribution coming from the disks of non-zero relative degree, thus the space of states are determined by the classical intersection points.\footnote{There is a subtlety when the Lagrangians are not transversal. It appears the lift of degeneracy of the ground states \cite{sb,gms} in quantum mechanics corresponds to such singularities.} In other words, the problem of quantization reduces to enumeration of the intersection of the variety of opers and a $\omega_K$-Lagrangian brane. The isolated intersection point defines a common eigenstate of the quantum Hamiltonians. The spectra of quantum Hamiltonians are the holomorphic functions on the variety of opers restricted to this locus.

\subsection{The NRS conjecture}
Since the variety of opers $\mathcal{O}_N [\EuScript{C}]$ is a complex Lagrangian submanifold of $\EuScript{M}_{\text{flat}} (SL(N), \EuScript{C})$, there exists the generating function $\mathcal{S}\left[{\mathcal{O}_N [\EuScript{C}]} \right]$ for $\mathcal{O}_N [\EuScript{C}]$,
\begin{align}
\boldsymbol{\beta}_i = \frac{\partial \mathcal{S} \left[ \mathcal{O}_N [\EuScript{C}] \right]} {\partial \boldsymbol{\alpha}_i}, \quad\quad\quad i=1, \cdots, \frac{1}{2} \dim{\EuScript{M}_{\text{flat}} (SL(N), \EuScript{C})},
\end{align}
for any Darboux coordinate system $\{ \boldsymbol{\alpha}_i , \boldsymbol{\beta}_j \} = \delta_{ij}$ on $\EuScript{M}_{\text{flat}} (SL(N), \EuScript{C})$. In \cite{nrs}, it was suggested that there exists a specific Darboux coordinate system (which we refer to as \textit{the NRS coordinate system}), in which the generating function for the variety of opers is identified with the effective twisted superpotential, up to a contribution from the boundary at the infinity which is independent of the gauge coupling, namely,
\begin{align} \label{nrsc}
\mathcal{S} \left[ \mathcal{O}_N [\EuScript{C}]\right] = \frac{1}{\varepsilon_1}\left( \widetilde{\EuScript{W}} \left[ \mathcal{T}[A_{N-1}, \EuScript{C}] \right] - \widetilde{\EuScript{W}}_{\infty} \right).
\end{align}
In the $N=2$ case, the NRS coordinate system on the moduli space of $SL(2,{\BC})$-flat connections essentially restricts to the coordinate systems proposed in \cite{gold2, fn, km} for the $SU(2)$ flat connections, Teichm{\"u}ller space (which is a component of the moduli space of $SL(2,{\BR})$-flat connections) and the $SO(1,2)$-flat connections, respectively. The intuition behind the above equivalence is that as we vary the complex structure of $\EuScript{C}$, the corresponding variation of $\mathcal{O}_2 [\EuScript{C}]$ is represented by a closed holomorphic one-form on $\mathcal{O}_2 [\EuScript{C}]$, which is a derivative of a holomorphic function since $\mathcal{O}_2 [\EuScript{C}]$ is simply-connected. As we noted earlier, the holomorphic functions on $\mathcal{O}_2 [\EuScript{C}]$ are the spectra of the quantum Hamiltonians, which are, in the spirit of the Bethe/gauge correspondence,
\begin{align}
u = \mathfrak{q} \frac{\partial \widetilde{\EuScript{W}} \left[ \mathcal{T}[A_{1}, \EuScript{C}] \right] }{\partial \mathfrak{q}}.
\end{align}
Since the complex structure of $\EuScript{C}$ is controlled by the gauge coupling $\qe$, this motivated \cite{nrs} to identify the generating function for the variety of opers with the effective twisted superpotential, and thereby with the Yang-Yang functional. As a result, the classical symplectic geometry (which operates with symplectic manifolds and their Lagrangian subvarieties), the $\EuScript{N}=2$ gauge theory, and quantum integrable system (which belongs to the domain of noncommutative algebras, their commutative subalgebras, and representation theory) are nicely interconnected through the equality \eqref{nrsc}.  Note that this is a finite-dimensional version of the quantum/classical duality studied at some examples in \cite{lukzam}, which connects the integrable quantum field theories to the classical nonlinear differential equations.

There were many questions that remain unanswered. Some of them are:
\begin{enumerate}
\item Can one precisely describe the variety of opers $\mathcal{O}_N [\EuScript{C}]$  as of a deformation of the Coulomb moduli space $\EuScript{B}$ (of course, the first order deformation is simply the WKB approximation)? In particular, how the meromorphic coefficients $\left( t_k (z) \right)_{k=2} ^N$ in \eqref{oper} are related to the expectation values  $\left( u_k \right)_{k=2} ^N$ of the chiral observables in \eqref{spec}?
\item  How is the NRS coordinate system generalized to the higher rank case, at least for $\mathfrak{g} = A_{N-1}$?\footnote{In the genus one case it was done in \cite{Gorsky:1993dq, fr}.}

\item How should the equality \eqref{nrsc} be understood? Specifically, the left hand side is written in the NRS coordinates, while the right hand side is written in the gauge theoretic terms. How do we match these parameters?\footnote{Some of these questions are addressed in \cite{hol} from a geometric point of view.}
\item Most importantly, derive the equality \eqref{nrsc} from the first principles of the gauge theory (to all orders in the gauge coupling $\mathfrak{q}$)? 
\end{enumerate}
We address these questions  below:

\subsection{Outline}
The key players of the work are the half-BPS codimension two (surface) defects  in the four-dimensional $\EuScript{N}=2$ gauge theories. The surface defects can be constructed in several ways \cite{N1bis, gukwit,aggtv}. The exact computation of their partition functions became accessible in part by \cite{dgh,afkmy, kantachi}, and in a more general setting in \cite{nek7}. In particular, the explicit forms of the surface defects as the observables in the underlying gauge theory were written down in \cite{nek7}.

Meanwhile, the analysis of the analytic properties of the $\EuScript{N}=2$ partition functions became available since \cite{nek2}. The $qq$-characters were introduced as gauge theory observables, which can be constructed out of the spiked instanton configurations \cite{nek3, nek4, nekprab}. The crucial property of these observables is the regularity of their expectation values \cite{nek2}, which follows from the compactness theorem  \cite{nek3}. From the regularity of $qq$-characters follows the vanishing theorem for the non-regular parts of the expectation values, thereby constraining the partition functions. We call these vanishing equations \textit{the non-perturbative Dyson-Schwinger equations} \cite{nek2}.

In section \ref{surfdef}, we recall two independent constructions of surface defects: the quiver and the orbifold. In section \ref{ssnpds}, we describe the fundamental $qq$-character for the surface defects, and derive the non-perturbative Dyson-Schwinger equations for their partition functions. We show that the final equations satisfied by the surface defect partition functions can be regarded as a quantized version of the opers, in the sense that they reduce to the differential equations for the opers in the NS limit $\varepsilon_2 \to 0$. The relations of the expectation values of the chiral observables to the holomorphic coordinates on the variety of opers are naturally revealed through this procedure, clarifying in what sense the variety of opers is a quantization of the Coulomb moduli space.

Being solutions to the non-perturbative Dyson-Schwinger equations, in the NS limit the asymptotics $\bf\chi$ of the appropriately normalized surface defect partition function  becomes the oper solution ${\widehat{\mathfrak{D}}}{\bf\chi}=0$. Consequently, the monodromy of the solutions of the oper can be obtained by first computing the monodromy of the surface defect partition functions and then taking the NS limit. However, each surface defect partition function has its own convergence domain, and to compute the monodromy we need \textit{the connection matrix} which links the surface defect partition functions lying on different domains. This is the subject of the section \ref{congl}. Namely, we present how the surface defect partition function is analytically continued to another convergence domain, and how they can be \textit{glued} together. In fact, the analytically continued quiver surface defect partition function is shown to be identical to a specific orbifold surface defect partition function, suggesting the equivalence of the two distinct types of surface defects. It may be regarded as an independent nontrivial result in itself, realizing the duality between the surface defects \cite{frenguktes} at the level of the partition functions.

In relating the gauge effective theory twisted superpotential to the generating function of the variety of opers, we need to specify the Darboux coordinate system on the moduli space of flat connections relevant to the correspondence. More precisely, we need at least the coordinates on the patch of the moduli space, in which the theory has a weak coupling description (the twisted superpotential is defined, of course, everywhere, however we can only compute it directly in quantum field theory in that region). It may appear that the coupling constant of the theory, being the complex moduli of the underlying Riemann surface, has nothing to do with the coordinate charts on the moduli space of flat connections in the $J$-complex structure, as the latter depends only on the topology of $\EuScript{C}$. The explanation is the following. The continuous dependence on the couplings $\qe$ is indeed absent. However, the universality classes of the Lagrangians describing the theory depend on the type of the degeneration of the Riemann surface $\EuScript{C}$, the so-called pair-of-pants decomposition. The latter is determined by the choice of a handlebody (together with an embedded graph) whose boundary is $\EuScript{C}$ (with the punctures being the end-points of the graph edges). 

With this understood,  in section \ref{sec:gennrs}, we propose  Darboux coordinates on a
particular patch of the moduli space of flat $SL(N)$-connections on the $r+3$-punctured sphere. Our coordinates agree (up to a simple shift) with the NRS coordinates \cite{nrs} restricted to the corresponding patch of the $SL(2)$-moduli space.  We verify the canonical Poisson relations for the proposed coordinate system by using the geometric representation of Poisson brackets between the Wilson loops in the classical Chern-Simons theory. We compute explicitly the invariants of the holonomies of flat connections in our main $r=1$ example.\footnote{There are alternative approaches to the construction of Darboux coordinates from spectral networks \cite{gmn2}, motivated by the work of A. Voros \cite{voros} on the exact WKB approximation, and from symplectic doubles \cite{fg3}, motivated by the work of W. Thurston on the measured laminations. The spectral networks were used in \cite{gmn2,  hn, fg3, hol} to generalize Fock-Goncharov \cite{fg1}, NRS \cite{nrs}, Goldman \cite{gold} and Fenchel-Nielsen \cite{fn} coordinates. We stress that we only work on an open subset of the moduli space, so the subtleties discussed in \cite{fg1,fg2, fg3}, forcing one to work on  certain covers of the moduli space, are not visible at the level we are working.}

Finally, the monodromy data of opers is computed in section \ref{mono}. More precisely, we compute the analytic continuation of the surface defect partition functions, using the results of section \ref{congl}. Then we take the NS limit of the resulting transfer matrices to reduce them to the monodromies of the opers. Then we express those data in terms of the generalized NRS coordinates proposed in the section \ref{sec:gennrs}. This procedure reveals that the effective twisted superpotential is naturally identified with the generating function of the variety of opers. The conclusions and discussions are presented in the section \ref{dis}. The appendices contain some computational details. 

\acknowledgments

SJ is grateful to Xinyu Zhang for useful discussions. NN thanks Andrei Okounkov and Alexei Rosly for numerous discussions. The work was supported in part by the NSF grant PHY 1404446. The work of SJ was also supported by the Samsung Scholarship, and by the generous support of the Simons Center for Geometry and Physics. Research of NN is partially supported by Laboratory of Mirror Symmetry NRU HSE, RF Government grant, ag. N\textsuperscript{\underline{\scriptsize o}} 14.641.31.0001.

\section{Surface defects} \label{surfdef}
We start on the Hitchin system side. We will mainly consider the four-punctured Riemann sphere $\EuScript{C} = \mathbb{P}^1 \backslash \{ 0, {\qe} , 1, {\infty} \}$. All the punctures are assumed to be \textit{regular}. That is, we only allow a simple pole for the Higgs field $\varphi$ at each puncture. Moreover, we call a puncture \textit{maximal} when the residue of $\varphi$ at the puncture belongs to a generic semisimple conjugacy class of $\mathfrak{g} = A_{N-1}$, and \textit{minimal} when the residue is in a maximally degenerate semisimple conjugacy class (as in \cite{Krichever:1980, Gorsky:1994dj}). We assume the punctures at $0$ and $\infty$ are maximal (this is the typical limit of a Hitchin system on a stably degenerate curve, see \cite{Nekrasov:1995nq}), while the punctures at $\mathfrak{q}$ and $1$ are minimal. In what follows in listing the punctures we underline the minimal ones,  as in $\{ 0, \underline{\qe} , \underline{1}, {\infty} \}$. We shall also denote by $\uCalC$ the punctured Riemann surface together with the assignment of the minimal and maximal punctures, e.g. ${\uCalC} = \mathbb{P}^1 \backslash \{ 0, \underline{\qe} , \underline{1}, {\infty} \}$. There is no distinction between the maximal and the minimal punctures in the $N=2$ case.  For $N>2$ the difference is significant. The corresponding class-$\EuScript{S}$ theory $\mathcal{T} [A_{N-1} ,  {\uCalC}]$ is the superconformal $\EuScript{N}=2$ gauge theory with the gauge group $SU(N)$ and the $2N$ hypermultiplets, whose gauge coupling is $\qe$ and the masses of the hypermultiplets are determined by specific combinations of the eigenvalues of the residue of $\varphi$ \cite{gai1}. 

A half-BPS surface defect on $\mathcal{T} [A_{N-1} , {\uCalC}]$ can be constructed in several ways. Here we present two constructions relevant to our study. It is convenient to treat the gauge group formally  as $U(N)$, by making an overall shift in the masses of the hypermultiplets, as we do throughout the discussion. The $SU(N)$ gauge theory parameters can be easily recovered by shifting back the Coulomb moduli and the masses of hypermultiplets.

\subsection{The quiver construction} \label{ssquivsurf}
The construction starts with the superconformal $A_2$-quiver $U(N)$ gauge theory. As reviewed in appendix \ref{appA} in detail, the equivariant localization reduces the instanton partition function of the theory to that of a grand canonical ensemble on the $2N$-tuples of Young diagrams $\boldsymbol{\lambda} = \{ \lambda^{(\mathbf{i} , \alpha)} \; \vert \; \mathbf{i}=1,2, \; \alpha = 1, \cdots, N  \} $. It can be conveniently written as
\begin{align} \label{parta2}
\EuScript{Z}_{A_2} \left(\mathbf{a}_{0}; \mathbf{a}_{1}; \mathbf{a}_{2}; \mathbf{a}_{3} \vert {\ve}_1, {\ve}_2 \vert {\qe}_1,{\qe}_2 \right) = \sum_{\boldsymbol{\lambda}} \prod_{\mathbf{i}=1,2} \mathfrak{q}_{\mathbf{i}} ^{\vert \boldsymbol{\lambda^{(\mathbf{i})}} \vert} \; \epsilon \left[ \EuScript{T}_{A_2} [\boldsymbol{\lambda}] \right]
\end{align}
where the character $\EuScript{T}_{A_2}$ is
\begin{align} \label{chara2}
\begin{split}
\EuScript{T}_{A_2} &= \sum_{\mathbf{i} =1,2} \left( N_{\mathbf{i}} K_{\mathbf{i}} ^* + q_{12} N_{\mathbf{i}} ^* K_{\mathbf{i}} - P_{12} K_{\mathbf{i}}  K_{\mathbf{i}} ^* \right) - M_0 K_1 ^* - q_{12} M_3 ^* K_2 \\
& \quad\quad\quad\quad\quad\quad\quad\quad\quad\quad\quad\quad\quad -N_1 K_2 ^* - q_{12} N_2 ^* K_1 + P_{12} K_1 K_2 ^*,
\end{split}
\end{align}
and the $\epsilon$-operation\footnote{Not to be confused with the $\Omega$-deformation parameters ${\ve}_{1}, {\ve}_{2}$.}, also known as the plethystic exponent, converts a character into the product of weights,
\begin{align}
\epsilon(R) = \frac{\prod_{w\in R^-} w(\theta)}{\prod_{w \in R^+ } w(\theta)} \quad \text{for} \quad \theta \in \text{Lie}(T_H), \quad R = \sum_{w \in R^+} e^{w(\theta)} - \sum_{w \in R^-} e^{w(\theta)}.
\end{align}
Let us choose $\beta \in \{1, \cdots, N \}$, and 
tune the Coulomb moduli of the first gauge node as
\begin{align} \label{a2const}
\begin{split}
\begin{cases} a_{1,\beta} = a_{0,\beta} -\varepsilon_2  \\
	a_{1,\alpha} = a_{0,\alpha} \quad\quad\quad \text{for } \alpha \neq \beta \end{cases}
\end{split}
\end{align}
 We define the defect partition function  as $\EuScript{Z}_{A_2}$ with the constrained Coulomb parameters:
\begin{align} \label{zl}
\EuScript{Z}^L _{\beta} \equiv \EuScript{Z}_{A_2} \left( \mathbf{a}_{0}; \, a_{1,\alpha} = a_{0,\alpha} - {\ve}_2 {\delta}_{\alpha, \beta}  ;\, \mathbf{a}_{2}; \, \mathbf{a}_{3} \; \vert \; \varepsilon_{1}, \varepsilon_2\; \vert \; \mathfrak{q}_1 = z^{-1}, \mathfrak{q}_2 = \mathfrak{q} \right).
\end{align}
The constraints can be succinctly expressed as the relation between the characters
\begin{align}
M_0 = N_1 - P_2 \mu,
\end{align}
where we have defined $\mu = e^{\beta (a_{0,\beta}-\varepsilon_2)}$. Note that due to the constraints, almost all the Young diagrams for the first gauge node have vanishing contributions to the partition function, except the ones of the form
\begin{align}
\boldsymbol{\lambda}^{(1)} = \left( \varnothing, \cdots, \varnothing, \begin{rcases} \ytableausetup
{mathmode, boxsize=1em, centertableaux}
\begin{ytableau}
 \\ \\ \\ \none [\vdots] \\ \\
\end{ytableau} \end{rcases} k \;, \varnothing, \cdots, \varnothing  \right) ,  \label{singcol}
\end{align}
which is empty $\lambda^{(1,\alpha)} = \varnothing$ except the single-columned $\lambda^{(1, \beta)}$. 

We can view the constraint \eqref{a2const} as adding an extra equation in the ADHM construction for the quiver instanton moduli space, as we now recall. First, the $A_2$-quiver $U(N)$ theory can be obtained by the $\mathbb{Z}_4$-orbifold procedure from the $\EuScript{N}=2^*$ $U(4N)$ theory. The ADHM data for the $\EuScript{N}=2^*$ $U(4N)$ gauge theory is the following collection of linear maps between complex vector spaces:
\begin{align}
\begin{split}
&B_{1,2,3,4} : K \longrightarrow K \\
&I : N \longrightarrow K \\
&J: K \longrightarrow N,
\end{split}
\end{align}
where $N = \mathbb{C}^{4N}$ and $K=\mathbb{C}^{k_1+k_2}$. The reason for strange dimensions of these spaces will become clear momentarily. The extended ADHM equations are written as \cite{nek3}
\begin{align} \label{adhm}
\begin{split}
&[B_1, B_2]+I J +[B_3,B_4]^\dagger =0 \\
&[B_1,B_3]+[B_4,B_2]^\dagger =0\\
&[B_1,B_4]+[B_2,B_3]^\dagger=0\\
&s^+ \equiv B_3 I + (J B_4)^\dagger =0 \\
&s^- \equiv B_4I-(JB_3)^\dagger =0.
\end{split}
\end{align} 
We also impose the stability condition (cf. \eqref{adhmmod})
\begin{align}
\mathbb{C}[B_1, B_2, B_3, B_4] I(N) = K.
\end{align}
Upon the $\mathbb{Z}_4$-orbifolding, the spaces $N$ and $K$ become $\mathbb{Z}_4$-modules, and therefore can be decomposed according to the $\mathbb{Z}_4$-representations
\begin{align}
N = \bigoplus_{\omega \in \mathbb{Z}_4} N_\omega \otimes \EuScript{R}_\omega, \quad K = \bigoplus_{\omega \in \mathbb{Z}_4} K_\omega \otimes \EuScript{R}_\omega.
\end{align}
The coupling constant is also fractionalized accordingly, $\mathfrak{q}_\omega$ for $\omega \in \mathbb{Z}_4$. We manually set $\mathfrak{q}_0 = \mathfrak{q}_3=0$, then we are restricted to $K_0 = K_3 =0$ due to the measure factor $\mathfrak{q}_\omega ^{\vert K_\omega \vert}$. Let $\vert K_1 \vert = k_1$ and $\vert K_2 \vert =k_2$. Also, we impose the $\mathbb{Z}_4$-weights to the space $N$ in such a way that $N_\omega = \mathbb{C}^N$ for each $\omega \in \mathbb{Z}_4$. Let the maps
\begin{align}
\begin{split}
\Omega_N : N \longrightarrow N, \quad \Omega_K : K \longrightarrow K,
\end{split}
\end{align} 
be defined by the diagonal action of $i^\omega$ to the elements in $N_\omega$ and $K_\omega$. Then we impose the conditions for the ADHM data
\begin{align}
\begin{split}
&\Omega_K ^{-1} \: B_{1,2} \: \Omega_K = B_{1,2} \\
&\Omega_K ^{-1} \: B_3 \: \Omega_K = i B_3 \\
&\Omega_K  ^{-1}\: B_4\: \Omega_K = -i B_4 \\
&\Omega_K ^{-1} \: I \: \Omega_N = I \\
&\Omega_N ^{-1} \:J \: \Omega_K =J,
\end{split}
\end{align}
which fractionalize these matrices as
\begin{align}
\begin{split}
&B_{\omega,1} : K_\omega \longrightarrow K_\omega \\
&B_{\omega,2} : K_\omega \longrightarrow K_\omega \\
&B_{\omega,3} : K_\omega \longrightarrow K_{\omega+1} \\
&B_{\omega,4} : K_\omega \longrightarrow K_{\omega-1} \\
&I_{\omega} : N_\omega \longrightarrow K_\omega \\
&J_{\omega} : K_\omega \longrightarrow N_\omega.
\end{split}
\end{align}
Note that many of these maps are identically zero due to the restriction $K_0 =K_3=0$. Hence only the following equations survive among the ADHM equations \eqref{adhm},
\begin{align} \label{adhmred}
\begin{split}
&[B_{1,1},B_{1,2}] +I_1 J_1 -B_{1,3}^\dagger B_{2,4} ^\dagger =0 \\
&[B_{2,1},B_{2,2}]+I_2 J_2 -B_{2,4}^\dagger B_{1,3}^\dagger =0 \\
&B_{2,1} B_{1,3} -B_{1,3} B_{1,1} +B_{2,2} ^\dagger B_{2,4} ^\dagger -B_{2,4} ^\dagger B_{1,2} ^\dagger =0 \\
&B_{1,1} B_{2,4} -B_{2,4} B_{2,1} +B_{1,3} ^\dagger B_{2,2} ^\dagger -B_{1,2} ^\dagger B_{1,3} ^\dagger =0 \\
&s^+ _1 \equiv B_{1,3} I_1 +B_{2,4} ^\dagger J_1 ^\dagger =0 \\
&s^- _2 \equiv B_{2,4} I_2 - B_{1,3} ^\dagger J_2 ^\dagger =0.
\end{split}
\end{align}
The stability condition also becomes
\begin{align}
\begin{split}
\mathbb{C}[B_{1,1},B_{1,2},B_{2,1},B_{2,2},B_{1,3},B_{2,4}] I(N) = K.
\end{split}
\end{align}
We find that the sum of the squares of the norms of the first two equations of \eqref{adhmred} can be simplified, using the other four equations, into a sum of squares,
\begin{align} \label{normsq}
\begin{split}
0&=\vert\vert [B_{1,1},B_{1,2}] +I_1 J_1 \vert\vert^2 + \vert\vert [B_{2,1},B_{2,2}] +I_2 J_2 \vert\vert^2 + \vert\vert B_{1,3} I_1 \vert\vert^2 + \vert \vert B_{2,4} I_2 \vert\vert^2 \\
&\quad+ \vert\vert B_{2,1} B_{1,3}-B_{1,3} B_{1,1} \vert\vert^2 + \vert \vert B_{1,1} B_{2,4} -B_{2,4}B_{2,1} \vert\vert^2.
\end{split}
\end{align}
Applying the last two equations to the stability condition, we can commute $B_{1,3}$ and $B_{2,4}$ through all the way to hit $I_1 (N_1)$ or $I_2 (N_2)$, respectively. This vanishes as a result of the third and the fourth equations. Hence, the stability condition is reduced to
\begin{align}\label{redstab}
\begin{split}
&\mathbb{C}[B_{\mathbf{i},1},B_{\mathbf{i},2}] I_{\mathbf{i}} (N_\mathbf{i}) = K_\mathbf{i}, \quad \mathbf{i}=1,2.
\end{split}
\end{align}
This implies $B_{1,3} = B_{2,4} =0$. The first and the second equations of \eqref{normsq} provide the reduced ADHM equations
\begin{align}
[B_{\mathbf{i},1}, B_{\mathbf{i},2}] +I_{\mathbf{i}} J_{\mathbf{i}} =0, \quad \mathbf{i}=1,2,
\end{align}
which are precisely the ADHM equations for the instanton moduli space of the $A_2$-quiver $U(N)$ theory. 

In this construction of the $A_2$-theory, the constraint \eqref{a2const} can be understood as adding an equation $\text{``}s^+ _0 \text{''}: N_0 \longrightarrow K_1$. Note that we neglected the equation
\begin{align}
s^+ _0 \equiv B_{0,3} I_0 + B_{1,4} ^\dagger J_0 ^\dagger : N_0 \longrightarrow K_1,
\end{align}
since it is identically zero by $B_{0,3} = I_0 = B_{1,4} = J_0 =0$ due to the restriction $K_0=0$. However, we can avoid this restriction if we first set
\begin{align}
N_0 = \widetilde{N} \oplus L, \quad N_1 = \widetilde{N} \oplus q_2 L,
\end{align}
where we have chosen an one-dimensional subspace $L \subset N_0$, which corresponds the choice of $\beta \in \{1, \cdots, N\}$ in the constraint \eqref{a2const}. Then we may define a non-vanishing map
\begin{align}
s^+ _0 = I_1 \big\vert_{\widetilde{N}} \oplus B_{1,2} I_1 \big\vert_{L} : N_0 \longrightarrow K_1.
\end{align}
Adding the equation $s^+ _0 =0$ to the ADHM construction, we find that the space $K_1$ is further restricted by the stability condition \eqref{redstab},
\begin{align}
K_1 = \mathbb{C} [B_{1,1}, B_{1,2}] \: I_1 (N_1) = \mathbb{C}[B_{1,1}] \: I_1 (L).
\end{align}
In other words, the Young diagram that denotes the space $K_1$ only grows in one direction from the chosen basis vector $I_1 (L)$. This exactly manifests the single-columnedness expressed in \eqref{singcol}.
\begin{figure}\label{a2quiverfig}
\centering
\begin{tikzpicture}[square/.style={regular polygon,regular polygon sides=4}]
\node[square, draw=black,fill=white, inner sep=3.9pt] at (-3,0) (N0) {$N_0$};
\node[circle, draw=black,fill=white, inner sep=5pt] at (-1,0) (N1) {$N_1$};
\node[circle, draw=black,fill=white, inner sep=5pt] at (1,0) (N2) {$N_2$};
\node[square, draw=black,fill=white, inner sep=3.9pt] at (3,0) (N3) {$N_3$};
\node[circle, draw=black,fill=white, inner sep=5pt] at (-1,-2) (K1) {$K_1$};
\node[circle, draw=black,fill=white, inner sep=5pt] at (1,-2) (K2) {$K_2$};
\draw (N0) -- (N1);
\draw (N1) -- (N2);
\draw (N2) -- (N3);
\draw[very thick,decoration={markings, mark=at position 0.65 with {\arrow[scale=1]{latex}}}, postaction={decorate}] (-1.2,-0.5) -- (-1.2,-1.5);
\draw[very thick,decoration={markings, mark=at position 0.65 with {\arrow[scale=1]{latex}}}, postaction={decorate}] (-0.8,-1.5) -- (-0.8,-0.5);
\draw[very thick,decoration={markings, mark=at position 0.65 with {\arrow[scale=1]{latex}}}, postaction={decorate}] (0.8,-0.5) -- (0.8,-1.5);
\draw[very thick,decoration={markings, mark=at position 0.65 with {\arrow[scale=1]{latex}}}, postaction={decorate}] (1.2,-1.5) -- (1.2,-0.5);
\path[very thick] (K1) [loop below] edge (K1);
\path[very thick] (K2) [loop below] edge (K2);
\node at (-1.5,-1) {$I_1$};
\node at (-0.45,-1) {$J_1$};
\node at (1.55,-1) {$J_2$};
\node at (0.5,-1) {$I_2$};
\node at (-1,-3.5) {$B_{1,1}$, $B_{1,2}$};
\node at (1.2,-3.5) {$B_{2,1}$, $B_{2,2}$};
\draw[dotted,very thick,decoration={markings, mark=at position 0.55 with {\arrow[scale=1]{latex}}}, postaction={decorate}] (N0) -- (K1);
\node at (-2.3,-1.3) {$s^+ _0$};
\end{tikzpicture}
\caption{The ADHM data for the $A_2$-quiver gauge theory, with the extra map $s^+ _0$.}
\end{figure}
The physics picture of what is happening is the following. The constraint \eqref{a2const} makes $N$ hypermultiplets nearly massless (exactly massless in the absence of $\Omega$-deformation). The theory can then go to the Higgs branch, where the gauge group is partially Higgsed to a subgroup, by the expectation values of the hypermultiplet scalars. Now, the theory allows for the half-BPS field configurations where the gauge group is restored along a codimension two defect, essentially a vortex string. 
Consequently, the gauge field configuration of the first gauge node is squeezed into a two-dimensional plane ($\varepsilon_1$-plane), effectively forming a vortex. The resulting two-dimensional supersymmetric sigma model couples to the remaining four-dimensional $A_1$-theory, generating a surface defect in the four-dimensional point of view.

We can confirm that the 2d-4d coupled system arises at the level of the partition function.  First we have the simplified expression for
\begin{align}
K_1 = \mu \frac{1-q_1 ^{k}}{1-q_1}.
\end{align}
Therefore, the character \eqref{chara2} can also be simplified into
\begin{align}
\EuScript{T}_{A_2} = \left [N_2 K_2 ^* +q_{12} N_2 ^* K_2 - P_{12} K_2 K_2 ^* - N_1 K_2 ^* - q_{12} M_3 ^* K_2 \right] + \left[ P_2 \mu q_1 ^k K_1 ^* + q_{12} K_1 (N_1 ^* - S_2 ^*) \right].
\end{align}
Accordingly, the partition function \eqref{parta2} of the $A_2$-quiver gauge theory is reduced to the expectation value of an observable in the $A_1$-quiver gauge theory
\begin{align} \label{a2partred}
\EuScript{Z}_\beta ^L = \sum_{\boldsymbol{\lambda}^{(2)}} \mathfrak{q}_2 ^{\vert \boldsymbol{\lambda}^{(2)} \vert} \mathcal{I}_{\beta} ^L [\boldsymbol{\lambda}^{(2)}] \; \epsilon\left[ \EuScript{T}_{A_1}[\boldsymbol{\lambda} ^{(2)} ] \right] = \langle \mathcal{I}_{\beta} ^L \rangle \; \EuScript{Z}_{A_1},
\end{align}
where we have defined the character for the $A_1$-theory
\begin{align}
\EuScript{T}_{A_1} \equiv N_2 K_2 ^* + q_{12} N_2 ^* K_2 - P_{12} K_2 K_2^* -N_1 K_2 ^*- q_{12} M_3 ^* K_2,
\end{align}
which defines the instanton partition function of the $A_1$-theory by
\begin{align} \label{a1}
\EuScript{Z}_{A_1} \equiv \sum_{\boldsymbol{\lambda}^{(2)}} \mathfrak{q}_2 ^{\vert \boldsymbol{\lambda}^{(2)} \vert} \epsilon \left[ \EuScript{T}_{A_1}[\boldsymbol{\lambda} ^{(2)} ] \right],
\end{align}
and the surface defect as an element of the chiral ring
\begin{align} \label{surfquiv}
\begin{split}
\mathcal{I}_{\beta} ^L [\boldsymbol{\lambda}^{(2)}] &\equiv \sum_{k=0} ^\infty \mathfrak{q}_1 ^k \; \epsilon \left[ P_2 \mu q_1 ^k K_1 ^* + q_{12} K_1 (N_1 ^* - S_2 ^*) \right]  \\
&= \sum_{k=0} ^{\infty} \mathfrak{q}_1 ^k \; \epsilon \left[ \sum_{l=1}^k q_1 ^l  (P_2 + \mu q_2 (N_1 ^* -S_2 ^*))  \right] \\
&= \sum_{k=0} ^{\infty}  \mathfrak{q}_1 ^k  \; \prod_{l=1} ^k \frac{\EuScript{Y}_2 (a_{0,\beta} + l \varepsilon_1)[\boldsymbol{\lambda}^{(2)}] }{P_0 (a_{0,\beta} + l \varepsilon_1)}.
\end{split}
\end{align}
Here, we have used the $\EuScript{Y}$-observable \eqref{yobs} for the second gauge node, and $P_0 (x) \equiv \prod_{\alpha=1} ^N (x-a_{0,\alpha})$ by definition. Let us focus on the zero bulk instanton sector, $\vert \boldsymbol{\lambda}^{(2)} \vert =0$. The contribution of this sector is the vortex partition function of a two-dimensional gauged linear sigma model. This sigma model generates the surface defect, when coupled to the four-dimensional bulk \cite{N1bis, dgh}. The $\EuScript{Y}$-observable in this sector simply reduces to a polynomial $\EuScript{Y}_2 (x) \to A_2 (x) \equiv \prod_{\alpha=1} ^N (x-a_{2,\alpha}) $. The partition function \eqref{surfquiv} is exactly that of the gauged linear sigma model on the $\text{Hom} (\mathcal{O}(-1), \mathbb{C}^N)$-bundle over $\mathbb{P}^{N-1}$ whose K\"{a}hler modulus is $\mathfrak{q}_1$ \cite{nek7}. For the non-trivial sectors of the four-dimension, the two-dimensional sigma model couples to the four-dimensional gauge theory through the non-perturbative corrections to the $\EuScript{Y}$-observable. Thus, the full partition function \eqref{a2partred} represents the 2d-4d coupled system in this manner.

It is instructive to cast the surface defect partition function \eqref{surfquiv} into the form relevant to our study. Recall that the $\EuScript{Y}$-observable \eqref{yobs} can be written as a ratio
\begin{align}
\EuScript{Y}_{\mathbf{i}} (x) [\boldsymbol{\lambda}] = \prod_{\alpha=1} ^N \frac{ \prod_{\square \in \partial_+ \lambda^{(\mathbf{i},\alpha)}} (x-c_{\square} ) }{\prod_{\square \in \partial_- \lambda^{(\mathbf{i}, \alpha)}} (x-  c_{\square} -\varepsilon) }.
\end{align}
This suggests to represent the $\EuScript{Y}$-observable as a ratio of two entire functions \cite{nek2},
\begin{align}
\EuScript{Y}_{\mathbf{i}} (x) = \frac{\EuScript{Q}_{\mathbf{i}} (x)}{\EuScript{Q}_{\mathbf{i}} (x-\varepsilon_1)},
\end{align}
where we have defined the $\EuScript{Q}$-observable
\begin{align} \label{qobs}
\EuScript{Q}_{\mathbf{i}} (x) [\boldsymbol{\lambda}] \equiv \prod_{\alpha=1} ^N \left( \frac{(-\varepsilon_1 )^{\frac{x-a_{\mathbf{i},\alpha}}{\varepsilon_1}}}{\Gamma \left( -\frac{x-a_{\mathbf{i},\alpha}}{\varepsilon_1} \right)}  \prod_{\square \in \lambda^{(\mathbf{i}, \alpha)}} \frac{x- c_{\square} -\varepsilon_2}{x- c_{\square}} \right).
\end{align}
Therefore, the surface defect \eqref{surfquiv} can be understood as an infinite sum of $\EuScript{Q}$-observables,
\begin{align} \label{quivsurfq}
\mathcal{I}_{\beta} ^L [\boldsymbol{\lambda}^{(2)}] = \sum_{k=0} ^{\infty} \mathfrak{q}_1 ^k \; \left( \prod_{\alpha=1} ^N \frac{\Gamma \left( 1+ \frac{a_{0,\beta} - a_{0, \alpha}}{\varepsilon_1} \right)}{\varepsilon_1 ^k \; \Gamma \left( k+1 + \frac{a_{0, \beta} - a_{0, \alpha}}{\varepsilon_1} \right)} \right) \frac{\EuScript{Q}_2 (a_{0, \beta} + k \varepsilon_1 ) [\boldsymbol{\lambda}^{(2)}]}{\EuScript{Q}_2 (a_{0,\beta}) [\boldsymbol{\lambda}^{(2)}]}.
\end{align}
We will observe that the $\EuScript{Q}$-observable reduces to the so-called Baxter $\EuScript{Q}$-function in the NS limit. It will be more apparent in section \ref{ssnpds} that this representation is useful for our purpose. 

Likewise, we can similarly impose the constraints for the Coulomb moduli in the second gauge node,
\begin{align} \label{constdiff}
\begin{split}
\begin{cases}a_{2,\beta} = a_{3,\beta} -\varepsilon - \varepsilon_2 \\
a_{2, \alpha} = a_{3, \alpha} - \varepsilon \quad\quad \quad  \alpha \neq \beta, \end{cases}
\end{split}
\end{align}
for some chosen $\beta \in \{1, \cdots, N\}$. Here, we are using the abbreviated notation $\varepsilon \equiv \varepsilon_1 + \varepsilon_2$. The constraint can also be written as
\begin{align} \label{constr}
q_{12} ^{-1} M_3 = N_2 - P_2 \mu,
\end{align}
where now $\mu = e^{\beta (a_{3, \beta} - \varepsilon - \varepsilon_2)}$. For a reason that will be clarified in section \ref{glue}, we make the following re-definition for the parameters after imposing the constraints \eqref{constr},
\begin{align} \label{redef}
\begin{split}
&a_{0,\alpha} \longrightarrow -a_{0,\alpha} -\varepsilon \\
&a_{1,\alpha} \longrightarrow -a_{1,\alpha} \quad\quad \quad\quad\quad \alpha = 1, \cdots, N \\
&a_{3,\alpha} \longrightarrow -a_{3,\alpha} +2\varepsilon,
\end{split}
\end{align}
The corresponding partition function,
\begin{align} \label{zr}
\EuScript{Z}^R _\beta \equiv \EuScript{Z}_{A_2} \left( -a_{0,\alpha}-{\ve} ; \, -a_{1,\alpha}; \, -a_{3,\alpha} + \varepsilon - {\ve}_{2}{\delta}_{\alpha,\beta}; \, -a_{3, \alpha}+2\varepsilon \; \vert \; \varepsilon_1, \varepsilon_2 \; \vert \; \mathfrak{q}_1 = \mathfrak{q} , \mathfrak{q}_2 = {\qe}^{-1} \, z  \right),
\end{align}
can be likewise simplified to:
\begin{align} \label{a2diff}
\EuScript{Z}_\beta ^R = \sum_{\boldsymbol{\lambda}^{(1)}} \mathfrak{q}_1 ^{\vert \boldsymbol{\lambda}^{(1)} \vert } \; \mathcal{I} ^R _\beta [\boldsymbol{\lambda}^{(1)}] \; \epsilon \left[ \EuScript{T}_{A_1} [\boldsymbol{\lambda}^{(1)}] \right],
\end{align}
where the character for the $A_1$-theory is now
\begin{align}
\EuScript{T}_{A_1} \equiv N_1 K_1 ^* +q_{12} N_1 ^* K_1 - P_{12} K_1 K_1 ^* - M_0 K_1 ^* - q_{12} N_2 ^* K_1,
\end{align}
and the surface defect is
\begin{align}
\begin{split}
\mathcal{I}_{\beta} ^R [\boldsymbol{\lambda}^{(1)}] &\equiv \sum_{k=0} ^{\infty} \mathfrak{q}_2 ^k \; \prod_{l=1} ^k \frac{\EuScript{Y}_1 ( -a_{3,\beta} + l \varepsilon_1)}{P_3 (-a_{3, \beta}+ 2\varepsilon + l \varepsilon)} \\
&= \sum_{k=0} ^{\infty} \mathfrak{q}_2 ^k\; \prod_{\alpha=1} ^N \left( \frac{\Gamma \left( 1+ \frac{a_{3,\alpha}-a_{3,\beta}}{\varepsilon_1} \right)}{\varepsilon_1 ^k \; \Gamma\left( k+1 + \frac{a_{3,\alpha}-a_{3,\beta}}{\varepsilon_1} \right)} \right) \frac{\EuScript{Q}_1 (-a_{3,\beta}+ k \varepsilon_1)}{\EuScript{Q}_1 (-a_{3,\beta} )},
\end{split}
\end{align}
where we have used the $\EuScript{Y}$-observable for the first gauge node and $P_3 (x) \equiv \prod_{\alpha=1} ^N (x+a_{3, \alpha} -2\varepsilon)$ (Be cautious about the re-definition of the parameters). Also, the $\EuScript{Y}$-observable has been replaced by a ratio of $\EuScript{Q}$-observables in the second line. Note that the bulk coupling is now $\mathfrak{q}_1$, while the K\"{a}hler modulus for the two-dimensional sigma model is $\mathfrak{q}_2$. Thus it is natural to expect that the $\mathfrak{q}_2$ of $\EuScript{Z}_\beta ^L$ would correspond to $\mathfrak{q}_1$ of $\EuScript{Z}_\beta ^R$, when we try to connect these partition functions. The issue will be clarified in section \ref{congl}.

\subsection{The orbifold construction} \label{ssorbsurf}
We construct a surface defect by placing the gauge theory on an orbifold. We first form an orbifold $\mathbb{C}_{\varepsilon_1} \times \left( \mathbb{C}_{\varepsilon_2} / \mathbb{Z}_p \right) $ by the following $\mathbb{Z}_p$-action on $\mathbb{C}_{\varepsilon_1} \times \mathbb{C}_{\varepsilon_2}$
\begin{align}
\zeta: (z_1, z_2) \longmapsto (z_1, \zeta z_2), \quad \zeta \equiv \exp \left( \frac{2 \pi i}{p} \right) \in \mathbb{Z}_p.
\end{align}
Here, $\mathbb{C}_{\varepsilon_i}$ denotes the complex plane with the equivariant parameter $\varepsilon_i$ for the $\mathbb{C}^\times$-action. Then the surface defect is constructed as a prescription of performing the path integral only over the $\mathbb{Z}_p$-invariant field configurations. Indeed, under the map $(z_1 , z_2) \mapsto (\widetilde{z}_1 \equiv z_1 , \widetilde{z}_2 \equiv z_2 ^p)$, the orbifold $\mathbb{C}_{\varepsilon_1} \times \left( \mathbb{C}_{\varepsilon_2} / \mathbb{Z}_p \right) $ is mapped to $\mathbb{C}_{\varepsilon_1} \times \mathbb{C}_{p \varepsilon_2}$, and the field configurations are allowed to be singular along the surface $\widetilde{z}_2 =0$. Therefore, the resulting theory on $\mathbb{C}_{\varepsilon_1} \times \mathbb{C}_{p \varepsilon_2}$ can be interpreted as a surface defect inserted upon the underlying gauge theory.

To fully characterize the surface defect, we have to specify how the field configurations are projected out by the $\mathbb{Z}_p$-action. We present here how this is done for the $A_1$-theory. Let us introduce the coloring function
\begin{align}
c: [N] \longrightarrow \mathbb{Z}_p.
\end{align}
Then the space $N$ is decomposed according to the $\mathbb{Z}_p$-representations,
\begin{align}
N = \bigoplus_{\omega \in \mathbb{Z}_p} N_{\omega} \otimes \EuScript{R}_{\omega} , \quad N_{\omega} \equiv \sum_{\alpha \in c^{-1} (\omega)} e^{\beta a_{\alpha}}.
\end{align}
Also,
\begin{align}
K = \bigoplus_{\omega \in \mathbb{Z}_p} K_{\omega} \otimes \EuScript{R}_{\omega} , \quad K_{\omega} \equiv  \sum_{\alpha =1}^N e^{\beta a_{\alpha}  } \sum_{i=1} ^{l \left(\lambda^{(\alpha)} \right)} q_1 ^{i-1} \sum_{\substack{ 1 \leq j \leq \lambda_i ^{(\alpha)} \\ c(\alpha) + j-1 \equiv \omega \text{ mod }p}} q_2 ^{j-1},
\end{align}
where $\EuScript{R}_\omega$ is the one-dimensional irreducible representation of $\mathbb{Z}_p$ with the weight $\omega$, and $l\left( \lambda^{(\alpha)} \right) = \lambda^{(\alpha) \: t} _{1}$ is the number of rows in the Young diagram $\lambda^{(\alpha)}$. It is straightforward to include the fundamental matter fields, namely,
\begin{align}
M = \bigoplus_{\omega \in \mathbb{Z}_p} M_{\omega} \otimes \EuScript{R}_{\omega}.
\end{align}
Now as explained, we perform the path integral only for the $\mathbb{Z}_p$-invariant field configurations, projecting out the non-invariant contributions. At the level of the character (see \eqref{1loopinst}), this is to pick up the $\mathbb{Z}_p$-invariant piece, namely,
\begin{align}\label{orbchar}
\EuScript{T}^{\mathbb{Z}_p} \equiv \left[  \frac{1}{P_{12}} \left( -S S^* + M^* S  \right)\right] ^{\mathbb{Z}_p},
\end{align}
from which the partition function is given by
\begin{align} \label{orbpart}
\EuScript{Z}^{\mathbb{Z}_p,c} \equiv \sum_{\boldsymbol{\lambda}} \mathfrak{q}_\omega ^{\vert K_\omega \vert} \; \epsilon \left[ \EuScript{T}^{\mathbb{Z}_p} [\boldsymbol{\lambda} ]\right].
\end{align}
Though providing a concrete formula, it is not so obvious from \eqref{orbpart} that the partition function can be interpreted as an insertion of an observable in the $A_1$-theory. Thus it is important to properly construct the projection of the set of $N$-tuples of Young diagrams
\begin{align}
\rho: \boldsymbol{\lambda}  \longmapsto \boldsymbol{\Lambda},
\end{align}
where $\boldsymbol{\Lambda}$ is supposed to enumerates the fixed points of the instanton moduli space of the $A_1$-theory on $\mathbb{C}_{\varepsilon_1} \times \mathbb{C}_{p\varepsilon_2} = \mathbb{C}^2/ \mathbb{Z}_p$.

The construction of the map $\rho$ can be done as follows. Let us first re-define the Coulomb moduli by the shift
\begin{align}
\widetilde{a}_\alpha \equiv a_\alpha - \varepsilon_2 \: c(\alpha),
\end{align}
so that
\begin{align}
\widetilde{N}_\omega \equiv \sum_{\alpha \in c^{-1} (\omega)} e^{\beta \widetilde{a}_\alpha}, \quad \widetilde{N} \equiv \sum_{\omega =0} ^{p-1} \widetilde{N}_\omega.
\end{align}
and
\begin{align}
\widetilde{K}_\omega \equiv K_\omega q_2 ^{-\omega} = \sum_{\alpha=1} ^N e^{\beta \widetilde{a}_\alpha} \sum_{i=1} ^{l \left(\lambda^{(\alpha)} \right)} q_1 ^{i-1} \sum_{j = 1\:\text{or}\:2} ^{l_{i,\alpha, \omega}} \widetilde{q}_2 ^{j-1},
\end{align}
where 
\begin{align}
& l_{i,\alpha, \omega} = \left[ \frac{\lambda_i ^{(\alpha)} +c(\alpha) - \omega +p-1}{p} \right] \\
& \widetilde{q}_2 \equiv q_2 ^p \quad (\widetilde{\varepsilon}_2 \equiv p \varepsilon_2),
\end{align}
and the lower limit of the sum over $j$ is equal to 1 for $c(\alpha) \leq \omega$ and 2 for $c (\alpha)> \omega$. In particular, for $\omega = p-1$,
\begin{align} \label{spacek}
\widetilde{K} \equiv \widetilde{K}_{p-1} =\sum_{\alpha=1} ^N e^{\beta \widetilde{a}_\alpha} \sum_{i=1} ^{l\left( \Lambda^{(\alpha)} \right)} q_1 ^{i-1} \; \sum_{j=1} ^{\Lambda_i ^{(\alpha)}} \widetilde{q}_2 ^{j-1},
\end{align}
where we have defined a new $N$-tuple of Young diagrams $\boldsymbol{\Lambda} = \left( \Lambda ^{(\alpha)} \right) _{\alpha=1} ^N$ by
\begin{align} \label{lambda}
\Lambda_i ^{(\alpha)} = l_{i, \alpha, p-1} \equiv \left[ \frac{\lambda_i ^{(\alpha)} +c(\alpha)}{p} \right].
\end{align}
The map $\rho$ is defined by this relation, $\rho(\boldsymbol{\lambda}) = \boldsymbol{\Lambda}$.

The partition function \eqref{orbpart} is a weighted sum over $\boldsymbol{\lambda}$, which can be first summed over $\rho^{-1}(\boldsymbol{\Lambda})$ for fixed $\boldsymbol{\Lambda}$ and then summed over $\boldsymbol{\Lambda}$. We will show the sum over $\rho^{-1} (\boldsymbol{\Lambda})$ provides an observable insertion to the $A_1$-theory whose measure for the partition function is given by $\boldsymbol{\Lambda}$. First, the vector multiplet contribution in the measure \eqref{orbchar} is
\begin{align} \label{orbcharvec}
- \sum_{\omega, \omega', \omega'' =0} ^{p-1} \frac{S_\omega S_{\omega'}^*}{P_1 (1- \widetilde{q}_2)} q_2 ^{\omega''} \delta_{\omega - \omega' +\omega''} ^{\mathbb{Z}_p},
\end{align}
where we have used the identity
\begin{align}
\frac{1}{1- q_2 \EuScript{R}_1} = \frac{1}{1- \widetilde{q}_2} \sum_{\omega=0} ^{p-1} q_2 ^{\omega} \: \EuScript{R}_\omega.
\end{align}
After defining
\begin{align} 
\widetilde{S}_\omega \equiv S_\omega q_2 ^{-\omega}, \quad \widetilde{S} \equiv \sum_{\omega=0} ^{p-1} \widetilde{S}_\omega = \widetilde{N} - P_1 \widetilde{P}_2 \widetilde{K},
\end{align}
the character \eqref{orbcharvec} can be written as
\begin{align}\label{charred}
-\frac{\widetilde{S} \widetilde{S}^*}{\widetilde{P}_{12}} + \frac{1}{P_1} \sum_{0 \leq \omega < \omega' <p} \widetilde{S}_{\omega'} \widetilde{S}_\omega ^*.
\end{align}
Note that the first term is precisely the vector multiplet contribution to the partition function of the $A_1$-theory on $\mathbb{C}_{\varepsilon_1} \times \mathbb{C}_{\widetilde{\varepsilon}_2}$ in the $\widetilde{k}$-instanton sector. The second term is interpreted as an observable insertion to this theory. We can further simplify it by introducing
\begin{align}
\begin{split}
&\Sigma_{\omega} \equiv \widetilde{N}_0 + \cdots + \widetilde{N}_{\omega-1} - P_1 \widetilde{K}_{\omega-1} + \widetilde{q}_2 P_1 \widetilde{K}, \quad \omega = 1, \cdots, p.
\end{split}\label{sigma}
\end{align}
The second term in \eqref{charred} is now given by
\begin{align}
\frac{1}{P_1} \; \sum_{\omega=1} ^{p-1} (\Sigma_{\omega+1}-\Sigma_\omega)\Sigma_\omega ^*.
\end{align}

Similarly, the matter contribution in the measure \eqref{orbchar} can be written as
\begin{align}
\sum_{\omega, \omega', \omega'' =0} ^{p-1} \frac{M_{\omega'}^*  S_{\omega}}{P_1 (1-\widetilde{q}_2) } q_2 ^{\omega''} \delta_{\omega-\omega' + \omega''} ^{\mathbb{Z}_p}.
\end{align}
After defining $\widetilde{M}_\omega \equiv M_\omega q_2 ^{-\omega}$ and $\widetilde{M} \equiv \widetilde{q}_2 ^{-1} \sum_{\omega=0} ^{p-1} \widetilde{M}_\omega$, it can be re-expressed as
\begin{align}
\frac{\widetilde{M}^* \widetilde{S}}{\widetilde{P}_{12}} + \frac{1}{P_1} \sum_{\omega=0} ^{p-1} \widetilde{M} ^* _{\omega} \Sigma_{\omega+1} 
\end{align}
Note that the first term is the usual fundamental matter contribution to the measure of the $A_1$-theory on $\mathbb{C}_{\varepsilon_1} \times \mathbb{C}_{\widetilde{\varepsilon}_2}$ in the $\widetilde{k}$-instanton sector. The second term is interpreted as an observable insertion to the theory.

We also introduce the auxiliary variables $(\mathpzc{z}_\omega)$ and $\mathfrak{q}$ to express the fractionalized couplings,
\begin{align} \label{fraccoupling}
\begin{split}
&\mathfrak{q}_\omega \equiv \frac{\mathpzc{z}_{\omega+1}}{\mathpzc{z}_\omega}, \quad \omega = 0, \cdots, p-2, \\
&\mathfrak{q}_{p-1} = \mathfrak{q} \frac{\mathpzc{z}_0}{\mathpzc{z}_{p-1}},
\end{split}
\end{align}
so that
\begin{align}
\prod_{\omega=0} ^{p-1} \mathfrak{q}_\omega = \mathfrak{q}.
\end{align}
Note that $\mathfrak{q}$ is weighted with the power $\widetilde{k}$ in the measure and therefore is the bulk coupling of the $A_1$-theory on $\mathbb{C}_{\varepsilon_1} \times \mathbb{C}_{\widetilde{\varepsilon}_2}$. 

As a result, the full partition function \eqref{orbpart} can be written as
\begin{align} \label{simppart}
\EuScript{Z}^{\mathbb{Z}_p, c} = \sum_{\boldsymbol{\Lambda}} \mathfrak{q} ^{\vert \boldsymbol{\Lambda} \vert} \; \mathcal{I}_c [\boldsymbol{\Lambda}] \; \epsilon \left[ \frac{1}{\widetilde{P}_{12}} \left( - \widetilde{S} \widetilde{S} ^* +\widetilde{M} ^* \widetilde{S} \right) \right] = \left\langle \mathcal{I}_c \right\rangle  \EuScript{Z}_{A_1},
\end{align}
where the surface defect is expressed as a chiral ring element
\begin{align} \label{orbsurfdef}
\mathcal{I}_c [\boldsymbol{\Lambda}] \equiv \sum_{\boldsymbol{\lambda} \in \rho^{-1} (\boldsymbol{\Lambda})} \; \prod_{\omega=0} ^{p-1} \mathpzc{z}_\omega ^{k_{\omega-1} - k_\omega}\; \epsilon \left[ \frac{1}{P_1} \left( \sum_{\omega=1} ^{p-1} \left(  (\widetilde{M}_{\omega-1} - \Sigma_\omega)^* \Sigma_\omega +\Sigma_{\omega+1} \Sigma_\omega ^* \right) + \widetilde{M}_{p-1} ^* \widetilde{S} \right) \right].
\end{align}
Let us focus on the zero-instanton sector, $\vert \boldsymbol{\Lambda} \vert = \widetilde{k}=0$. An element of the inverse image $\lambda = \left( \lambda^{(\alpha)} \right)_{\alpha=1} ^N \in \rho^{-1} (\varnothing)$ is of the form
\begin{align}
\lambda^{(\alpha)}\;\; = \;\; \ytableausetup{mathmode,boxsize=2.5em, centertableaux}
\begin{ytableau}\scriptstyle c(\alpha)&\scriptstyle c(\alpha)+1 &\dots& \scriptstyle p-2\\
\vdots &\vdots & \vdots & \vdots \\
\vdots & \vdots & \vdots & \\ 
\vdots & \vdots & \\ 
\vdots & \\
{}
\end{ytableau},
\end{align}
where the number in each box denotes its color. We may define the length of the column of color $\omega$ to be $d_{\omega+1, \alpha}$ ($c(\alpha) \leq \omega < p-1$). Note that $k_{\omega-1} = \sum_{\alpha} d_{\omega, \alpha}$. Consequently, we have 
\begin{align}
\widetilde{K}_\omega = \sum_{c(\alpha) \leq \omega }  e^{\beta \widetilde{a}_\alpha} \sum_{i=1} ^{d_{\omega+1, \alpha}} q_1 ^{i-1},
\end{align}
from which we simplify \eqref{sigma} as
\begin{align}
\Sigma_\omega = \sum_{c(\alpha) < \omega } e^{\beta \widetilde{a}_\alpha} q_1 ^{d_{\omega, \alpha}}.
\end{align}
Therefore, the partition function \eqref{orbsurfdef} is reduced to a sum over the non-negative integers
\begin{align}
\begin{cases} \;\; d_{\omega, \alpha} \geq 0 \;\;\;\; \Bigg\vert \end{cases}  \begin{rcases} \omega = 1, \cdots, p-1, \;\; c(\alpha) < \omega \\ d_{\omega, \alpha} \geq d_{\omega+1,\alpha} \quad\quad\quad\; \end{rcases},
\end{align}
with the simplified $\Sigma_\omega$ given above. This is precisely the partition function of the gauged linear sigma model on the $\bigoplus_{\omega=1} ^{p-1} \text{Hom} (\mathcal{E}_\omega , \widetilde{M}_{\omega-1})$-bundle over the partial flag variety $\text{Flag} (l_1, l_2, \cdots, N)$, with 
\begin{align}
l_{\omega} \equiv \vert \{ \alpha \; \vert \; c(\alpha) < \omega \}\vert,
\end{align}
under certain stability condition \cite{nek7}. Here, $\mathcal{E}_\omega$ is the $\omega$-th tautological bundle with $\text{rk}\: \mathcal{E}_\omega = l_\omega$. The K\"{a}hler moduli are precisely $\{ \mathfrak{q}_{\omega-1} = \mathpzc{z}_{\omega}/\mathpzc{z}_{\omega-1} \; \vert \; \omega =1, \cdots, p-1 \}$.

In the non-zero instanton sector of the four-dimensional theory, the sigma model couples to the four-dimensional gauge theory through \eqref{orbsurfdef} in a non-trivial way, generating a surface defect. In this way, the full partition function \eqref{simppart} represents the 2d-4d coupled system.

The investigations in this paper mainly utilize the special case, the $(N-1, 1)$-type $\mathbb{Z}_2$-orbifold. That is, we set $p=2$ and assign the coloring function
\begin{align}
c_\beta (\alpha) \equiv \begin{cases} 1  \quad \text{for } \alpha= \beta \\ 0 \quad \text{otherwise} \end{cases},
\end{align}
for some chosen $\beta \in \{ 1, \cdots, N \}$. We also set $\widetilde{M}_0 , \widetilde{M}_1 = \mathbb{C} ^N$. For later use, it is instructive to separate out the instanton part in the partition function \eqref{simppart},
\begin{align} \label{orbpartition}
\EuScript{Z}^{\mathbb{Z}_2} _\beta = \sum_{\boldsymbol{\Lambda}}\mathfrak{q}  ^{\vert \boldsymbol{\Lambda} \vert} \; \mathcal{I}_\beta [\boldsymbol{\Lambda}] \; \epsilon \left[ \widetilde{N} ^* \widetilde{K} +q_{12} ^{-1} \widetilde{N} \widetilde{K} ^* - P_{12} ^* \widetilde{K} \widetilde{K} ^* - \widetilde{M} ^* \widetilde{K} \right],
\end{align}
where the instanton part of the surface defect is
\begin{align}
\begin{split}
\mathcal{I}_\beta [\boldsymbol{\Lambda}] &= \sum_{\boldsymbol{\lambda} \in \rho^{-1} (\boldsymbol{\Lambda})} \; z ^{k_0 -k_1} \; \epsilon \left[ (\widetilde{K}_0 -\widetilde{K}_1) (\widetilde{N}_0 - P_1 \widetilde{K}_0 +\widetilde{q}_2 P_1 \widetilde{K}_1)^* \right. \\
&\quad\quad\quad\quad\quad\quad\quad\quad\quad\quad\quad \left. + q_1 \widetilde{N}_1 (\widetilde{K}_0 - \widetilde{q}_2 \widetilde{K}_1)^* - \widetilde{M}_0 ^* (\widetilde{K}_0 -\widetilde{q}_2 \widetilde{K}_1) - \widetilde{P}_2 \widetilde{M}_1 ^* \widetilde{K}_1 \right].
\end{split}\end{align}
In this special case, the target space of the two-dimensional sigma model that generates the surface defect is the $\text{Hom} (\mathcal{O}(-1), \mathbb{C}^N)$-bundle over $\mathbb{P}^{N-1}$, which is exactly the same with that of the quiver surface defect in section \ref{ssquivsurf}. Thus it is natural to expect the two distinct types of surface defects are actually related to each other. However, it is not so obvious from the explicit expressions for their partition functions, \eqref{a2partred} and \eqref{simppart}, how they can really be associated. In particular, the combinatorics that define these partition functions are quite different; one involves a simple sum over non-negative integers while the other involves the non-trivial mapping $\rho$ between $N$-tuples of Young diagrams. We come back to this problem in section \ref{glue}.

\section{Non-perturbative Dyson-Schwinger equations} \label{ssnpds}
We investigate the non-perturbative Dyson-Schwinger equations satisfied by the surface defect partition functions that we constructed in the previous section. The primary object of this investigation is the $qq$-character, which is a gauge theory observable formed as a certain Laurent polynomial of $\EuScript{Y}$-observables \cite{nek2}. The most general $qq$-characters were constructed in \cite{nek2, nek4} from the spiked instanton configurations, by integrating out the degrees of freedom orthogonal to the four-dimensional gauge theory. The compactness theorem for the spiked instanton moduli space proven in \cite{nek3} provided the crucial property of the $qq$-character, the holomorphicity of its expectation value. Schematically,
\begin{align}
\bigg\langle \EuScript{X} (\EuScript{Y}(x)) \bigg\rangle = \frac{1}{\EuScript{Z}^{\text{inst}}} \sum_{\boldsymbol{\lambda}} \EuScript{X} (\EuScript{Y}(x)[\boldsymbol{\lambda}]) \; \mathfrak{q}^{\vert \boldsymbol{\lambda} \vert} \; \boldsymbol{\mu}_{\boldsymbol{\lambda}} = T(x),
\end{align}
where $T(x)$ is a polynomial in $x$ of certain degree. Therefore, the $qq$-character generates an infinite number of constraints that the partition function satisfies, from the expectation values of its non-regular parts
\begin{align}
[x^{-n}] \; \bigg\langle \EuScript{X}(\EuScript{Y}(x))  \bigg\rangle = 0 , \quad n \geq 1,
\end{align}
which we call \textit{the non-perturbative Dyson-Schwinger equations}. 

In this section, we present the fundamental $qq$-characters relevant to each surface defect, and study the consequences of their non-perturbative Dyson-Schwinger equations. For other analysis on the non-perturbative Dyson-Schwinger equations, see \cite{nek8, sbxin} in the context of the BPS/CFT correspondence, and \cite{sb} in the context of the Bethe/gauge correspondence.

\subsection{The quiver} \label{npdsquiver}
As in section \ref{ssquivsurf}, we start with the $A_2$-quiver gauge theory with the $U(N)$ gauge group. The fundamental $qq$-characters for this theory is given by \cite{nek2}
\begin{subequations} \label{a2qq}
\begin{align}
\EuScript{X}_1 (x) &= \EuScript{Y}_1 (x + \varepsilon) + \mathfrak{q}_1 \frac{\EuScript{Y}_0 (x) \EuScript{Y}_2 (x+\varepsilon)}{\EuScript{Y}_1 (x)} + \mathfrak{q}_1 \mathfrak{q}_2 \frac{\EuScript{Y}_0 (x) \EuScript{Y}_3 (x+ \varepsilon) }{\EuScript{Y}_2 (x)}, \\
\EuScript{X}_2 (x) &= \EuScript{Y}_2 (x+ \varepsilon) + \mathfrak{q}_2 \frac{\EuScript{Y}_1 (x) \EuScript{Y}_3 (x+\varepsilon)}{\EuScript{Y}_2 (x)} + \mathfrak{q}_1 \mathfrak{q}_2 \frac{\EuScript{Y}_0 (x-\varepsilon) \EuScript{Y}_3 (x+ \varepsilon)}{\EuScript{Y}_1 (x - \varepsilon)},
\end{align}
\end{subequations}
where $\EuScript{Y}_0 (x) \equiv \prod_{\alpha=1} ^N (x- a_{0, \alpha})$ and $\EuScript{Y}_3 (x) \equiv \prod_{\alpha=1} ^N (x-a_{3, \alpha})$ by definition.\footnote{Be cautious about the re-definition of parameters \eqref{redef} when we deal with the case \eqref{constdiff}. The expressions for $\EuScript{Y}_0$ and $\EuScript{Y}_3$ also change correspondingly.} We construct the surface defect by imposing the constraints \eqref{a2const} or \eqref{constdiff} for the Coulomb moduli. In each case, the $\EuScript{Y}$-observable of the first or the second gauge node is simplified to
\begin{subequations} \label{a1y1}
\begin{align}
&\EuScript{Y}_1(x)[\boldsymbol{\lambda}^{(1)} (k_1)] = \EuScript{Y}_0 (x) \frac{x-a_{0,\beta}+\varepsilon_2 - k_1 \varepsilon_1}{x-a_{0,\beta}- k_1 \varepsilon_1}, \quad \text{for } \eqref{a2const} \\
&\EuScript{Y}_2(x)[\boldsymbol{\lambda}^{(2)} (k_2)] = \EuScript{Y}_3 (x+\varepsilon) \frac{x+a_{3,\beta} -\varepsilon_1 - k_2 \varepsilon_1}{x+a_{3,\beta}- \varepsilon - k_2 \varepsilon_1}, \quad \text{for } \eqref{constdiff}
\end{align}
\end{subequations}
It is now straightforward to plug \eqref{a1y1} back into \eqref{a2qq} and compute their expectation values of the non-regular parts. However, it is convenient to follow the systematic procedure established in \cite{sbxin}. First let us define
\begin{align}
\EuScript{G} (x;t) \equiv \frac{1}{\EuScript{Y}_0 (x) \prod_{i=0} ^2 (1+t z_i)} \sum_{l=0} ^3 z_0 z_1 \cdots z_{l-1} \; t^l \; \EuScript{X}_l (x-\varepsilon (1-l)) = \sum_{n=0} ^{\infty} \frac{\EuScript{G} ^{(-n)}(t)}{x^n},
\end{align}
where we have defined the parameters $z_i$ by $\mathfrak{q}_i \equiv \frac{z_i}{z_{i-1}}$ ($z_{-1} = \infty$ and $z_3=0$ by definition), and $t$ is an auxiliary parameter. The non-perturbative Dyson-Schwinger equations imply
\begin{align} \label{npdssimp}
[x^{-n}] \; \Big\langle \EuScript{Y}_0 (x) \EuScript{G} (x;t) \Big\rangle =0 , \quad n\geq 1,
\end{align}
for any value of $t$. In appendix \ref{appB}, we summarize the systematic approach for computing $\EuScript{G}^{(-n)} (t)$. We focus on presenting the results below.

\subsubsection{$N=2$} \label{npdsn2case}
We observe that $\EuScript{Y}_0 (x) = \prod_{\alpha} ^N (x-a_{0, \alpha})$ is a polynomial of degree $N$. Hence in the case of $N=2$, the $x^{-1}$-term in \eqref{npdssimp} is
\begin{align} \label{n2case}
0 = \Big\langle \EuScript{G} ^{(-3)} (t) \Big\rangle -(a_{0,1} +a_{0,2})  \Big\langle \EuScript{G} ^{(-2)} (t) \Big\rangle +a_{0,1} a_{0,2}  \Big\langle \EuScript{G} ^{(-1)} (t) \Big\rangle.
\end{align}
Recall that with the constraints \eqref{a2const} the Young diagram $\boldsymbol{\lambda}^{(1)}$ for the first gauge node is restricted to be single-columned. Thus we can simplify
\begin{align} \label{c1simp}
\begin{split}
\Bigg\langle \sum_{\Box \in \boldsymbol{\lambda}^{(1)}} c_{\square} \Bigg\rangle &= (a_{0,\beta} -\varepsilon) \Big\langle k_1 \Big\rangle +  \frac{\varepsilon_1}{2} \Big\langle k_1 (k_1+1) \Big\rangle \\
&=\frac{1}{\EuScript{Z}_\beta ^L} \left[ \left( a_{0,\beta} -\varepsilon -\frac{\varepsilon_1}{2} \right) \mathfrak{q}_1  \frac{\partial}{\partial \mathfrak{q}_1}  + \frac{\varepsilon_1}{2} \left(  \mathfrak{q}_1  \frac{\partial}{\partial \mathfrak{q}_1}\right)^2 \right] \EuScript{Z}_\beta ^L.
\end{split}
\end{align}
Using this relation, the residue of \eqref{n2case} at $t=-z_0 ^{-1}$ can be written as the following second order differential equation
\begin{align} \label{a2caseinter}
\begin{split}
0 &=\left[  \varepsilon_1 ^2 \left( z_0 \frac{\partial}{\partial z_0} \right)^2 - \varepsilon_1  \left( \sum_{i=1}^2 \frac{z_i}{z_i -z_0} \mathcal{A}_i ^{(1)} + 2 a_{0,\beta}- a_{0,1} -a_{0,2} \right)  \left( z_0 \frac{\partial}{\partial z_0} \right) \right. \\
&\quad\quad + \sum_{i=1} ^2 \frac{z_i}{z_i -z_0} \left( \frac{1}{2} \left(\mathcal{A}_i ^{(2)} + \left(\mathcal{A}_1 ^{(1)}\right)^2\right) -\varepsilon_1 \varepsilon_2 z_i \frac{\partial}{\partial z_i} - ( a_{0,1}+ a_{0,2} -a_{0,\beta}) \mathcal{A}_i ^{(1)} \right)  \\
& \quad\quad \left. +  \frac{z_1 z_2}{(z_1 -z_0)(z_2 -z_0)} \left(\mathcal{A}_1 ^{(1)} -\varepsilon\right) \mathcal{A}_2 ^{(1)} \right] \EuScript{Z}_\beta ^L,
\end{split}
\end{align}
where we have introduced
\begin{align} \label{vara}
\mathcal{A}_{i} ^{(n)} \equiv \sum_{\alpha=1} ^N \left( a_{i,\alpha}^n - (a_{i +1, \alpha} -\varepsilon)^n  \right), \quad \quad i=1,2.
\end{align}
Here $N=2$ but we will also extend to the higher $N$ by the same expression. In particular, for $n=1$ we can write
\begin{align}
\mathcal{A}_i ^{(1)} = N (\bar{a}_i -\bar{a}_{i+1} +\varepsilon),
\end{align}
where we have defined $\bar{a}_i \equiv \frac{1}{N} \sum_{\alpha=1} ^N a_{i,\alpha}$. For our purpose of investigating the relations with the opers, it is important to re-define the partition function as
\begin{align} \label{n2prefactor}
\widetilde{\EuScript{Z}}_{A_2} \equiv \prod_{i=0} ^2 z_i ^{L_i} \prod_{0\leq i<j \leq 2} \left( 1- \frac{z_j}{z_i} \right)^{T_{ij}} \EuScript{Z}_{A_2},
\end{align}
where we have multiplied the prefactors with the exponents,
\begin{align} \label{n2expnents}
\begin{split}
&L_i \equiv \frac{(a_{i+1,1}-a_{i+1,2})^2 - (a_{i,1}-a_{i,2})^2}{4 \varepsilon_1 \varepsilon_2} + \frac{ (\bar{a}_i -\bar{a}_{i+1} +\varepsilon )(\bar{a}_i -\bar{a}_{i+1})}{\varepsilon_1 \varepsilon_2} , \quad i=0,1,2, \\
&T_{ij} = \frac{2(\bar{a}_j -\bar{a}_{j+1} + \varepsilon) (\bar{a}_i -\bar{a}_{i+1} ) }{\varepsilon_1 \varepsilon_2}, \quad i,j=0,1,2.
\end{split}
\end{align}
With the constraints \eqref{a2const} imposed on these prefactors, the modification for the surface defect partition function $\EuScript{Z}_\beta ^L$ is simpler than the most generic case. Let us set $z_0 = z$, $z_1=1$, and $z_2 = \mathfrak{q}$ by using the redundancy of overall scaling of $z_i$'s. Then we find the prefactors (with the overall constant that we choose at our convenience) for $\EuScript{Z}_\beta ^L$ can be written as
\begin{align}
\begin{split}
&\left(-\frac{1}{z}\right)^{-r_{L,\beta}} \; \mathfrak{q} ^{-\Delta_{\mathfrak{q}} -\Delta_0 +\frac{\varepsilon^2 -(a_{2,1}-a_{2,2})^2}{4 \varepsilon_1 \varepsilon_2}} \\
& \quad\left( 1-\frac{1}{z} \right)^{ \frac{2\bar{a}_0 - 2\bar{a}_2 + 2\varepsilon_1 + \varepsilon_2}{2 \varepsilon_1} } \; \left( 1-\frac{\mathfrak{q}}{z} \right)^{ \frac{\bar{a}_2 -\bar{a}_3 +\varepsilon}{\varepsilon_1} } \; \left( 1-\mathfrak{q} \right)^{ \frac{2(\bar{a}_2 -\bar{a}_3+\varepsilon)( 2\bar{a}_0 -2 \bar{a}_2  -\varepsilon_2 )}{\varepsilon_1 \varepsilon_2} },
\end{split}
\end{align}
where we have defined
\begin{align} \label{critexpinf}
\left(r_{L,\beta} \right)_{\beta=1,2} = \left( \frac{-a_{0,1}+a_{0,2}+\varepsilon + \varepsilon_2}{2\varepsilon_1} , \;  \frac{a_{0,1}-a_{0,2}+\varepsilon + \varepsilon_2}{2\varepsilon_1} \right),
\end{align} 
and
\begin{align} \label{delta}
\begin{split}
&\Delta_0 \equiv \frac{\varepsilon^2-(a_{3,1}-a_{3,2})^2}{4\varepsilon_1 \varepsilon_2} \\
&\Delta_{\mathfrak{q}} \equiv - \frac{(\bar{a}_2 -\bar{a}_3 )(\bar{a}_2 -\bar{a}_3 +\varepsilon) }{\varepsilon_1 \varepsilon_2} \\
&\Delta_1 \equiv -\frac{(2\bar{a}_0 -2\bar{a}_2 +2\varepsilon_1 + \varepsilon_2)(2\bar{a}_0 -2\bar{a}_2 -\varepsilon_2)}{4\varepsilon_1 \varepsilon_2} \\
&\Delta_\infty \equiv \frac{\varepsilon^2 -(a_{0,1}-a_{0,2})^2}{4 \varepsilon_1 \varepsilon_2}.
\end{split}
\end{align}
For the re-defined partition function $\widetilde{\EuScript{Z}}_\beta ^L$, the differential equation \eqref{a2caseinter} becomes 
\begin{align} \label{a2npdsfinal}
\begin{split}
0=& \left[  \varepsilon_1 ^2 \partial^2 - \varepsilon_1\varepsilon_2 \frac{2z-1}{z (z-1)} \partial + \varepsilon_1 \varepsilon_2 \frac{\mathfrak{q}-1}{z(z-1)(z-\mathfrak{q})} \mathfrak{q} \frac{\partial}{\partial \mathfrak{q}} \right. \\ 
& \left. \quad+ \varepsilon_1 \varepsilon_2 \left( \frac{\Delta_0}{z ^2} + \frac{\Delta_1}{(z-1)^2} + \frac{\Delta_{\mathfrak{q}}}{(z-\mathfrak{q})^2} - \frac{ -\frac{2\varepsilon +\varepsilon_2}{4\varepsilon_1} +  \Delta_1 + \Delta_{\mathfrak{q}} +\Delta_0 - \Delta_{\infty}}{z (z-1)}  \right)  \right] \widetilde{\EuScript{Z}}_\beta ^L.
\end{split}
\end{align}
We can view this differential equation as the second-order differential operator $\widehat{\widehat{\mathfrak{D}}}_2$ annihilating the modified partition function $\widetilde{\EuScript{Z}}_\beta ^L$. Note that the operator $\widehat{\widehat{\mathfrak{D}}}_2$ is independent of $\beta$, so that each choice of $\beta \in \{1,2 \}$ provides a solution to $\widehat{\widehat{\mathfrak{D}}}_2$. We may regard $\widehat{\widehat{\mathfrak{D}}}_2$ as the quantization of the $SL(2)$-oper $\widehat{\mathfrak{D}}_2$ for the four-punctured sphere $\mathbb{P}^1 \backslash \{ 0, \qe , 1, {\infty} \}$, as we now argue.

Under the NS limit ($\varepsilon_1 \neq 0, \varepsilon_2 \to 0$), the surface defect partition function \eqref{a2partred} is dominated by the limit shape \cite{nekokoun}. Viewed as the expectation value of $\mathcal{I}_\beta ^L$ in the $A_1$-theory, the surface defect partition function gets the singular contribution, or the effective twisted superpotential, from the bulk $A_1$-theory, while the observable $\mathcal{I}_\beta ^L$ only contributes regular terms. Therefore, we arrive at the following asymptotics of the partition function 
\begin{align}
\widetilde{\EuScript{Z}}_\beta ^L ( \mathbf{a}_2 \equiv \mathbf{a} ,z, \mathfrak{q}) =  e^ {\frac{\widetilde{\EuScript{W}}( \mathbf{a} ,\mathfrak{q})}{\varepsilon_2}} \left( \chi_\beta (\mathbf{a},z,\mathfrak{q}) + \mathcal{O}(\varepsilon_2 ) \right),
\end{align}
where we have omitted the subscript for the Coulomb moduli $\mathbf{a}_2$ since it precisely becomes the Coulomb moduli $\mathbf{a}$ of the $A_1$-theory. $\widetilde{\EuScript{W}}$ is a part of the effective twisted superpotential of the underlying $A_1$-gauge theory,
\begin{align} \label{efftwpoten}
\begin{split}
\widetilde{\EuScript{W}} &\equiv \lim_{\varepsilon_2 \to 0} \varepsilon_2  \log \widetilde{\EuScript{Z}}_\beta ^L \\
&= \widetilde{\EuScript{W}}^{\text{classical}} + \widetilde{\EuScript{W}}^{\text{inst}} + \widetilde{\EuScript{W}}^{\text{extra}},
\end{split}
\end{align}
where we have defined
\begin{subequations} \label{n2twsupo}
\begin{align}
&\widetilde{\EuScript{W}}^{\text{classical}} \equiv -\frac{\left( a_{1}-a_{2} \right)^2}{4\varepsilon_1 } \log \mathfrak{q} \\
&\widetilde{\EuScript{W}}^{\text{inst}} \equiv \lim_{\varepsilon_2 \to 0} \varepsilon_2 \log \EuScript{Z}_{A_1} ^{\text{inst}} \\
&\widetilde{\EuScript{W}}^{\text{extra}} \equiv \varepsilon_1 \left( \frac{1}{4} -\delta_{\mathfrak{q}} -\delta_0 \right) \log\mathfrak{q} + \frac{2(\bar{a}_0 - \bar{a})(\bar{a} -\bar{a}_3 + \varepsilon_1 )}{\varepsilon_1}  \log(1-\mathfrak{q}),
\end{align}
\end{subequations}
where the instanton partition function $\EuScript{Z}_{A_1} ^{\text{inst}}$ for the $A_1$-theory is given by \eqref{a1}. In particular, $\widetilde{\EuScript{W}}^{\text{inst}}$ is fully determined by the Young diagram expansions reviewed in appendix \ref{appA}. Also, we have defined the limit,
\begin{align}
\varepsilon_2 \Delta_i \xrightarrow{\varepsilon_2 \to 0} \varepsilon_1 \delta_i, \quad i= 0, \mathfrak{q}, 1, \infty.
\end{align}
We have emphasized that \eqref{efftwpoten} is only a part of the full effective twisted superpotential, since we are missing the 1-loop term. This is because the 1-loop term is independent of the gauge coupling and therefore ignorant of the differential equation that the partition function satisfies. The missing 1-loop part will re-combine in section \ref{mono}.

Thus, under the NS limit the equation for the differential operator $\widehat{\widehat{\mathfrak{D}}}_2$ becomes
\begin{align} \label{heun}
0 = \left[  \partial^2 +  \frac{\delta_0}{z^2} +\frac{\delta_1}{(z-1)^2} +\frac{\delta_{\mathfrak{q}}}{(z-\mathfrak{q})^2} - \frac{\delta_1 +\delta_{\mathfrak{q}} +\delta_0 - \delta_{\infty}}{z(z-1)}  +  \frac{H}{z(z-1)(z-\mathfrak{q})}  \right] \chi_\beta,
\end{align}
which is exactly the equation for \textit{the Heun's oper}, the Fuchsian differential operator $\widehat{\mathfrak{D}}_2$ of degree 2 with fixed conjugacy class of monodromy at each puncture of $\mathbb{P}^1 \backslash \{ 0, \qe , 1, {\infty} \}$. The variety $\mathcal{O}_2 [\mathbb{P} ^1 \backslash \{ 0,\mathfrak{q},1,\infty \} ]$ of these opers is spanned by \textit{the accessory parameter},
\begin{align} \label{coordi}
\begin{split}
H &\equiv -\mathfrak{q}(1-\mathfrak{q})  \frac{1}{\varepsilon_1} \frac{ \partial \widetilde{\EuScript{W}}}{\partial \mathfrak{q}} \\
&= (1-\mathfrak{q}) \left( \frac{1}{2\varepsilon_1 ^2} \lim_{\varepsilon_2 \to 0} \Big\langle \EuScript{O}_2 \Big\rangle_{A_1} -  \frac{1}{4} +\delta_{\mathfrak{q}} +\delta_0   \right) + \frac{2(\bar{a}_0 - \bar{a})(\bar{a} -\bar{a}_3 + \varepsilon_1 )}{\varepsilon_1^2} \mathfrak{q}.
\end{split}
\end{align}
All the terms are just some constants except the expectation value of chiral observable $\EuScript{O}_2 = \text{Tr} \phi_2 ^2$. Thus, a holomorphic coordinate on the variety $\mathcal{O}_2 [\mathbb{P} ^1 \backslash \{0,\mathfrak{q},1,\infty \} ] $ of opers is provided by the expectation value of the chiral observable $\EuScript{O}_2 $ in the limit $\varepsilon_2 \to 0$. The variety $\mathcal{O}_2 [\mathbb{P} ^1 \backslash \{0,\mathfrak{q},1,\infty \} ] $ of opers is a quantization of the Coulomb moduli space of $\mathcal{T} [A_1, \mathbb{P} ^1 \backslash \{0,\mathfrak{q},1,\infty \} ]$ in this sense. The expectation value $\lim_{\varepsilon_2 \to 0} \Big\langle \EuScript{O}_2 \Big\rangle_{A_1}$ is also identified with the off-shell spectrum of the quantum Hitchin system on $\mathbb{P} ^1 \backslash \{0,\mathfrak{q},1,\infty \}$ through the Bethe/gauge correspondence. Hence, we observe that the relation \eqref{coordi} establishes the connection between the accessory parameter $H$ of $\widehat{\mathfrak{D}}_2$ and the off-shell spectrum of quantum Hitchin Hamiltonian. A proper on-shell condition is expected to be introduced by a $\omega_K$-Lagrangian brane which intersects with $\mathcal{O}_2 [\mathbb{P} ^1 \backslash \{0,\mathfrak{q},1,\infty \} ] $ at isolated points. As we argued earlier, the holomorphic coordinate, i.e., the expectation value, \eqref{coordi} evaluated at these points gives the on-shell spectrum of the quantum Hitchin system.\paragraph{Remarks}
\begin{itemize}
\item It was checked in \cite{pm1, pm2} that the series expansion \eqref{coordi} for the accessory parameter $H$ matches with the direct computation in which $H$ is determined by fixing the monodromy of the oper $\widehat{\mathfrak{D}}_2$ along the $A$-cycle (see Figure \ref{fig2}), up to some low orders in the gauge coupling $\mathfrak{q}$. The derivation above is purely gauge theoretical and therefore guarantees the validity to all orders in $\mathfrak{q}$.
\item The series expansion for the instanton partition function is valid when $0< \vert \mathfrak{q}_1 \vert, \vert \mathfrak{q}_2 \vert<1$. This implies that the solutions $\widetilde{\EuScript{Z}}_\beta ^L$ for the operator $\widehat{\widehat{\mathfrak{D}}}_2$ are in the convergence domain $ 0<\vert \mathfrak{q} \vert <1 <\vert z \vert $.
\item The solution $\chi_\beta$ for the oper $\widehat{\mathfrak{D}}_2$ can be represented as a sum of the Baxter $\EuScript{Q}$-functions, by using \eqref{quivsurfq} and taking the limit $\varepsilon_2 \to 0$. This expression reflects that the equation for the oper is the Fourier transform of the Baxter $TQ$-equation.
\item The fact that \eqref{a2npdsfinal} coincides with well-known null-vector decoupling equation in two-dimensional CFT \cite{bpz}, see also \cite{tes1, vt,abdfjl}, confirms the paradigm of the BPS/CFT correspondence \cite{N1} at the example of the AGT correspondence \cite{agt}.
\end{itemize}

Similarly, it is not too difficult to derive a closed differential equation for $\EuScript{Z}_\beta ^R$. Again, we re-define partition function as in \eqref{n2prefactor} with the prefactors \eqref{n2expnents}, yet with the constraints \eqref{constdiff}. Also we need the re-definition of parameters \eqref{redef} for the prefactors this time. By setting $z_0=1$, $z_1=\mathfrak{q}$, and $z_2 = z$, the relevant prefactor for $\EuScript{Z}_\beta ^R$ is
\begin{align}\label{prefactor0}
\begin{split}
& \left(-\frac{\mathfrak{q}}{z}\right)^{-r_{R,\beta}} \; \mathfrak{q} ^{\frac{\varepsilon^2 -(a_{1,1}-a_{1,2})^2}{4\varepsilon_1 \varepsilon_2} -\Delta_0 -\Delta_{\mathfrak{q}}' + \frac{2\varepsilon + \varepsilon_2}{4\varepsilon_1} } \\
&\quad \left( 1-\mathfrak{q} \right)^{\frac{(\bar{a}_0 - \bar{a}_1 +\varepsilon)(2\bar{a}_1 -2\bar{a}_3 - \varepsilon_2)}{\varepsilon_1 \varepsilon_2} } \; \left( 1-z \right)^{\frac{\bar{a}_0 -\bar{a}_1 +\varepsilon}{\varepsilon_1} } \; \left( 1- \frac{z}{\mathfrak{q}} \right)^{\frac{2\bar{a}_1 -2\bar{a}_3 +2 \varepsilon_1 + \varepsilon_2}{2\varepsilon_1}}
\end{split}
\end{align}
where we have defined
\begin{align} \label{critexp0}
\left( r_{R,\beta} \right)_{\beta=1,2} \equiv \left( \frac{-a_{3,1}+a_{3,2}+\varepsilon}{2\varepsilon_1} ,\;  \frac{a_{3,1}-a_{3,2}+\varepsilon}{2\varepsilon_1}  \right),
\end{align}
and
\begin{align} \label{delta'}
\begin{split}
&\Delta_{\mathfrak{q}} ' \equiv - \frac{(2\bar{a}_1 - 2\bar{a}_3 -\varepsilon_2)(2\bar{a}_1 -2\bar{a}_3 +2\varepsilon_1 +\varepsilon_2)}{4\varepsilon_1 \varepsilon_2} \\
&\Delta_1 ' \equiv - \frac{(\bar{a}_0 -\bar{a}_1)(\bar{a}_0 - \bar{a}_1 + \varepsilon)}{\varepsilon_1 \varepsilon_2}.
\end{split}
\end{align}
Then the differential equation satisfied by the modified partition function $\widetilde{\EuScript{Z}}_\beta ^R$ is
\begin{align} \label{rightequation}
\begin{split}
0=& \left[  \varepsilon_1 ^2 \partial^2 - \varepsilon_1\varepsilon_2 \frac{2z-1}{z (z-1)} \partial + \varepsilon_1 \varepsilon_2 \frac{\mathfrak{q}-1}{z(z-1)(z-\mathfrak{q})} \mathfrak{q} \frac{\partial}{\partial \mathfrak{q}} \right. \\ 
& \left. \quad+ \varepsilon_1 \varepsilon_2 \left( \frac{\Delta_0}{z ^2} + \frac{\Delta_1 '}{(z-1)^2} + \frac{\Delta_{\mathfrak{q}} '}{(z-\mathfrak{q})^2} - \frac{ -\frac{2\varepsilon +\varepsilon_2}{4\varepsilon_1} +\Delta_1 ' + \Delta_{\mathfrak{q}} ' +\Delta_0 - \Delta_{\infty}}{z (z-1)}  \right)  \right] \widetilde{\EuScript{Z}}_\beta ^R.
\end{split}
\end{align}
Note that this differential equation is precisely the equation \eqref{a2npdsfinal} for $\widehat{\widehat{\mathfrak{D}}}_2$, except $\Delta_1 \to \Delta_1 ', \Delta_{\mathfrak{q}} \to \Delta_{\mathfrak{q}} '$. To equate these quantities to get the same equation, we have to clarify how the Coulomb moduli of the two theories are associated. This will be the subject of section \ref{shiftmatr}.

\paragraph{Remarks}
\begin{itemize}
\item This time, $0< \vert \mathfrak{q}_1 \vert ,\vert \mathfrak{q}_2 \vert <1$ implies the convergence domain $0 <\vert z \vert< \vert \mathfrak{q} \vert  <1 $. Thus the domains for the solutions $\widetilde{\EuScript{Z}}_\beta ^L$ and $\widetilde{\EuScript{Z}}_\beta ^R$ are disjoint.
\end{itemize}

\subsubsection{$N=3$}
Since $\EuScript{Y}_0 (x)$ is now a polynomial of degree 3, the $x^{-1}$-term of \eqref{npdssimp} can be written as
\begin{align} \label{n3dseq}
0 = \Big\langle \EuScript{G} ^{(-4)} (t) \Big\rangle - \sum_{\alpha=1} ^3 a_{0,\alpha}  \Big\langle \EuScript{G} ^{(-3)} (t) \Big\rangle + \sum_{1\leq \alpha<\beta \leq 3} a_{0,\alpha} a_{0,\beta}  \Big\langle \EuScript{G} ^{(-2)} (t) \Big\rangle - \prod_{\alpha=1} ^3 a_{0, \alpha} \Big\langle \EuScript{G} ^{(-1)} (t) \Big\rangle.
\end{align}
In addition to \eqref{c1simp}, we utilize the following relation from the single-columnedness of $\boldsymbol{\lambda}^{(1)}$
\begin{align} \label{c2simp}
\begin{split}
\Bigg\langle & \sum_{\Box \in \boldsymbol{\lambda}^{(1)}} c_{\square} ^2 \Bigg\rangle = (a_{0,\beta} -\varepsilon)^2 \Big\langle k_1 \Big\rangle + \varepsilon_1 (a_{0,\beta}-\varepsilon) \Big\langle k_1 (k_1+1) \Big\rangle +\frac{\varepsilon_1 ^2}{6} \Big\langle k_1 (k_1+1)(2k_1+1) \Big\rangle \\
&=\frac{1}{\EuScript{Z}_\beta ^L} \left[ \frac{\varepsilon_1 ^2}{3} \left( \mathfrak{q}_1 \frac{\partial}{\partial \mathfrak{q}_1} \right)^3 + \left( \frac{\varepsilon_1 ^2}{2} +\varepsilon_1 (a_{0,\beta}-\varepsilon) \right) \left( \mathfrak{q}_1 \frac{\partial}{\partial \mathfrak{q}_1} \right)^2 + \left( \frac{\varepsilon_1 ^2}{6}  + (a_{0,\beta}-\varepsilon)(a_{0,\beta}-\varepsilon_2) \right)\left( \mathfrak{q}_1 \frac{\partial}{\partial \mathfrak{q}_1} \right) \right]\EuScript{Z}_\beta ^L.
\end{split}
\end{align}
Using the relations \eqref{c1simp} and \eqref{c2simp}, the residue of \eqref{n3dseq} at $t=-z_0 ^{-1}$ can be written as the following third order differential equation
\begin{align}\label{3rdorder}
\begin{split}
0 &= \left[ -\varepsilon_1 ^3 \left( z_0 \frac{\partial}{\partial z_0}  \right)^3 + \varepsilon_1 ^2 \left( 3 a_{0,\beta} - \sum_{\alpha=1} ^3 a_{0,\alpha} - \varepsilon_2 \frac{z_1}{z_1-z_0} + \sum_{i=1} ^2 \frac{z_i}{z_i -z_0} \mathcal{A}_i ^{(1)} \right) \left( z_0 \frac{\partial}{\partial z_0} \right)^2 \right.  \\
& - \varepsilon_1 \left( \prod_{\alpha \neq \beta} (a_{0,\beta} -a_{0,\alpha})  + \frac{z_1 z_2}{(z_1 -z_0)(z_2 -z_0)} \mathcal{A}_2 ^{(1)} (\mathcal{A}_1 ^{(1)} -\varepsilon) - \varepsilon_2 \varepsilon \left( 2 a_{0,\beta} - \varepsilon -\varepsilon_1  \right) \frac{z_1}{z_1-z_0} \right.  \\
& \quad \quad \quad\left. + \sum_{i=1} ^2 \frac{z_i}{z_i -z_0} \left( \left(2a_{0,\beta}-\sum_{\alpha=1} ^3 a_{0,\alpha} \right)\mathcal{A}_i ^{(1)} -\varepsilon_1 \varepsilon_2 z_i \frac{\partial}{\partial z_i} + \frac{1}{2} \left(\mathcal{A}_i ^{(2)} + \left(\mathcal{A}_i ^{(1)}\right)^2\right) \right) \right) \left( z_0 \frac{\partial}{\partial z_0} \right) \\
& + \sum_{i=1} ^2 \frac{z_i}{z_i -z_0} \left( \frac{1}{6}(\mathcal{A}_i ^{(1)})^3 + \frac{1}{3} \mathcal{A}_i ^{(3)} + \frac{1}{2} \mathcal{A}_i ^{(1)} \mathcal{A}_i ^{(2)} -\frac{1}{2} \left(\mathcal{A}_i ^{(2)} +\left(\mathcal{A}_i ^{(1)}\right)^2\right)\left(\sum_{\alpha=1} ^3a_{0,\alpha} -a_{0,\beta}\right) \right. \\
& \quad \quad \quad \quad \left. + \varepsilon_1 \varepsilon_2 \left(\sum_{\alpha=1} ^3 a_{0,\alpha} -a_{0,\beta} -\mathcal{A}_i ^{(1)}\right) \left( z_i \frac{\partial}{\partial z_i} \right) + \prod_{\alpha\neq\beta} a_{0,\alpha} \; \mathcal{A}_i ^{(1)}  - \varepsilon_1 \varepsilon_2 \varepsilon \; z_2 \partial_2 \right)  \\
& - 2 \varepsilon_1 \varepsilon_2  \frac{z_0 (z_1-z_2)}{(z_0-z_1)(z_0-z_2)} \Bigg\langle \sum_{\square \in \boldsymbol{\lambda}^{(2)}} c_{\square}  \Bigg\rangle_{A_2} \\
&+  \frac{z_1 z_2}{(z_1 -z_0)(z_2 -z_0)} \left( \frac{1}{2} \left(\mathcal{A}_1 ^{(1)} - 2\varepsilon\right)\left(\mathcal{A}_2 ^{(2)} + \left(\mathcal{A}_2 ^{(1)}\right)^2\right) + \frac{1}{2}\mathcal{A}_2 ^{(1)} \left( \mathcal{A}_1 ^{(2)} + \left(\mathcal{A}_1 ^{(1)}\right)^2\right) \right. \\
&\quad\quad \quad \quad\quad \left. \left. -\mathcal{A}_2 ^{(1)} \left(\mathcal{A}_1 ^{(1)} -\varepsilon\right)\left(\sum_{\alpha=1} ^3 a_{0,\alpha} -a_{0,\beta} +\varepsilon\right) -\varepsilon_1 \varepsilon_2 \left( \mathcal{A}_2 ^{(1)} z_1 \frac{\partial}{\partial z_1} +(\mathcal{A}_1 ^{(1)} -2\varepsilon) z_2 \frac{\partial}{\partial z_2} \right) \right) \right]\EuScript{Z}_\beta ^L,
\end{split}
\end{align}
where we have used \eqref{vara}. We modify the partition function by multiplying the prefactors,
\begin{align}\label{modify3}
\widetilde{\EuScript{Z}}_3 \equiv \prod_{i=0} ^2 z_i ^{L_i} \prod_{0 \leq i <j \leq 2} \left( 1-\frac{z_j}{z_i} \right)^{T_{ij}} \EuScript{Z}_3,
\end{align}
where
\begin{align} \label{prefactors3}
\begin{split}
&L_i \equiv \frac{(a_{i+1,1}-a_{i+1,2})^2 +(a_{i+1,1}-a_{i+1,3})^2 -(a_{i+1,1}-a_{i+1,2})(a_{i+1,1}-a_{i+1,3})}{3\varepsilon_1 \varepsilon_2} \\
&\quad\quad - \frac{(a_{i,1}-a_{i,2})^2 +(a_{i,1}-a_{i,3})^2 -(a_{i,1}-a_{i,2})(a_{i,1}-a_{i,3})}{3\varepsilon_1 \varepsilon_2}  \\
&\quad\quad + \frac{3(\bar{a}_i-\bar{a}_{i+1}+\varepsilon)(\bar{a}_i-\bar{a}_{i+1})}{\varepsilon_1\varepsilon_2}, \quad\quad\quad\quad\quad\quad i=0,1,2, \\
&T_{ij} \equiv \frac{3(\bar{a}_j -\bar{a}_{j+1} +\varepsilon)(\bar{a}_i -\bar{a}_{i+1})}{\varepsilon_1 \varepsilon_2}, \quad i,j=0,1,2.
\end{split}
\end{align}
With the constraints \eqref{a2const}, the prefactors simplify. We also set $z_0=z$, $z_1=1$, and $z_2=\mathfrak{q}$. Then the prefactor for $\EuScript{Z}_\beta ^L$ becomes
\begin{align}
\begin{split}
&\left(-\frac{1}{z} \right)^{-r_{L,\beta}} \; \mathfrak{q}^{-\Delta_{\mathfrak{q}} -\Delta_0 + \frac{1}{\varepsilon_1 \varepsilon_2} \left( \varepsilon^2 -\frac{(a_{2,1}-a_{2,2})^2 +(a_{2,1}-a_{2,3})^2 -(a_{2,1}-a_{2,2})(a_{2,1}-a_{2,3})}{3}  \right)  } \\
& \quad \left( 1-\frac{1}{z} \right)^{ \frac{3\bar{a}_0 -3\bar{a}_2 + 3\varepsilon-\varepsilon_2 }{3\varepsilon_1}} \; \left( 1- \frac{\mathfrak{q}}{z} \right)^{ \frac{\bar{a}_2 -\bar{a}_3 +\varepsilon}{\varepsilon_1} } \; \left( 1-\mathfrak{q} \right)^{ \frac{(\bar{a}_2 -\bar{a}_3+\varepsilon)(3\bar{a}_ -3\bar{a}_2 -\varepsilon_2) }{\varepsilon_1 \varepsilon_2} },
\end{split}
\end{align}
where the exponents are
\begin{align} \label{critexp3inf}
\begin{split}
&\left( r_{L,\beta} \right)_{\beta=1} ^3  \equiv  \left( \frac{-3a_{0,\beta} +\sum_{\gamma=1} ^3 a_{0,\gamma} + 3\varepsilon_1 + 5\varepsilon_2}{3\varepsilon_1}\right)_{\beta=1} ^3,
\end{split}
\end{align}
and
\begin{align} \label{delta3}
\begin{split}
&\Delta_0 \equiv \frac{1}{\varepsilon_1 \varepsilon_2} \left( \varepsilon^2 - \frac{(a_{3,1}-a_{3,2})^2 +(a_{3,1}-a_{3,3})^2 -(a_{3,1}-a_{3,2})(a_{3,1}-a_{3,3})}{3} \right) \\
&\Delta_{\mathfrak{q}} \equiv - \frac{3(\bar{a}_2 -\bar{a}_3)(\bar{a}_2 -\bar{a}_3+\varepsilon)}{\varepsilon_1 \varepsilon_2} \\ 
&\Delta_{1} \equiv - \frac{(3\bar{a}_0 -3\bar{a}_2 -\varepsilon_2)(3\bar{a}_0 -3\bar{a}_2 +3\varepsilon -\varepsilon_2)}{3\varepsilon_1 \varepsilon_2}\\
&\Delta_{\infty} \equiv \frac{1}{\varepsilon_1 \varepsilon_2} \left( \varepsilon^2 -\frac{(a_{0,1}-a_{0,2})^2 +(a_{0,1}-a_{0,3})^2 -(a_{0,1}-a_{0,2})(a_{0,1}-a_{0,3})}{3} \right).
\end{split}
\end{align}
It is also convenient to define the quantities
\begin{align} \label{lambda3}
\begin{split}
&\Lambda_0 \equiv \frac{(2a_{3,1}-a_{3,2}-a_{3,3}) (-a_{3,1}+2a_{3,2}-a_{3,3})(-a_{3,1}-a_{3,2}+2a_{3,3})}{27\varepsilon_1 ^3} \\
&\Lambda_{\mathfrak{q}} \equiv \frac{(\bar{a}_2 -\bar{a}_3)(\bar{a}_2-\bar{a}_3 +\varepsilon)(2\bar{a}_2 -2\bar{a}_3 +\varepsilon) }{\varepsilon_1 ^3}\\
&\Lambda_1 \equiv \frac{1}{\varepsilon_1 ^3} \left(\bar{a}_0 -\bar{a}_2 - \frac{\varepsilon_2}{3}\right)\left(\bar{a}_0-\bar{a}_2 +\varepsilon - \frac{\varepsilon_2}{3}\right)\left(2\bar{a}_0 -2\bar{a}_2 +\varepsilon -\frac{2\varepsilon_2}{3} \right) \\
&\Lambda_\infty \equiv \frac{(2a_{0,1}-a_{0,2}-a_{0,3}) (-a_{0,1}+2a_{0,2}-a_{0,3})(-a_{0,1}-a_{0,2}+2a_{0,3})}{27 \varepsilon_1 ^3} .
\end{split}
\end{align}
Then under the modification, the differential equation \eqref{3rdorder} defines an operator $\widehat{\widehat{\mathfrak{D}}}_3$ annihilating the partition function $\widetilde{\EuScript{Z}}_\beta ^L$,
\begin{align} \label{d3}
\begin{split}
0&= \left[ \varepsilon_1 ^3 \partial^3  - \varepsilon_1 ^2 \varepsilon_2 \frac{5z -3}{z(z-1)}  \partial^2 + \varepsilon_1  \widehat{t}_2 (z, \mathfrak{q}) \partial +  \widehat{t}_3 (z, \mathfrak{q})  \right] \widetilde{\EuScript{Z}}_\beta ^L,
\end{split}
\end{align}
where we have defined the meromorphic operators,
\begin{subequations}
\begin{align}
&\widehat{t}_2 (z,\mathfrak{q}) \equiv  \varepsilon_1 \varepsilon_2 \left( \frac{ \Delta_0}{z^2} + \frac{\Delta_{\mathfrak{q}}}{(z-\mathfrak{q})^2} + \frac{\Delta_1}{(z-1)^2} +\frac{\Delta_\infty -\Delta_1 -\Delta_{\mathfrak{q}} -\Delta_0 + \frac{3\varepsilon+\varepsilon_2}{3\varepsilon_1} }{z(z-1)} +\frac{\widehat{H}_1}{z(z-\mathfrak{q})(z-1)}  \right) \nonumber \\
&\quad\quad\quad\quad + \varepsilon_2 \left( \frac{3\varepsilon_1 +2\varepsilon_2}{z^2} -\frac{ 2( \bar{a}_2-\bar{a}_3 +\varepsilon)}{(z-\mathfrak{q})^2} +\frac{\bar{a}_0 -\bar{a}_2 +\varepsilon -\frac{\varepsilon_2}{3}}{(z-1)^2} \right. \nonumber \\
& \quad\quad\quad\quad\quad\quad\quad \left. -\frac{3\bar{a}_0 - 9\bar{a}_2 +6\bar{a}_3 - 16\varepsilon_2}{3z(z-1)} - \frac{2(1-\mathfrak{q})(\bar{a}_2-\bar{a}_3+\varepsilon)}{z(z-\mathfrak{q})(z-1)} \right) \\
&\widehat{t}_3 (z, \mathfrak{q}) \equiv  \frac{\varepsilon_1 ^3 \Lambda_0}{z^3} +  \frac{\varepsilon_1 ^3 \Lambda_{\mathfrak{q}}}{(z-\mathfrak{q})^3}  +  \frac{\varepsilon_1 ^3 \Lambda_1}{(z-1)^3} +  \frac{\varepsilon_1 ^3 ( \Lambda_\infty -\Lambda_0 -\Lambda_{\mathfrak{q}} -\Lambda_1 - \frac{\varepsilon_2 (3\varepsilon+\varepsilon_2)(3\varepsilon +2\varepsilon_2)}{27 \varepsilon_1^3} ) }{z(z-\mathfrak{q})(z-1)}  \\
&\quad\quad\quad+ \frac{(1-\mathfrak{q})(6\bar{a}_0 -6\bar{a}_2 +3\varepsilon_1 +\varepsilon_2)}{6z(z-\mathfrak{q})(z-1)^2} \varepsilon_1 \varepsilon_2 \left( -\Delta_\infty +\Delta_0 + \Delta_{\mathfrak{q}} + \Delta_1 -\frac{3\varepsilon + \varepsilon_2}{3\varepsilon_1} \right)  \nonumber \\
&\quad\quad\quad - \frac{1}{2z (z-\mathfrak{q})(z-1)} \varepsilon_1 \varepsilon_2 \left( \frac{2\bar{a}_2 -2\bar{a}_3 +\varepsilon}{z-\mathfrak{q}} +\frac{ 6\bar{a}_0 -6\bar{a}_2 +3\varepsilon_1 +\varepsilon_2 }{3(z-1)} \right) \widehat{H}_1 \nonumber \\
& \quad\quad\quad +\frac{\widehat{H}_2}{z^2 (z-\mathfrak{q})(z-1)} + \frac{\varepsilon_1}{2} \partial \left( \widehat{t}_2 (z,\mathfrak{q}) \right) + \varepsilon_2 \left( \cdots \right). \nonumber
\end{align}
\end{subequations}
We have omitted the last term in $\widehat{t}_3 (z,\mathfrak{q})$ which is rather lengthy but is constant and subleading in $\varepsilon_2$. This term decouples in the limit $\varepsilon_2 \to 0$. Also, we have defined
\begin{subequations}
\begin{align}
&\widehat{H}_1 \equiv -\mathfrak{q}(1-\mathfrak{q}) \frac{\partial}{\partial \mathfrak{q}} \\
&\widehat{H}_2 \equiv - (1-\mathfrak{q}) \left( \frac{1}{3} \Big\langle \EuScript{O}_3 \Big\rangle_{A_2} +  \varepsilon_1 \varepsilon_2 \left( 3\bar{a}_2 -\bar{a}_3 +\frac{3\varepsilon_1 + 2\varepsilon_2}{2} \right) \mathfrak{q} \frac{\partial}{\partial \mathfrak{q}} + \cdots \right) \\
&  \quad\quad + \mathfrak{q} \left( \varepsilon_1 \varepsilon_2 (3\bar{a}_0 -6\bar{a}_2 +3\bar{a}_3 -2\varepsilon -\varepsilon_2) \mathfrak{q} \frac{\partial}{\partial \mathfrak{q}} + \cdots  \right) \nonumber \\
& \quad \quad +  \frac{\mathfrak{q}^2}{1-\mathfrak{q}} (3\bar{a}_0 -6\bar{a}_2 +3\bar{a}_3-2\varepsilon -\varepsilon_2)(3\bar{a}_0-3\bar{a}_2-\varepsilon_2)(\bar{a}_2-\bar{a}_3 +\varepsilon) \nonumber.
\end{align}
\end{subequations}
It is not very instructive to write down the full lengthy expression of $\widehat{H}_2$ here, but we emphasize that it is fully expressed in gauge theoretical terms. In particular, it includes the expectation value of the chiral observable
\begin{align}
\begin{split}
\EuScript{O}_3 &= \text{Tr} \phi_2 ^3 = \sum_{\alpha=1 } ^3 a_{2,\alpha} ^3 - 3\varepsilon_1 \varepsilon_2 \varepsilon k_2 - 6 \varepsilon_1 \varepsilon_2 \sum_{\square \in \boldsymbol{\lambda}^{(2)}} c_\square
\end{split}
\end{align}
of the $A_2$-theory. We present the full expression for $\widehat{H}_2$ in the appendix \ref{H2}.

{}In the NS limit, the partition function exhibits the asymptotics:
\begin{align}
\widetilde{\EuScript{Z}}_\beta ^L ( \mathbf{a}_2 \equiv \mathbf{a} ,z, \mathfrak{q}) =  e^ {\frac{\widetilde{\EuScript{W}}( \mathbf{a} ,\mathfrak{q})}{\varepsilon_2}} \left( \chi_\beta (\mathbf{a},z,\mathfrak{q}) + \mathcal{O}(\varepsilon_2 ) \right),
\end{align}
where $\widetilde{\EuScript{W}}$ is a part of the effective twisted superpotential of the underlying $A_1$-gauge theory,
\begin{align}
\begin{split}
\widetilde{\EuScript{W}} &\equiv \lim_{\varepsilon_2 \to 0} \varepsilon_2  \log \widetilde{\EuScript{Z}}_\beta ^L \\
&= \widetilde{\EuScript{W}}^{\text{classical}} + \widetilde{\EuScript{W}}^{\text{inst}} + \widetilde{\EuScript{W}}^{\text{extra}}.
\end{split}
\end{align}
Each piece is given as
\begin{subequations} \label{efftwpoten3}
\begin{align}
&\widetilde{\EuScript{W}}^{\text{classical}} \equiv -\frac{(a_1-a_2)^2 +(a_1-a_3)^2 -(a_1-a_2)(a_1-a_3)}{3\varepsilon_1 } \log \mathfrak{q} \\
&\widetilde{\EuScript{W}}^{\text{inst}} \equiv \lim_{\varepsilon_2 \to 0} \varepsilon_2 \log \EuScript{Z}_{A_1} ^{\text{inst}} \\
&\widetilde{\EuScript{W}}^{\text{extra}} \equiv \varepsilon_1 \left( 1 -\delta_{\mathfrak{q}} -\delta_0 \right) \log\mathfrak{q} +\frac{3(\bar{a} -\bar{a}_3+\varepsilon)(\bar{a}_0 -\bar{a}) }{\varepsilon_1 }  \log(1-\mathfrak{q}),
\end{align}
\end{subequations}
where $\widetilde{\EuScript{W}}^{\text{inst}}$ is the is fully determined by the Young diagram expansions. Also we have defined the limit,
\begin{align}
\varepsilon_2 \Delta_i \xrightarrow{\varepsilon_2 \to 0} \varepsilon_1 \delta_i, \quad i= 0, \mathfrak{q}, 1, \infty.
\end{align}
It is convenient to define also
\begin{align}
\Lambda_i \xrightarrow{\varepsilon_2 \to 0}  \lambda_i, \quad i= 0, \mathfrak{q}, 1, \infty.
\end{align}
It is clear that $\delta_i$ and $\lambda_i$ are written in gauge theoretical terms by their definitions. Now, the equation \eqref{d3} for the operator $\widehat{\widehat{\mathfrak{D}}}_3$ becomes
\begin{align} \label{heun3}
0 = \left[ \partial^3 + t_2 (z) \partial + t_3 (z) \right] \chi_\beta,
\end{align}
where the meromorphic functions $t_i (z)$ are obtained by taking the limit to the meromorphic operators $\widehat{t}_i (z,\mathfrak{q})$,
\begin{subequations}
\begin{align}
&t_2 (z) \equiv \frac{\delta_0}{z^2} + \frac{\delta_{\mathfrak{q}}}{(z-\mathfrak{q})^2} + \frac{\delta_1}{(z-1)^2} + \frac{\delta_\infty -\delta_1 -\delta_{\mathfrak{q}} -\delta_0}{z(z-1)} + \frac{H_1}{z(z-\mathfrak{q})(z-1)}, \\
&t_3(z)\equiv \frac{\lambda_0}{z^3} + \frac{\lambda_{\mathfrak{q}}}{(z-\mathfrak{q})^3} + \frac{\lambda_1}{(z-1)^3} + \frac{\lambda_\infty -\lambda_0 -\lambda_{\mathfrak{q}} -\lambda_1}{z(z-\mathfrak{q})(z-1)} \\
& \quad - \frac{H_1}{2 z (z-\mathfrak{q})(z-1)} \frac{1}{\varepsilon_1 } \left( \frac{2\bar{a} -2\bar{a}_3 +\varepsilon_1}{z-\mathfrak{q}} + \frac{2\bar{a}_0 -2\bar{a} +\varepsilon_1}{z-1} \right)  + \frac{H_2}{z^2 (z-\mathfrak{q})(z-1)} \nonumber \\
&\quad + \frac{(1-\mathfrak{q}) (2\bar{a}_0-2\bar{a}_2 +\varepsilon_1)}{2z (z-\mathfrak{q})(z-1)} \frac{1}{\varepsilon_1} \left( -\delta_\infty +\delta_0 + \delta_{\mathfrak{q}} + \delta_1 \right) +\frac{1}{2} t_2 ' (z).  \nonumber 
\end{align}
\end{subequations}
This is exactly the equation for the $SL(3)$-oper $\widehat{\mathfrak{D}}_3$ on the four-punctured sphere $\mathbb{P}^1 \backslash \{ 0, \underline{\mathfrak{q}} , \underline{1}, \infty \}$.\footnote{The equation \eqref{heun3} matches exactly the one for the
generalized Heun oper in \cite{hol}, where it is derived from the constraints for the minimal punctures.} In particular, the monodromies of $\widehat{\mathfrak{D}}_3$ around the punctures exhibit the desired semi-simplicity and degeneracy of the eigenvalues, as verified by the analytic properties of the solutions $\chi$ obtained from the surface defect partition functions (see section \ref{mono}). The variety $\mathcal{O}_3 [\mathbb{P}^1 \backslash \{ 0, \underline{\mathfrak{q}} , \underline{1}, \infty\}]$ of such opers is parametrized by the accessory parameters,
\begin{subequations} \label{accessory3exp}
\begin{align}
&H_1 \equiv -\mathfrak{q}(1-\mathfrak{q}) \frac{1}{\varepsilon_1} \frac{\partial \widetilde{\EuScript{W}}}{\partial \mathfrak{q}}, \\
& \quad =  (1-\mathfrak{q}) \left( \frac{1}{2\varepsilon_1 ^2} \lim_{\varepsilon_2 \to 0} \Big\langle \EuScript{O}_2 \Big\rangle_{A_1} - 1 +\delta_{\mathfrak{q}} +\delta_0 \right) + \frac{3(\bar{a}-\bar{a}_3+\varepsilon_1)(\bar{a}_0-\bar{a})}{\varepsilon_1 ^2} \mathfrak{q} \nonumber \\
&H_2 \equiv -(1-\mathfrak{q}) \left( \frac{1}{3\varepsilon_1 ^3} \lim_{\varepsilon_2 \to 0} \Big\langle \EuScript{O}_3 \Big\rangle_{A_1} + \frac{1}{\varepsilon_1^2} \left( 3\bar{a} -\bar{a}_3 +\frac{3\varepsilon_1}{2} \right) \mathfrak{q}\frac{\partial \widetilde{\EuScript{W}}}{\partial \mathfrak{q}} + \cdots \right) \label{eq:h2exp} \\
&  \quad\quad + \mathfrak{q} \left( \frac{1}{\varepsilon_1 ^2}  (3\bar{a}_0 -6\bar{a} +3\bar{a}_3 -2\varepsilon_1 ) \mathfrak{q} \frac{\partial \widetilde{\EuScript{W}}}{\partial \mathfrak{q}} + \cdots  \right) \nonumber \\
& \quad \quad +  \frac{\mathfrak{q}^2}{1-\mathfrak{q}} \frac{3(\bar{a}_0-\bar{a})(\bar{a}-\bar{a}_3 +\varepsilon_1) (3\bar{a}_0 -6\bar{a} +3\bar{a}_3-2\varepsilon_1)}{\varepsilon_1 ^3} \nonumber.
\end{align}
\end{subequations}
We present the full expression for $H_2$ in appendix \ref{H2}. Notice that the accessory parameters are expanded as series in $\mathfrak{q}$ whose coefficients are completely determined in gauge theoretical terms. In particular, the series begin with 
\begin{align} \label{accessory3}
\begin{split}
&H_1 =  \frac{(a_1-a_2)^2 +(a_1-a_3)^2 -(a_1-a_2)(a_1-a_3)}{3\varepsilon_1 ^2} -1 + \delta_{\mathfrak{q}} + \delta_0 + \mathcal{O}(\mathfrak{q}) \\
&H_2 = \lambda_0  - \frac{\lambda_{\mathfrak{q}}}{2} -\frac{(2a_1-a_2-a_3)(-a_1+2a_2-a_3)(-a_1-a_2+2a_3)}{27 \varepsilon_1 ^3} \\
& \quad +\frac{2\bar{a}-2\bar{a}_3+\varepsilon_1}{2\varepsilon_1} \left(\delta_0-  1+\frac{(a_1-a_2)^2+(a_1-a_3)^2-(a_1-a_2)(a_1-a_3)}{3\varepsilon_1 ^2} \right)  + \mathcal{O}(\mathfrak{q}).
\end{split}
\end{align}
Thus holomorphic coordinates on the variety $\mathcal{O}_3 [\mathbb{P}^1 \backslash \{ 0, \underline{\mathfrak{q}} , \underline{1}, \infty \}]$ of opers are given by the expectation values of the chiral observables in the $A_1$-theory, $\EuScript{O}_2$ and $\EuScript{O}_3$ \footnote{Here, we are using the fact that \begin{align} \lim_{\varepsilon_2 \to 0} \Big\langle \EuScript{O}_3 \Big\rangle_{A_2} = \lim_{\varepsilon_2 \to 0} \frac{ \Big\langle \mathcal{I}_\beta ^L \: \EuScript{O}_3 \Big\rangle_{A_1}}{\Big\langle \mathcal{I}_\beta ^L \Big\rangle_{A_1}} = \lim_{\varepsilon_2 \to 0} \Big\langle \EuScript{O}_3 \Big\rangle_{A_1}  \nonumber, \end{align} since the expectation value is dominated by the limit shape when $\varepsilon_2 \to 0$.}, in the limit $\varepsilon_2 \to 0$. Hence we observe that the variety $\mathcal{O}_3 [\mathbb{P}^1 \backslash \{ 0, \underline{\mathfrak{q}} , \underline{1}, \infty \}]$ of opers gives a quantization of the Coulomb moduli space of $\mathcal{T}[A_2 , \mathbb{P}^1 \backslash \{ 0, \underline{\mathfrak{q}} , \underline{1}, \infty \}]$. The Bethe/gauge correspondence identifies the NS limits of the expectation values of $\EuScript{O}_2$ and $\EuScript{O}_3$ with the off-shell spectra of the Hamiltonians of the quantum Hitchin system on $\mathbb{P}^1 \backslash \{ 0, \underline{\mathfrak{q}} , \underline{1}, \infty \}$. Thus, the relations \eqref{accessory3exp} establish the connection between the holomorphic functions on the variety $\mathcal{O}_3 [\mathbb{P}^1 \backslash \{ 0, \underline{\mathfrak{q}} , \underline{1}, \infty \}]$ of opers and the off-shell spectra of the quantum Hitchin Hamiltonians.

\paragraph{Remarks}
\begin{itemize}
\item The gauge theoretical derivation of the series expansions \eqref{accessory3exp} for the accessory parameters guarantee their validity to all orders in the gauge coupling $\mathfrak{q}$. It would be nice to mimic the procedure in \cite{pm1, pm2} and check the series expansions by directly computing the monodromy of the oper $\widehat{\mathfrak{D}}_3$ \eqref{heun3} along the $A$-cycle on $\mathbb{P}^1 \backslash \{ 0, \underline{\mathfrak{q}} , \underline{1}, \infty \}$ (see Figure \ref{fig2}).

\item From the point of view of the AGT correspondence \cite{agt}, the expectation value of the higher chiral observable $\EuScript{O}_3$ corresponds to the conformal block with a $\mathcal{W}$-descendant (we briefly mention this issue in section \ref{dis}). It is not very obvious how we should relate the semi-classical conformal block with a $\mathcal{W}$-descendant to the off-shell spectrum of the higher quantum Hitchin Hamiltonian. In the gauge theoretical perspective, the Bethe/gauge correspondence immediately establishes the relation between the expectation value of $\EuScript{O}_3$ and the off-shell spectrum of the higher quantum Hitchin Hamiltonian. Thus, the relation between the accessory parameter $H_2$ and the off-shell spectrum of the higher quantum Hitchin Hamiltonian is also revealed through \eqref{eq:h2exp}.

\end{itemize}

Similarly, we can start by imposing other constraints, e.g. \eqref{constdiff} on the $A_2$-theory. Hence we consider the partition function $\EuScript{Z}_\beta ^R$ \eqref{zr}. Again, we modify the partition function as \eqref{modify3} with the prefactors \eqref{prefactors3}, yet this time under the constraint \eqref{constdiff} and the re-definition \eqref{redef}. The final form of the prefactor is
\begin{align} \label{prefactor3diff}
\begin{split}
&\left( - \frac{\mathfrak{q}}{z} \right)^{-r_{R,\beta}} \mathfrak{q} ^{\frac{1}{\varepsilon_1 \varepsilon_2}\left( \varepsilon^2 - \frac{(a_{1,1}-a_{1,2})^2 +(a_{1,1}-a_{1,3})^2 -(a_{1,1}-a_{1,2})(a_{1,1}-a_{1,3})}{3} \right) -\Delta_{\mathfrak{q}} ' -\Delta_0 +\frac{3\varepsilon +\varepsilon_2}{3\varepsilon_1}  } \\
&\quad (1-\mathfrak{q})^{\frac{(\bar{a}_0 -\bar{a}_1 +\varepsilon)(3\bar{a}_1 -3\bar{a}_3 -\varepsilon_2)}{\varepsilon_1 \varepsilon_2}} (1-z) ^{\frac{\bar{a}_0 -\bar{a}_1 + \varepsilon}{\varepsilon_1} }  \left(1-\frac{z}{\mathfrak{q}}\right)^{\frac{3\bar{a}_1 -3\bar{a}_3 +3\varepsilon -\varepsilon_2}{3\varepsilon_1}},
\end{split}
\end{align}
where we have defined
\begin{align}\label{critexp3zero}
\begin{split}
\left( r_{R,\beta} \right)_{\beta=1} ^3 \equiv \left( \frac{- 3a_{3,\beta} + \sum_{\gamma=1} ^3 a_{3,\gamma} +3\varepsilon}{3\varepsilon_1} \right)_{\beta=1} ^3
\end{split}
\end{align}
and
\begin{align} \label{delta3'}
\begin{split}
&\Delta_{\mathfrak{q}} ' \equiv -\frac{(3\bar{a}_1 -3\bar{a}_3 -\varepsilon_2)(3\bar{a}_1 -3\bar{a}_3 +3\varepsilon -\varepsilon_2)}{3\varepsilon_1 \varepsilon_2} \\
&\Delta_1 ' \equiv - \frac{(\bar{a}_0 -\bar{a}_1)(\bar{a}_0-\bar{a}_1 +\varepsilon)}{\varepsilon_1 \varepsilon_2}.
\end{split}
\end{align}
Let us also define
\begin{align} \label{lambda3'}
\begin{split}
&\Lambda_{\mathfrak{q}} ' \equiv \frac{(3\bar{a}_1-3\bar{a}_3 +3\varepsilon -\varepsilon_2)(3\bar{a}_1 -3\bar{a}_3 -\varepsilon_2)(6\bar{a}_1-6\bar{a}_3 -3\varepsilon +2\varepsilon_2)}{27 \varepsilon_1 ^3} \\
&\Lambda_1 ' \equiv \frac{(\bar{a}_0 -\bar{a}_1)(\bar{a}_0 -\bar{a}_1 +\varepsilon)(2\bar{a}_0 -2\bar{a}_1 +\varepsilon)}{\varepsilon_1 ^3}.
\end{split}
\end{align}
Then the modified partition function $\widetilde{\EuScript{Z}}_\beta ^R$ satisfies the equation of the form \eqref{d3}, after substituting $\Delta_{\mathfrak{q}, 1} \to \Delta_{\mathfrak{q},1} '$ and $\Lambda_{\mathfrak{q},1} \to \Lambda_{\mathfrak{q},1} '$.

\subsection{The $(N-1,1)$-type $\mathbb{Z}_2$-orbifold} \label{z2orbifold}
We construct the surface defect on the $A_1$-theory by placing it on $\mathbb{Z}_p$-orbifold. Due to the orbifolding, the bulk $\EuScript{Y}$-observable fractionalizes into $p$ observables,
\begin{align}
\EuScript{Y}_\omega (x)[\boldsymbol{\lambda}] = \prod_{\alpha \in c^{-1} (\omega)} (x - a_\alpha ) \prod_{\Box \in K_\omega} \frac{x-c_{\Box} -\varepsilon_1}{x-c_{\Box}} \prod_{\Box \in K_{\omega-1} } \frac{x-c_{\Box} - \varepsilon_2}{x-c_{\Box}-\varepsilon} . \label{refy}
\end{align}
The fundamental refined $qq$-characters are given by \cite{nek4}
\begin{align} \label{refqq}
\EuScript{X}_\omega (x) = \EuScript{Y}_{\omega+1} (x+\varepsilon) + \mathfrak{q}_\omega \frac{P_\omega (x)}{\EuScript{Y}_\omega (x)}.
\end{align}
It is often possible to derive a useful equation for the partition function for specific $p$ and the coloring function $c$ from the non-perturbative Dyson-Schwinger equations of \eqref{refqq}. We now describe how this is be done for the $(N-1,1)$-type $\mathbb{Z}_2$-orbifold. The details of the computation for the non-regular parts of $\EuScript{X}_\omega$ is given in the appendix \ref{appC}. Below we focus on the results.

\subsubsection{$N=2$}
For $N=2$, we consider $(1,1)$-type $\mathbb{Z}_2$-orbifold. This case is special since the coloring function is one-to-one. Let us define
\begin{align}
c^{-1} (1)= \beta, \quad c^{-1}(0) =\bar{\beta},
\end{align}
without any loss of generality. Each of the non-perturbative Dyson-Schwinger equations
\begin{align}
[x^{-1}] \Big\langle \EuScript{X}_0 (x) \Big\rangle = [x^{-1}] \Big\langle \EuScript{X}_1 (x) \Big\rangle =0 
\end{align}
involves the unwanted term
\begin{align}
\Bigg\langle \sum_{\square \in K_0 }c_{\square}  -\sum_{\square \in K_1} c_{\square}  \Bigg\rangle,
\end{align}
but they can be combined to cancel this term and to yield the following closed equation
\begin{align} \label{npdsz2}
\begin{split}
0 =& \left[ \varepsilon_1 ^2 (z \partial )^2 - \varepsilon_1 \left\{-2 \widetilde{a}_{\bar{\beta}} + \sum_{\alpha=1,2} \widetilde{m}_{+, \alpha} -\frac{\mathfrak{q}}{z-\mathfrak{q}} \sum_{\alpha=1,2} (\widetilde{a}_{\alpha} - \widetilde{m}_{-,\alpha}) + \frac{1}{1-z} \sum_{\alpha=1,2} (\widetilde{a}_\alpha -\widetilde{m}_{+,\alpha})  \right\} (z \partial) \right. \\
&+ \varepsilon_1 \widetilde{\varepsilon}_2 \frac{z (1-\mathfrak{q})}{(1-z)(z-\mathfrak{q})} \mathfrak{q} \frac{\partial}{\partial \mathfrak{q}} + \frac{1}{2} \left( \widetilde{a}_{\bar{\beta}} -\sum_{\alpha=1,2} \widetilde{m}_{+,\alpha}  \right)^2 + \frac{1}{2}  \widetilde{a}_{\bar{\beta}} ^2 -\frac{1}{2}\sum_{\alpha=1,2} \widetilde{m}_{+,\alpha}  ^2 \\
& - \frac{1}{2(1-z)} \left[  \left( \widetilde{a}_{\bar{\beta}} -\sum_{\alpha=1,2} \widetilde{m}_{+,\alpha} \right)^2 +\widetilde{a}_{\bar{\beta}} ^2 - \sum_{\alpha=1,2} \widetilde{m}_{+,\alpha} ^2  \right] \\
&\left. -\frac{\mathfrak{q}}{2(z-\mathfrak{q})} \left[ \left( \widetilde{a}_{\beta} - \sum_{\alpha=1,2} \widetilde{m}_{-,\alpha} -\frac{\widetilde{\varepsilon}_2}{2} \right)^2 +\left( \widetilde{a}_{\beta} +\frac{\widetilde{\varepsilon}_2}{2} \right)^2 - \sum_{\alpha=1,2} \left( \widetilde{m}_{-,\alpha} +\frac{\widetilde{\varepsilon}_2}{2} \right)^2 \right] \right] \EuScript{Z}^{\mathbb{Z}_2} _\beta,
\end{split}
\end{align}
where we have re-defined the couplings as in \eqref{fraccoupling}, $\mathfrak{q}_0 = -z$ and $\mathfrak{q}_1 = -\frac{\mathfrak{q}}{z}$ (up to the sign which is not very important). Now, let us also re-define the parameters as
\begin{align}
\widetilde{a}_\alpha = a_{2,\alpha}, \quad \widetilde{m}_{+,\alpha} = a_{0,\alpha}, \quad \widetilde{m}_{-,\alpha} = a_{3,\alpha} -\varepsilon_1 -\widetilde{\varepsilon}_2, \quad \alpha=1,2.
\end{align}
Then we decouple multiplicative prefactors
\begin{align}
\begin{split}
\widetilde{\EuScript{Z}}^{\mathbb{Z}_2} _\beta &\equiv -\left(-\frac{1}{z}\right) ^{-r^{\mathbb{Z}_2} _{\beta}} \; \mathfrak{q} ^{\frac{\varepsilon^2 -(a_{2,1}-a_{2,2})^2}{4 \varepsilon_1 \varepsilon_2} - \Delta_0 -\Delta_{\mathfrak{q}}} \\
& \quad (1-z) ^{\frac{\varepsilon (2\bar{a}_0 -2\bar{a}_2 -\varepsilon_2)}{2\varepsilon_1 \varepsilon_2} } (1-\mathfrak{q})^{\frac{(2\bar{a}_0 -2\bar{a}_2-\varepsilon_2) (\bar{a}_2 -\bar{a}_3 + \varepsilon) }{\varepsilon_1 \varepsilon_2}} \left( 1-\frac{\mathfrak{q}}{z} \right)^{\frac{\bar{a}_2 -\bar{a}_3  + \varepsilon}{\varepsilon_1}} \EuScript{Z}^{\mathbb{Z}_2} _\beta,
\end{split}
\end{align}
where
\begin{align} \label{critexpz2}
\left( r^{ \mathbb{Z}_2} _{\beta}\right)_{\beta=1,2} \equiv \left( \frac{-a_{2,1}+a_{2,2}+\varepsilon}{2\varepsilon_1}, \; \frac{a_{2,1}-a_{2,2}+\varepsilon }{2\varepsilon_1} \right),
\end{align}
and the other exponents have been defined in the previous section. The differential equation \eqref{npdsz2} then becomes
\begin{align}
\begin{split}
0=& \left[  \varepsilon_1 ^2 \partial ^2 - \varepsilon_1 \varepsilon_2 \frac{2z-1}{z(z-1)}  \partial + \varepsilon_1 \varepsilon_2 \frac{\mathfrak{q}-1}{z(z-1)(z-\mathfrak{q})} \mathfrak{q} \frac{\partial}{\partial \mathfrak{q}} \right. \\ 
& \left. \quad+ \varepsilon_1 \varepsilon_2 \left( \frac{\Delta_0}{z ^2} + \frac{\Delta_1}{(z-1)^2} + \frac{\Delta_{\mathfrak{q}} }{(z-\mathfrak{q})^2} - \frac{ -\frac{2\varepsilon + \varepsilon_2}{4\varepsilon_1} +\Delta_1  + \Delta_{\mathfrak{q}}  +\Delta_0 - \Delta_{\infty}}{z (z-1)}  \right)  \right] \widetilde{\EuScript{Z}} ^{\mathbb{Z}_2} _\beta,
\end{split}
\end{align}
which is precisely the differential equation \eqref{a2npdsfinal} for $\widehat{\widehat{\mathfrak{D}}}_2$.

\paragraph{Remarks}
\begin{itemize}
\item The convergence domain for the partition function is $0< \vert \mathfrak{q}_0 \vert, \vert \mathfrak{q}_1 \vert <1$. This implies the solutions $\widetilde{\EuScript{Z}}_\beta ^{\mathbb{Z}_2}$ are in yet another \textit{intermediate} domain $0 < \vert \mathfrak{q} \vert < \vert z \vert<1$. 
\end{itemize}

\subsubsection{$N=3$}
For $N=3$, the computation is more involved. First, recall that the $(2,1)$-type $\mathbb{Z}_2$-orbifold surface defect partition function \eqref{orbpartition} is split into the underlying $A_1$-theory part and the surface defect part. The fixed points of the instanton moduli space of the underlying $A_1$-theory are enumerated by the Young diagrams $\boldsymbol{\Lambda}$ \eqref{lambda}, whose weights are encoded in the space $\widetilde{K} =\widetilde{K}_1$ \eqref{spacek}. Thus the observables in the underlying $A_1$-theory descends from the observables in the space $K_1$ of the original theory on the $\mathbb{Z}_2$-orbifold . In particular, we have
\begin{align}
\sum_{\square \in K_1} c_{\square} = \sum_{\square \in \boldsymbol{\Lambda}} \widetilde{c}_{\square} + \frac{1}{2} \widetilde{\varepsilon}_2 k_1,
\end{align}
where
\begin{align}
\widetilde{c}_{\square} \equiv \widetilde{a}_{\alpha} + (i-1)\varepsilon_1 + (j-1)\widetilde{\varepsilon}_2, \quad \text{for} \quad \square_{(i,j)} \in \Lambda^{(\alpha)}.
\end{align}
We will reduce the non-perturbative Dyson-Schwinger equations so that the final equation only involves the expectation value of this observable, since it comprises the chiral observable
\begin{align}
\EuScript{O}_3 [\boldsymbol{\Lambda}] = \sum_{\alpha=1} ^3 \widetilde{a}_\alpha ^3 -3 \varepsilon_1 \widetilde{\varepsilon}_2 (\varepsilon_1 + \widetilde{\varepsilon}_2) k_1 - 6 \varepsilon_1 \widetilde{\varepsilon}_2 \sum_{\square \in \boldsymbol{\Lambda}} \widetilde{c}_{\square},
\end{align}
of the underlying $A_1$-theory. The non-perturbative Dyson-Schwinger equations that we utilize are
\begin{align}
[x^{-1}] \Big\langle \EuScript{X}_1 (x) \Big\rangle = [x^{-1}] \Big\langle \EuScript{X}_0 (x) \Big\rangle =  [x^{-2}] \Big\langle \EuScript{X}_0 (x) \Big\rangle =0.
\end{align}
The second equation can be used to cancel the unwanted terms
\begin{align}
\Bigg\langle \sum_{\square \in K_0 }c_{\square}  \Bigg\rangle, \quad \Bigg\langle (k_0 -k_1) \left( \sum_{\square \in K_0} c_{\square} -\sum_{\square\in K_1} c_{\square} \right) \Bigg\rangle,
\end{align}
while the first and the third equations can be combined to cancel the unwanted term
\begin{align}
\Bigg\langle \sum_{\square \in K_0 }c_{\square} ^2  -\sum_{\square \in K_1} c_{\square} ^2 \Bigg\rangle,
\end{align}
The final equation only involves the partition function itself and the expectation value $\Bigg\langle \sum_{\square \in \boldsymbol{\Lambda
} }\widetilde{c}_{\square}  \Bigg\rangle$:
\footnotesize
\begin{align}
\begin{split}
&0 = \left[ -\varepsilon_1 ^3 (z \partial)^3 +\varepsilon_1 ^2 \left( 3\widetilde{a}_\beta -3\bar{\widetilde{a}} +\frac{3\mathfrak{q}}{z-\mathfrak{q}} \left( -\bar{\widetilde{a}} + {\bar{\widetilde{m}}}_{-} \right) -\frac{6 z}{z-1} \left( {\bar{\widetilde{a}}} -{\bar{\widetilde{m}}}_{+} \right) \right) (z\partial)^2 \right. \\
& -\varepsilon_1 \left\{ -\frac{z}{1-z} \left( \left( \sum_{\bar{\beta}\neq\beta} {\widetilde{a}}_{\bar{\beta}} -\sum_{\alpha=1} ^3  {\widetilde{m}}_{+,\alpha}  \right)^2 + \sum_{\bar{\beta}\neq\beta} {\widetilde{a}}_{\bar{\beta}}^2 -\sum_{\alpha=1} ^3  {\widetilde{m}}_{+,\alpha}^2 \right) -\frac{6\varepsilon_1 \left( \bar{ a } - \bar{\widetilde{m}}_+ \right) z}{(1-z)^2} +\frac{\widetilde{\varepsilon}_2}{2} \left( \varepsilon_1 +\frac{\widetilde{\varepsilon}_2}{2} \right) \right. \\
& + \frac{\prod_{\bar{\beta} \neq \beta} \left( \varepsilon_1 + \frac{\widetilde{\varepsilon}_2}{2} -{\widetilde{a}}_{\bar{\beta}} \right) z}{z-\mathfrak{q}} +\frac{\varepsilon_1 (\varepsilon_1 -{\widetilde{a}}_\beta)}{2 (1-z)} - \frac{ z (\varepsilon_1 +\widetilde{\varepsilon}_2) \left(2\varepsilon_1 +\widetilde{\varepsilon}_2 -\sum_{\bar{\beta}\neq\beta} {\widetilde{a}}_{\bar{\beta}}  \right) }{2(z-\mathfrak{q})}  \\
&-\frac{\mathfrak{q}}{2(z-\mathfrak{q})} \left( \left( {\widetilde{a}}_\beta - 3 \bar{\widetilde{m}}_- -\widetilde{\varepsilon}_2 \right)^2 + (\varepsilon_1 +\widetilde{\varepsilon}_2) ({\widetilde{a}}_\beta - 3 \bar{\widetilde{m}}_- -\widetilde{\varepsilon}_2 ) + \left( {\widetilde{a}}_\beta + \frac{\widetilde{\varepsilon}_2}{2} \right) ^2 -\sum_{\alpha=1} ^3  \left( {\widetilde{m}}_{-,\alpha} + \frac{\widetilde{\varepsilon}_2}{2} \right) ^2 \right) \\
& + \frac{z}{2(1-z)} \left( \left( \sum_{\bar{\beta}\neq\beta} {\widetilde{a}}_{\bar{\beta}} - 3 \bar{\widetilde{m}}_+  \right)^2 + \sum_{\bar{\beta}\neq\beta} {\widetilde{a}}_{\bar{\beta}} ^2 -\sum_{\alpha=1} ^3  {\widetilde{m}}_{+,\alpha} ^2 +\varepsilon_1 \left( \sum_{\bar{\beta}\neq\beta} {\widetilde{a}}_{\bar{\beta}} -\sum_{\alpha=1} ^3  {\widetilde{m}}_{+,\alpha} \right) \right) \\
& \left. -\varepsilon_1 \widetilde{\varepsilon}_2 \frac{z(1-\mathfrak{q})}{(z-1)(z-\mathfrak{q})} \mathfrak{q} \frac{\partial}{\partial \mathfrak{q}} + \left( \sum_{\bar{\beta}\neq\beta} \widetilde{a}_{\bar{\beta}} -3\bar{\widetilde{m}}_+ +\frac{\varepsilon_1}{2} +\frac{3(\bar{\widetilde{a}}-\bar{\widetilde{m}}_+)}{z-1}  \right) \left( 2 \sum_{\bar{\beta}\neq\beta} {\widetilde{a}}_{\bar{\beta}} -3 \bar{\widetilde{m}}_+ +  \frac{3(\bar{\widetilde{a}}-\bar{\widetilde{m}}_+) +\varepsilon_1}{z-1} +\frac{3\mathfrak{q} (\bar{\widetilde{a}}-\bar{\widetilde{m}}_-)}{z-\mathfrak{q}} \right)\right\} (z\partial) \\
& + \left\{ -\frac{2 \varepsilon_1 z}{(1-z)^2} +\frac{1+z}{2(1-z)} \left( -2 \sum_{\bar{\beta} \neq\beta} \widetilde{a}_{\bar{\beta}} +3 \bar{\widetilde{m}}_+ - \frac{3(\bar{\widetilde{a}} -\bar{\widetilde{m}}_+ )+\varepsilon_1 }{z-1} - \frac{3 \mathfrak{q} ( \bar{\widetilde{a}} -\bar{\widetilde{m}}_- )}{z-\mathfrak{q}}  \right) -\varepsilon_1 \frac{z(1-\mathfrak{q})}{(1-z)(z-\mathfrak{q})}  \right. \\
&\left. -\varepsilon_1 - \frac{\widetilde{\varepsilon}_2}{2}  + \frac{\varepsilon_1 - \widetilde{a}_\beta }{2(1-z)} +\frac{z(2\varepsilon_1 +\widetilde{\varepsilon}_2 -\sum_{\bar{\beta}\neq\beta} \widetilde{a}_{\bar{\beta}})}{2(z-\mathfrak{q})} +\frac{z \left( -\sum_{\bar{\beta} \neq \beta} \widetilde{a}_{\bar{\beta}} + 3 \bar{\widetilde{m}}_+ \right)}{2(z-1)} + \frac{(\widetilde{a}_\beta -\widetilde{\varepsilon}_2 -3\bar{\widetilde{m}}_- )\mathfrak{q}}{2(z-\mathfrak{q})} \right\}\varepsilon_1 \widetilde{\varepsilon}_2 \mathfrak{q} \frac{\partial}{\partial \mathfrak{q}}  \\
&   + \frac{z(1-\mathfrak{q})}{(1-z)(z-\mathfrak{q})} \left( 2\varepsilon_1 \widetilde{\varepsilon}_2 \Bigg\langle \sum_{\square \in \boldsymbol{\Lambda}} \widetilde{c}_\square \Bigg\rangle +\varepsilon_1 \widetilde{\varepsilon}_2 (\varepsilon_1 + \widetilde{\varepsilon}_2) \mathfrak{q}\frac{\partial}{\partial \mathfrak{q}} \right) +\frac{z \varepsilon_1}{(1-z)^2} \left( \left( \sum_{\bar{\beta}\neq\beta} \widetilde{a}_{\bar{\beta}} - 3\bar{\widetilde{m}}_+ \right)^2 +\sum_{\bar{\beta}\neq\beta} \widetilde{a}_{\bar{\beta}} ^2 -\sum_{\alpha=1} ^3 \widetilde{m}_{+,\alpha} ^2 \right) \\
& +\frac{z}{2(1-z)} \left( \left( \sum_{\bar{\beta}\neq\beta} \widetilde{a}_{\bar{\beta}} -3\bar{\widetilde{m}}_+ \right)^2 +\sum_{\bar{\beta}\neq\beta} \widetilde{a}_{\bar{\beta}}^2 - \sum_{\alpha=1} ^3 \widetilde{m}_{+,\alpha} ^2 \right) \left( 2\sum_{\bar{\beta}\neq\beta} \widetilde{a}_{\bar{\beta}} -3\bar{\widetilde{m}}_+ +\frac{3 (\bar{\widetilde{a}}-\bar{\widetilde{m}}_+)+\varepsilon_1}{z-1} +\frac{3\mathfrak{q}(\bar{\widetilde{a}} -\bar{\widetilde{m}}_-)}{z-\mathfrak{q}}  \right) \\
&-\frac{\mathfrak{q}}{z-\mathfrak{q}} \left( \frac{\left( \widetilde{a}_{\beta} -3\bar{\widetilde{m}}_- -\widetilde{\varepsilon}_2 \right)^3}{6} +\frac{\left( \widetilde{a}_{\beta} +\frac{\widetilde{\varepsilon}_2}{2} \right) ^3 -\sum_{\alpha=1} ^3 \left( \widetilde{m}_{-,\alpha} +\frac{\widetilde{\varepsilon}_2}{2} \right)^3 }{3} \right. \\
&\left. \quad\quad\quad\quad\quad\quad\quad\quad\quad +\frac{\left( \widetilde{a}_{\beta} -3\bar{\widetilde{m}}_- -\widetilde{\varepsilon}_2 \right) \left( \left( \widetilde{a}_{\beta} +\frac{\widetilde{\varepsilon}_2}{2} \right) ^2 -\sum_{\alpha=1} ^3 \left( \widetilde{m}_{-,\alpha} +\frac{\widetilde{\varepsilon}_2}{2} \right) ^2 \right) }{2} \right) \\
& \left. -\frac{z}{1-z} \left( \frac{\left( \sum_{\bar{\beta}\neq\beta} \widetilde{a}_{\bar{\beta}} -3\bar{\widetilde{m}}_+ \right)^3}{6} +\frac{\sum_{\bar{\beta}\neq\beta} \widetilde{a}_{\bar{\beta}} ^3 -\sum_{\alpha=1} ^3 \widetilde{m}_{+,\alpha} ^3 }{3} +\frac{\left( \sum_{\bar{\beta}\neq\beta} \widetilde{a}_{\bar{\beta}} -3\bar{\widetilde{m}}_+ \right) \left( \sum_{\bar{\beta}\neq\beta} \widetilde{a}_{\bar{\beta}} ^2 -\sum_{\alpha=1} ^3 \widetilde{m}_{+,\alpha} ^2 \right) }{2} \right) \right]\EuScript{Z}_\beta ^{\mathbb{Z}_2},
\end{split}
\end{align}
\normalsize
where we have re-defined the couplings as $\mathfrak{q}_0 = -z$ and $\mathfrak{q}_1 = -\frac{\mathfrak{q}}{z}$. Let us also re-define the other parameters as
\begin{align}
\widetilde{a}_\alpha = a_{2,\alpha}, \quad \widetilde{m}_{+,\alpha} = a_{0,\alpha}, \quad \widetilde{m}_{-,\alpha} = a_{3,\alpha} -\varepsilon_1 -\widetilde{\varepsilon}_2, \quad \alpha=1,2,3.
\end{align}
and modify the partition function by the prefactors,
\begin{align}
\begin{split}
	\widetilde{\EuScript{Z}}_\beta ^{\mathbb{Z}_2} &\equiv -\left(-\frac{1}{z}\right) ^{-r^{\mathbb{Z}_2} _\beta } \mathfrak{q}^{\frac{1}{\varepsilon_1 \varepsilon_2} \left( \varepsilon^2 -\frac{(a_{2,1}-a_{2,2})^2 +(a_{2,1}-a_{2,3})^2-(a_{2,1}-a_{2,2})(a_{2,1}-a_{2,3})}{3} \right) -\Delta_{\mathfrak{q}} -\Delta_0 } \\
& \quad  \left( 1-z \right)^{ - \frac{2(3\bar{a}_0 -3\bar{a}_2-\varepsilon_2)}{3\varepsilon_1}} \;\left( 1-\mathfrak{q} \right) ^{\frac{(3\bar{a}_0 -3\bar{a}_2 -\varepsilon_2)(\bar{a}_2-\bar{a}_3+\varepsilon)}{\varepsilon_1 \varepsilon_2}} \; \left( 1-\frac{\mathfrak{q}}{z} \right) ^{\frac{\bar{a}_2 -\bar{a}_3+\varepsilon}{\varepsilon_1}} \EuScript{Z}_\beta ^{\mathbb{Z}_2}.
\end{split}
\end{align}
Here, we have defined the critical exponent for $z$ as
\begin{align} \label{critexpinterz2}
\left( r_\beta ^{\mathbb{Z}_2} \right)_{\beta=1} ^3 = \left( \frac{-3a_{2,\beta}+ \sum_{\gamma=1} ^3 a_{2,\gamma} +3\varepsilon}{3\varepsilon_1}\right)_{\beta=1} ^3 .
\end{align}
Then the equation satisfied by the modified partition function $\widetilde{\EuScript{Z}}_\beta ^{\mathbb{Z}_2}$ becomes of the form \eqref{d3}.

\section{Analytic continuation and gluing} \label{congl}
To compute the monodromies of the solutions to the quantized opers, it is necessary to know how to connect the solutions in different convergence domains. We accomplish this by analytically continuing the surface defect partition functions to different convergence domains, and gluing those continuations in the intermediate regime.

\subsection{Analytic continuation}

We use the duality transformation similar to the one described on p.13 of \cite{Losev:1999nt}. There, one traded the sum over the fluxes of the two dimensional abelian gauge field (magnetic fluxes) for the sum over a dual integral variable (electric flux), which could be viewed as the label enumerating the sheets of the (possibly disconnected) effective target space.

\subsubsection{Gauged linear sigma model}
Let us begin with the two-dimensional gauged linear sigma model (GLSM), which would generate the surface defect when coupled to the four-dimensional $A_1$-theory. In section \ref{surfdef}, we have shown that the 2d GLSM responsible for the quiver surface defect and the $(N-1,1)$-type $\mathbb{Z}_2$-orbifold surface defect is the one which flows to the non-linear sigma model on the $\text{Hom}(\mathcal{O}(-1),\mathbb{C}^N)$-bundle over $\mathbb{P}^{N-1}$. This theory is the $\EuScript{N}=(2,2)$ supersymmetric $U(1)$ gauge theory with the field contents
\begin{align}
\begin{split}
\text{Twisted chiral :}&\quad \Sigma = (\sigma, A) \\
\text{Fundamental chiral :}&\quad Q_\alpha \quad \alpha = 1, \cdots, N, \\
\text{Anti-fundamental chiral :}&\quad \widetilde{Q}_\alpha \quad \alpha=1, \cdots, N,
\end{split}
\end{align}
where we have only denoted the bosonic component fields. By weakly gauging the $\left(U(N)\times U(N)\right)/U(1)$ flavor symmetry, the fundamental and the anti-fundamental acquire the twisted masses which we denote as $\left(a_{0,\alpha} \right)_{\alpha=1} ^N $ and $\left( a_{2,\alpha} \right)_{\alpha=1} ^N$ respectively, for the reason to be clarified soon. Note that we may re-define $\sigma$ by a constant amount so that the twisted masses appear as if weakly gauging the full $U(N) \times U(N)$ symmetry. Due to the twisted masses all the chiral multiplets can be integrated out. The resulting effective theory is the $\EuScript{N}=(2,2)$ $U(1)$ gauge theory with the effective twisted superpotential
\begin{align}
\widetilde{\EuScript{W}} (\sigma) = -t \sigma - \sum_{\alpha=1} ^N  (\sigma - a_{0,\alpha})  \left( \log(\sigma - a_{0,\alpha} )-1 \right) - \sum_{\alpha=1} ^N (-\sigma +a_{2,\alpha} )\left( \log(-\sigma +a_{2,\alpha} )-1 \right),
\end{align}
where we have introduced the complex coupling $t=r-i \theta$ from the Fayet-Illiopoulos parameter $r$ and the two-dimensional $\theta$-angle. Hence the vacuum equation reads
\begin{align} \label{vaceqs}
\prod_{\alpha=1} ^N \frac{-\sigma +a_{2,\alpha}}{\sigma -a_{0,\alpha}} = e^{t} = z,
\end{align}
with the K\"{a}hler modulus defined by $z\equiv e^{t}$. Note that the Fayet-Illiopoulos parameter $r$ is not renormalized since the total charge of the chiral multiplets is zero, and we can imagine flowing from the region $r \gg 0$ to the region $r \ll 0$. The GLSM in both regions gives rise to the non-linear sigma model on the $\text{Hom}(\mathcal{O}(-1), \mathbb{C}^N)$-bundle over $\mathbb{P}^{N-1}$, yet with the base and the fiber exchanged with each other as we cross $r=0$. The classical singularity at $r=0$ is actually shifted by the quantum effect, leaving only a single point $\theta=N \pi$ (mod $2\pi$) singular. Hence the flow can be smoothly continued to the other region, connecting the two sigma models. The vacuum equation \eqref{vaceqs} implies that the $N$-vacua continuously flow from $\sigma \sim a_{0,\alpha}$ at $r \gg 0$ to $\sigma \sim a_{2,\alpha}$ at $r \ll 0$.

Upon the $\Omega$-deformation on the two-dimensional plane, the partition function of the GLSM can be exactly computed by the equivariant localization. The effective twisted superpotential only exhibits the leading singular term in the partition function, so we investigate how the flow of $z$ appears at the level of the partition function. The partition function localizes on the generalized vortex configurations,
\begin{align}
\begin{split}
&D_{\bar{z}} Q \equiv \partial_{\bar{z}} Q + A_{\bar{z}} Q = 0 \\
&D_{\bar{z}} \widetilde{Q} = 0 \\
&F_{z \bar{z}} + \vert Q \vert^2 -\vert \widetilde{Q} \vert^2 = r.
\end{split}
\end{align}
Depending on the sign of $r$, we are forced to localize on either vortices or anti-vortices. Let us assume $r>0$ for now. The asymptotics of the D-term equation forbids the anti-fundamental $\widetilde{Q}$ to generate any bosonic moduli, and only allows its fermionic zero-modes \cite{vortexcounting}. The final form of the partition function is precisely the expression \eqref{surfquiv} without the coupling to the four-dimension,
\begin{align} \label{genhyp}
\begin{split}
\EuScript{Z}^{\text{GLSM}} _\beta &= \sum_{k=0} ^{\infty} \frac{z ^{-k}}{k!}  \; \frac{ \prod_{\alpha=1} ^N \left( 1+ \frac{a_{0,\beta} -a_{2,\alpha}}{\varepsilon_1} \right)_k }{ \prod_{\alpha \neq \beta} \left( 1+ \frac{a_{0,\beta}-a_{0,\alpha}}{\varepsilon_1} \right)_k} \\
&= {}_{N}F_{N-1} \left( \left( 1+\frac{a_{0,\beta}-a_{2,\alpha}}{\varepsilon_1} \right)_{\alpha=1,\cdots, N} ; \: \left( 1+\frac{a_{0,\beta}-a_{0,\alpha}}{\varepsilon_1} \right)_{\alpha\neq \beta}; \: z^{-1} \right),
\end{split}
\end{align} 
where we have chosen the vacuum at the infinity as $\sigma = a_{0,\beta}$. The effective twisted superpotential evaluated at this vacuum can be obtained by taking the asymptotics of the partition function,
\begin{align}
\EuScript{Z}_\beta ^{\text{GLSM}} = e^{\frac{\widetilde{\EuScript{W}}_\beta}{\varepsilon_1}} \left( 1+ \mathcal{O}(\varepsilon_1) \right).
\end{align}
Once we flow to the region $r<0$, the above series expansion is no longer valid. However, we can still study the asymptotics of the partition function, i.e., the effective twisted superpotential, in this region by applying the Picard-Lefschetz theory to the integral representation of the partition function \cite{wit}. To illustrate the idea, let us consider the case $N=2$. Also let us assume $\text{Re}\left( 1+\frac{a_{0,1}-a_{0,2}}{\varepsilon_1} \right) > \text{Re} \left( 1+ \frac{a_{0,1}-a_{2,2}}{\varepsilon_1} \right) >0$ for simplicity. Then the Euler integral representation for the hypergeometric function gives
\begin{align}
\EuScript{Z}_1 ^{\text{GLSM}} = \frac{\Gamma\left( 1+\frac{a_{0,1}-a_{0,2}}{\varepsilon_1} \right)}{\Gamma\left( 1+\frac{a_{0,1}-a_{2,2}}{\varepsilon_1} \right) \Gamma\left( \frac{a_{2,2}-a_{0,2}}{\varepsilon_1} \right)} \int_0 ^1 dt \; t^{\frac{a_{0,1}-a_{0,2}}{\varepsilon_1}} (1-t)^{-1+\frac{a_{2,2}-a_{0,2}}{\varepsilon_1}} (1-z^{-1} t)^{-1-\frac{a_{0,1}-a_{2,1}}{\varepsilon_1}}.
\end{align}
We now promote the real integral to an integral on the complex $t$-plane. We can represent the integral as
\begin{align}
\begin{split}
&\int_{C=[0,1]} dt \; g(t) e^{\frac{S(t)}{\varepsilon_1}}, \\ \\
&g(t)=(1-t)^{-1} (1-z^{-1} t)^{-1} \\
&S(t) = (a_{0,1}-a_{0,2})\log t +(a_{2,2}- a_{0,2})\log (1-t) -(a_{0,1}-a_{2,1})\log (1-z^{-1} t).
\end{split}
\end{align}
The critical points of $S(t)$ are at
\begin{align}
S'(t) = \frac{a_{0,1}-a_{2,2}}{t} -\frac{a_{2,2}-a_{0,2}}{1-t} + \frac{(a_{0,1}-a_{2,1})z^{-1}}{1- z^{-1} t} =0.
\end{align}
Let us denote the critical points as $t_\pm$, namely, $S'(t_\pm)=0$. Let us assume that the masses are generic enough so that the critical points $t_\pm$ are distinct. We would like to deform the integration contour $C$ into a union of paths, in which each path passes through one of the critical points and the imaginary part $\text{Im}S(t)$ is constant along the path. Such paths are called \textit{the Lefschetz thimbles}, and can be obtained by treating the imaginary part of $S(t)$ as a Hamiltonian
\begin{align}
H(t) \equiv \text{Im}S(t) = \frac{1}{2 i} (S(t) -\bar{S}(t)),
\end{align}
which defines the gradient flow by the equation
\begin{align}\label{flow}
\dot{\bar{t}} = \{H, \bar{t} \} = \omega^{ab} \partial_a H \partial_b \bar{t} = - \frac{\partial S(t)}{\partial t},
\end{align}
where the symplectic form on the $t$-plane is given by $\omega = \frac{1}{2 i} dt \wedge d \bar{t}$. The Lefschetz thimble $\mathcal{J}_\pm$ is defined as the union of these paths emanating from the critical points $t_\pm$. Note that $\text{Re}S(t)$ monotonically decreases along the flow \eqref{flow}, so that the integral along $\mathcal{J}_\pm$ would show good convergence. Now the problem is decomposing the contour $C$ into a union of those Lefschetz thimbles, and this procedure can be done as follows. Note that the integration contour $C$ defines an element of the relative homology $H_1 (\mathbb{C},\mathbb{C}_{-T};\mathbb{Z})$, where 
\begin{align}
\mathbb{C}_{-T} \equiv \{t \in \mathbb{C} \: \vert \: \text{Re}S(t) \leq -T \},
\end{align}
for $T \gg 1$. The Lefschetz thimbles are defined as the paths emanating from the critical points, in which $\text{Re}S(t)$ decreases along the flow. Hence the Lefschetz thimbles also define elements of the relative homology, $\mathcal{J}_\pm \in H_1 (\mathbb{C},\mathbb{C}_{-T};\mathbb{Z})$, and moreover they actually form a basis of this relative homology. Thus we can express $C$ as a linear combination of the basis elements $\mathcal{J}_\pm$, say, $C = \sum_{\pm} n_\pm \mathcal{J}_\pm$. Then the integral in the partition function can be expressed as
\begin{align}
\sum_\pm n_\pm \int_{\mathcal{J}_\pm} dt \; g(t) e^{\frac{S(t)}{\varepsilon_1}}.
\end{align}
The remaining problem is to find the number $n_\pm$. For this, let us consider the relative homology $H_1 (\mathbb{C},\mathbb{C}^T ; \mathbb{Z})$, where
\begin{align}
\mathbb{C}^T \equiv \{ t\in \mathbb{C} \; \vert \; \text{Re}S(t) \geq T \},
\end{align}
for $T \gg 1$. This relative homology is generated by \textit{the dual Lefschetz thimbles}, $\mathcal{K}_\pm$, which are defined as the union of the paths \eqref{flow} converging to the critical point $t_\pm$. Note that we have the intersection pairing
\begin{align}
\langle \mathcal{J}_\tau , \mathcal{K}_{\tau'} \rangle = \delta_{\tau, \tau'}, \quad \tau, \tau' = \pm,
\end{align}
under an appropriate orientation on these thimbles, since $\mathcal{J}_\pm$ and $\mathcal{K}_\pm$ intersect at $t_\pm$ and $\text{Re}S(t)$ only decreases or increases along these thimbles. Therefore, we derive
\begin{align}
n_\pm = \langle C, \mathcal{K}_\pm \rangle,
\end{align}
and the final form of the integral is
\begin{align}
\sum_\pm \langle C, \mathcal{K}_\pm \rangle \int_{\mathcal{J}_\pm} dt \; g(t) e^{\frac{S(t)}{\varepsilon_1}}.
\end{align}

When $r>0$ ($\vert z \vert >1$), it can be checked that only one dual thimble, say, $\mathcal{K}_+$, intersects with the original contour $C=[0,1]$. Hence the integral can be performed in the WKB sense as
\begin{align}
\sqrt{-\frac{\pi \varepsilon_1}{S''(t_+)}} \: g(t_+) \: e^{\frac{S(t_+)}{\varepsilon_1}} \left(1+ \sum_{k=1} ^\infty c_{+,k} \: \varepsilon_1 ^k \right)
\end{align}
In particular, the effective twisted superpotential is essentially $S(t_+)$. This confirms that we have a contribution from the single vacuum $\sigma = a_{0,\beta}$. However, when $r<0$ ($\vert z \vert <1$) the topology of thimbles change so that both dual thimbles $\mathcal{K}_\pm$ intersect with the contour $C=[0,1]$. Hence the integral is rather performed as
\begin{align}
\sqrt{-\frac{\pi \varepsilon_1}{S''(t_+)}} \: g(t_+) \: e^{\frac{S(t_+)}{\varepsilon_1}} \left( 1+\sum_{k=1} ^\infty c_{+,k}\: \varepsilon_1 ^k \right) + \sqrt{-\frac{\pi \varepsilon_1}{S''(t_-)}} \: g(t_-) \: e^{\frac{S(t_-)}{\varepsilon_1}} \left( 1+\sum_{k=1} ^\infty c_{-,k} \: \varepsilon_1 ^k \right) ,
\end{align}
In other words, we start to get a contribution from the other vacuum, represented by the thimble $\mathcal{J}_-$. The continuous flow the the vacua \eqref{vaceqs} only exhibits the leading contribution from $\mathcal{J}_+$, but the Picard-Lefschetz analysis shows that the contribution from the other vacuum also emerges as we flow to the region $r<0$.

For higher ranks $N \geq 3$, we have to deal with the Euler integral representation for the generalized hypergeometric function ${}_N F_{N-1}$ which is $N-1$-complex dimensional. It is more difficult to visualize, but the basic idea is the same. When we fix a vacuum in the region $r>0$ and flow to the region $r<0$, the exponentially suppressed contributions from the other $N-1$-vacua start to emerge. It can be also understood as the manifestation of the analytic continuation of the generalized hypergeometric function. In the domain $\vert z \vert <1$, the generalized hypergeometric function \eqref{genhyp} is still well-defined by the analytic continuation, and the proper series expansion for this analytic continuation is simply obtained by the connection formula,
\begin{align}
\begin{split}
&{}_{N}F_{N-1} \left( \left( 1+\frac{a_{0,\beta}-a_{2,\gamma}}{\varepsilon_1} \right)_{\gamma=1,\cdots, N} ; \: \left( 1+\frac{a_{0,\beta}-a_{0,\beta'}}{\varepsilon_1} \right)_{\beta'\neq \beta}; \: z^{-1} \right) \\
& = - \sum_{\alpha=1} ^N \prod_{\beta'\neq\beta} \frac{\Gamma \left( 1+ \frac{a_{0,\beta}-a_{0,\beta'}}{\varepsilon_1} \right)}{\Gamma\left( \frac{a_{2,\alpha} -a_{0,\beta'}}{\varepsilon_1} \right)} \prod_{\alpha' \neq\alpha} \frac{\Gamma\left( \frac{a_{2,\alpha}-a_{2,\alpha'}}{\varepsilon_1} \right)}{\Gamma \left( 1+\frac{a_{0,\beta}-a_{2,\alpha'}}{\varepsilon_1} \right)} (-z)^{1+ \frac{a_{0,\beta}-a_{2,\alpha}}{\varepsilon_1}} \\
&\quad\quad {}_N F_{N-1} \left( \left(1+\frac{a_{0,\gamma}-a_{2,\alpha}}{\varepsilon_1} \right)_{\gamma=1,\cdots,N} ; \left( 1+\frac{a_{2,\alpha'}-a_{2,\alpha}}{\varepsilon_1} \right)_{\alpha'\neq \alpha} ; \; z \right).
\end{split}
\end{align}
The Picard-Lefschetz analysis provides a physical interpretation of this formula, i.e., the emergence of other $N-1$-vacua as a consequence of the flow from $r>0$ to $r<0$.

\subsubsection{Four-dimensional theory with surface defect}
The analytic continuation along the flow of the K\"{a}hler modulus can be conducted in a more general setting: the two-dimensional gauged linear sigma model coupled to the four-dimensional gauge theory. Let us start with the quiver surface defect partition function \eqref{zl} with the constraints \eqref{a2const}, namely,
\begin{align} 
\EuScript{Z}^L _{\beta} = \EuScript{Z}_{A_2} \left( \mathbf{a}_{0}; \, a_{1,\alpha} = a_{0,\alpha} - {\ve}_2 {\delta}_{\alpha, \beta}  ; \, \mathbf{a}_{2} ;\, \mathbf{a}_{3} \; \vert \; \varepsilon_{1}, \varepsilon_2\; \vert \; \mathfrak{q}_1 = z^{-1}, \mathfrak{q}_2 = \mathfrak{q} \right).
\end{align}
We recall that this can be expressed in terms of the $\EuScript{Q}$-observables \eqref{qobs}. Thus \eqref{zl} can be written as
\begin{align}
\begin{split}
\EuScript{Z}^L _\beta &= \sum_{\boldsymbol{\lambda}^{(2)}} \mathfrak{q}_2 ^{\vert \boldsymbol{\lambda}^{(2)}  \vert } \boldsymbol{\mu}_{\boldsymbol{\lambda}^{(2)}} \sum_{k=0} ^\infty \mathfrak{q}_1 ^{k}  \; \prod_{\alpha=1} ^N \frac{ (-1)^k \; \Gamma \left( 1+ \frac{a_{0,\beta} - a_{0, \alpha}}{\varepsilon_1} \right) \Gamma \left( -\frac{a_{0,\beta}-a_{2,\alpha}}{\varepsilon_1} \right)}{ \; \Gamma \left( k+1 + \frac{a_{0, \beta} - a_{0, \alpha}}{\varepsilon_1}  \right) \Gamma\left( -k -\frac{a_{0,\beta}-a_{2,\alpha}}{\varepsilon_1} \right)} \\
& \quad\quad\quad\quad\quad \quad \quad \quad \quad \quad \quad  \prod_{\square \in K_2} \frac{a_{0,\beta}+k \varepsilon_1 -c_\square -\varepsilon_2}{a_{0,\beta}+k \varepsilon_1 -c_\square} \;\; \frac{a_{0,\beta}-c_\square}{a_{0,\beta}-c_\square -\varepsilon_2} \\ 
&= \sum_{\boldsymbol{\lambda}^{(2)}} \mathfrak{q}_2 ^{\vert \boldsymbol{\lambda}^{(2)}  \vert } \boldsymbol{\mu}_{\boldsymbol{\lambda}^{(2)}} \sum_{k=0} ^\infty \mathfrak{q}_1 ^{k}  \; \prod_{\alpha=1} ^N \frac{  \Gamma \left( 1+ \frac{a_{0,\beta} - a_{0, \alpha}}{\varepsilon_1} \right) \Gamma \left(k+1 +\frac{a_{0,\beta}-a_{2,\alpha}}{\varepsilon_1} \right)}{ \; \Gamma \left( k+1 + \frac{a_{0, \beta} - a_{0, \alpha}}{\varepsilon_1}  \right) \Gamma\left( 1+ \frac{a_{0,\beta}-a_{2,\alpha}}{\varepsilon_1} \right)} \\
& \quad\quad\quad\quad\quad \quad \quad \quad \quad \quad \quad  \prod_{\square \in K_2} \frac{a_{0,\beta}+k \varepsilon_1 -c_\square -\varepsilon_2}{a_{0,\beta}+k \varepsilon_1 -c_\square} \;\; \frac{a_{0,\beta}-c_\square}{a_{0,\beta}-c_\square -\varepsilon_2},
\end{split}
\end{align}
where we have used the reflection formula $\Gamma(x) \Gamma(1-x) = \frac{\pi}{\sin \pi x}$ in the second equality. It is crucial to notice that the partition function now can be represented as a contour integral
\begin{align} \label{partcont}
\begin{split}
\EuScript{Z}^L _\beta &= - \prod_{\alpha=1} ^N \frac{\Gamma \left( 1+ \frac{a_{0,\beta} -a_{0,\alpha}}{\varepsilon_1} \right)}{\Gamma \left( 1+\frac{a_{0,\beta}-a_{2,\alpha}}{\varepsilon_1} \right)} (-\mathfrak{q}_1)^{-\frac{a_{0,\beta}}{\varepsilon_1}} \sum_{\boldsymbol{\lambda}^{(2)}}  \mathfrak{q}_2 ^{\vert \boldsymbol{\lambda}^{(2)}  \vert } \widetilde{\boldsymbol{\mu}}_{\boldsymbol{\lambda}^{(2)}} \\
&  \oint_{\mathcal{C}} dx \; (-\mathfrak{q}_1)^{\frac{x}{\varepsilon_1}} \frac{\Gamma \left( -\frac{x-a_{0,\beta}}{\varepsilon_1} \right) \prod_{\alpha=1} ^N \Gamma \left( 1+\frac{x-a_{2,\alpha}}{\varepsilon_1} \right)}{\prod_{\alpha \neq \beta} \Gamma\left(1+ \frac{x-a_{0,\alpha}}{\varepsilon_1} \right)} \prod_{\alpha=1} ^N \prod_{i=1} ^{l\left(\lambda^{(2,\alpha)}\right)} \frac{x-a_{2,\alpha}-(i-1)\varepsilon_1 -\lambda_i ^{(2,\alpha)} \varepsilon_2 }{x-a_{2,\alpha}-(i-1)\varepsilon_1},
\end{split}
\end{align}
where we have defined
\begin{align}
\begin{split}
\widetilde{\boldsymbol{\mu}}_{\boldsymbol{\lambda}^{(2)}} &\equiv \boldsymbol{\mu}_{\boldsymbol{\lambda}^{(2)}} \prod_{\square \in {\boldsymbol{\lambda}^{(2)}}} \frac{a_{0,\beta}-c_\square}{a_{0,\beta}-c_\square-\varepsilon_2} \\
&= \epsilon \left[ N_2 K_2 ^* +q_{12} N_2 ^* K_2 -P_{12} K_2 K_2 ^* -M_0 K_2 ^* -q_{12} M_3 ^* K_2 \right].
\end{split}
\end{align}
The contour $\mathcal{C}$ is described in Figure \ref{fig1}. Here, we are assuming the Coulomb moduli $\mathbf{a}_2 = \left(a_{2,\alpha}\right)_{\alpha=1} ^N$ and the masses of hypermultiplets $\mathbf{a}_0 = \left(a_{0,\alpha}\right)_{\alpha=1} ^N $ (and $\mathbf{a}_3 = \left(a_{3,\alpha}\right)_{\alpha=1} ^N$ for $\EuScript{Z}_\beta ^R$) are generic, so that the simple poles do not overlap with each other.
\begin{figure}
\centering
\resizebox{4 in}{4 in}{
\begin{tikzpicture}
\path[black] (0,-9) edge[->] (0,9)  (-9,0) edge[->] (9,0);
\node[cross out,draw=black] at (0,0) {}; \node[cross out,draw=black] at (1,0) {}; \node[cross out,draw=black] at (2,0) {};\node[cross out,draw=black] at (3,0) {};\node[cross out,draw=black] at (4,0) {}; \node[cross out,draw=black] at (5,0) {};\node[cross out,draw=black] at (6,0) {};

\node[cross out,draw=black] at (3.2,5.6) {}; \node[cross out,draw=black] at (2.2,5.6) {};\node[cross out,draw=black] at (1.2,5.6) {};\node[cross out,draw=black] at (0.2,5.6) {};\node[cross out,draw=black] at (-0.8,5.6) {};\node[cross out,draw=black] at (-1.8,5.6) {};\node[cross out,draw=black] at (-2.8,5.6) {};\node[cross out,draw=black] at (-3.8,5.6) {};\node[cross out,draw=black] at (-4.8,5.6) {}; 
\node[circle,fill,inner sep=1pt] at (-2,4.6) {};
\node[circle,fill,inner sep=1pt] at (-2,3.8) {};
\node[circle,fill,inner sep=1pt] at (-2,3.0) {};
\node[cross out,draw=black] at (2.1,2) {};\node[cross out,draw=black] at (1.1,2) {};\node[cross out,draw=black] at (0.1,2) {}; \node[cross out,draw=black] at (-0.9,2) {}; \node[cross out,draw=black] at (-1.9,2) {};\node[cross out,draw=black] at (-2.9,2) {};\node[cross out,draw=black] at (-3.9,2) {}; \node[cross out,draw=black] at (-4.9,2) {};

\node[cross out,draw=black] at (4.6,-2.4) {}; \node[cross out,draw=black] at (3.6,-2.4) {};\node[cross out,draw=black] at (2.6,-2.4) {};\node[cross out,draw=black] at (1.6,-2.4) {}; \node[cross out,draw=black] at (0.6,-2.4) {}; \node[cross out,draw=black] at (-0.4,-2.4) {}; \node[cross out,draw=black] at (-1.4,-2.4) {};\node[cross out,draw=black] at (-2.4,-2.4) {};\node[cross out,draw=black] at (-3.4,-2.4) {};\node[cross out,draw=black] at (-4.4,-2.4) {};

\node[circle,fill,inner sep=1pt] at (-2,-3.2) {};
\node[circle,fill,inner sep=1pt] at (-2,-3.8) {};
\node[circle,fill,inner sep=1pt] at (-2,-4.4) {};
\node[cross out,draw=black] at (5.9,-5) {}; \node[cross out,draw=black] at (4.9,-5) {};\node[cross out,draw=black] at (3.9,-5) {};\node[cross out,draw=black] at (2.9,-5) {};\node[cross out,draw=black] at (1.9,-5) {};\node[cross out,draw=black] at (0.9,-5) {}; \node[cross out,draw=black] at (-0.1,-5) {}; \node[cross out,draw=black] at (-1.1,-5) {};\node[cross out,draw=black] at (-2.1,-5) {};\node[cross out,draw=black] at (-3.1,-5) {};\node[cross out,draw=black] at (-4.1,-5) {};

\draw[blue,very thick,decoration={markings, mark=at position 0.6 with {\arrow[scale=2]{latex}}}, postaction={decorate}] (0,-9) -- (0,-5.3); \draw[blue,very thick] (0,-5.3) -- (5.9,-5.3);
\draw[blue,very thick] (5.9,-5.3) arc (-90:90:0.3) -- (5.9,-4.7); \draw[blue,very thick] (5.9,-4.7) -- (0,-4.7);\draw[blue,very thick] (0,-4.7) -- (0,-4.2); 
\node[blue,circle,fill,inner sep=1pt] at (0,-4.0) {}; \node[blue,circle,fill,inner sep=1pt] at (0,-3.75) {};\node[blue,circle,fill,inner sep=1pt] at (0,-3.5) {};
\draw[blue,very thick] (0,-3.3) -- (0,-2.7); \draw[blue,very thick] (0,-2.7) -- (4.6,-2.7);
\draw[blue,very thick] (4.6,-2.7) arc (-90:90:0.3) -- (4.6,-2.1);\draw[blue,very thick] (4.6,-2.1) -- (0,-2.1);  \draw[blue,very thick,decoration={markings, mark=at position 0.6 with {\arrow[scale=2]{latex}}}, postaction={decorate}] (0,-2.1) -- (0,-0.3);  \draw[blue,very thick] (0,-0.3) arc (270:90:0.3) -- (0,0.3);  \draw[blue,very thick,decoration={markings, mark=at position 0.6 with {\arrow[scale=2]{latex}}}, postaction={decorate}] (0,0.3) -- (0,1.7); \draw[blue,very thick] (0,1.7) -- (2.1,1.7); 
\draw[blue, very thick] (2.1,1.7) arc (-90:90:0.3) -- (2.1,2.3); \draw[blue,very thick] (2.1,2.3) -- (0,2.3); \draw[blue,very thick] (0,2.3) -- (0,3);
\node[blue,circle,fill,inner sep=1pt] at (0,3.4) {}; \node[blue,circle,fill,inner sep=1pt] at (0,3.8) {};\node[blue,circle,fill,inner sep=1pt] at (0,4.2) {}; 
\draw[blue,very thick] (0,4.6) -- (0,5.3); \draw[blue,very thick] (0,5.3) -- (3.2,5.3);
\draw[blue,very thick] (3.2,5.3) arc (-90:90:0.3) -- (3.2,5.9);  \draw[blue,very thick] (3.2,5.9) -- (0,5.9); \draw[blue, very thick,decoration={markings, mark=at position 0.6 with {\arrow[scale=2]{latex}}}, postaction={decorate}] (0,5.9) -- (0,9); 

\draw[red, very thick,decoration={markings, mark=at position 0.3 with {\arrow[scale=2]{latex}}, mark=at position 0.7 with {\arrow[scale=2]{latex}}  }, postaction={decorate}] (0,9) arc (90:-90:9) -- (0,-9);
\node at (8,6) {\scalebox{2.5}{$\textcolor{red}{\mathcal{R}_+}$}};

\draw[teal,very thick,decoration={markings, mark=at position 0.3 with {\arrow[scale=2]{latex}}, mark=at position 0.7 with {\arrow[scale=2]{latex}}  }, postaction={decorate}] (0,9) arc (90:270:9) -- (0,-9);
\node at (-8,6) {\scalebox{2.5}{$\textcolor{teal}{\mathcal{R}_-}$}};

\node[circle,fill,inner sep=1pt] at (6.6,0) {};\node[circle,fill,inner sep=1pt] at (7.2,0) {};\node[circle,fill,inner sep=1pt] at (7.8,0) {};
\node[circle,fill,inner sep=1pt] at (-5.4,5.6) {};\node[circle,fill,inner sep=1pt] at (-5.8,5.6) {};\node[circle,fill,inner sep=1pt] at (-6.2,5.6) {};
\node[circle,fill,inner sep=1pt] at (-5.5,2) {};\node[circle,fill,inner sep=1pt] at (-5.9,2) {};\node[circle,fill,inner sep=1pt] at (-6.3,2) {};
\node[circle,fill,inner sep=1pt] at (-5,-2.4) {};\node[circle,fill,inner sep=1pt] at (-5.4,-2.4) {};\node[circle,fill,inner sep=1pt] at (-5.8,-2.4) {};
\node[circle,fill,inner sep=1pt] at (-4.7,-5) {};\node[circle,fill,inner sep=1pt] at (-5.1,-5) {};\node[circle,fill,inner sep=1pt] at (-5.5,-5) {};

\node at (-0.7,-1.2) {\scalebox{2.5}{$\textcolor{blue}{\mathcal{C}}$}};
\node at (4.8,4.9) {$\frac{a_{2,1}-a_{0,\beta}}{\varepsilon_1}+ l\left( \lambda^{(2,1)} \right) -1 $};
\node at (7.4,-5.8) {$\frac{a_{2,N}-a_{0,\beta}}{\varepsilon_1}+ l\left( \lambda^{(2,N)} \right) -1 $};
\node at (0,-0.5) {$0$}; \node at (1,-0.5) {$1$};\node at (2,-0.5) {$2$};
\end{tikzpicture}
}
\caption{The contour $\mathcal{C}$ on the $\frac{x-a_{0,\beta}}{\varepsilon_1}$-plane.} \label{fig1}
\end{figure}
Note that this contour integral is analogous to the famous \textit{Barnes integral}. It is straightforward to prove that the integral \eqref{partcont} uniformly converges as long as $\text{Arg}(-\mathfrak{q}_1) < \pi$, i.e., $\mathfrak{q}_1 \notin \mathbb{R}^+$, using the asymptotics of the $\Gamma$-functions. The equality in \eqref{partcont} is obtained as we close the contour by adding the semi-circle $\mathcal{R}_+$ at the infinity, picking only the poles at $x=a_{0,\beta} + k\varepsilon_1$, $k \in \mathbb{Z}^{\geq 0}$. It can be shown that the integral along $\mathcal{R}_+$ uniformly converges to zero in the regime $\vert \mathfrak{q}_1 \vert <1$, and therefore it is safe to add $\mathcal{R}_+$ to the contour $\mathcal{C}$.

Now, we take the contour integral representation \eqref{partcont} as the analytic continuation of the partition function $\EuScript{Z}_\beta ^L$. In particular, the partition function assumes a different series expansion in the regime $\vert \mathfrak{q}_1 \vert >1$, and it can be computed as we close the contour by adding a semi-circle $\mathcal{R}_-$ on the opposite side. It is possible to show that the integral along $\mathcal{R}_-$ uniformly converges to zero in the regime $\vert \mathfrak{q}_1 \vert >1$, and hence it is safe to add $\mathcal{R}_-$ to the contour $\mathcal{C}$. The resulting contour encloses the rest of the poles, i.e., $x = a_{2,\alpha} + \left(l \left(\lambda^{(2,\alpha)} \right) -k-1\right)\varepsilon_1$ where $\alpha = 1, \cdots, N$ and $k \in \mathbb{Z}^{\geq 0}$. First note that the denominator in the contour integral can be absorbed into the $\Gamma$-functions, yielding
\begin{align}
\begin{split}
\oint_{\mathcal{C}} dx& \; (-\mathfrak{q}_1)^{\frac{x}{\varepsilon_1}} \frac{ \Gamma\left( -\frac{x-a_{0,\beta}}{\varepsilon_1} \right) \prod_{\alpha=1} ^N \Gamma \left( -l \left( \lambda^{(2,\alpha)} \right) +1 + \frac{x-a_{2,\alpha}}{\varepsilon_1} \right)}{\prod_{\alpha \neq \beta} \Gamma \left( 1+ \frac{x-a_{0,\alpha}}{\varepsilon_1} \right)} \\
&\quad\quad\quad\quad\quad \prod_{\alpha=1} ^N \prod_{i=1} ^{l \left( \lambda^{(2,\alpha)} \right)} \frac{x-a_{2,\alpha} -(i-1)\varepsilon_1 - \lambda_i ^{(2,\alpha)} \varepsilon_2}{\varepsilon_1}.
\end{split}
\end{align}
Then we can pick up the residues of the $N$-rays of poles at $x = a_{2,\alpha} + \left(l\left(\lambda^{(2,\alpha)} \right) -k-1\right)\varepsilon_1$, $\alpha=1, \cdots, N$ and $k\in \mathbb{Z}^{\geq 0}$. We can write the resulting series expansion for the analytically continued partition function as a sum over these $N$-rays,
\begin{align} \label{analcont1}
\EuScript{Z}^L _\beta = \sum_{\alpha=1} ^N \prod_{\beta' \neq \beta} \frac{\Gamma\left( 1+ \frac{a_{0,\beta}-a_{0,\beta'}}{\varepsilon_1} \right)}{\Gamma \left( \frac{a_{2,\alpha} -a_{0,\beta'}}{\varepsilon_1} \right)} \prod_{\alpha' \neq \alpha} \frac{\Gamma\left( \frac{a_{2,\alpha} -a_{2,\alpha'}}{\varepsilon_1} \right)}{\Gamma\left( 1+\frac{a_{0,\beta} -a_{2,\alpha'}}{\varepsilon_1} \right)} \mathfrak{q}_1 ^{-1} (-\mathfrak{q}_1)^{ \frac{a_{2,\alpha} - a_{0,\beta}}{\varepsilon_1}} \EuScript{Z}_\alpha ^{L \to M},
\end{align}
where we have defined the basis function in the regime $\vert \mathfrak{q}_1 \vert >1$, which is independent of the choice of $\beta$ in the constraints \eqref{a2const}, by
\begin{align} \label{contbase}
\begin{split}
\EuScript{Z}_\alpha ^{L \to M} \left( \mathbf{a}_{2} \right) &\equiv \sum_{\boldsymbol{\lambda} ^{(2)}} \mathfrak{q}_2 ^{\vert \boldsymbol{\lambda}^{(2)}  \vert } \widetilde{\boldsymbol{\mu}}_{\boldsymbol{\lambda}^{(2)}} \sum_{k=0} ^{\infty} \mathfrak{q}_1 ^{ -k +l \left( \lambda^{(2,\alpha)} \right) } \frac{(-1)^{k}}{k!} \\
&\prod_{\alpha' \neq \alpha} \frac{ \Gamma \left( -k +l \left( \lambda^{(2,\alpha)} \right) -l \left( \lambda^{(2,\alpha')} \right) +\frac{a_{2,\alpha}-a_{2,\alpha'}}{\varepsilon_1} \right)}{\Gamma\left( \frac{a_{2,\alpha}-a_{2,\alpha'}}{\varepsilon_1} \right)} \prod_{\gamma=1 } ^N \frac{\Gamma \left( \frac{a_{2,\alpha} - a_{0,\gamma}}{\varepsilon_1} \right)}{\Gamma \left( -k + l \left( \lambda^{(2,\alpha)} \right) +\frac{a_{2,\alpha} -a_{0,\gamma}}{\varepsilon_1} \right)} \\
&\prod_{\gamma=1} ^N \prod_{i=1} ^{l \left( \lambda^{(2,\gamma)} \right)} \frac{a_{2,\alpha} -a_{2,\gamma} +\left( l \left( \lambda^{(2,\alpha)} \right)-k-i \right)\varepsilon_1 - \lambda_i ^{(2,\gamma)} \varepsilon_2  }{\varepsilon_1},
\end{split}
\end{align}
so that the choice of $\beta$ only affects the coefficients of the continuation formula \eqref{analcont1}. We will explicitly write the argument of $\EuScript{Z}_\alpha ^{L \to M} $ only when we emphasize its Coulomb moduli, but otherwise we omit it. Note that the basis function can be expressed as the expectation value of an infinite sum of $\EuScript{Q}$-observables \eqref{qobs},
\begin{align} \label{contq}
\begin{split}
\EuScript{Z}_\alpha ^{L \to M} &= \sum_{\boldsymbol{\lambda} ^{(2)}} \mathfrak{q}_2  ^{\vert \boldsymbol{\lambda}^{(2)}  \vert } \widetilde{\boldsymbol{\mu}}_{\boldsymbol{\lambda}^{(2)}} \prod_{\square \in {\boldsymbol{\lambda}^{(2)}}} \frac{a_{2,\alpha} -\varepsilon -c_\square }{a_{2,\alpha} -\varepsilon_1 -c_\square} \; \sum_{k=0} ^{\infty} \mathfrak{q}_1  ^{-k+  l \left( \lambda^{(2,\alpha)} \right) }\;  \varepsilon_1 ^{ N\left(k-l \left( \lambda^{(2,\alpha)} \right)\right)}
  \\
&\quad\quad \prod_{\gamma =1} ^N \frac{\Gamma \left( \frac{a_{2,\alpha} - a_{0,\gamma}}{\varepsilon_1} \right)}{\Gamma \left( -k + l \left( \lambda^{(2,\alpha)} \right) +\frac{a_{2,\alpha} -a_{0,\gamma}}{\varepsilon_1} \right)} \; \frac{ \EuScript{Q}_2 \left(a_{2,\alpha} + \left( l \left( \lambda^{(2,\alpha)} \right) -k-1 \right)\varepsilon_1  \right)}{\EuScript{Q}_2 \left( a_{2,\alpha}- \varepsilon_1 \right)}.
\end{split}
\end{align}
\paragraph{Remarks}
\begin{itemize}
\item The ratios of the $\Gamma$-functions in \eqref{contbase} and \eqref{contq} can be expressed as Pochhammer symbols, but they may appear either in the numerator or in the denominator depending on $k$ and $l \left( \lambda^{(2,\alpha)} \right)$'s.
\item While the exponent of $\mathfrak{q}_2$ is always positive, the exponent of $\mathfrak{q}_1$ can either be positive or negative depending on $k$ and $l \left( \lambda^{(2,\alpha)} \right)$. The convergence regime is $0<\vert \mathfrak{q}_2 \vert < \vert \mathfrak{q}_1 ^{-1} \vert<1$. We may introduce new coupling constants
\begin{align}
\mathfrak{q}_1 \equiv {\mathfrak{q}_1 '}^{-1}, \quad \mathfrak{q}_2 \equiv \mathfrak{q}_1 ' \mathfrak{q}_2 ',
\end{align}
so that the the convergence regime becomes $0< \vert \mathfrak{q}_1 ' \vert, \vert \mathfrak{q}_2 ' \vert<1$. Indeed, the exponent of the new coupling constant $\mathfrak{q} '_1$ is $k+ \vert \boldsymbol{\lambda} ^{(2)} \vert - l \left( \lambda^{(2,\alpha)} \right) \geq k \geq 0$, i.e., bounded below. The first inequality is saturated if and only if $\boldsymbol{\lambda}^{(2)}$ is single-columned, namely, $\lambda^{(2,\alpha')} = \varnothing$ for all $\alpha' \neq \alpha$ and $\lambda^{(2,\alpha)}$ is single-columned. This suggests the basis function $\EuScript{Z}_\alpha$ is related to the $A_2$-theory in which the Coulomb moduli of the two gauge nodes are subject to certain constraints. We come back to this question in section \ref{conmats}.
\item The reparametrization of the couplings $\mathfrak{q}_1 = z^{-1}$ and $\mathfrak{q}_2= \mathfrak{q}$ of \eqref{zl} were introduced to be consistent with the convention in \eqref{a2npdsfinal}. Note that $\mathfrak{q}' _1 = z$ and $\mathfrak{q}_2 '  = \frac{\mathfrak{q}}{z}$ under the reparametrization.
\item Let $\EuScript{O}$ be an observable lying only on the second gauge node, i.e., $\EuScript{O} [\boldsymbol{\lambda}] = \EuScript{O} [\boldsymbol{\lambda}^{(2)}]$. The expectation value of such observables can similarly be analytically continued. We simply need to insert the observable inside \eqref{contbase}, along with the measure $\widetilde{\boldsymbol{\mu}}_{\boldsymbol{\lambda}^{(2)}}$.
\end{itemize}

Similarly, we can analytically continue the quiver surface defect partition function \eqref{a2diff}. After imposing the constraints \eqref{constdiff} and the re-definition of parameters \eqref{redef} we consider
\begin{align} 
\EuScript{Z}^R _\beta = \EuScript{Z}_{A_2} \left( -a_{0,\alpha}-{\ve} ; \, -a_{1,\alpha}; \, -a_{3,\alpha} + \varepsilon - {\ve}_{2}{\delta}_{\alpha,\beta}; \, -a_{3, \alpha}+2\varepsilon \; \vert \; \varepsilon_1, \varepsilon_2 \; \vert \; \mathfrak{q}_1 = \mathfrak{q} , \mathfrak{q}_2 = {\qe}^{-1} \, z  \right),
\end{align}
The partition function can be analytically continued in the same way,
\begin{align} \label{analcont2}
\EuScript{Z}^R _\beta = \sum_{\alpha=1} ^N \prod_{\beta' \neq \beta} \frac{ \Gamma \left( 1+\frac{a_{3,\beta'} -a_{3,\beta}}{\varepsilon_1} \right)}{\Gamma \left( \frac{a_{3,\beta'}-a_{1,\alpha}}{\varepsilon_1} \right)} \prod_{\alpha' \neq\alpha} \frac{\Gamma \left( \frac{a_{1,\alpha'} -a_{1,\alpha}}{\varepsilon_1} \right)}{\Gamma \left( 1+\frac{a_{1,\alpha'}-a_{3,\beta}}{\varepsilon_1} \right)} \mathfrak{q}_2 ^{-1} (-\mathfrak{q}_2)^{\frac{a_{3,\beta} -a_{1,\alpha}}{\varepsilon_1}}\EuScript{Z}_\alpha ^{R \to M},
\end{align}
where
\begin{align} \label{contright}
\begin{split}
\EuScript{Z}_\alpha ^{R \to M} (\mathbf{a}_{1}) &= \sum_{\boldsymbol{\lambda} ^{(1)}} \mathfrak{q}_1 ^{\vert \boldsymbol{\lambda}^{(1)}  \vert } \widetilde{\boldsymbol{\mu}}_{\boldsymbol{\lambda}^{(1)}} \sum_{k=0} ^{\infty} \mathfrak{q}_2 ^{ -k +l \left( \lambda^{(1,\alpha)} \right) } \frac{(-1)^{ k}}{k!} \\
&\quad\quad\prod_{\alpha' \neq \alpha} \frac{ \Gamma \left( -k +l \left( \lambda^{(1,\alpha)} \right) -l \left( \lambda^{(1,\alpha')} \right) +\frac{a_{1,\alpha'}-a_{1,\alpha}}{\varepsilon_1} \right)}{\Gamma\left( \frac{a_{1,\alpha'}-a_{1,\alpha}}{\varepsilon_1} \right)} \prod_{\gamma=1 }^N \frac{\Gamma \left( \frac{a_{3,\gamma} - a_{1,\alpha}}{\varepsilon_1} \right)}{\Gamma \left( -k + l \left( \lambda^{(1,\alpha)} \right) +\frac{a_{3,\gamma} -a_{1,\alpha}}{\varepsilon_1} \right)} \\
&\quad\quad\prod_{\gamma=1} ^N \prod_{i=1} ^{l \left( \lambda^{(1,\gamma)} \right)} \frac{ - a_{1,\alpha}+a_{1,\gamma} +\left( l \left( \lambda^{(1,\alpha)} \right)-k-i \right)\varepsilon_1 - \lambda_i ^{(1,\gamma)} \varepsilon_2  }{\varepsilon_1}. \\
&= \sum_{\boldsymbol{\lambda} ^{(1)}} {\mathfrak{q}_1 } ^{\vert \boldsymbol{\lambda}^{(1)}  \vert } \widetilde{\boldsymbol{\mu}}_{\boldsymbol{\lambda}^{(1)}} \prod_{\boldsymbol{\lambda}^{(1)}} \frac{-a_{1,\alpha} -c_\square -\varepsilon}{-a_{1,\alpha} -c_\square -\varepsilon_1}  \; \sum_{k=0} ^{\infty} \mathfrak{q}_2  ^{-k+  l \left( \lambda^{(1,\alpha)} \right) }\;  \varepsilon_1 ^{ N\left(k-l \left( \lambda^{(1,\alpha)} \right)\right)}
  \\
&\quad\quad \prod_{\gamma=1} ^N \frac{\Gamma \left( \frac{a_{3,\gamma} - a_{1,\alpha}}{\varepsilon_1} \right)}{\Gamma \left( -k + l \left( \lambda^{(1,\alpha)} \right) +\frac{a_{3,\gamma} -a_{1,\alpha}}{\varepsilon_1} \right)} \frac{ \EuScript{Q}_1 \left(-a_{1,\alpha} + \left( l \left( \lambda^{(1,\alpha)} \right) -k-1 \right)\varepsilon_1  \right)}{\EuScript{Q}_1 \left(- a_{1,\alpha}- \varepsilon_1 \right)},
\end{split}
\end{align}
We have defined the modified measure
\begin{align}
\begin{split}
\widetilde{\boldsymbol{\mu}}_{\boldsymbol{\lambda}^{(1)}}  &\equiv \boldsymbol{\mu}_{\boldsymbol{\lambda}^{(1)}} \prod_{\square \in K_1} \frac{-a_{3,\beta} -c_\square}{-a_{3,\beta} -c_\square -\varepsilon_2} \\
&= \epsilon \left[ N_1 K_1 ^* +q_{12} N_1 ^* K_1 -P_{12} K_1 K_1 ^* - M_0 K_1 ^* - q_{12} ^2 M_3 ^* K_1  \right].
\end{split}
\end{align}
\paragraph{Remarks}
\begin{itemize}
\item The convergence regime of \eqref{contright} is $0< \vert \mathfrak{q}_1 \vert <\vert \mathfrak{q}_2 ^{-1} \vert<1$. We may define new coupling constants
\begin{align}
\mathfrak{q}_1 \equiv \mathfrak{q}_1 ' \mathfrak{q}_2 ' , \quad \mathfrak{q}_2 \equiv {\mathfrak{q}_2 '} ^{-1},
\end{align}
so that the convergence regime becomes $0< \vert \mathfrak{q}_1 ' \vert, \vert \mathfrak{q}_2 ' \vert <1$.
\item The reparametrizations of coupling constants \eqref{zr} were introduced to be consistent with the convention in \eqref{rightequation}. Note that the new coupling constants $\mathfrak{q}_1 ' = z$ and $\mathfrak{q}_2 ' = \frac{\mathfrak{q}}{z}$ match with the previous ones. Thus, both analytically continued partition functions lie in the intermediate domain, $0< \vert \mathfrak{q} \vert < \vert z \vert<1$.
\end{itemize}

\subsection{Gluing the partition functions} \label{glue}
\subsubsection{The connection matrix} \label{conmats}
Recall that the surface defect partition functions are annihilated by the operators $\widehat{\widehat{\mathfrak{D}}}$ obtained in section \ref{ssnpds}. The uniqueness of the analytic continuation guarantees that the continued functions satisfy the same differential equations. Therefore we may regard the analytically continued partition functions as the extensions of the solutions to other convergence domains. Motivated by the analytic continuation formulas \eqref{analcont1} and \eqref{analcont2}, let us define \textit{the connection matrices}
\begin{subequations} \label{contmat}
\begin{align}
\left( \mathbf{C}_\infty  \right)_{\alpha \beta} &\equiv \prod_{\alpha' \neq\alpha} \frac{\Gamma \left( 1+\frac{a_{0,\alpha}-a_{0,\alpha'}}{\varepsilon_1} \right)}{\Gamma \left( \frac{a_{2,\beta} -a_{0,\alpha'}}{\varepsilon_1} \right)} \prod_{\beta' \neq\beta} \frac{\Gamma \left( \frac{a_{2,\beta}-a_{2,\beta'}}{\varepsilon_1} \right)}{\Gamma \left( 1+ \frac{a_{0,\alpha} -a_{2,\beta'}}{\varepsilon_1} \right)}, \label{conmat1} \\
\left(\mathbf{C}_0\right)_{\alpha\beta} &\equiv \prod_{\alpha' \neq\alpha} \frac{\Gamma \left( 1+\frac{a_{3,\alpha'}-a_{3,\alpha}}{\varepsilon_1} \right)}{\Gamma \left( \frac{a_{3,\alpha'} -a_{1,\beta}}{\varepsilon_1} \right)} \prod_{\beta' \neq\beta} \frac{\Gamma \left( \frac{a_{1,\beta'}-a_{1,\beta}}{\varepsilon_1} \right)}{\Gamma \left( 1+ \frac{a_{1,\beta'} -a_{3,\alpha}}{\varepsilon_1} \right)} . \label{conmat2}
\end{align}
\end{subequations}
We will scrutinize below how the connection matrices associate the solutions to $\widehat{\widehat{\mathfrak{D}}}$ in different convergence domains, for each $N\geq 2$.
\paragraph{$\boldsymbol{N=2}$} We have shown in section \ref{npdsn2case} that the modified surface defect partition functions,
\begin{align}
\boldsymbol{\widetilde{\EuScript{Z}}}^L \equiv \left( \; \widetilde{\EuScript{Z}}_{\alpha} ^L \; \right)_{\alpha=1,2},
\end{align}
solve the differential equation \eqref{a2npdsfinal} given by $\widehat{\widehat{\mathfrak{D}}}_2$, with the prefactors in \eqref{n2expnents}. These functions provided the solutions of the form
\begin{align}
\sum_{k_1, k_2 =0} ^\infty c_{k_1, k_2 } \;z^{r_L -k_1} \; \mathfrak{q} ^{L_2 +k_2},
\end{align}
in the domain $0<\vert \mathfrak{q} \vert < 1 < \vert z \vert$. It is not so difficult to show that they are the only solutions once the critical exponent $L_2$ is given. Indeed, by directly acting $\widehat{\widehat{\mathfrak{D}}}_2$ to the ansatz and expanding in $z^{-1}$ and $\mathfrak{q}$, we get a recursive relations for the coefficients $c_{k_1, k_2}$. In particular, the zeroth order equation is
\begin{align}
0 = \left(\varepsilon_1 ^2 r_L ^2 - \varepsilon_1 (\varepsilon_1 +2 \varepsilon_2)r_L + \varepsilon_1 \varepsilon_2 \left( \frac{2\varepsilon+\varepsilon_2}{4\varepsilon_1} + \Delta_1 \right)\right)c_{0,0}.
\end{align}
The existence of the solution $(c_{0,0} \neq 0)$ implies that we are restricted to only two choices for the critical exponent $r_L$,
\begin{align} \label{critexpinf2}
\left(r_{L,\alpha} \right)_{\alpha=1,2} = \left( \frac{-a_{0,1}+a_{0,2}+\varepsilon + \varepsilon_2}{2\varepsilon_1} , \;  \frac{a_{0,1}-a_{0,2}+\varepsilon + \varepsilon_2}{2\varepsilon_1} \right),
\end{align}
which are precisely \eqref{critexpinf}. Once $r_L$ is chosen, the recursive relations fully determine all the coefficients $c_{k_1, k_2}$. Since the partition functions $\boldsymbol{\widetilde{\EuScript{Z}}}^L$ already provide two solutions, we conclude that the surface defect partition functions  $\boldsymbol{\widetilde{\EuScript{Z}}}^L$ provide all solutions to $\widehat{\widehat{\mathfrak{D}}}_2$ in the domain $0<\vert \mathfrak{q} \vert < 1 < \vert z \vert$, for each fixed $L_2$.

With the modification of the partition function by the multiplication of the prefactors \eqref{n2expnents}, the analytic continuation formula \eqref{analcont1} becomes
\begin{align}
\begin{split}
\widetilde{\EuScript{Z}}^L _\alpha &= -\sum_{\beta=1,2 }  \left(  \mathbf{C}_\infty \right)_{\alpha \beta} \left(-\frac{1}{z}\right)^{-r_{L,\alpha} + \frac{2\bar{a}_0 - 2\bar{a}_2  + \varepsilon_2}{2 \varepsilon_1}  + \frac{a_{2,\beta} -a_{0,\alpha} }{\varepsilon_1} } \;  \mathfrak{q}^{-\Delta_{\mathfrak{q}} -\Delta_0 +\frac{\varepsilon^2 -(a_{2,1}-a_{2,2})^2}{4 \varepsilon_1 \varepsilon_2}} \\
& \quad\quad\quad \left( 1-z\right)^{ \frac{2\bar{a}_0 - 2\bar{a}_2 + 2\varepsilon_1 + \varepsilon_2}{2 \varepsilon_1} }\left( 1-\frac{\mathfrak{q}}{z} \right)^{ \frac{\bar{a}_2 -\bar{a}_3 +\varepsilon}{\varepsilon_1} } \; \left( 1-\mathfrak{q} \right)^{ \frac{2(\bar{a}_2 -\bar{a}_3+\varepsilon)( 2\bar{a}_0 -2 \bar{a}_2  -\varepsilon_2 )}{\varepsilon_1 \varepsilon_2} } \EuScript{Z}_\beta ^{L \to M}.
\end{split}
\end{align}
Note that the critical exponent of $z$ is independent of $\alpha$, namely,
\begin{align} \label{critexpinf1}
\begin{split}
\left( r_{L \to M,\beta} \right)_{\beta=1,2} &\equiv \left( r_{L,\alpha} - \frac{2\bar{a}_0 - 2\bar{a}_2  + \varepsilon_2}{2 \varepsilon_1}   - \frac{a_{2,\beta} -a_{0,\alpha} }{\varepsilon_1}\right)_{\beta=1,2} \\
&= \left( \frac{-a_{2,1}+a_{2,2}+\varepsilon}{2\varepsilon_1}, \;  \frac{a_{2,1}-a_{2,2}+\varepsilon}{2\varepsilon_1}\right).\
\end{split}
\end{align}
Finally, we define the modified basis functions as
\begin{align}
\begin{split}
&\widetilde{\EuScript{Z}}_\beta ^{L\to M} \equiv -\left(-\frac{1}{z}\right)^{-r_{L \to M,\beta}  } \; \mathfrak{q}^{-\Delta_{\mathfrak{q}} -\Delta_0 +\frac{\varepsilon^2 -(a_{2,1}-a_{2,2})^2}{4 \varepsilon_1 \varepsilon_2}} \\
&\quad \quad\quad\quad \left( 1-z\right)^{ \frac{2\bar{a}_0 - 2\bar{a}_2 + 2\varepsilon_1 + \varepsilon_2}{2 \varepsilon_1} }\left( 1-\frac{\mathfrak{q}}{z} \right)^{ \frac{\bar{a}_2 -\bar{a}_3 +\varepsilon}{\varepsilon_1} } \; \left( 1-\mathfrak{q} \right)^{ \frac{2(\bar{a}_2 -\bar{a}_3+\varepsilon)( 2\bar{a}_0 -2 \bar{a}_2  -\varepsilon_2 )}{\varepsilon_1 \varepsilon_2} } \EuScript{Z}_\beta ^{L \to M}, \\
&\boldsymbol{\widetilde{\EuScript{Z}}}^{L\to M} \equiv \left( \widetilde{\EuScript{Z}}_\beta ^{L\to M}  \right)_{\beta=1,2}.
\end{split}
\end{align}
The uniqueness of analytic continuation guarantees that $\boldsymbol{\widetilde{\EuScript{Z}}}^{L\to M} $ also provides solutions to $\widehat{\widehat{\mathfrak{D}}}_2$. Therefore, the analytic continuation formula,
\begin{align}
\boldsymbol{\widetilde{\EuScript{Z}}}^L = \mathbf{C}_\infty  \boldsymbol{\widetilde{\EuScript{Z}}}^{L \to M},
\end{align}
connects the solutions to the differential operators $\widehat{\widehat{\mathfrak{D}}}_2$ in different convergence domains, through the connection matrix defined in \eqref{conmat1}.
\paragraph{Remarks}
\begin{itemize}
\item In the limit $\varepsilon_2 \to 0$, the modified functions $\boldsymbol{\EuScript{Z}} ^{L \to M}$ produce solutions to the oper $\widehat{\mathfrak{D}}$. It is evident from the expression \eqref{contq} that the solutions are again expressed as sums of the Baxter $\EuScript{Q}$-functions.
\item We observe that the critical exponent $r_{L \to M,\beta} $ of $z$ for $\widetilde{\EuScript{Z}}_\beta ^{L \to M}$ is precisely the $L_1$ in \eqref{n2expnents} subject to the constraint
\begin{align}
\begin{cases} a_{1,\beta} = a_{2,\beta} +\varepsilon_2 \\ a_{1,\alpha} = a_{2,\alpha} \;\; (\alpha \neq \beta) \end{cases}.
\end{align}
This strongly indicates the identity,
\begin{align}
\EuScript{Z}_{\beta} ^{L \to M} = \left( 1-z \right)^{- \frac{2\varepsilon}{\varepsilon_1 \varepsilon_2} \left(  \bar{a}_{0} -\bar{a}_{2}  -\varepsilon_2 \right)} \EuScript{Z}_{A_2} \left(  a_{1,\alpha} = a_{2,\alpha} +\varepsilon_2  \delta_{\alpha,\beta}  \right),
\end{align} 
between the two seemingly distinct partition functions. Even though this identity is rather obvious in the point of view of AGT \cite{agt}, its rigorous proof in the gauge theory side is not. As is clear from the definition of each side, this identity implies a lot of non-trivial combinatoric identities. It would be nice to directly prove the identity, perhaps by using the non-perturbative Dyson-Schwinger equations, but it is not necessary for our study so we do not attempt it here.
\end{itemize} 

Similarly, we have shown that the modified surface defect partition functions,
\begin{align}
\boldsymbol{\widetilde{\EuScript{Z}}}^ R \equiv \left( \widetilde{\EuScript{Z}}_\alpha ^R \right)_{\alpha=1,2},
\end{align}
give solutions to $\widehat{\widehat{\mathfrak{D}}}_2$ in the domain $0<\vert z\vert<\vert \mathfrak{q} \vert<1$, which are now of the form
\begin{align}
\sum_{k_1, k_2=0} ^{\infty} c_{k_1, k_2} \; z^{r_R +k_2} \; \mathfrak{q} ^{L_1 +k_1 -k_2} =  \sum_{k_1 , k_2=0} ^\infty c_{k_1, k_2} \; \mathfrak{q} ^{L_1 +r_R+k_1} \left( \frac{z}{\mathfrak{q}} \right)^{r_R +k_2}.
\end{align}
We can act with $\widehat{\widehat{\mathfrak{D}}}_2$ on this series and expand in $\mathfrak{q}$ and $\frac{z}{\mathfrak{q}}$, to find the indicial equation,
\begin{align}
0= \varepsilon_1 ^2 r_R ^2 -\varepsilon_1 \varepsilon \: r_R + \varepsilon_1 \varepsilon_2 \Delta_0,
\end{align}
whose solutions are precisely \eqref{critexp0}, namely,
\begin{align} \label{critexp02}
\left( r_{R,\alpha} \right)_{\alpha=1,2} \equiv \left( \frac{-a_{3,1}+a_{3,2}+\varepsilon}{2\varepsilon_1} ,\;  \frac{a_{3,1}-a_{3,2}+\varepsilon}{2\varepsilon_1}  \right).
\end{align}
Once $r_R$ is chosen, all the coefficients $c_{k_1,k_2}$ are determined recursively. Thus we conclude that $\boldsymbol{\widetilde{\EuScript{Z}}}^R$ provide the only two solutions to $\widehat{\widehat{\mathfrak{D}}}_2$ in the domain $0<\vert z\vert<\vert \mathfrak{q} \vert<1$, for each fixed $L_1 + r_R$. With the prefactors \eqref{prefactor0}, the analytic continuation formula \eqref{analcont2} becomes
\begin{align}
\begin{split}
\widetilde{\EuScript{Z}}_\alpha ^R &= -\sum_{\beta=1,2} \left( \mathbf{C}_0 \right)_{\alpha \beta} \; \left( - \frac{\mathfrak{q}}{z} \right)^{-r_{R,\alpha}-\frac{2\bar{a}_1 -2\bar{a}_3  +\varepsilon_2}{2\varepsilon_1} -\frac{a_{3,\alpha} -a_{1,\beta}}{\varepsilon_1}} \mathfrak{q} ^{\frac{\varepsilon^2 -(a_{1,1}-a_{1,2})^2}{4\varepsilon_1 \varepsilon_2} -\Delta_0 -\Delta_{\mathfrak{q}}' + \frac{2\varepsilon + \varepsilon_2}{4\varepsilon_1} } \\
&\quad\quad \left( 1-\mathfrak{q} \right)^{\frac{(\bar{a}_0 - \bar{a}_1 +\varepsilon)(2\bar{a}_1 -2\bar{a}_3 - \varepsilon_2)}{\varepsilon_1 \varepsilon_2} } \; \left( 1-z \right)^{\frac{\bar{a}_0 -\bar{a}_1 +\varepsilon}{\varepsilon_1} } \; \left(1-\frac{\mathfrak{q}}{z} \right)^{\frac{2\bar{a}_1 -2\bar{a}_3 +2 \varepsilon_1 + \varepsilon_2}{2\varepsilon_1}} \EuScript{Z}_\beta ^{R \to M} .
\end{split}
\end{align}
Note that the critical exponent for $z$ becomes, again, independent of $\alpha$,
\begin{align} \label{critexp01}
\begin{split}
\left(r_{R\to M , \beta} \right)_{\beta=1,2} &\equiv \left( r_{R,\alpha}+\frac{2\bar{a}_1 -2\bar{a}_3  +\varepsilon_2}{2\varepsilon_1} +\frac{a_{3,\alpha} -a_{1,\beta}}{\varepsilon_1} \right)_{\beta=1,2} \\
&= \left( \frac{-a_{1,1}+a_{1,2}+\varepsilon+\varepsilon_2}{2\varepsilon_1}, \; \frac{a_{1,1}-a_{1,2}+\varepsilon+\varepsilon_2}{2\varepsilon_1}\right).
\end{split}
\end{align}
Hence we define the modified basis function by
\begin{align}
\begin{split}
&\widetilde{\EuScript{Z}}_\beta ^{R \to M} \equiv -\left( -\frac{\mathfrak{q}}{z} \right)^{-r_{R \to M , \beta}} \mathfrak{q} ^{\frac{\varepsilon^2 -(a_{1,1}-a_{1,2})^2}{4\varepsilon_1 \varepsilon_2} -\Delta_0 -\Delta_{\mathfrak{q}}' + \frac{2\varepsilon + \varepsilon_2}{4\varepsilon_1} } \\
&\quad\quad\quad\quad \left( 1-\mathfrak{q} \right)^{\frac{(\bar{a}_0 - \bar{a}_1 +\varepsilon)(2\bar{a}_1 -2\bar{a}_3 - \varepsilon_2)}{\varepsilon_1 \varepsilon_2} } \; \left( 1-z \right)^{\frac{\bar{a}_0 -\bar{a}_1 +\varepsilon}{\varepsilon_1} } \; \left(1-\frac{\mathfrak{q}}{z}  \right)^{\frac{2\bar{a}_1 -2\bar{a}_3 +2 \varepsilon_1 + \varepsilon_2}{2\varepsilon_1}} \EuScript{Z}_\beta ^{R \to M}, \\
&\boldsymbol{\widetilde{\EuScript{Z}}} ^{R \to M} \equiv \left( \widetilde{\EuScript{Z}}_\beta ^{R \to M} \right)_{\beta=1,2}.
\end{split}
\end{align}
By the uniqueness of the analytic continuation, we conclude that $\boldsymbol{\widetilde{\EuScript{Z}}} ^{R \to M}$ gives the solutions to $\widehat{\widehat{\mathfrak{D}}}_2$ in the domain $0<\vert \mathfrak{q} \vert< \vert z\vert<1$. The analytic continuation formula,
\begin{align}
\boldsymbol{\widetilde{\EuScript{Z}}}^R = \mathbf{C}_0 \; \boldsymbol{\widetilde{\EuScript{Z}}}^{R \to M},
\end{align}
connects the solutions in different convergence domains, through the connection matrix defined in \eqref{conmat2}.

\paragraph{$\boldsymbol{N=3}$} Under the modification \eqref{modify3} with the prefactors \eqref{prefactors3}, the analytic continuation formula \eqref{analcont1} becomes
\begin{align}
\widetilde{\EuScript{Z}}_\alpha ^L &= -\sum_{\beta=1} ^3 \left( \mathbf{C}_\infty \right)_{\alpha\beta} \; \left(-\frac{1}{z}\right)^{-r_{L,\alpha} + \frac{3\bar{a}_0 -3\bar{a}_2 + 2\varepsilon_2 }{3\varepsilon_1} + \frac{a_{2,\beta}-a_{0,\alpha}}{\varepsilon_1} } \\
&\quad \mathfrak{q}^{-\Delta_{\mathfrak{q}} -\Delta_0 + \frac{1}{\varepsilon_1 \varepsilon_2} \left( \varepsilon^2 -\frac{(a_{2,1}-a_{2,2})^2 +(a_{2,1}-a_{2,3})^2 -(a_{2,1}-a_{2,2})(a_{2,1}-a_{2,3})}{3}  \right)  } \\
& \quad \left( 1-z\right)^{ \frac{3\bar{a}_0 -3\bar{a}_2 + 3\varepsilon-\varepsilon_2 }{3\varepsilon_1}} \; \left( 1- \frac{\mathfrak{q}}{z} \right)^{ \frac{\bar{a}_2 -\bar{a}_3 +\varepsilon}{\varepsilon_1} } \; \left( 1-\mathfrak{q} \right)^{ \frac{(\bar{a}_2 -\bar{a}_3+\varepsilon)(3\bar{a}_ -3\bar{a}_2 -\varepsilon_2) }{\varepsilon_1 \varepsilon_2} }  \EuScript{Z}_{\beta} ^{L \to M}.
\end{align}
Again, the critical exponent of $z$ is independent of $\alpha$,
\begin{align} \label{critexpinter}
\begin{split}
\left( r_{L \to M,\beta} \right)_{\beta=1} ^3 &\equiv \left( r_{L,\alpha} - \frac{3\bar{a}_0 -3\bar{a}_2 + 2\varepsilon_2 }{3\varepsilon_1} - \frac{a_{2,\beta}-a_{0,\alpha}}{\varepsilon_1} \right)_{\beta=1} ^3 \\
& = \left( \frac{-3a_{2,\beta}+ \sum_{\gamma=1} ^3 a_{2,\gamma} +3\varepsilon}{3\varepsilon_1}\right)_{\beta=1} ^3.
\end{split}
\end{align}
Hence, we define the modified basis functions as
\begin{align}
\begin{split}
&\widetilde{\EuScript{Z}}_\beta ^{L \to M} \equiv -\left(-\frac{1}{z}\right)^{-r_{L \to M,\beta}} \; \mathfrak{q}^{-\Delta_{\mathfrak{q}} -\Delta_0 + \frac{1}{\varepsilon_1 \varepsilon_2} \left( \varepsilon^2 -\frac{(a_{2,1}-a_{2,2})^2 +(a_{2,1}-a_{2,3})^2 -(a_{2,1}-a_{2,2})(a_{2,1}-a_{2,3})}{3}  \right)  } \\
& \quad\quad\quad\quad \left( 1-z\right)^{ \frac{3\bar{a}_0 -3\bar{a}_2 + 3\varepsilon-\varepsilon_2 }{3\varepsilon_1}} \; \left( 1- \frac{\mathfrak{q}}{z} \right)^{ \frac{\bar{a}_2 -\bar{a}_3 +\varepsilon}{\varepsilon_1} } \; \left( 1-\mathfrak{q} \right)^{ \frac{(\bar{a}_2 -\bar{a}_3+\varepsilon)(3\bar{a}_ -3\bar{a}_2 -\varepsilon_2) }{\varepsilon_1 \varepsilon_2} }  \EuScript{Z}_{\beta} ^{L \to M}, \\
&\boldsymbol{\widetilde{\EuScript{Z}}}  ^{L \to M} \equiv \left( \widetilde{\EuScript{Z}}_\beta ^{L \to M}\right)_{\beta=1} ^3.
\end{split}
\end{align}
Then the analytic continuation formula,
\begin{align}
\boldsymbol{\widetilde{\EuScript{Z}}}^{L} = \mathbf{C}_\infty \boldsymbol{\widetilde{\EuScript{Z}}}^{L \to M},
\end{align}
connects the solutions to $\widehat{\widehat{\mathfrak{D}}}_3$ in different converence domains.

Likewise, under the multiplication of the prefactors \eqref{prefactor3diff}, the analytic continuation formula \eqref{analcont2} becomes 
\begin{align}
\begin{split}
&\widetilde{\EuScript{Z}}_\alpha ^R = -\sum_{\beta=1} ^3 \left( \mathbf{C}_0 \right)_{\alpha \beta} \left( -\frac{\mathfrak{q}}{z} \right)^{-r_{R,\alpha}-\frac{3\bar{a}_1 -3\bar{a}_3 + 2\varepsilon_2}{3\varepsilon_1} -\frac{a_{3,\alpha}-a_{1,\beta}}{\varepsilon_1} }  (1-\mathfrak{q})^{\frac{(\bar{a}_0 -\bar{a}_1 +\varepsilon)(3\bar{a}_1 -3\bar{a}_3 -\varepsilon_2)}{\varepsilon_1 \varepsilon_2}} (1-z) ^{\frac{(\bar{a}_0 -\bar{a}_1 + \varepsilon)}{\varepsilon_1} }  \\
& \quad\quad\quad \mathfrak{q} ^{\frac{1}{\varepsilon_1 \varepsilon_2}\left( \varepsilon^2 - \frac{(a_{1,1}-a_{1,2})^2 +(a_{1,1}-a_{1,3})^2 -(a_{1,1}-a_{1,2})(a_{1,1}-a_{1,3})}{3} \right) -\Delta_{\mathfrak{q}} ' -\Delta_0 +\frac{3\varepsilon +\varepsilon_2}{3\varepsilon_1}  } \left(1-\frac{\mathfrak{q}}{z}\right)^{\frac{3\bar{a}_1 -3\bar{a}_3 +3\varepsilon -\varepsilon_2}{3\varepsilon_1}} \\
&\quad\quad\quad \EuScript{Z}_\beta ^{R \to M}.
\end{split}
\end{align}
Note that the critical exponent of $z$ becomes independent of $\alpha$, namely,
\begin{align} \label{critexpinter'}
\begin{split}
\left( r_{R \to M, \beta} \right)_{\beta=1} ^3&\equiv \left( r_{R,\alpha} +\frac{3\bar{a}_1 -3\bar{a}_3 +2\varepsilon_2}{3\varepsilon_1} +\frac{a_{3,\alpha}-a_{1,\beta}}{\varepsilon_1} \right)_{\beta=1} ^3 \\
&= \left( \frac{-3a_{1,\beta}+\sum_{\gamma=1} ^3 a_{1,\gamma}+3\varepsilon+2\varepsilon_2}{3\varepsilon_1}\right)_{\beta=1} ^3.
\end{split}
\end{align}
Therefore we modify the basis function by
\begin{align}
\begin{split}
&\widetilde{\EuScript{Z}}_\beta ^{R \to M} \equiv -\left( -\frac{\mathfrak{q}}{z} \right)^{-r_{R \to M, \beta}}  (1-\mathfrak{q})^{\frac{(\bar{a}_0 -\bar{a}_1 +\varepsilon)(3\bar{a}_1 -3\bar{a}_3 -\varepsilon_2)}{\varepsilon_1 \varepsilon_2}} (1-z) ^{\frac{\bar{a}_0 -\bar{a}_1 + \varepsilon}{\varepsilon_1} } \left(1-\frac{\mathfrak{q}}{z} \right)^{\frac{3\bar{a}_1 -3\bar{a}_3 +3\varepsilon -\varepsilon_2}{3\varepsilon_1}} \\
& \quad\quad\quad\quad\quad \mathfrak{q} ^{\frac{1}{\varepsilon_1 \varepsilon_2}\left( \varepsilon^2 - \frac{(a_{1,1}-a_{1,2})^2 +(a_{1,1}-a_{1,3})^2 -(a_{1,1}-a_{1,2})(a_{1,1}-a_{1,3})}{3} \right) -\Delta_{\mathfrak{q}} ' -\Delta_0 +\frac{3\varepsilon +\varepsilon_2}{3\varepsilon_1}  } \EuScript{Z}_\beta ^{R \to M}  \\
&\boldsymbol{\widetilde{\EuScript{Z}}} ^{R \to M} \equiv \left(\widetilde{\EuScript{Z}}_\beta ^{R \to M}\right)_{\beta=1} ^3 .
\end{split}
\end{align}
We conclude that the connection formula,
\begin{align}
\boldsymbol{\widetilde{\EuScript{Z}}} ^R = \mathbf{C}_0 \: \boldsymbol{\widetilde{\EuScript{Z}}} ^{R \to M},
\end{align}
associate the solutions in different domains, through the connection matrix \eqref{conmat2}.

\subsubsection{The shift matrix} \label{shiftmatr}
We have verified in section \ref{conmats} that the analytically continued partition functions $\boldsymbol{\widetilde{\EuScript{Z}}}^{L \to M}$ and $\boldsymbol{\widetilde{\EuScript{Z}}}^{R \to M}$ provide the solutions to the operator $\widehat{\widehat{\mathfrak{D}}}$ in the intermediate domain, $0<\vert \mathfrak{q} \vert <\vert z \vert <1$. Moreover, we have found in section \ref{z2orbifold} that the $(N-1,1)$-type $\mathbb{Z}_2$-orbifold surface defect partition functions $\boldsymbol{\EuScript{Z}}^{\mathbb{Z}_2}$ also provide the solutions to $\widehat{\widehat{\mathfrak{D}}}$ in the same domain. The question arises on how these solutions are associated to each other. Exact identities between these partition functions are established with the help of \textit{the shift matrix}
\begin{align} \label{shift}
\mathbf{S} _{\alpha\beta} \equiv  e^{\varepsilon_2 \frac{\partial}{\partial a_\alpha}} \; \delta_{\alpha \beta},
\end{align}
which is introduced to facilitate shifting the Coulomb moduli of the underlying $A_1$-theory. We proceed below with the derivation of the identities, for each $N \geq 2$.

\paragraph{$\boldsymbol{N=2}$} Let us consider the generic ansatz for $\widehat{\widehat{\mathfrak{D}}}_2$ in the intermediate domain $0<\vert \mathfrak{q} \vert <\vert z \vert <1$,
\begin{align}
\sum_{k_1, k_2 =0} ^{\infty} c_{k_1, k_2} \; z^{r_M +k_1 - k_2 } \mathfrak{q}^{L_2 +k_2} = \sum_{k_1, k_2 =0} ^\infty c_{k_1,k_2} \; z^{r_M + L_2 +k_1} \left( \frac{\mathfrak{q}}{z} \right)^{L_2 + k_2}.
\end{align}
By acting $\widehat{\widehat{\mathfrak{D}}}_2$ to the ansatz and expanding in $z$ and $\frac{\mathfrak{q}}{z}$, we find the indicial equation
\begin{align} \label{indicial}
0 = \varepsilon_1 ^2 r_M ^2 - \varepsilon_1 \varepsilon\: r_M +\varepsilon_1 \varepsilon_2 (\Delta_{\mathfrak{q}} + \Delta_0) +\varepsilon_1 \varepsilon_2 L_2.
\end{align}
Once the critical exponents $r_1$ and $L_2$ are chosen to satisfy the indicial equation, all the coefficients $c_{k_1,k_2}$ are determined recursively. The solution is unique in this sense.

We have seen that $\boldsymbol{\widetilde{\EuScript{Z}}}^{L \to M}$, $\boldsymbol{\widetilde{\EuScript{Z}}}^{R \to M}$, and $\boldsymbol{\widetilde{\EuScript{Z}}}^{\mathbb{Z}_2}$ are annihilated by $\widehat{\widehat{\mathfrak{D}}}_2$, and therefore their critical exponents evidently satisfy the indicial equation \eqref{indicial}. Moreover, we observe from \eqref{critexpinf1}, \eqref{critexp01}, \eqref{critexpz2}, \eqref{delta}, and \eqref{delta'} that
\begin{align}
&\left(r_{L \to M ,\alpha} \right)_{\alpha=1,2} = \left( r_{R \to M, \alpha} \; \Big\vert_{a_{1,\alpha} \to a_{2,\alpha} +\varepsilon_2 }  \; \right)_{\alpha=1,2} = \left( r^{\mathbb{Z}_2} _{ \alpha}  \right)_{\alpha=1,2}, \\
&\Delta_{\mathfrak{q}} = \Delta_{\mathfrak{q}} ' \; \Big\vert_{a_{1,\alpha} \to a_{2,\alpha} +\varepsilon_2 },
\end{align}
so that the indicial equation guarantees that those solutions are identical under the shift of the Coulomb moduli, namely,
\begin{align}
\boldsymbol{\widetilde{\EuScript{Z}}}^{L \to M} (\mathbf{a}) = \mathbf{S} \; \boldsymbol{\widetilde{\EuScript{Z}}}^{R \to M} (\mathbf{a}) =\boldsymbol{\widetilde{\EuScript{Z}}}^{\mathbb{Z}_2} (\mathbf{a}).
\end{align}
Note that the re-definitions \eqref{redef} of the Coulomb moduli and the masses of the hypermultiplets for $\boldsymbol{\widetilde{\EuScript{Z}}}^R $ were carefully designed to yield this equality. Consequently, we conclude that the analytically continued partition functions agree in the intermediate domain, and this is also identical to the orbifold surface defect partition function.

\paragraph{$\boldsymbol{N=3}$} From \eqref{delta3}, \eqref{lambda3}, \eqref{delta3'}, and \eqref{lambda3'}, we observe that
\begin{align} \label{condition1}
\begin{split}
&\Delta_{\mathfrak{q,1}} = \Delta_{\mathfrak{q,1}} ' \; \Big\vert_{a_{1,\alpha} \to a_{2,\alpha} +\varepsilon_2 } \\
&\Lambda_{\mathfrak{q,1}} = \Lambda_{\mathfrak{q,1}} ' \; \Big\vert_{a_{1,\alpha} \to a_{2,\alpha} +\varepsilon_2 } 
\end{split}
\end{align}
Also, from \eqref{critexpinter}, \eqref{critexpinter'}, and \eqref{critexpinterz2}, we have
\begin{align} \label{condition2}
\left(r_{L \to M ,\alpha} \right)_{\alpha=1} ^3 = \left( r_{R \to M, \alpha} \; \Big\vert_{a_{1,\alpha} \to a_{2,\alpha} +\varepsilon_2 }  \; \right)_{\alpha=1}^3 = \left( r^{\mathbb{Z}_2} _{ \alpha} \right)_{\alpha=1} ^3.  
\end{align}
Although these relations look promising, they do not guarantee the equality of the partition functions this time. The problem is that the equation for $\widehat{\widehat{\mathfrak{D}}}_3$ involves the expectation value $\Big\langle \EuScript{O}_3 \Big\rangle$, which is an object independent of the partition function itself. Without an additional information on equating the expectation values analytically continued from different domains, the single equation of $\widehat{\widehat{\mathfrak{D}}}_3$ is not enough to fully determine the partition function. Nevertheless, in the limit $\varepsilon_2 \to 0$ the equation is reduced to the oper $\widehat{\mathfrak{D}}_3$ on $\mathbb{P}^1 \backslash \{0,\underline{\mathfrak{q}}, \underline{1}, \infty \}$, and the relations \eqref{condition1} and \eqref{condition2} are indeed enough to guarantee that the solutions agree with each other. This is because, as we have seen earlier, the expectation value  $\Big\langle \EuScript{O}_3 \Big\rangle$ is dominated by the limit shape and becomes a series only in $\mathfrak{q}$, comprising an accessory parameter for the oper $\widehat{\mathfrak{D}}_3$ which is unambiguously determined once the monodromy along the $A$-cycle is fixed.

We furthermore suspect that even for generic values of $\varepsilon_2$, there is a proper matching between the analytically continued expectation values in the intermediate domain, so that the identities,
\begin{align} \label{identify}
\boldsymbol{\widetilde{\EuScript{Z}}}^{L \to M} (\mathbf{a}) = \mathbf{S} \; \boldsymbol{\widetilde{\EuScript{Z}}}^{R \to M} (\mathbf{a}) =\boldsymbol{\widetilde{\EuScript{Z}}}^{\mathbb{Z}_2} (\mathbf{a}),
\end{align}
persist to be true. We have checked the identities at low orders in the gauge couplings $z$ and $\mathfrak{q}$. We discuss more on this issue in section \ref{dis}.

\paragraph{Remarks}
\begin{itemize}
\item The duality between the quiver-type and the orbifold-type surface defects was realized in \cite{frenguktes} as the M-theory brane transition, for the $A_1$-theories. It would be interesting to study the relation between the higher rank generalization of the duality in \cite{frenguktes} and the exact identification of the partition functions \eqref{identify}.
\end{itemize}

\section{Darboux coordinates} \label{sec:gennrs}
Recall that the main assertion of \cite{nrs} is that the generating function for the variety of opers with respect to the NRS coordinate system is identical to the effective twisted superpotential of a class-$\EuScript{S}$ theory:
\begin{align}
\mathcal{S} \left[ \mathcal{O}_N [{\uCalC}]\right] = \frac{1}{\varepsilon_1}\left( \widetilde{\EuScript{W}} \left[ \mathcal{T}[A_{N-1}, {\uCalC}] \right] - \widetilde{\EuScript{W}}_{\infty} \right).
\end{align}
We need a generalization of the NRS coordinates for $N >2$ to give any meaning to  the  left hand side of the correspondence. 

Here, we propose a Darboux coordinate system on the moduli space of flat $SL(N)$-connections on the $r+3$-punctured sphere $\mathbb{P}^{1}_{ 2, \underline{r+1}}$ with two maximal and $r+1$ minimal punctures, for the arbitrary higher rank $N-1$. The proposed coordinates reduce to the usual NRS coordinate system in $N=2$ on a specific patch of the moduli space of flat connections.

In this section ${\uCalC}_{r}$ denotes $\mathbb{P}^{1}_{ 2, \underline{r+1}} = \mathbb{P}^{1} \backslash \{ z_{-1}, \underline{z_{0}, \ldots , z_{r}} , z_{r+1} \}$. We often set 
$z_{-1} = \infty$, $z_{r+1} = 0$, and $z_0 = 1$. 

\subsection{Construction of Darboux coordinates} \label{subsec:constgennrs}
\subsubsection{Definition}
We construct  Darboux coordinates on a patch of the moduli space of flat $SL(N)$-connections on the $r+3$-punctured sphere ${\uCalC}_{r}$, which reduces to the NRS coordinates in the $N=2$ case. Our main example of the four-punctured sphere is the case $r=1$. As in \eqref{flatmoduli}, the moduli space $\EuScript{M}_{\text{flat}} (SL(N), {\uCalC}_{r})$ is the space of (stable) equivalence classes of the homomorphisms of the fundamental group of the punctured Riemann sphere to $SL(N)$, in which the loops encircling each puncture are mapped to the prescribed conjugacy classes in $SL(N)$. In particular, the two maximal punctures correspond to generic semisimple conjugacy classes in $SL(N)$, while the $r+1$ minimal punctures correspond to semisimple conjugacy classes in $SL(N)$ with maximally degenerate eigenvalues. We fix the conjugacy classes by specifying the eigenvalues of the holonomy matrices $g_{i}$, $i = -1, 0, 1, \ldots, r+1$. The moduli space is given by:
\small
\begin{align}
\begin{split}
&\EuScript{M}_{\text{flat}} (SL(N), {\uCalC}_{r})  \\
&= \begin{cases} \\
\\ \; \left( g_{i} \right)_{i=-1}^{r+1} \\   \\ \\
\end{cases} \Biggl\vert \;\; \begin{rcases}   g_{i} \in SL(N) \, , \\
 {\rm Det}(g_{i} - x) = ({\fm}_{i} - x)^{N-1} ({\fm}_{i}^{1-N} - x) \, , \ i = 0, \ldots , r \\
\\
{\rm Det}(g_{i} - x) = \prod\limits_{\alpha = 1}^{N} \left( \left( {\fm}_{i}^{({\alpha})}\right)^{-{\rm sgn}(i)} - x \right) \, , \ i = -1, r+1 \\
g_{-1}  g_0 \cdots g_{r+1} = \mathds{1}_N \\ 
\\ \end{rcases}^{\rm stable} \Bigg/ SL(N).
\end{split}
\end{align}
\normalsize
The stability condition chooses an open subset in the set of matrices $g_{i}$ obeying all of the conditions above. We shall not need to specify the stability condition since we are going to work on an open patch of the moduli space which belongs to the stable subset. 

The holonomies  $g_i \sim \text{diag}(\fm_i,\cdots ,\fm_i,\fm_i ^{-N+1})$ around the minimal punctures require more detailed notation. We can form such an element of $SL(N)$ by setting
\begin{align}
g_i = \fm_i \left( \mathds{1}_N + \left(\fm_i ^{-N} -1 \right) E_i \otimes \tilde{E}_i \right),
\end{align}
where
\begin{align} \label{eq:egvec}
\begin{split}
&E_i \in \mathbb{C}^N,\; \tilde{E} _i \in \left(\mathbb{C}^N \right)^* \text{ (dual space)} \\
&\tilde{E}_i (E_i) =1,
\end{split}
\end{align}
which are defined up to rescaling $\left( E_i, \tilde{E}_i \right) \mapsto \left( t_i E_i, t_i ^{-1} \tilde{E}_i \right)$, $t_i \in \mathbb{C}^\times$. For fixed $\tilde{E}_i$, its null subspace in $\mathbb{C}^N$ is $N-1$-dimensional. Hence we have $N-1$-dimensional eigenspace of $g_i$ with the eigenvalue $\fm_i$. The one last eigenvector is given by $E_i$, with the eigenvalue $\fm_i ^{-N+1}$ fixed by the normalization condition. The number of degrees of freedom in such a $g_i$ is equal to
\begin{align}
2N \text{ (from } E \text{ and } \tilde{E} )- 1 \text{ (normalization)} - 1 \text{ (rescaling)} = 2(N-1). 
\end{align}
Therefore, a simple dimension count gives
\begin{align}
\begin{split}
\dim{\EuScript{M}_{\text{flat}} (SL(N), {\mathbb{P}} ^1 _{2 , \underline{r+1}})} &= 2 \left( (N^2-1) - (N-1) \right) + (r+1) \left( 2(N-1) \right) - 2(N^2-1) \\
&= 2r (N-1).
\end{split}
\end{align}

\begin{figure} 
\centering
\begin{tikzpicture}
\node[inner sep=0pt] (figure) at (0,0) {\includegraphics[width=0.75\textwidth]{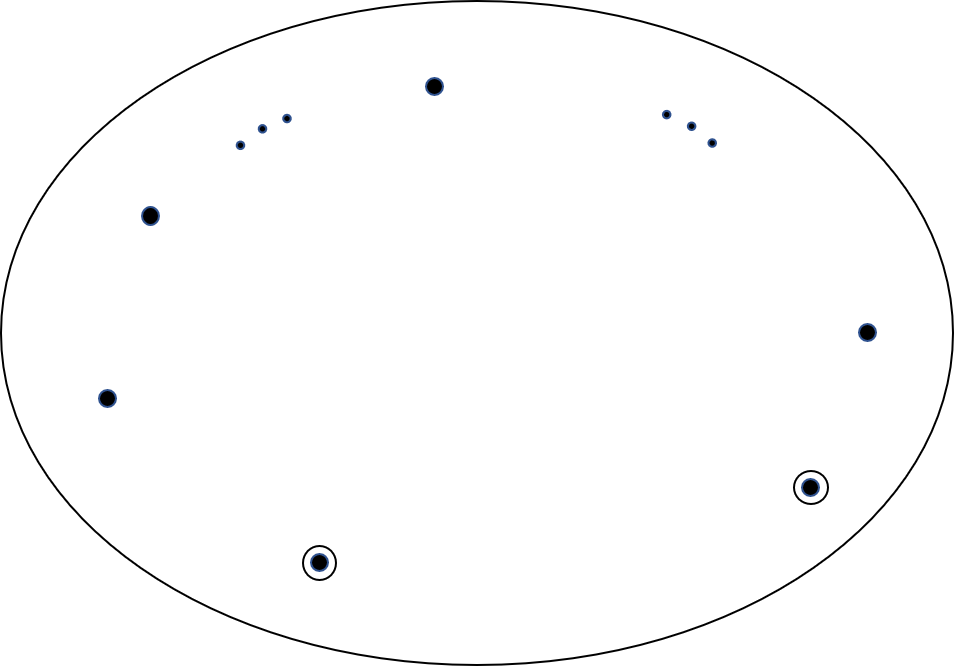}};
\node at (-2,-3.4) {\scalebox{1.0}{$-1$}};
\node at (-4.5,-1.3) {\scalebox{1.0}{$0$}};
\node at (-3.98,1) {\scalebox{1.0}{$1$}};
\node at (-0.5,2.6) {\scalebox{1.0}{$i$}};
\node at (4.8,-0.5) {\scalebox{1.0}{$r$}};
\node at (4.1,-2.4) {\scalebox{1.0}{$r+1$}};

\draw[blue,very thick,decoration={markings, mark=at position 0.8 with {\arrow[scale=1.5]{latex}}}, postaction={decorate}] (-0.5,2.8) circle (0.7cm);
\node at (-0.5,1.6) {\scalebox{1.2}{$\textcolor{blue}{g_i}$}};

\draw [red,very thick,decoration={markings, mark=at position 0.95 with {\arrow[scale=1.5]{latex}}}, postaction={decorate}] (0,0) to[out=45,in=-40] (-0.1,3.7) to[out=160,in=40] (-2.5,3.2) to[out=210,in=70] (-5,1) to[out=260,in=110] (-5.1,-1.6) to[out=310,in=190] (-1,-3.6) to[out=35,in=210] (0,0);
\node at (0,-1) {\scalebox{1.2}{$\textcolor{red}{M_i}$}};

\end{tikzpicture}
\caption{The $r+3$-punctured sphere $\mathbb{P}_{2, \underline{r+1}} ^1$. The $(-1)$-th puncture (located at $z=\infty$) and the $(r+1)$-th puncture (located at $z=0$) are maximal, denoted by double circles, while all the other punctures are minimal, denoted by simple dots. The holonomy along the loop encircling each puncture is represented by $g_i$ (blue line), while the holonomy along the loop enclosing $i+2$ punctures is represented by $M_i$ (red line).} \label{fig:sphere}
\end{figure}

We need to define $r(N-1)$-pairs of coordinates which are canonical under the Poisson bracket. For this, it is convenient to parametrize the moduli space as follows. Let us define the projection operators
\begin{align}
\begin{split}
&\Pi_i = E_i \otimes \tilde{E}_i, \quad i = 0,1, \cdots, r \\
& \Pi_i ^2 = \Pi_i,
\end{split}
\end{align}
formed by the eigenvector $E_i$ \eqref{eq:egvec} of $g_i$ and its dual-vector. Then $g_i$ is expressed as
\begin{align}
& g_{i} = {\fm}_{i} \left( \mathds{1}_N  + \left( {\fm}_{i}^{-N}-1 \right) {\Pi}_{i} \right) \, , \qquad i = 0, 1, \ldots, r.
\end{align}
For later use, we also give the expression for its inverse:
\begin{align}
g_i ^{-1} = \fm_i ^{-1} \left( \mathds{1}_N + (\fm_i ^N-1) \Pi_i \right) \, , \qquad i = 0, 1, \ldots, r.
\end{align}
Let us also define
\begin{align}
& M_{i} \equiv g_{-1}g_{0} \ldots g_{i} \in SL(N), \quad i =-1, 0, 1, \cdots, r+1.
\end{align}
These matrices represent the holonomies along the curves on the $r+3$-punctured sphere enclosing $i+2$ punctures (see Figure \ref{fig:sphere}).
In particular, it is immediate that we have $M_{-1} = g_{-1}$, $M_{r} = g_{r+1} ^{-1}$, and $M_{r+1} = \mathds{1}_N$. We can express these matrices as
\beq
\begin{aligned}
& M_{i} = \sum_{\alpha=1}^{N} \, {\fm}_{i}^{(\alpha)}\, {\Pi}_{i}^{(\alpha)} \, , \\
\end{aligned}
\label{eq:gmrel}
\eeq
with the projection operators $\Pi_i ^{(\alpha)}$ obeying
\beq
{\Pi}_{i}^{(\alpha)}{\Pi}^{(\beta)}_{i} = {\delta}_{\alpha,\beta} \, {\Pi}^{(\alpha)}_{i},
\eeq
each having rank one. Using the eigenbasis $E_{i}^{(\alpha)} \in \mathbb{C}^N$, $i = 0, 1, \ldots , r-1$ of $M_i$, and its dual-basis ${\tilde E}_{i}^{(\beta)} \in \left( \mathbb{C} ^N \right)^{*}$, ${\tilde E}_{i}^{(\alpha)}(E_{i}^{(\beta)}) = {\delta}_{\alpha,\beta}$, we can write
\beq
{\Pi}_{i}^{(\alpha)} = E_{i}^{(\alpha)} \otimes {\tilde E}_{i}^{(\alpha)}.
\eeq
The basis vectors are defined up to rescalings $\left( E_{i}^{(\alpha)} , {\tilde E}_{i}^{(\alpha)} \right) \mapsto
\left( t_{i}^{(\alpha)} E_{i}^{(\alpha)}, \left(t_{i}^{(\alpha)}\right)^{-1} {\tilde E}_{i}^{(\alpha)}\right)$, $t_{i}^{(\alpha)} \in {\BC}^{\times}$, and reorderings
$E_{i}^{(\alpha)} \mapsto E_{i}^{({\sigma}_{i}(\alpha))}$, ${\sigma}_{i} \in {\CalS}(N)$.

Now we are ready to propose a Darboux coordinate system. We define the coordinates ${\boldsymbol\alpha}_{i}^{(\alpha)}$, ${\tilde{\boldsymbol\beta}}_{i}^{(\alpha)}$, 
$i = 0, 1, \ldots , r-1$, $\alpha = 1, \ldots , N$, subject to the constraints
\beq \label{eq:slnconst}
\sum_{\alpha=1}^{N} {\boldsymbol\alpha}_{i}^{(\alpha)}  = 0
\eeq
and defined up to the shifts 
\beq
{\tilde{\boldsymbol\beta}}_{i}^{(\alpha)} \mapsto {\tilde{\boldsymbol\beta}}_{i}^{(\alpha)} + b_{i} \, , \qquad b_{i} \in \BC
\label{eq:betamb}
\eeq 
via
\beq
M_{i}\, \, E_{i}^{(\alpha)} = e^{2\pi\ii {\boldsymbol\alpha}_{i}^{(\alpha)}} \, E_{i}^{(\alpha)}
\label{eq:alphaco}
\eeq
and
\beq
e^{-{\tilde{\boldsymbol\beta}}_{i}^{(\alpha)}+{\tilde{\boldsymbol\beta}}_{i}} =\frac{{\tilde E}_{i}^{(\alpha)} (E_{i+1})} {{\tilde E}_{i}^{(\alpha)} (E_{i})} {\tilde E}_{i+1}
(E_{i}) = \frac{\text{Tr}_N \Pi_i \Pi_i ^{(\alpha)} \Pi_{i+1}}{\text{Tr}_N \Pi_i \Pi_i ^{(\alpha)}},
\label{eq:betaii}
\eeq
where
$\tilde{\boldsymbol\beta}_i$ is defined by:
\beq \label{eq:betaover}
e^{\tilde{\boldsymbol\beta}_i} = \sum_{\alpha=1}^{N} e^{{\tilde{\boldsymbol\beta}}_{i}^{(\alpha)}} {\Tr}_{N} \left( 
{\Pi}_{i+1}{\Pi}_{i}^{(\alpha)} \right).
\eeq
Due to the constraint \eqref{eq:slnconst} and the ambiguity \eqref{eq:betamb}, the coordinates ${\boldsymbol\alpha}_{i}^{(\alpha)}$, ${\tilde{\boldsymbol\beta}}_{i}^{(\alpha)}$ are redundant. Thus we refine the coordinates by choosing mutually independent $r(N-1)$-pairs
\begin{align} \label{eq:coordref}
\left(\boldsymbol\alpha_i ^{(\alpha)}, \; \boldsymbol\beta_i ^{(\alpha)} \equiv \tilde{\boldsymbol\beta}_i ^{(\alpha)} - \tilde{\boldsymbol\beta}_i ^{(N)} \right), \quad i=0,1, \cdots, r-1, \, \alpha =1, \cdots, N-1.
\end{align}
to form a proper coordinate system on $\EuScript{M}_\text{flat} (SL(N), \mathbb{P}_{2, \underline{r+1}} ^1)$.

\subsubsection{Canonical Poisson relations}
To show that $\left\{ \boldsymbol\alpha_i ^{(\alpha)}, \boldsymbol\beta_i ^{(\alpha)} \;\vert\; i=0,1, \cdots, r-1, \, \alpha=1, \cdots, N-1 \right\}$ forms a Darboux coordinate system on $\EuScript{M}_\text{flat} (SL(N), \mathbb{P}_{2, \underline{r+1}} ^1)$, we have to verify that the Poisson brackets \cite{fr} are canonical\footnote{It is clear that the Poisson brackets for the refined coordinates \eqref{eq:coordref} are also canonical once \eqref{eq:canpoi} is proven.}
\begin{align} \label{eq:canpoi}
\begin{split}
&\Biggl\{ \,  {\tilde{\boldsymbol\beta}}_{i}^{(\alpha)}    \, , \,  {\boldsymbol\alpha}_{j}^{(\beta)}  \, \Biggr\} = {\delta}_{i,j} {\delta}_{\alpha,\beta} \\
&\Biggl\{ \boldsymbol\alpha_i ^{(\alpha)} , \boldsymbol\alpha_j ^{(\beta)} \Biggl\} = \Biggl\{ \tilde{\boldsymbol\beta}_i ^{(\alpha)} , \tilde{\boldsymbol\beta}_j ^{(\beta)} \Biggl\} =0,
\end{split}  \begin{split} &i, j=0, 1, \cdots, r-1, \\ &\alpha, \beta=1, \cdots, N. \end{split}
\end{align}
The Poisson bracket on the space of all gauge fields
\beq
\Biggl\{ \mathcal{A}^{a} (x), \mathcal{A}^{b} (y) \Biggr\} = {\delta}^{ab}  {\delta}^{(2)}(x,y) 
\eeq
(the $\delta^{(2)}$ is a two-form on $\mathbb{P}_{2,\underline{r+1}} ^1$) has a simple geometric description when represented on the holonomies. To illustrate, consider two distinct elements of the fundamental group $\left[\gamma_{1,2} \right] \in \pi_1 \left( \mathbb{P}_{2, \underline{r+1}} ^1 \right)$. We can choose their representatives $\gamma_{1,2}$ to intersect transversally. We assign to each intersection point $x \in \gamma_1 \cap \gamma_2$ a sign
\begin{align}
\begin{split}
&s:\gamma_{1} \cap \gamma_2 \longrightarrow \{\pm\} \\
\end{split}
\end{align}
according to the orientation of the curves $\gamma_{1,2}$ at $x$ relative to the orientation of the sphere (see Figure \ref{fig:resolution}).
\begin{figure}
\centering
\begin{tikzpicture}
\draw[very thick,blue,decoration={markings, mark=at position 0.8 with {\arrow[scale=1]{latex}}}, postaction={decorate}] (-6,0) -- (-3,0);
\draw[very thick,red,decoration={markings, mark=at position 0.8 with {\arrow[scale=1]{latex}}}, postaction={decorate}] (-4.5,-1.5) -- (-4.5,1.5);
\draw [->] (-2.5,0) -- (-2,0);
\node at (-3.7,-0.4) {\scalebox{1.0}{$\textcolor{blue}{\gamma_1}$}};
\node at (-4.9,0.7) {\scalebox{1.0}{$\textcolor{red}{\gamma_2}$}};
\node at (-4.75,-0.3) {\scalebox{1.0}{$+$}};
\draw[very thick,decoration={markings, mark=at position 0.5 with {\arrow[scale=1]{latex}}}, postaction={decorate}] (-0.2,-1.3) arc (-180:-270:1.8cm);
\draw[very thick,decoration={markings, mark=at position 0.5 with {\arrow[scale=1]{latex}}}, postaction={decorate}] (-1.3,-0.2) arc (270:360:1.8cm);

\draw[very thick,red,decoration={markings, mark=at position 0.8 with {\arrow[scale=1]{latex}}}, postaction={decorate}] (2.5,0) -- (5.5,0);
\draw[very thick,blue,decoration={markings, mark=at position 0.8 with {\arrow[scale=1]{latex}}}, postaction={decorate}] (4,-1.5) -- (4,1.5);
\draw [->] (6,0) -- (6.5,0);
\draw[very thick,decoration={markings, mark=at position 0.5 with {\arrow[scale=1]{latex}}}, postaction={decorate}] (8,-1.3) arc (-180:-270:1.8cm);
\draw[very thick,decoration={markings, mark=at position 0.5 with {\arrow[scale=1]{latex}}}, postaction={decorate}] (6.9,-0.2) arc (270:360:1.8cm);
\node at (4.8,-0.4) {\scalebox{1.0}{$\textcolor{red}{\gamma_2}$}};
\node at (3.65,0.7) {\scalebox{1.0}{$\textcolor{blue}{\gamma_1}$}};
\node at (3.75,-0.3) {\scalebox{1.0}{$-$}};
\end{tikzpicture} \caption{The sign assignment to intersection points and the resolution of the union of curves.}\label{fig:resolution}
\end{figure}
Then we define
\begin{align}
\left(\gamma_1 \cap \gamma_2 \right)^\pm \equiv \left\{ x \in \gamma_{1} \cap \gamma_2 \; \vert \; s(x) = \pm \right\}.
\end{align}
At each intersection $x$, we compose a resolution $\left( \gamma_1 \cup \gamma_2 \right)_x$ of the union of the curves as described in Figure \ref{fig:resolution}. Now the Poisson structure on the moduli space of flat connections can be represented on the holonomies $\rho$ along $\gamma_{1,2}$ by
\begin{align} \label{eq:poigeo}
\Biggl\{ \rho ([\gamma_1]) , \rho ([\gamma_2]) \Biggr\} = \sum_{x\in (\gamma_1 \cap \gamma_2)^+} \rho \left([(\gamma_1 \cup \gamma_2)_x]\right) - \sum_{x\in (\gamma_1 \cap \gamma_2)^-} \rho \left([(\gamma_1 \cup \gamma_2)_x]\right).
\end{align}

Using the geometric description of the Poisson structure, we can show that the coordinates defined in \eqref{eq:alphaco} and \eqref{eq:betaii} satisfy the canonical Poisson relations \eqref{eq:canpoi}. Let us package \eqref{eq:alphaco} into the generating function:
\beq
{\BA}_{i}(x) \equiv {\Tr}_{N} \, \left( x - M_{i}\right)^{-1} = \sum_{l=0} ^\infty \frac{1}{x^{l+1}} \Tr_N \, M_i ^l, \eeq
which has a simple geometric meaning as the generating function of the loops which wind along the same curve (whose holonomy is represented by $M_i$) multiple times. Since there is no intersection among these curves, it is clear that we have
\begin{align}
\Biggl\{ {\BA}_{i}(x) , {\BA}_{j}(y) \Biggr\} = 0,
\end{align}
for any $i,j = 0, 1, \cdots, r-1$. Thus we derive
\begin{align}
\Biggl\{ \boldsymbol\alpha_i ^{(\alpha)}, \boldsymbol\alpha_j ^{(\beta)} \Biggr\} = 0,
\end{align}
for any $i,j = 0,1, \cdots, r-1$, $\alpha, \beta =1, \cdots, N$.

We can also package \eqref{eq:betaii} into
\beq
{\BB}_{i}(x) \equiv {\Tr}_{N}\, {\Pi}_{i} \left( x- M_{i}\right)^{-1} {\Pi}_{i+1} = e^{\tilde{\boldsymbol\beta}_i} \sum_{\alpha=1}^{N} e^{-{\tilde{\boldsymbol\beta}}_{i}^{(\alpha)}} \, \frac{{\Tr}_{N} \, {\Pi}_{i}{\Pi}_{i}^{(\alpha)}}{x-m_{i}^{(\alpha)}}.
\eeq
We can re-express this via:
\begin{align} \label{eq:dfunction}
\begin{split}
{\BD}_{i}(x) &\equiv {\Tr}_{N}\, g_{i} \left( x- M_{i}\right)^{-1} g_{i+1} \\
&= {\fm}_{i}{\fm}_{i+1} ({\fm}_{i}^{-N}-1) ({\fm}_{i+1}^{-N}-1) 
{\BB}_{i}(x) + \fm_i \fm_{i+1} x^{-1} \left( \frac{P_{i-1} (\fm_i ^{-1} x)}{P_i (x)} -1 \right) \\
& \quad - \fm_i \fm_{i+1} ^{1-N} x^{-1} \left( \frac{P_{i+1} (\fm_{i+1} x)}{P_i (x)} -1 \right) + \fm_i \fm_{i+1} \BA_i (x),
\end{split}
\end{align}
where $P_{i}(x)$ is the characteristic polynomial of $M_i$:
\begin{align}
\begin{split}
P_{i}(x) = {\Det}(x - M_{i}) = \prod_{\alpha=1}^{N} \left( x - {\fm}_{i}^{(\alpha)} \right).
\end{split}
\end{align}
In deriving the second equality of \eqref{eq:dfunction}, we had simple manipulations on the determinants\footnote{Use that for any rank one projector $\Pi$, and any operator $A$, ${\Det}(1+A {\Pi} ) = 1 + {\Tr} (A {\Pi})$} and \eqref{eq:gmrel}:
\beq
\frac{P_{i-1}({\fm}_{i}^{-1}x)}{P_{i}(x)} - 1 = x \left( 1 - {\fm}_{i}^{-N}\right) {\Tr} (M_{i}-x)^{-1} {\Pi}_{i} 
\label{eq:piimi}
\eeq
\beq
\frac{P_{i}({\fm}_{i}x)}{P_{i-1}(x)} - 1 = x \left( 1 - {\fm}_{i}^{N} \right) {\Tr} (M_{i-1}-x)^{-1} {\Pi}_{i}.
\label{eq:piimipo}
\eeq 
The function ${\BD}_{i}(x)$ has a simple geometric 
meaning:
\beq
{\BD}_{i}(x) = \sum_{l=0}^{\infty} \frac{1}{x^{l+1}} {\Tr}_{N} \, g_{i} M_{i}^{l} g_{i+1}
\eeq
from which it is obvious that $\{ {\BD}_{i}(x) , {\BA}_{j}(y) \} = 0$ for $i \neq j$
(the corresponding loops on the $r+3$-punctured sphere do not intersect), as
well as that $\{ {\BD}_{i}(x), {\BD}_{j}(y) \} = 0$ for $|i-j| > 1$. From these, we derive
\begin{align}
\begin{split}
&\Biggl\{ \tilde{\boldsymbol\beta}_i ^{(\alpha)} , \boldsymbol\alpha_j ^{(\beta)} \Biggr\} = 0, \quad i\neq j, \; \alpha,\beta = 1, \cdots, N, \\
&\Biggl\{ \tilde{\boldsymbol\beta}_i ^{(\alpha)}, \tilde{\boldsymbol\beta}_j ^{(\beta)} \Biggr\} =0, \quad \vert i-j \vert >1, \; \alpha, \beta =1, \cdots, N.
\end{split}
\end{align}
It remains to compute:
\beq
\Biggl\{ {\BD}_{i}(x) , {\BA}_{i}(y) \Biggr\}, \quad  \Biggl\{ {\BD}_{i}(x), {\BD}_{i+1}(y) \Biggr\}, \quad \text{and} \quad \Biggl\{ {\BD}_i (x), {\BD}_{i}(y) \Biggr\},
\eeq
which are a bit more involved. As we elaborate in appendix \ref{appD} in detail, the rest of the canonical Poisson relations \eqref{eq:canpoi} are obtained out of these brackets, confirming that the proposed coordinate system is indeed Darboux.

\paragraph{Remarks}
\begin{itemize}
\item  Other constructions generalizing the NRS-type coordinates were proposed in the $SL(2)$ case in \cite{hn}, in the arbitrary group case in \cite{fg3}, and specifically in the $SL(3)$ case in \cite{hol}. In \cite{hn} and \cite{hol}, the spectral coordinates are defined as the holonomies of a (twisted) flat $GL(1)$-connection on a line bundle over the $N$-fold branched covering $\Sigma$ of the Riemann surface $\EuScript{C}$, which is a certain uplift (called \textit{abelianization}) of the flat $SL(N)$-connection on $\EuScript{C}$ \cite{gmn2}. To give some credit to our construction, as described above, it produces the Darboux coordinates for arbitrary $N$ in an elementary fashion, albeit only on a specific patch of the moduli space. 
\end{itemize}

\subsection{The four-punctured sphere} \label{subsec:fourpunc}
\begin{figure} 
\centering
\begin{tikzpicture}
\node[inner sep=0pt] (figure) at (0,0) {\includegraphics[width=0.8\textwidth]{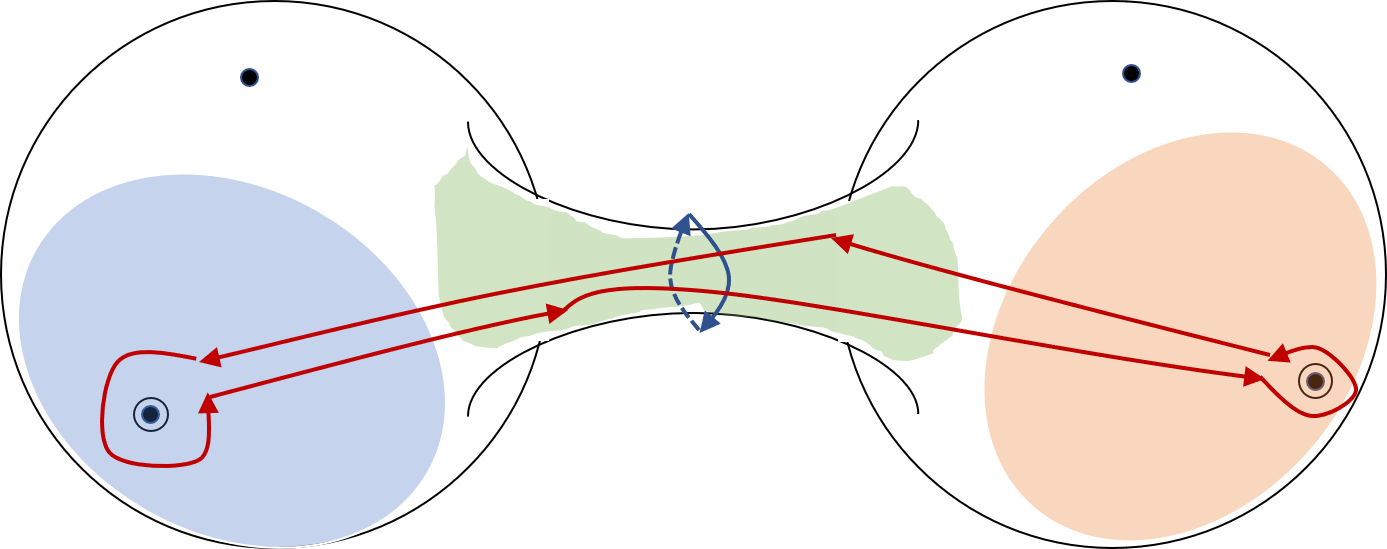}};
\node at (-4.9,-1.6) {\scalebox{1.0}{$\infty$}};
\node at (-4.2,2.05) {\scalebox{1.0}{$1$}};
\node at (4.15,2.05) {\scalebox{1.0}{$\mathfrak{q}$}};
\node at (5.5,-1.3) {\scalebox{1.0}{$0$}};

\node at (-4.9,0.3) {\scalebox{1.5}{$\textcolor{blue}{L}$}};
\node at (5,0.65) {\scalebox{1.5}{$\textcolor{red}{R}$}};
\node at (-1.8,0.4) {\scalebox{1.5}{$\textcolor{teal}{M}$}};
\node at (0,0.8) {\scalebox{1.2}{$\textcolor{navyblue}{A}$}};
\node at (2.65,0.18) {\scalebox{1.2}{$\textcolor{firebrick}{B}$}};
\end{tikzpicture} 
\caption{The $A$-cycle and the $B$-cycle on the four-punctured sphere $\mathbb{P}^1 \backslash \{0,\underline{\mathfrak{q}},\underline{1}, \infty \}$. The double circles represent the maximal punctures at $0$ and $\infty$, while the simple dots represent the minimal punctures at $\mathfrak{q}$ and $1$. The shaded regions represent the convergence domains $L$, $M$, and $R$, respectively. The $A$-cycle is represented by the dark blue line, while the $B$-cycle is represented by the dark red line. } \label{fig2}
\end{figure}

We consider our main example, the four-punctured sphere $\mathbb{P} ^1 \backslash \{0,\underline{\mathfrak{q}},\underline{1},\infty \} $. The generalized NRS coordinate system on the moduli space $\EuScript{M}_\text{flat} (SL(N), \mathbb{P} ^1 \backslash \{0,\underline{\mathfrak{q}},\underline{1},\infty\} )$ is just a special case $r=1$ of the one defined in the previous section. In this special case, it is convenient to express the generalized NRS coordinates in terms of the trace invariants of the holonomies, and take these expressions as equivalent definitions for those coordinates. Since the dimension of the moduli space is $\dim \EuScript{M}_\text{flat} (SL(N), \mathbb{P} ^1 \backslash \{0,\underline{\mathfrak{q}},\underline{1},\infty\} ) = 2(N-1)$, it is enough to consider two independent cycles on $\mathbb{P} ^1 \backslash \{0,\underline{\mathfrak{q}},\underline{1},\infty \} $ which we choose to be the $A$-cycle and the $B$-cycle in Figure \ref{fig2}. We describe how the traces of the holonomies $M_{A,B}$ along these cycles are expressed in terms of the generalized NRS coordinates, for $N=2$ and $N=3$.
\subsubsection{$SL(2)$}
We start with the $A$-cycle. It is clear that we have
\begin{align}
M_A = M_0 ^{-1} = \sum_{\alpha=1} ^2 \left( \fm_0 ^{(\alpha)} \right)^{-1} \Pi_0 ^{(\alpha)}.
\end{align}
Thus we find
\begin{align}
\text{Tr}\, M_A = \left( \fm_0 ^{(1)} \right)^{-1} + \left(  \fm_0 ^{(2)} \right)^{-1} = e^{-2\pi i \boldsymbol\alpha_0 ^{(1)}} + e^{-2\pi i \boldsymbol\alpha_0 ^{(2)}}.
\end{align}
It is convenient to omit the superscript and write $\boldsymbol\alpha \equiv \boldsymbol\alpha_0 ^{(1)} = - \boldsymbol\alpha_0 ^{(2)}$. Thus we have
\begin{align} \label{eq:tra2}
\text{Tr}\, M_A = 2 \cos 2 \pi \boldsymbol\alpha.
\end{align}
Next, we can express the holonomy along the $B$-cycle as
\begin{align}
\begin{split}
M_B &= g_2 g_{-1} = g_1 ^{-1} g_0 ^{-1} \\
&= \fm_0 ^{-1} \fm_1 ^{-1} \left(\mathds{1}_2 +(\fm_1 ^2 -1) \Pi_1\right) \left(\mathds{1}_2 +(\fm_0 ^2 -1)\Pi_0\right).
\end{split}
\end{align}
Thus we find
\begin{align}
\begin{split}
&\text{Tr}\, M_B = \fm_0 \fm_1 ^{-1} + \fm_0 ^{-1} \fm_1 + (\fm_0 - \fm_0 ^{-1})(\fm_1 -\fm_1 ^{-1}) \, \text{Tr}\, \Pi_0 \Pi_1
\end{split}
\end{align}
Note that we can express the trace in the last term using the $\boldsymbol\beta$ coordinates,
\begin{align}
\begin{split}
\text{Tr} \, \Pi_0 \Pi_1 &= \sum_{\alpha=1} ^2 \text{Tr}\, \Pi_0 \Pi_0 ^{(\alpha)} \Pi_1 \\
&= \sum_{\alpha=1} ^2 e^{-\tilde{\boldsymbol\beta}_0 ^{(\alpha)} + \tilde{\boldsymbol\beta}_0} \, \text{Tr} \, \Pi_0 \Pi_0 ^{(\alpha)} \\
&= \sum_{\alpha=1} ^2 \text{Tr}\, \Pi_0 \Pi_0 ^{(\alpha)} \, \text{Tr}\, \Pi_1 \Pi_0 ^{(\alpha)} \\
& \quad + e^{\tilde{\boldsymbol\beta}_0 ^{(1)} - \tilde{\boldsymbol\beta}_0 ^{(2)}} \text{Tr}\, \Pi_0 \Pi_0 ^{(2)}\, \text{Tr}\, \Pi_1 \Pi_0 ^{(1)} + e^{\tilde{\boldsymbol\beta}_0 ^{(2)} - \tilde{\boldsymbol\beta}_0 ^{(1)}} \text{Tr}\, \Pi_0 \Pi_0 ^{(1)} \, \text{Tr}\, \Pi_1 \Pi_0 ^{(2)},
\end{split}
\end{align}
where we have used \eqref{eq:betaover} in the third equality. Using \eqref{eq:piimi} and \eqref{eq:piimipo}, we can express the rest of the traces in terms of the $\boldsymbol\alpha$ coordinates. For simplicity, let us define $\fm_{-1} \equiv \fm_{-1} ^{(1)}$ and $\fm_2 \equiv \fm_1 ^{(2)}$.\footnote{Not to be confused with the eigenvalue $\fm_2$ of $g_2$ which appears when $r>1$. Here, we restrict ourselves only to the case $r=1$ and there would be no confusion in notation.} Also we refine the $\boldsymbol\beta$ coordinate according to the definition \eqref{eq:coordref}: $\boldsymbol\beta \equiv \tilde{\boldsymbol\beta}_0 ^{(1)} - \tilde{\boldsymbol\beta}_0 ^{(2)}$. Then the final expression for the trace of the holonomy along the $B$-cycle is
\begin{align} \label{eq:trb2}
\begin{split}
&\text{Tr}\, M_B = \frac{\left( \fm_2 + \fm_2 ^{-1} -\fm_1 -\fm_1 ^{-1} \right) \left( \fm_{-1} + \fm_{-1} ^{-1} -\fm_0 -\fm_0 ^{-1} \right) }{8 \sin ^2  \pi \boldsymbol\alpha} \\
&\quad \quad\quad + \frac{\left( \fm_2 + \fm_2 ^{-1} +\fm_1 +\fm_1 ^{-1} \right) \left( \fm_{-1} + \fm_{-1} ^{-1} +\fm_0 +\fm_0 ^{-1} \right) }{8 \cos^2  \pi \boldsymbol\alpha} \\
& \;\; - \sum_{\pm} \frac{ \left( e^{\mp 2\pi i \boldsymbol\alpha} -\fm_0 \fm_{-1} \right) \left( \fm_0 ^{-1} e^{\mp 2\pi i \boldsymbol\alpha} - \fm_{-1} ^{-1} \right)\left( e^{ \pm 2\pi i \boldsymbol\alpha} - \fm_1 ^{-1} \fm_2 ^{-1} \right) \left( \fm_1 e^{\pm 2\pi i \boldsymbol\alpha} - \fm_2 \right) }{4 \sin ^2 2\pi \boldsymbol\alpha} e^{\pm \boldsymbol\beta}.
\end{split}
\end{align}
Thus we observe that the coordinates $(\boldsymbol\alpha, \boldsymbol\beta)$ determine the traces of the holonomies along the $A$-cycle and $B$-cycle on $\mathbb{P} ^1 \backslash \{0,\mathfrak{q},1,\infty \} $ by \eqref{eq:tra2} and \eqref{eq:trb2}. Conversely, we may take these formulas as defining equations for the coordinates $(\boldsymbol\alpha, \boldsymbol\beta)$.
\paragraph{Remarks}
\begin{itemize}
\item For a direct comparison with the coordinates defined in \cite{nrs}, let us define
\begin{align}
x_1 = \fm_{-1} + \fm_{-1} ^{-1}, \quad x_2 = \fm_0 + \fm_0 ^{-1}, \quad  x_3 = \fm_2 + \fm_2 ^{-1}, \quad x_4 = \fm_1+\fm_1 ^{-1}.
\end{align}
Also we use the abbreviation
\begin{align}
A = \text{Tr}\, M_A, \quad B= \text{Tr}\, M_B.
\end{align}
Then we find that \eqref{eq:trb2} becomes
\begin{align} \label{eq:nrsbeq}
\begin{split}
B(A^2-4) &= 2(x_1 x_4 +x_2 x_3) - A(x_1 x_3 +x_2 x_4) \\
& \quad + \left( e^{\boldsymbol\beta} + e^{-\boldsymbol\beta} \right) \sqrt{c_{12} (A) c_{34} (A)},
\end{split}
\end{align}
where
\begin{align}
c_{ij} (A) \equiv A^2 - A \, x_i x_j + x_i ^2 +x_j ^2 -4,
\end{align}
under the canonical transformation
\begin{align} \label{eq:cantransnrs}
\begin{split}
\boldsymbol\beta \rightarrow \boldsymbol\beta &+ \frac{1}{2} \log \left( e^{- 2\pi i \boldsymbol\alpha} -\fm_0 \fm_{-1} \right) \left( \fm_0 ^{-1} e^{- 2\pi i \boldsymbol\alpha} - \fm_{-1} ^{-1} \right)\left( e^{  2\pi i \boldsymbol\alpha} - \fm_1 ^{-1} \fm_2 ^{-1} \right) \left( \fm_1 e^{ 2\pi i \boldsymbol\alpha} - \fm_2 \right)\\
&- \frac{1}{2} \log \left( e^{ 2\pi i \boldsymbol\alpha} -\fm_0 \fm_{-1} \right) \left( \fm_0 ^{-1} e^{ 2\pi i \boldsymbol\alpha} - \fm_{-1} ^{-1} \right)\left( e^{ - 2\pi i \boldsymbol\alpha} - \fm_1 ^{-1} \fm_2 ^{-1} \right) \left( \fm_1 e^{- 2\pi i \boldsymbol\alpha} - \fm_2 \right).
\end{split}
\end{align}
The relation \eqref{eq:nrsbeq} is precisely the defining equation for the NRS coordinate $\boldsymbol\beta$ for the four-punctured sphere. Thus we confirm that the Darboux coordinate system proposed in section \ref{subsec:constgennrs} is a higher-rank generalization of the NRS coordinate system.

As we will see in section \ref{mono}, the canonical transformation \eqref{eq:cantransnrs} amounts to change the boundary contribution to the effective twisted superpotential. Although the transformed coordinates may be natural in some context, we will find in section \ref{mono} that our original definition is more natural in the gauge theoretical context. Therefore we stick to our original definition of the generalized NRS coordinates in section \ref{subsec:constgennrs} without making additional canonical transformation throughout the discussion.
\end{itemize}

\subsubsection{$SL(3)$}
We begin with the $A$-cycle holonomy which is clearly conjugate to
\begin{align}
M_A = M_0 ^{-1} = \sum_{\alpha=1} ^3 \left(\fm_0 ^{(\alpha)} \right)^{-1} \Pi_0 ^{(\alpha)}.
\end{align}
Thus we find
\begin{align}
\text{Tr}\, M_A ^{\pm 1} = \sum_{\alpha=1} ^3 \left(\fm_0 ^{(\alpha)} \right)^{\mp 1} = e^{\mp 2\pi i \boldsymbol\alpha_0 ^{(1)}} + e^{\mp 2\pi i \boldsymbol\alpha_0 ^{(2)}} + e^{\pm 2\pi i (\boldsymbol\alpha_0 ^{(1)} + \boldsymbol\alpha_0 ^{(2)})}.
\end{align}
For notational convenience, let us define the coordinates without superscripts,
\begin{align}
\begin{split}
&\boldsymbol\alpha_\alpha \equiv \boldsymbol\alpha_0 ^{(\alpha)}, \quad \alpha=1,2,3. \\
&\boldsymbol\alpha_3 = -\boldsymbol\alpha_1-\boldsymbol\alpha_2,
\end{split}
\end{align}
so that we have
\begin{align} \label{eq:tra3}
\text{Tr}\, M_A ^{\pm 1} = e^{\mp 2\pi i \boldsymbol\alpha_1} +e^{\mp 2\pi i \boldsymbol\alpha_2} +e^{\pm 2\pi i( \boldsymbol\alpha_1 + \boldsymbol\alpha_2)}.
\end{align}

The expressions for the holonomy along the $B$-cycle and its inverse are
\begin{align}
\begin{split}
M_B ^{\pm 1} &= \left( g_1 ^{-1} g_0 ^{-1} \right) ^{\pm 1} \\
&= \fm_0 ^{\mp 1} \fm_1 ^{\mp 1} \left(\mathds{1}_3 +(\fm_1 ^{\pm 3} -1) \Pi_1\right) \left(\mathds{1}_3 +(\fm_0 ^{\pm 3} -1)\Pi_0\right).
\end{split}
\end{align}
Thus we obtain
\begin{align}
\begin{split}
&\text{Tr}\, M_B ^{\pm 1} = \fm_0 ^{\mp 1} \fm_1 ^{\mp 1} (1+\fm_0 ^{\pm 3}+\fm_1 ^{\pm 3}) + (\fm_0 ^{\pm 2} - \fm_0 ^{\mp 1})(\fm_1 ^{\pm 2} -\fm_1 ^{\mp 1}) \, \text{Tr}\, \Pi_0 \Pi_1.
\end{split}
\end{align}
Again, we can express the trace in the last term using the $\boldsymbol\beta$ coordinates,
\begin{align}
\begin{split}
\text{Tr} \, \Pi_0 \Pi_1 &= \sum_{\alpha=1} ^3 \text{Tr}\, \Pi_0 \Pi_0 ^{(\alpha)} \Pi_1 \\
&= \sum_{\alpha=1} ^3 e^{-\tilde{\boldsymbol\beta}_0 ^{(\alpha)} +\tilde{\boldsymbol\beta}_0} \, \text{Tr} \, \Pi_0 \Pi_0 ^{(\alpha)} \\
&= \sum_{\alpha=1} ^3 \text{Tr}\, \Pi_0 \Pi_0 ^{(\alpha)} \, \text{Tr}\, \Pi_1 \Pi_0 ^{(\alpha)} + \sum_{\alpha\neq\beta} e^{\tilde{\boldsymbol\beta}_0 ^{(\alpha)} -\tilde{\boldsymbol\beta}_0 ^{(\beta)}} \text{Tr}\, \Pi_0 \Pi_0 ^{(\beta)} \, \text{Tr}\, \Pi_1 \Pi_0 ^{(\alpha)}.
\end{split}
\end{align}
The rest of the traces can be expressed in terms of the $\boldsymbol\alpha$ coordinates by using \eqref{eq:piimi} and \eqref{eq:piimipo}. We also write the refined $\boldsymbol\beta$ coordinates without superscripts,
\begin{align}
\boldsymbol\beta_\alpha \equiv \tilde{\boldsymbol\beta}_0 ^{(\alpha)} - \tilde{\boldsymbol\beta}_0 ^{(3)}, \quad \alpha=1,2.
\end{align}
Then the final expressions for the traces of the holonomies along the $B$-cycle are
\begin{align} \label{eq:trb3}
\begin{split}
\text{Tr}\, M_B ^{\pm1} &= B_0 ^{\pm} + B_{12} ^\pm e^{\boldsymbol\beta_1 -\boldsymbol\beta_2} + B_{13} ^\pm e^{\boldsymbol\beta_1} + B_{23} ^\pm e^{\boldsymbol\beta_2} \\
&\quad\quad\;\;\, + B_{21} ^\pm e^{-\boldsymbol\beta_1 +\boldsymbol\beta_2} + B_{31} ^\pm e^{-\boldsymbol\beta_1} + B_{32} ^\pm e^{-\boldsymbol\beta_2},
\end{split}
\end{align}
where
\begin{align}
\begin{split}
&B_0 ^+ = \fm_0 ^{-1} \fm_1 ^{-1} +\fm_0 ^2 \fm_1 ^{-1} +\fm_0 ^{-1} \fm_1 ^2 \\
&\quad\quad\; - \frac{\fm_0 ^2 \fm_1 ^{-1}}{16} \sum_{\alpha=1} ^3 \frac{\prod_{\gamma=1} ^3 \left( \fm_0 ^{-1} e^{\pi i \boldsymbol\alpha_\alpha} -\fm_{-1} ^{(\gamma)} e^{-\pi i \boldsymbol\alpha_\alpha} \right) \left( \fm_1 e^{\pi i \boldsymbol\alpha_\alpha} -\fm_1 ^{(\gamma)} e^{-\pi i \boldsymbol\alpha_\alpha} \right) }{ \prod_{\alpha' \neq\alpha} \sin^2 \pi (\boldsymbol\alpha_\alpha -\boldsymbol\alpha_{\alpha'} ) } \\
&B_0 ^- = \fm_0  \fm_1  +\fm_0 ^{-2} \fm_1  +\fm_0  \fm_1 ^{-2} \\
&\quad\quad\; - \frac{\fm_0 \fm_1 ^{-2}}{16} \sum_{\alpha=1} ^3 \frac{\prod_{\gamma=1} ^3 \left( \fm_0 ^{-1} e^{\pi i \boldsymbol\alpha_\alpha} -\fm_{-1} ^{(\gamma)} e^{-\pi i \boldsymbol\alpha_\alpha} \right) \left( \fm_1 e^{\pi i \boldsymbol\alpha_\alpha} -\fm_1 ^{(\gamma)} e^{-\pi i \boldsymbol\alpha_\alpha} \right) }{ \prod_{\alpha' \neq\alpha} \sin^2 \pi (\boldsymbol\alpha_\alpha -\boldsymbol\alpha_{\alpha'} ) }
\end{split}
\end{align}
and
\begin{align}
\begin{split}
&B_{\alpha \beta} ^+ = - \frac{\fm_0 ^{2} \fm_1 ^{-1}}{16} \, \frac{\prod_{\gamma=1} ^3 \left( \fm_0 ^{-1} e^{\pi i \boldsymbol\alpha_\beta} - \fm_{-1} ^{(\gamma)}  e^{-\pi i \boldsymbol\alpha_\beta} \right)\left( \fm_1 e^{\pi i \boldsymbol\alpha_\alpha} -\fm_1 ^{(\gamma)} e^{-\pi i \boldsymbol\alpha_\alpha} \right) }{\prod_{\alpha' \neq \alpha} \sin \pi (\boldsymbol\alpha_{\alpha'} -\boldsymbol\alpha_\alpha ) \prod_{\beta' \neq \beta}  \sin \pi (\boldsymbol\alpha_\beta-\boldsymbol\alpha_{\beta'})   } \\
&B_{\alpha \beta} ^- = - \frac{\fm_0  \fm_1 ^{-2}}{16} \, \frac{\prod_{\gamma=1} ^3 \left( \fm_0 ^{-1} e^{\pi i \boldsymbol\alpha_\beta} - \fm_{-1} ^{(\gamma)}  e^{-\pi i \boldsymbol\alpha_\beta} \right)\left( \fm_1 e^{\pi i \boldsymbol\alpha_\alpha} -\fm_1 ^{(\gamma)} e^{-\pi i \boldsymbol\alpha_\alpha} \right) }{\prod_{\alpha' \neq \alpha} \sin \pi (\boldsymbol\alpha_{\alpha'} -\boldsymbol\alpha_\alpha ) \prod_{\beta' \neq \beta}  \sin \pi (\boldsymbol\alpha_\beta-\boldsymbol\alpha_{\beta'})   }.
\end{split}
\end{align}
Therefore, we obtain the traces of the holonomies along the $A$-cycle and $B$-cycle on $\mathbb{P} ^1  \backslash \{0, \underline{\mathfrak{q}}, \underline{1}, \infty \}$ expressed in terms of the generalized NRS coordinates in \eqref{eq:tra3} and \eqref{eq:trb3}. We can conversely regard these formulas as the defining equations for the generalized NRS coordinates $\{\boldsymbol\alpha_\alpha, \boldsymbol\beta_\alpha \, \vert\, \alpha=1,2 \}$.

\paragraph{Remarks}
\begin{itemize}
\item After our work has been completed and submitted to the arXiv, we were informed that the generalized Fenchel-Nielsen spectral coordinates constructed in \cite{hol} are equivalent to the ones obtained here in \eqref{eq:tra3} and \eqref{eq:trb3}, up to some simple shifts for the $\boldsymbol\beta$ coordinates. Since our construction is elementary and does not use the auxiliary constructs such as the Seiberg-Witten curve disguised in the form of the spectral network, we may hope that more general spectral network constructions of Darboux coordinates could be simplified as well.  In this way we expect our coordinates to match with the (generalized) Fenchel-Nielsen spectral coordinates in \cite{hn, hol} and their higher-rank analogues \cite{fg3}, possibly up to some simple shifts.
\end{itemize}

\section{Monodromies and generating functions of opers} \label{mono}
Finally, we compute the monodromies of opers to find the expressions for the generalized NRS coordinates restricted to the variety of opers. Since the variety of opers is a Lagrangian submanifold in the moduli space of flat connections and the generalized NRS coordinates form a Darboux coordinate system, there exists generating function $\mathcal{S}\left[\mathcal{O}_N [\mathbb{P}_{2,\underline{r+1}} ^1]\right]$ for the variety $\mathcal{O}_N [\mathbb{P}_{2,\underline{r+1}} ^1]$ of opers with respect to the generalized NRS coordinates:
\begin{align}
\boldsymbol\beta_i ^{(\alpha)} = \frac{\partial \mathcal{S}\left[\mathcal{O}_N [\mathbb{P}_{2,\underline{r+1}} ^1]\right]}{\partial \boldsymbol\alpha_i ^{(\alpha)}}, \quad i=0,1, \cdots r-1,\, \alpha=1,\cdots, N-1.
\end{align}
We verify that the generating function for the variety of opers is identified with the effective twisted superpotential of the corresponding class-$\EuScript{S}$ theory $\mathcal{T}\left[A_{N-1}, \mathbb{P}_{2,\underline{r+1}} ^1 \right]$, for the example of the four-punctured sphere $\mathbb{P} ^1 \backslash \{0, \underline{\mathfrak{q}},\underline{1},\infty\} $.

The strategy to compute the monodromies of the oper $\widehat{\mathfrak{D}}_N$ is to study the holonomy of the operator $\widehat{\widehat{\mathfrak{D}}}_N$, and then take the limit $\varepsilon_2 \to 0$. The monodromy along the $A$-cycle is easy to compute: as noted in the section \ref{congl}, the solution $\boldsymbol{\widetilde{\EuScript{Z}}}^{L \to M}$ is defined in the domain $0<\vert \mathfrak{q} \vert <\vert z \vert <1$ (it is easy to estimate the growth of coefficients of $z$-expansion to conclude it converges there). Thus we simply continue along the path
\begin{align} \label{shiftz}
z \longrightarrow z\, e^{it} \quad \text{with} \quad0 \leq  t \leq 2\pi,
\end{align}
to enclose the punctures at $0$ and $\mathfrak{q}$, thereby making the $A$-cycle. In this fashion we pick up the multiplicative factors from the non-integral part of the exponent of $z$ in the perturbative prefactor, and thereby obtain the holonomy $\mathbf{M}_A (\widehat{\widehat{\mathfrak{D}}})$. The monodromy of the oper $\widehat{\mathfrak{D}}$ is then computed by taking the limit:
\begin{align}
M_A (\widehat{\mathfrak{D}}) = \lim_{\varepsilon_2 \to 0} \mathbf{M}_A (\widehat{\widehat{\mathfrak{D}}}).
\end{align}

The monodromy along the $B$-cycle is more involved. First we need \textit{the rotation matrices} $\mathbf{R}_0$ and $\mathbf{R}_\infty$ which are the monodromy matrices for the $2\pi$-rotations around the punctures at $0$ and $\infty$. As noted in section \ref{ssnpds}, the solutions $\boldsymbol{\widetilde{\EuScript{Z}}}^L$ and $\boldsymbol{\widetilde{\EuScript{Z}}}^R$ are defined by gauge theory as series expansions in the domains $0 <\vert \mathfrak{q} \vert <1 <\vert z\vert $ and $0< \vert z \vert < \vert \mathfrak{q} \vert < 1$, respectively. Thus we get $\mathbf{R}_0$ by following $\boldsymbol{\widetilde{\EuScript{Z}}}^R$ along the circle $z \mapsto z \; e^{2 \pi i}$, and we get $\mathbf{R}_\infty$ by following $\boldsymbol{\widetilde{\EuScript{Z}}}^L$  along the circle  $z \mapsto z \; e^{-2 \pi i}$. For completeness, we give the expressions for the rotation matrices for the punctures at $\mathfrak{q}$ and $1$ also: 
\begin{align} 
\begin{split}
&\mathbf{R}_{\mathfrak{q}} = \mathbf{M}_A \, \mathbf{C}_0 ^{-1}\, \mathbf{R}_0 ^{-1}\, \mathbf{C}_0 \\ 
&\mathbf{R}_1 = \mathbf{M}_A ^{-1}\, \mathbf{C}_\infty ^{-1}\, \mathbf{R}_\infty ^{-1}\, \mathbf{C}_\infty.  
\end{split}
\end{align}
It is immediate to see, in the $N=3$ case for example, that the eigenvalues of $\lim_{\varepsilon_2 \to 0} \mathbf{R}_{\mathfrak{q}}$ and $\lim_{\varepsilon_2 \to 0} \mathbf{R}_1$ are maximally degenerate, which means they correspond to minimal punctures. Now for the $B$-monodromy matrix, we start from the solution $\boldsymbol{\widetilde{\EuScript{Z}}}^L $. By concatenating the connection matrices, the shift matrices, and the rotation matrices, we construct the following sequence of continuations of the solutions
\begin{align}
\boldsymbol{\widetilde{\EuScript{Z}}}^L \xrightarrow{\mathbf{C}_\infty} \boldsymbol{\widetilde{\EuScript{Z}}}^{L \to M}  \xrightarrow{\mathbf{S}} \boldsymbol{\widetilde{\EuScript{Z}}}^{R \to M} \xrightarrow{\mathbf{C}_0 ^{-1}} \boldsymbol{\widetilde{\EuScript{Z}}}^R \xrightarrow{\mathbf{R}_0 ^{-1}} \boldsymbol{\widetilde{\EuScript{Z}}}^R \xrightarrow{\mathbf{C}_0} \boldsymbol{\widetilde{\EuScript{Z}}}^{R \to M} \xrightarrow{\mathbf{S}^{-1}}  \boldsymbol{\widetilde{\EuScript{Z}}}^{L \to M} \xrightarrow{\mathbf{C}_\infty ^{-1}} \boldsymbol{\widetilde{\EuScript{Z}}}^L \xrightarrow{\mathbf{R}_\infty ^{-1}} \boldsymbol{\widetilde{\EuScript{Z}}}^L.
\end{align}
Hence the corresponding holonomy is
\begin{align} \label{bmono1}
\mathbf{M}_B ( \widehat{\widehat{\mathfrak{D}}} ) = \mathbf{R}_\infty \; \mathbf{C}_\infty \; \mathbf{S} \; \mathbf{C}_0 ^{-1} \; \mathbf{R}_0  \; \mathbf{C}_0 \; \mathbf{S}^{-1} \; \mathbf{C}_\infty ^{-1}.
\end{align}
We have seen in section \ref{npdsquiver} that under the NS limit, the solutions $\widetilde{\boldsymbol{\EuScript{Z}}}^L $ for $\widehat{\widehat{\mathfrak{D}}}$ behaves as
\begin{align}
\widetilde{\boldsymbol{\EuScript{Z}}}^L = e^{\frac{\widetilde{\EuScript{W}}}{\varepsilon_2}}\; \left( \boldsymbol{\chi} + \mathcal{O}(\varepsilon_2 ) \right),
\end{align}
which leads to the equation for the oper, $\widehat{\mathfrak{D}} \boldsymbol{\chi}=0$. Therefore, we compute the $B$-monodromy for the oper $\widehat{\mathfrak{D}}$ as
\begin{align} \label{bmono2}
\begin{split}
M_B (\widehat{\mathfrak{D}}) &= \lim_{\varepsilon_2 \to 0}  \mathbf{M}_B ( \widehat{\widehat{\mathfrak{D}}} ) \; e^{\frac{\widetilde{\EuScript{W}}}{\varepsilon_2}} \\
&= \lim_{\varepsilon_2 \to 0}  \mathbf{R}_\infty \; \mathbf{C}_\infty \; \mathbf{S} \; \mathbf{C}_0 ^{-1} \; \mathbf{R}_0  \; \mathbf{C}_0 \; \mathbf{S}^{-1} \; \mathbf{C}_\infty ^{-1} \; e^{\frac{\widetilde{\EuScript{W}}}{\varepsilon_2}}.
\end{split}
\end{align}
Now we exhibit in detail how these computations can actually be done.

\subsection{$SL(2)$-oper}
We obtain the $A$-monodromy matrix for $\widehat{\widehat{\mathfrak{D}}}_2$ by letting $z$ make the full circle $z \mapsto z \; e^{2 \pi i}$ in the expression for $\boldsymbol{\widetilde{\EuScript{Z}}}^{L \to M}$. Since the critical exponent for $z$ is given as \eqref{critexpinf1}, we find that
\begin{align}
\mathbf{M}_A (\widehat{\widehat{\mathfrak{D}}}_2) = \text{diag} \left( e^{2 \pi i \frac{-a_1 + a_2 + \varepsilon}{2\varepsilon_1} }, \; e^{2 \pi i \frac{a_1 - a_2 + \varepsilon}{2\varepsilon_1} } \right).
\end{align}
Then by taking the NS limit ($\varepsilon_1 \neq 0, \varepsilon_2 \to 0$), we obtain the $A$-monodromy matrix for the oper $\widehat{\mathfrak{D}}_2$,
\begin{align}
\begin{split}
M_A (\widehat{\mathfrak{D}}_2) &= \lim_{\varepsilon_2 \to 0} \mathbf{M}_A (\widehat{\widehat{\mathfrak{D}}}_2) \\
&=  \text{diag} \left( e^{2 \pi i \frac{-a_1 + a_2 +\varepsilon_1 }{2\varepsilon_1} }, \; e^{2 \pi i \frac{a_1 - a_2 +\varepsilon_1 }{2\varepsilon_1} } \right),
\end{split}
\end{align}
so that
\begin{align} \label{sl2operamono}
\text{Tr}\; M_A (\widehat{\mathfrak{D}}_2) =  2 \cos 2 \pi \left( \frac{a_1-a_2+\varepsilon_1}{2\varepsilon_1} \right).
\end{align}
Comparing this with \eqref{eq:tra2}, we obtain
\begin{align}
\boldsymbol\alpha = \frac{a_1-a_2+\varepsilon_1}{2\varepsilon_1}.
\end{align}

Next, we find the expression for the $\boldsymbol{\beta}$ coordinate by computing the $B$-monodromy matrix. It is necessary to compute the rotation matrices first, by shifting $z \mapsto z \; e^{-2 \pi i}$ and $z \mapsto z \; e^ {2 \pi i}$ for $\widetilde{\boldsymbol{\EuScript{Z}}}^L $ and $\widetilde{\boldsymbol{\EuScript{Z}}}^R$, respectively. Since their critical exponents are given as \eqref{critexpinf2} and \eqref{critexp02}, we immediately compute
\begin{align}
\begin{split}
&\mathbf{R}_\infty = \text{diag}\left( e^{ \pi i \frac{a_{0,1}-a_{0,2} - \varepsilon_1 -2 \varepsilon_2}{ \varepsilon_1}}, \;e^{ \pi i \frac{-a_{0,1}+a_{0,2} - \varepsilon_1 -2 \varepsilon_2}{ \varepsilon_1}}   \right), \\
&\mathbf{R}_0 = \text{diag}\left( e^{ \pi i \frac{ -a_{3,1}+a_{3,2} + \varepsilon }{ \varepsilon_1}}, \;e^{ \pi i \frac{a_{3,1}-a_{3,2} + \varepsilon}{ \varepsilon_1}} \right).
\end{split}
\end{align}
The connection matrices and the shift matrices are given by \eqref{contmat} and \eqref{shift}. For $N=2$, it is easier to write these matrices explicitly. In particular, the connection matrices are
\begin{align}
\begin{split}
&\mathbf{C}_\infty = \begin{bmatrix} \frac{\Gamma \left( 1+ \frac{a_{0,1} -a_{0,2}}{\varepsilon_1} \right)\; \Gamma \left( \frac{a_1 -a_2}{\varepsilon_1} \right) }{\Gamma \left( 1+\frac{a_{0,1}-a_2}{\varepsilon_1} \right) \; \Gamma \left(  \frac{a_1-a_{0,2}}{\varepsilon_1} \right) }  & \frac{\Gamma \left( 1+ \frac{a_{0,1} -a_{0,2}}{\varepsilon_1} \right)\; \Gamma \left( \frac{a_2 -a_1}{\varepsilon_1} \right) }{\Gamma \left( 1+\frac{a_{0,1}-a_1}{\varepsilon_1} \right) \; \Gamma \left(  \frac{a_2-a_{0,2}}{\varepsilon_1} \right) }   \\[15pt] \frac{\Gamma \left( 1+ \frac{a_{0,2} -a_{0,1}}{\varepsilon_1} \right)\; \Gamma \left( \frac{a_1 -a_2}{\varepsilon_1} \right) }{\Gamma \left( 1+\frac{a_{0,2}-a_2}{\varepsilon_1} \right) \; \Gamma \left(  \frac{a_1-a_{0,1}}{\varepsilon_1} \right) }   &  \frac{\Gamma \left( 1+ \frac{a_{0,2} -a_{0,1}}{\varepsilon_1} \right)\; \Gamma \left( \frac{a_2 -a_1}{\varepsilon_1} \right) }{\Gamma \left( 1+\frac{a_{0,2}-a_1}{\varepsilon_1} \right) \; \Gamma \left(  \frac{a_2-a_{0,1}}{\varepsilon_1} \right) }  \end{bmatrix}, \\
&\mathbf{C}_0 = \begin{bmatrix} \frac{\Gamma \left( 1+ \frac{a_{3,2} -a_{3,1}}{\varepsilon_1} \right)\; \Gamma \left( \frac{a_2 -a_1}{\varepsilon_1} \right) }{\Gamma \left( 1+\frac{a_2-a_{3,1}}{\varepsilon_1} \right) \; \Gamma \left(  \frac{a_{3,2}-a_1}{\varepsilon_1} \right) }  & \frac{\Gamma \left( 1+ \frac{a_{3,2} -a_{3,1}}{\varepsilon_1} \right)\; \Gamma \left( \frac{a_1 -a_2}{\varepsilon_1} \right) }{\Gamma \left( 1+\frac{a_1-a_{3,1}}{\varepsilon_1} \right) \; \Gamma \left(  \frac{a_{3,2}-a_2}{\varepsilon_1} \right) }  \\[15pt] \frac{\Gamma \left( 1+ \frac{a_{3,1} -a_{3,2}}{\varepsilon_1} \right)\; \Gamma \left( \frac{a_2 -a_1}{\varepsilon_1} \right) }{\Gamma \left( 1+\frac{a_2-a_{3,2}}{\varepsilon_1} \right) \; \Gamma \left(  \frac{a_{3,1}-a_1}{\varepsilon_1} \right) }  &  \frac{\Gamma \left( 1+ \frac{a_{3,1} -a_{3,2}}{\varepsilon_1} \right)\; \Gamma \left( \frac{a_1 -a_2}{\varepsilon_1} \right) }{\Gamma \left( 1+\frac{a_1-a_{3,2}}{\varepsilon_1} \right) \; \Gamma \left(  \frac{a_{3,1}-a_2}{\varepsilon_1} \right) } \end{bmatrix}.
\end{split}
\end{align}
Their inverses can also be computed directly as
\begin{align}
\begin{split}
&\mathbf{C}_\infty ^{-1} = \frac{a_1-a_2}{a_{0,1}-a_{0,2}} \begin{bmatrix}  \frac{\Gamma \left( 1+ \frac{a_{0,2} -a_{0,1}}{\varepsilon_1} \right)\; \Gamma \left( \frac{a_2 -a_1}{\varepsilon_1} \right) }{\Gamma \left( 1+\frac{a_{0,2}-a_1}{\varepsilon_1} \right) \; \Gamma \left(  \frac{a_2-a_{0,1}}{\varepsilon_1} \right) }   & -\frac{\Gamma \left( 1+ \frac{a_{0,1} -a_{0,2}}{\varepsilon_1} \right)\; \Gamma \left( \frac{a_2 -a_1}{\varepsilon_1} \right) }{\Gamma \left( 1+\frac{a_{0,1}-a_1}{\varepsilon_1} \right) \; \Gamma \left(  \frac{a_2-a_{0,2}}{\varepsilon_1} \right) } \\[15pt] -\frac{\Gamma \left( 1+ \frac{a_{0,2} -a_{0,1}}{\varepsilon_1} \right)\; \Gamma \left( \frac{a_1 -a_2}{\varepsilon_1} \right) }{\Gamma \left( 1+\frac{a_{0,2}-a_2}{\varepsilon_1} \right) \; \Gamma \left(  \frac{a_1-a_{0,1}}{\varepsilon_1} \right) } &   \frac{\Gamma \left( 1+ \frac{a_{0,1} -a_{0,2}}{\varepsilon_1} \right)\; \Gamma \left( \frac{a_1 -a_2}{\varepsilon_1} \right) }{\Gamma \left( 1+\frac{a_{0,1}-a_2}{\varepsilon_1} \right) \; \Gamma \left(  \frac{a_1-a_{0,2}}{\varepsilon_1} \right) }  \end{bmatrix}, \\
&\mathbf{C}_0 ^{-1} =  \frac{a_1-a_2}{a_{3,1}-a_{3,2}}  \begin{bmatrix}\frac{\Gamma \left( 1+ \frac{a_{3,1} -a_{3,2}}{\varepsilon_1} \right)\; \Gamma \left( \frac{a_1 -a_2}{\varepsilon_1} \right) }{\Gamma \left( 1+\frac{a_1-a_{3,2}}{\varepsilon_1} \right) \; \Gamma \left(  \frac{a_{3,1}-a_2}{\varepsilon_1} \right) }    & - \frac{\Gamma \left( 1+ \frac{a_{3,2} -a_{3,1}}{\varepsilon_1} \right)\; \Gamma \left( \frac{a_1 -a_2}{\varepsilon_1} \right) }{\Gamma \left( 1+\frac{a_1-a_{3,1}}{\varepsilon_1} \right) \; \Gamma \left(  \frac{a_{3,2}-a_2}{\varepsilon_1} \right) }  \\[15pt] - \frac{\Gamma \left( 1+ \frac{a_{3,1} -a_{3,2}}{\varepsilon_1} \right)\; \Gamma \left( \frac{a_2 -a_1}{\varepsilon_1} \right) }{\Gamma \left( 1+\frac{a_2-a_{3,2}}{\varepsilon_1} \right) \; \Gamma \left(  \frac{a_{3,1}-a_1}{\varepsilon_1} \right) }  &  \frac{\Gamma \left( 1+ \frac{a_{3,2} -a_{3,1}}{\varepsilon_1} \right)\; \Gamma \left( \frac{a_2 -a_1}{\varepsilon_1} \right) }{\Gamma \left( 1+\frac{a_2-a_{3,1}}{\varepsilon_1} \right) \; \Gamma \left(  \frac{a_{3,2}-a_1}{\varepsilon_1} \right) }  \end{bmatrix}.
\end{split}
\end{align}
Now it is straightforward to evaluate the $B$-monodromy matrix for $\widehat{\widehat{\mathfrak{D}}}_2$ by
\begin{align}
\mathbf{M}_B (\widehat{\widehat{\mathfrak{D}}}_2) = \mathbf{R}_\infty \; \mathbf{C}_\infty \; \mathbf{S} \; \mathbf{C}_0 ^{-1} \; \mathbf{R}_0  \; \mathbf{C}_0 \; \mathbf{S}^{-1} \; \mathbf{C}_\infty ^{-1}.
\end{align}
We need to evaluate the trace of the $B$-monodromy matrix for the $SL(2)$-oper $\widehat{\mathfrak{D}}_2$, which is obtained by taking the limit,
\begin{align}
\begin{split}
\text{Tr}\: M_B (\widehat{\mathfrak{D}}_2) = \text{Tr} \left(  \lim_{\varepsilon_2 \to 0}  \mathbf{R}_\infty \; \mathbf{C}_\infty \; \mathbf{S} \; \mathbf{C}_0 ^{-1} \; \mathbf{R}_0  \; \mathbf{C}_0 \; \mathbf{S}^{-1} \; \mathbf{C}_\infty ^{-1} \; e^{\frac{\widetilde{\EuScript{W}}}{\varepsilon_2}} \right).
\end{split}
\end{align}
A few pages of computation shows that
\begin{align}
\begin{split}
&\text{Tr}\: M_B (\widehat{\mathfrak{D}}_2) \\
 &= \frac{\left(\cos \pi \frac{a_{3,1}-a_{3,2}}{\varepsilon_1} - \cos \pi \frac{2(\bar{a}_3 -\bar{a})}{\varepsilon_1} \right) \left( \cos \pi \frac{a_{0,1}-a_{0,2}}{\varepsilon_1} - \cos \pi \frac{2(\bar{a}_0 -\bar{a}) }{\varepsilon_1} \right) }{2 \sin^2 \pi \frac{a_1-a_2}{2\varepsilon_1}} \\
& + \frac{\left(\cos \pi \frac{a_{3,1}-a_{3,2}}{\varepsilon_1} + \cos \pi \frac{2(\bar{a}_3 -\bar{a})}{\varepsilon_1} \right) \left( \cos \pi \frac{a_{0,1}-a_{0,2}}{\varepsilon_1} + \cos \pi \frac{2(\bar{a}_0 -\bar{a}) }{\varepsilon_1} \right) }{2 \cos^2 \pi \frac{a_1-a_2}{2\varepsilon_1}} \\
& -4  \frac{\prod_{\gamma=1,2} \sin \pi \frac{a_2-a_{0,\gamma}}{\varepsilon_1}\; \sin \pi \frac{a_{3,\gamma} -a_1}{\varepsilon_1} }{\sin ^2 \pi \frac{a_1-a_2}{\varepsilon_1}}   \frac{\Gamma \left( \frac{a_1-a_2}{\varepsilon_1} \right)^2}{\Gamma \left( \frac{a_2-a_1}{\varepsilon_1} \right)^2} \prod_{\gamma=1,2} \frac{\Gamma\left( \frac{a_{3,\gamma} -a_1}{\varepsilon_1} \right) \; \Gamma \left( \frac{a_2-a_{0,\gamma}}{\varepsilon_1} \right) }{\Gamma \left( \frac{a_1-a_{0,\gamma}}{\varepsilon_1} \right) \; \Gamma \left( \frac{a_{3,\gamma} -a_2}{\varepsilon_1} \right) } e^{  \left( \frac{\partial}{\partial a_1} - \frac{\partial}{\partial a_2} \right) \widetilde{\EuScript{W}}} \\
&-4  \frac{\prod_{\gamma=1,2} \sin \pi \frac{a_1-a_{0,\gamma}}{\varepsilon_1}\; \sin \pi \frac{a_{3,\gamma} -a_2}{\varepsilon_1} }{\sin ^2 \pi \frac{a_1-a_2}{\varepsilon_1}}   \left( \frac{\Gamma \left( \frac{a_1-a_2}{\varepsilon_1} \right)^2}{\Gamma \left( \frac{a_2-a_1}{\varepsilon_1} \right)^2} \prod_{\gamma=1,2} \frac{\Gamma\left( \frac{a_{3,\gamma} -a_1}{\varepsilon_1} \right) \; \Gamma \left( \frac{a_2-a_{0,\gamma}}{\varepsilon_1} \right) }{\Gamma \left( \frac{a_1-a_{0,\gamma}}{\varepsilon_1} \right) \; \Gamma \left( \frac{a_{3,\gamma} -a_2}{\varepsilon_1} \right) }  \right)^{-1}  e^{- \left( \frac{\partial}{\partial a_1} - \frac{\partial}{\partial a_2} \right) \widetilde{\EuScript{W}}}
\end{split}
\end{align}
It is crucial to note that the products of $\Gamma$-functions in the third and the fourth lines can be absorbed as the 1-loop part of the effective twisted superpotential of the $A_1$-theory computed under the $\zeta$-function regularization (see \eqref{1loopformula} and its derivation above), namely,
\begin{align}
\left( \frac{\partial}{\partial a_1} -\frac{\partial}{\partial a_2} \right) \widetilde{\EuScript{W}}^{\text{1-loop}} =  \log \; \frac{\Gamma \left( \frac{a_1-a_2}{\varepsilon_1} \right)^2}{\Gamma \left( \frac{a_2-a_1}{\varepsilon_1} \right)^2} \prod_{\gamma=1,2} \frac{\Gamma\left( \frac{a_{3,\gamma} -a_1}{\varepsilon_1} \right) \; \Gamma \left( \frac{a_2-a_{0,\gamma}}{\varepsilon_1} \right) }{\Gamma \left( \frac{a_1-a_{0,\gamma}}{\varepsilon_1} \right) \; \Gamma \left( \frac{a_{3,\gamma} -a_2}{\varepsilon_1} \right) }.
\end{align}
Hence we define the full effective twisted superpotential by
\begin{align}
\widetilde{\EuScript{W}}^{\text{full}} \equiv \widetilde{\EuScript{W}}^{\text{classical}} + \widetilde{\EuScript{W}}^{\text{1-loop}}+\widetilde{\EuScript{W}}^{\text{inst}}+\widetilde{\EuScript{W}}^{\text{extra}},
\end{align}
where the 1-loop part is given in \eqref{1looppart} and the other parts have been obtained in \eqref{n2twsupo},
\begin{subequations} 
\begin{align}
&\widetilde{\EuScript{W}}^{\text{classical}} = -\frac{\left( a_{1}-a_{2} \right)^2}{4\varepsilon_1 } \log \mathfrak{q} \\
&\widetilde{\EuScript{W}}^{\text{1-loop}} = \lim_{\varepsilon_2 \to 0} \varepsilon_2 \log \: \frac{\prod_{\alpha, \beta =1} ^2 \Gamma_2 (a_\alpha -a_\beta ; \varepsilon_1, \varepsilon_2)}{\prod_{\alpha,\beta=1} ^2 \Gamma_2 (a_{\alpha} -a_{0,\beta}; \varepsilon_1, \varepsilon_2) \Gamma_2 (a_{3,\alpha} -a_\beta; \varepsilon_1, \varepsilon_2)} \\
&\widetilde{\EuScript{W}}^{\text{inst}} = \lim_{\varepsilon_2 \to 0} \varepsilon_2 \log \EuScript{Z}_{A_1} ^{\text{inst}} \\
&\widetilde{\EuScript{W}}^{\text{extra}} = \varepsilon_1 \left( \frac{1}{4} -\delta_2 -\delta_3 \right) \log\mathfrak{q}  + \frac{2(\bar{a}_0 - \bar{a})(\bar{a} -\bar{a}_3 + \varepsilon_1 )}{\varepsilon_1}  \log(1-\mathfrak{q}). \label{twistedextra2}
\end{align}
\end{subequations}
Thus the expression for the trace of the $B$-monodromy matrix is simplified with the full effective twisted superpotential $\widetilde{\EuScript{W}}^{\text{full}}$. Let us also make an overall shift for the Coulomb moduli and the masses of the hypermultiplets to recover the $SU(2)$ parameters (see section \ref{surfdef}). The final form of the expression is
\begin{align} \label{trmb}
\begin{split}
\text{Tr}\: M_B (\widehat{\mathfrak{D}}_2) &= \frac{\left(\cos \pi \frac{a_{3,1}-a_{3,2}}{\varepsilon_1} - \cos \pi \frac{2\bar{a}_3 }{\varepsilon_1} \right) \left( \cos \pi \frac{a_{0,1}-a_{0,2}}{\varepsilon_1} - \cos \pi \frac{2\bar{a}_0  }{\varepsilon_1} \right) }{2 \cos^2 \pi \boldsymbol{\alpha} } \\
& + \frac{\left(\cos \pi \frac{a_{3,1}-a_{3,2}}{\varepsilon_1} + \cos \pi \frac{2\bar{a}_3 }{\varepsilon_1} \right) \left( \cos \pi \frac{a_{0,1}-a_{0,2}}{\varepsilon_1} + \cos \pi \frac{2 \bar{a}_0  }{\varepsilon_1} \right) }{2 \sin^2 \pi \boldsymbol{\alpha}} \\
&- \sum_{\pm} 4  \; \frac{\prod_{\gamma=1,2} \cos \pi \left( \mp \boldsymbol{\alpha} - \frac{a_{0,\gamma}}{\varepsilon_1} \right)\; \cos \pi \left( \frac{a_{3,\gamma}}{\varepsilon_1}  \mp \boldsymbol{\alpha} \right) }{\sin ^2 2 \pi \boldsymbol{\alpha}}  e^{\pm \frac{1}{\varepsilon_1} \frac{\partial \widetilde{\EuScript{W}} ^{\text{full}}}{\partial \boldsymbol{\alpha}} }.
\end{split}
\end{align}
This expression exactly matches with \eqref{eq:trb2} under the identification of parameters,
\begin{align}
\fm_{-1} = e^{\pi i \frac{a_{0,1}-a_{0,2}-\varepsilon_1}{\varepsilon_1}}, \;\; \fm_0 = e^{\pi i \frac{a_{0,1}+a_{0,2}}{\varepsilon_1}}, \;\; \fm_1 = e^{-\pi i \frac{a_{3,1}+a_{3,2}}{\varepsilon_1}}, \;\; \fm_2 = e^{\pi i \frac{-a_{3,1}+a_{3,2}+\varepsilon_1}{\varepsilon_1}}.
\end{align}
Most importantly, we observe that
\begin{align} \label{alphabeta}
\boldsymbol{\beta} = \frac{1}{\varepsilon_1} \frac{\partial \widetilde{\EuScript{W}} ^{\text{full}}}{\partial \boldsymbol{\alpha}}.
\end{align}
Consequently, the generating function for the variety $\mathcal{O}_2 [\mathbb{P}^1 \backslash \{ 0, \mathfrak{q} , 1, \infty \}] $ of opers is identified with the effective twisted superpotential, namely,
\begin{align} \label{equiv}
\mathcal{S}\left[ \mathcal{O}_2 [\mathbb{P}^1 \backslash \{ 0, \mathfrak{q} , 1, \infty \}] \right] = \frac{1}{\varepsilon_1} \widetilde{\EuScript{W}}^{\text{full}} \left[ \mathcal{T}[A_1, \mathbb{P}^1 \backslash \{ 0, \mathfrak{q} , 1, \infty \} ] \right],
\end{align}
by the relation \eqref{alphabeta}. This identification verifies the main assertion of \cite{nrs} for all orders in the gauge coupling $\mathfrak{q}$.
\paragraph{Remarks}
\begin{itemize}
\item The equivalence \eqref{equiv} involves an extra term in the effective twisted superpotential, $\widetilde{\EuScript{W}}^{\text{extra}}$. Note that $\widetilde{\EuScript{W}}^{\text{extra}}$ has been completely determined in gauge theoretical terms in \eqref{twistedextra2}.
\item The regularization scheme that has been used to define the 1-loop part $\widetilde{\EuScript{W}}^{\text{1-loop}}$ was the $\zeta$-function regularization, which is natural in the gauge theory context. Note that it is free to choose other schemes to regularize the infinite product, or the IR divergence, in the 1-loop contribution. The other choices would lead to the correction in the effective twisted superpotential of the form,
\begin{align}
\widetilde{\EuScript{W}}_\infty \sim \text{Li}_2 \; e^{l (\mathbf{a}, \mathbf{a}_0, \mathbf{a}_3)},
\end{align}
where $l$ is some linear function of the arguments. Physically, IR regulator corresponds to cutting the cigar $\mathcal{D}^2$ \eqref{cigar} at the infinity. Thus the correction $\widetilde{\EuScript{W}}_\infty$ to the effective twisted superpotential can be interpreted as the contribution from a three-dimensional theory coupled to the four-dimensional bulk theory at the boundary at infinity. Note that $\widetilde{\EuScript{W}}_\infty$ is independent of the coupling $\mathfrak{q}$. Hence this correction to the effective twisted superpotential corresponds to a canonical coordinate transformation.
\end{itemize}

\subsection{$SL(3)$-oper}
Using the critical exponent \eqref{critexpinter} of $z$, we see that under the $z \mapsto z \; e^ {2 \pi i}$ loop the partition function $\boldsymbol{\widetilde{\EuScript{Z}}}^{L \to M}$ transforms by
\begin{align}
\mathbf{M}_A (\widehat{\widehat{\mathfrak{D}}}_3) = \text{diag} \left( e^{2\pi i \: \frac{-2a_1 + a_2 + a_3 + 3\varepsilon}{3\varepsilon_1}}, \;e^{2\pi i \: \frac{a_1 -2 a_2 + a_3 + 3\varepsilon}{3\varepsilon_1}} , \; e^{2\pi i \: \frac{a_1 + a_2 -2 a_3 + 3\varepsilon}{3\varepsilon_1}} \right).
\end{align}
Hence, by taking the limit $\varepsilon_2 \to 0$, we obtain
\begin{align}
\begin{split}
M_A (\widehat{\mathfrak{D}}_3) &= \lim_{\varepsilon_2 \to 0} \mathbf{M}_A (\widehat{\widehat{\mathfrak{D}}}_3) \\
&=  \text{diag} \left( e^{2\pi i \: \frac{-2a_1 + a_2 + a_3 }{3\varepsilon_1}}, \;e^{2\pi i \: \frac{a_1 -2 a_2 + a_3 }{3\varepsilon_1}} , \; e^{2\pi i \: \frac{a_1 + a_2 -2 a_3 }{3\varepsilon_1}} \right).
\end{split}
\end{align}
Comparing with \eqref{eq:tra3}, we find
\begin{align}
\boldsymbol{\alpha}_\alpha = \frac{ 3a_\alpha - \sum_{\gamma=1} ^3 a_\gamma }{3\varepsilon_1}, \quad \alpha=1,2,
\end{align}
so that we have
\begin{align} \label{sl3tra}
\begin{split}
&\text{Tr} \:M_A (\widehat{\mathfrak{D}}_3) ^{\pm 1} = e^{\mp 2 \pi i \boldsymbol{\alpha}_1} + e^{\mp 2 \pi i \boldsymbol{\alpha}_2} + e^{\pm 2 \pi i (\boldsymbol{\alpha}_1  + \boldsymbol{\alpha}_2 )}  .
\end{split}
\end{align}
For notational convenience, let us also define $\boldsymbol{\alpha}_3 \equiv -\boldsymbol{\alpha}_1 - \boldsymbol{\alpha}_2$ as before. 

Next, we obtain the expression for the $\boldsymbol\beta$ coordinates restricted to the variety of opers by evaluating the $B$-monodromy matrix. First, we compute the rotation matrices by shifting $z \mapsto z \; e ^{-2 \pi i}$ and $z \mapsto z \; e^{2 \pi i}$ for $\widetilde{\boldsymbol{\EuScript{Z}}}^L $ and $\widetilde{\boldsymbol{\EuScript{Z}}}^R$, respectively. From their critical exponents \eqref{critexp3inf} and \eqref{critexp3zero} we get
\begin{align}
\begin{split}
&\mathbf{R}_\infty = \text{diag}\left( e^{2 \pi i \frac{2a_{0,1}-a_{0,2}-a_{0,3} - 3\varepsilon -2 \varepsilon_2}{3 \varepsilon_1}}, \;e^{2 \pi i \frac{-a_{0,1}+2a_{0,2}-a_{0,3} - 3\varepsilon -2 \varepsilon_2}{3 \varepsilon_1}} ,\; e^{2 \pi i \frac{-a_{0,1}-a_{0,2}+2a_{0,3} - 3\varepsilon -2 \varepsilon_2}{3 \varepsilon_1}}  \right), \\
&\mathbf{R}_0 = \text{diag}\left( e^{2 \pi i \frac{ -2a_{3,1}+a_{3,2}+a_{3,3} + 3\varepsilon }{3 \varepsilon_1}}, \;e^{2 \pi i \frac{a_{3,1}-2a_{3,2}+a_{3,3} + 3\varepsilon}{3 \varepsilon_1}} ,\; e^{2 \pi i \frac{a_{3,1}+a_{3,2}-2a_{3,3} + 3\varepsilon}{3 \varepsilon_1}}  \right).
\end{split}
\end{align}
Then the $B$-monodromy matrix for $\widehat{\widehat{\mathfrak{D}}}_3$ is obtained by \eqref{bmono1}
\begin{align}
\mathbf{M}_B (\widehat{\widehat{\mathfrak{D}}}_3) = \mathbf{R}_\infty \; \mathbf{C}_\infty \; \mathbf{S} \; \mathbf{C}_0 ^{-1} \; \mathbf{R}_0  \; \mathbf{C}_0 \; \mathbf{S}^{-1} \; \mathbf{C}_\infty ^{-1} ,
\end{align}
where the connection matrices and the shift matrices are given by \eqref{contmat} and \eqref{shift}. The $B$-monodromy matrix for the oper $\widehat{\mathfrak{D}}_3$ is obtained by taking the limit,
\begin{align}
\begin{split}
&M_B (\widehat{\mathfrak{D}}_3 ) = \lim_{\varepsilon_2 \to 0}  \mathbf{R}_\infty \; \mathbf{C}_\infty \; \mathbf{S} \; \mathbf{C}_0 ^{-1} \; \mathbf{R}_0  \; \mathbf{C}_0 \; \mathbf{S}^{-1} \; \mathbf{C}_\infty ^{-1} \; e^{\frac{\widetilde{\EuScript{W}}}{\varepsilon_2}}.
\end{split}
\end{align}
To determine the $\boldsymbol\beta$ coordinates, we need to compute the following traces of $M_B (\widehat{\mathfrak{D}}_3 ) $
\begin{align}
\begin{split}
&\text{Tr}\: M_B (\widehat{\mathfrak{D}}_3 ) = \text{Tr} \left(  \lim_{\varepsilon_2 \to 0}  \mathbf{R}_\infty \; \mathbf{C}_\infty \; \mathbf{S} \; \mathbf{C}_0 ^{-1} \; \mathbf{R}_0  \; \mathbf{C}_0 \; \mathbf{S}^{-1} \; \mathbf{C}_\infty ^{-1} \; e^{\frac{\widetilde{\EuScript{W}}}{\varepsilon_2}} \right) \\
&\text{Tr}\: M_B (\widehat{\mathfrak{D}}_3 )^{-1} = \text{Tr} \left(  \lim_{\varepsilon_2 \to 0} \mathbf{C}_\infty \; \mathbf{S} \; \mathbf{C}_0 ^{-1} \; \mathbf{R}_0 ^{-1} \; \mathbf{C}_0 \; \mathbf{S}^{-1} \; \mathbf{C}_\infty ^{-1} \; \mathbf{R}_\infty ^{-1} \; e^{\frac{\widetilde{\EuScript{W}}}{\varepsilon_2}}\right).
\end{split}
\end{align}
The computation of these traces can be broken in several steps. First note that
\begin{align}
\text{Tr}\: M_B (\widehat{\mathfrak{D}}_3 ) = \text{Tr} \left(  \lim_{\varepsilon_2 \to 0} ( \mathbf{C}_\infty ^{-1}\; \mathbf{R}_\infty \; \mathbf{C}_\infty ) \; \mathbf{S} \; (\mathbf{C}_0 ^{-1} \; \mathbf{R}_0  \; \mathbf{C}_0) \; \mathbf{S}^{-1} \;  e^{\frac{\widetilde{\EuScript{W}}}{\varepsilon_2}} \right),
\end{align}
due to the limit $\varepsilon_2 \to 0$. Then
\begin{align}
\text{Tr}\: M_B (\widehat{\mathfrak{D}}_3 ) = \sum_{\alpha, \beta=1} ^3 (\boldsymbol{\mathcal{C}}_\infty)_{\beta \alpha} \; (\boldsymbol{\mathcal{C}}_0)_{\alpha \beta}\; e^{ \left( \frac{\partial}{\partial a_\alpha} - \frac{\partial}{\partial a_\beta} \right) \widetilde{\EuScript{W}} },
\end{align}
where we have defined
\begin{align}
\begin{split}
&(\boldsymbol{\mathcal{C}}_\infty )_{\alpha \beta} \equiv \left( \lim_{\varepsilon_2 \to 0 }  \mathbf{C}_\infty ^{-1} \; \mathbf{R}_\infty  \; \mathbf{C}_\infty \right)_{\alpha \beta} \\
& = e^{\frac{i \pi}{\varepsilon_1} (a_\alpha+a_\beta -2 \bar{a}_0)} \\ 
& \left( \delta_{\alpha,\beta}  -2 i \; e^{\frac{ 3  \pi i}{\varepsilon_1} ( \bar{a}_0 -   \bar{a}  ) }   \frac{\prod_{\beta' \neq \beta} \Gamma \left( \frac{a_\beta-a_{\beta'}}{\varepsilon_1} \right) }{\prod_{\alpha' \neq \alpha} \Gamma \left( \frac{a_\alpha - a_{\alpha'}}{\varepsilon_1} \right)\; \sin \pi \frac{a_\alpha - a_{\alpha'}}{\varepsilon_1} } \; \prod_{\alpha'=1} ^3 \frac{\Gamma\left( \frac{a_\alpha - a_{0,\alpha'}}{\varepsilon_1} \right)\; \sin \pi \frac{a_\alpha - a_{0,\alpha'}}{\varepsilon_1}  }{\Gamma \left( \frac{a_\beta - a_{0,\alpha'}}{\varepsilon_1} \right)} \right) ,
\end{split}
\end{align}
and
\begin{align}
\begin{split}
&(\boldsymbol{\mathcal{C}}_0 )_{\alpha \beta} \equiv \left( \lim_{\varepsilon_2 \to 0 }  \mathbf{C}_0 ^{-1}\; \mathbf{R}_0  \; \mathbf{C}_0 \right)_{\alpha \beta} \\
&  = e^{\frac{i \pi}{\varepsilon_1} (-a_\alpha-a_\beta+2\bar{a}_3)} \\
& \left( \delta_{\alpha,\beta}  -2 i \; e^{\frac{3  \pi i}{\varepsilon_1} \left( -\bar{a}_3 + \bar{a}  \right) }   \frac{\prod_{\beta' \neq \beta} \Gamma \left( \frac{a_{\beta'}-a_{\beta}}{\varepsilon_1} \right) }{\prod_{\alpha' \neq \alpha} \Gamma \left( \frac{a_{\alpha'} - a_{\alpha}}{\varepsilon_1} \right)\; \sin \pi \frac{a_{\alpha'} - a_{\alpha}}{\varepsilon_1} } \; \prod_{\alpha'=1} ^3 \frac{\Gamma\left( \frac{a_{3,\alpha'} - a_{\alpha}}{\varepsilon_1} \right)\; \sin \pi \frac{a_{3,\alpha'} - a_{\alpha}}{\varepsilon_1}  }{\Gamma \left( \frac{a_{3,\alpha'} - a_{\beta}}{\varepsilon_1} \right)} \right) .
\end{split}
\end{align}
Similarly, we can write
\begin{align}
\begin{split}
\text{Tr}\: M_B (\widehat{\mathfrak{D}}_3 )^{-1} &=  \text{Tr} \left(  \lim_{\varepsilon_2 \to 0} \mathbf{S} \; (\mathbf{C}_0 ^{-1} \; \mathbf{R}_0 ^{-1}  \; \mathbf{C}_0) \; \mathbf{S}^{-1} \;  ( \mathbf{C}_\infty ^{-1}\; \mathbf{R}_\infty ^{-1} \; \mathbf{C}_\infty ) \;  e^{\frac{\widetilde{\EuScript{W}}}{\varepsilon_2}} \right), \\
&= \sum_{\alpha, \beta=1} ^3 (\boldsymbol{\mathcal{C}}_0 ^{-1})_{\alpha \beta} \; (\boldsymbol{\mathcal{C}}_\infty ^{-1})_{\beta \alpha}\; e^{ \left( \frac{\partial}{\partial a_\alpha} - \frac{\partial}{\partial a_\beta} \right) \widetilde{\EuScript{W}} },
\end{split}
\end{align}
where we have used
\begin{align}
\begin{split}
&(\boldsymbol{\mathcal{C}}_\infty ^{-1})_{\alpha \beta} = \left( \lim_{\varepsilon_2 \to 0 }  \mathbf{C}_\infty ^{-1}\; \mathbf{R}_\infty ^{-1} \; \mathbf{C}_\infty \right)_{\alpha \beta} \\
&= e^{\frac{i \pi}{\varepsilon_1} (-a_\alpha -a_\beta +2\bar{a}_0)} \\
&\left( \delta_{\alpha,\beta} +  2 i \; e^{\frac{3  \pi i}{\varepsilon_1} \left(  - \bar{a}_0 +   \bar{a}  \right) } \frac{\prod_{\beta' \neq \beta} \Gamma \left( \frac{a_\beta-a_{\beta'}}{\varepsilon_1} \right) }{\prod_{\alpha' \neq \alpha} \Gamma \left( \frac{a_\alpha - a_{\alpha'}}{\varepsilon_1} \right)\; \sin \pi \frac{a_\alpha - a_{\alpha'}}{\varepsilon_1} } \; \prod_{\alpha'=1} ^3 \frac{\Gamma\left( \frac{a_\alpha - a_{0,\alpha'}}{\varepsilon_1} \right)\; \sin \pi \frac{a_\alpha - a_{0,\alpha'}}{\varepsilon_1}  }{\Gamma \left( \frac{a_\beta - a_{0,\alpha'}}{\varepsilon_1} \right)} \right) ,
\end{split}
\end{align}
and
\begin{align}
\begin{split}
&(\boldsymbol{\mathcal{C}}_0 ^{-1})_{\alpha \beta} = \left( \lim_{\varepsilon_2 \to 0 }  \mathbf{C}_0 ^{-1}\; \mathbf{R}_0 ^{-1} \; \mathbf{C}_0 \right)_{\alpha \beta} \\
&= e^{\frac{i\pi}{\varepsilon_1} (a_\alpha+a_\beta-2\bar{a}_3) } \\
&\left( \delta_{\alpha,\beta} + 2 i \; e^{\frac{3  \pi i}{\varepsilon_1} \left( \bar{a}_3 -   \bar{a}  \right) } \frac{\prod_{\beta' \neq \beta} \Gamma \left( \frac{a_{\beta'}-a_{\beta}}{\varepsilon_1} \right) }{\prod_{\alpha' \neq \alpha} \Gamma \left( \frac{a_{\alpha'} - a_{\alpha}}{\varepsilon_1} \right)\; \sin \pi \frac{a_{\alpha'} - a_{\alpha}}{\varepsilon_1} } \; \prod_{\alpha'=1} ^3 \frac{\Gamma\left( \frac{a_{3,\alpha'} - a_{\alpha}}{\varepsilon_1} \right)\; \sin \pi \frac{a_{3,\alpha'} - a_{\alpha}}{\varepsilon_1}  }{\Gamma \left( \frac{a_{3,\alpha'} - a_{\beta}}{\varepsilon_1} \right)} \right) .
\end{split}
\end{align}
Therefore, the traces can be expressed as
\begin{align}
\text{Tr}\: M_B (\widehat{\mathfrak{D}}_3 )^{\pm 1} = B_0 ^{\pm} + \sum_{\alpha \neq \beta} \widetilde{B}_{\alpha\beta} ^{\pm} \; e^{ \left( \frac{\partial}{\partial a_\alpha} - \frac{\partial}{\partial a_\beta} \right) \widetilde{\EuScript{W}} },
\end{align}
where we have computed the coefficients as
\begin{align}
B_0 ^\pm \equiv \sum_{\alpha=1} ^3 \left(\boldsymbol{\mathcal{C}}_0 ^{\pm 1} \right)_{\alpha \alpha} \left(\boldsymbol{\mathcal{C}}_\infty ^{\pm 1} \right)_{\alpha \alpha},
\end{align}
and
\begin{align} \label{bab}
\begin{split}
\widetilde{B}_{\alpha\beta} ^\pm &\equiv \left( \boldsymbol{\mathcal{C}}_0 ^{\pm 1} \right)_{\alpha \beta} \left( \boldsymbol{\mathcal{C}}_\infty ^{\pm 1} \right)_{\beta \alpha} \\
&= -4 e^{ \pm i \pi  \frac{ \bar{a}_{0} -\bar{a}_{3} }{\varepsilon_1} } \; \frac{\prod_{\gamma=1} ^3 \sin \pi \frac{a_\beta -a_{0,\gamma}}{\varepsilon_1} \; \sin \pi \frac{ a_{3,\gamma} -a_{\alpha}}{\varepsilon_1} }{\prod_{\alpha' \neq \alpha} \sin \pi \frac{a_{\alpha'} -a_\alpha}{\varepsilon_1} \; \prod_{\beta' \neq \beta} \sin \pi \frac{a_\beta- a_{\beta'}}{\varepsilon_1}  } \\
& \quad\;  \prod_{\alpha' \neq \alpha} \frac{\Gamma \left( \frac{a_{\alpha}-a_{\alpha'}}{\varepsilon_1} \right)}{\Gamma \left( \frac{a_{\alpha'}-a_\alpha}{\varepsilon_1} \right)} \; \prod_{\beta' \neq\beta} \frac{\Gamma \left( \frac{a_{\beta'} -a_\beta}{\varepsilon_1} \right)}{\Gamma \left( \frac{a_\beta -a_{\beta'}}{\varepsilon_1} \right) } \; \prod_{\gamma=1} ^3 \frac{\Gamma \left( \frac{a_{3,\gamma} -a_\alpha}{\varepsilon_1} \right) \; \Gamma \left( \frac{a_\beta - a_{0,\gamma}}{\varepsilon_1} \right) }{\Gamma \left( \frac{a_\alpha - a_{0,\gamma}}{\varepsilon_1} \right) \; \Gamma \left( \frac{a_{3,\gamma}-a_\beta}{\varepsilon_1} \right) }  .
\end{split}
\end{align}
It is crucial to note that the last line of \eqref{bab} is precisely the contribution from 1-loop part of the effective twisted superpotential of the $A_1$-theory, under the $\zeta$-function regularization (see \eqref{1loopformula} and its derivation above),
\begin{align}
\left( \frac{\partial}{\partial a_\alpha} - \frac{\partial}{\partial a_\beta} \right) \widetilde{\EuScript{W}}^{\text{1-loop}} =\log \prod_{\alpha' \neq \alpha} \frac{\Gamma \left( \frac{a_{\alpha}-a_{\alpha'}}{\varepsilon_1} \right)}{\Gamma \left( \frac{a_{\alpha'}-a_\alpha}{\varepsilon_1} \right)} \; \prod_{\beta' \neq\beta} \frac{\Gamma \left( \frac{a_{\beta'} -a_\beta}{\varepsilon_1} \right)}{\Gamma \left( \frac{a_\beta -a_{\beta'}}{\varepsilon_1} \right) } \; \prod_{\gamma=1} ^3 \frac{\Gamma \left( \frac{a_{3,\gamma} -a_\alpha}{\varepsilon_1} \right) \; \Gamma \left( \frac{a_\beta - a_{0,\gamma}}{\varepsilon_1} \right) }{\Gamma \left( \frac{a_\alpha - a_{0,\gamma}}{\varepsilon_1} \right) \; \Gamma \left( \frac{a_{3,\gamma}-a_\beta}{\varepsilon_1} \right) }.
\end{align}
We define the full effective twisted superpotential by
\begin{align}
\widetilde{\EuScript{W}}^{\text{full}} \equiv \widetilde{\EuScript{W}}^{\text{classical}} + \widetilde{\EuScript{W}}^{\text{1-loop}}+\widetilde{\EuScript{W}}^{\text{inst}}+\widetilde{\EuScript{W}}^{\text{extra}}.
\end{align}
Here, the 1-loop part of the effective twisted superpotential is given in \eqref{1looppart} and the rest was obtained in \eqref{efftwpoten3},
\begin{subequations}
\begin{align}
&\widetilde{\EuScript{W}}^{\text{classical}} = -\frac{(a_1-a_2)^2 +(a_1-a_3)^2 -(a_1-a_2)(a_1-a_3)}{3\varepsilon_1 } \log \mathfrak{q} \\
&\widetilde{\EuScript{W}}^{\text{1-loop}} = \lim_{\varepsilon_2 \to 0} \varepsilon_2 \log \:  \frac{\prod_{\alpha, \beta =1} ^3 \Gamma_2 (a_\alpha -a_\beta ; \varepsilon_1, \varepsilon_2)}{\prod_{\alpha,\beta=1} ^3 \Gamma_2 (a_{\alpha} -a_{0,\beta}; \varepsilon_1, \varepsilon_2) \Gamma_2 (a_{3,\alpha} -a_\beta; \varepsilon_1, \varepsilon_2)} \\
&\widetilde{\EuScript{W}}^{\text{inst}} = \lim_{\varepsilon_2 \to 0} \varepsilon_2 \log \EuScript{Z}_{A_1} ^{\text{inst}} \\
&\widetilde{\EuScript{W}}^{\text{extra}} = \varepsilon_1 \left( 1 -\delta_{\mathfrak{q}} -\delta_0 \right) \log\mathfrak{q} +\frac{3(\bar{a} -\bar{a}_3+\varepsilon)(\bar{a}_0 -\bar{a}) }{\varepsilon_1 }  \log(1-\mathfrak{q}).
\end{align}
\end{subequations}
Again, the expression for the traces of the $B$-monodromy matrix are simplified with the full effective twisted superpotential $\widetilde{\EuScript{W}}^{\text{full}}$. Let us make an overall shift of the Coulomb moduli and the masses of the hypermultiplets to recover the $SU(3)$ parameters (see section \ref{surfdef}). Then we get the final expressions for the traces of the $B$-monodromy:
\begin{align} \label{sl3trb}
\begin{split}
&\text{Tr}\:  M_B (\widehat{\mathfrak{D}}_3 )^{\pm 1} \\ &= B_0 ^{\pm} + {B}_{12} ^{\pm} \; e^{ \frac{1}{\varepsilon_1} \left( \frac{\partial}{\partial \boldsymbol\alpha_1} - \frac{\partial}{\partial \boldsymbol\alpha_2} \right) \widetilde{\EuScript{W}}^{\text{full}} } + {B}_{13} ^{\pm} \; e^{\frac{1}{\varepsilon_1} \frac{\partial  \widetilde{\EuScript{W}}^{\text{full}} }{\partial \boldsymbol\alpha_1} } + {B}_{23} ^{\pm} \; e^{\frac{1}{\varepsilon_1}  \frac{\partial  \widetilde{\EuScript{W}}^{\text{full}}}{\partial \boldsymbol\alpha_2}  } \\
&\quad\quad\;\;\; +{B}_{21} ^{\pm} \; e^{-\frac{1}{\varepsilon_1} \left( \frac{\partial}{\partial \boldsymbol\alpha_1} - \frac{\partial}{\partial \boldsymbol\alpha_2} \right) \widetilde{\EuScript{W}}^{\text{full}} } +{B}_{31} ^{\pm} \; e^{-\frac{1}{\varepsilon_1}\frac{\partial  \widetilde{\EuScript{W}}^{\text{full}}}{\partial \boldsymbol\alpha_1}} + {B}_{32} ^{\pm} \; e^{-\frac{1}{\varepsilon_1}\frac{\partial  \widetilde{\EuScript{W}}^{\text{full}}}{\partial \boldsymbol\alpha_2}},
\end{split}
\end{align}
where the coefficients are computed as
\begin{align}
\begin{split}
B_0 ^{\pm} &= 3 e^{ \pm \frac{2 \pi i}{\varepsilon_1} (\bar{a}_3 -\bar{a}_0)} \pm 2 i\, e^{\pm \frac{i\pi }{\varepsilon_1} (2 \bar{a}_3 +\bar{a}_0)} \sin \pi \frac{ 3 \bar{a}_0}{\varepsilon_1} \mp 2 i\, e^{\mp \frac{i \pi  }{\varepsilon_1} (\bar{a}_3 +2\bar{a}_0)}  \sin \pi \frac{3 \bar{a}_3}{\varepsilon_1} \\
&\quad - 4 e^{\pm \frac{ i \pi }{\varepsilon_1} (\bar{a}_0 -\bar{a}_3)}  \sum_{\alpha=1} ^3 \frac{ \prod_{\gamma=1} ^3 \sin \pi \left( \boldsymbol\alpha_\alpha -\frac{ a_{0,\gamma}}{\varepsilon_1} \right) \sin \pi \left( \frac{a_{3,\gamma} }{\varepsilon_1} -\boldsymbol\alpha_\alpha \right) }{\prod_{\alpha' \neq \alpha} \sin ^2 \pi \left( \boldsymbol\alpha_\alpha -\boldsymbol\alpha_{\alpha'} \right)}
\end{split}
\end{align}
and
\begin{align}
B_{\alpha \beta} ^{\pm} =  -4 e^{ \pm i \pi  \frac{  \bar{a}_{0} -\bar{a}_{3} }{\varepsilon_1} } \; \frac{\prod_{\gamma=1} ^3 \sin \pi \left(  \boldsymbol{\alpha}_\beta -\frac{a_{0,\gamma}}{\varepsilon_1}\right) \; \sin \pi \left( \frac{ a_{3,\gamma} }{\varepsilon_1} - \boldsymbol{\alpha}_\alpha \right) }{\prod_{\alpha' \neq \alpha} \sin \pi \left( \boldsymbol{\alpha}_{\alpha'} -\boldsymbol{\alpha}_\alpha \right) \; \prod_{\beta' \neq \beta} \sin \pi \left( \boldsymbol{\alpha}_\beta- \boldsymbol{\alpha}_{\beta'} \right)  }.
\end{align}
We observe the precise agreement between \eqref{eq:trb3} and \eqref{sl3trb} under the identification of parameters,
\begin{align}
\begin{split}
&\fm_{-1} ^{(\alpha)} = e^{2\pi i \frac{a_{0,\alpha} - \bar{a}_0}{\varepsilon_1}}, \quad \fm_1 ^{(\alpha)} = e^{2\pi i \frac{ a_{3,\alpha}- \bar{a}_3}{\varepsilon_1}}, \quad \alpha =1,2, \\
&\fm_0 = e^{2\pi i \frac{\bar{a}_0}{\varepsilon_1}}, \quad \fm_1 = e^{-2\pi i \frac{\bar{a}_3}{\varepsilon_1}}.
\end{split}
\end{align}
Most importantly, we find
\begin{align} \label{genfunc}
\boldsymbol{\beta}_\alpha = \frac{1}{\varepsilon_1} \frac{\partial \widetilde{\EuScript{W}}^{\text{full}}}{\partial \boldsymbol{\alpha}_\alpha}, \quad \alpha=1,2.
\end{align}
Therefore, we verify that the generating function for the variety $\mathcal{O}_3 [\mathbb{P}^1 \backslash \{ 0, \underline{\mathfrak{q}} , \underline{1}, \infty \}]$ of opers with respect to the generalized NRS coordinate system is identical to the effective twisted superpotential, namely,
\begin{align} \label{equiv3}
\mathcal{S}\left[ \mathcal{O}_3 [\mathbb{P}^1 \backslash \{ 0, \underline{\mathfrak{q}} , \underline{1}, \infty\}] \right] = \frac{1}{\varepsilon_1} \widetilde{\EuScript{W}}^{\text{full}} \left[ \mathcal{T}[A_2, \mathbb{P}^1 \backslash \{ 0, \underline{\mathfrak{q}} , \underline{1}, \infty\} ] \right],
\end{align}
by the relation \eqref{genfunc}.
\paragraph{Remarks}
\begin{itemize}
\item The validity of the equivalence \eqref{equiv3} at the 1-loop level was checked in \cite{hol}.\footnote{In \cite{hol}, a different, the so-called Liouville/Toda regularization scheme was used. Although Liouville/Toda scheme is natural in the context of the AGT correspondence \cite{agt}, the $\zeta$-function regularization arises more naturally in the gauge theoretical context. Besides, the $\zeta$-function regularization has a notational advantage in that the defining equations for the generalized NRS coordinates \eqref{eq:tra3}, \eqref{eq:trb3} are  written more simply without any $\Gamma$-functions or square roots.} The gauge theoretical derivation of \eqref{equiv3} that we have shown guarantees its validity at all orders in the gauge coupling $\mathfrak{q}$.
\end{itemize}

\subsection{Higher $SL(N)$-oper}
It is straightforward to generalize the procedure to the higher $SL(N)$-opers $\widehat{\mathfrak{D}}_N$ on $ \mathbb{P}^1 \backslash \{ 0, \underline{\mathfrak{q}} , \underline{1}, \infty \}$. We schematically describe how we proceed. First, we need to express the traces of the holonomies of the flat $SL(N)$-connections in terms of the generalized NRS coordinates, as we did for $N=2$ and $N=3$ in section \ref{subsec:fourpunc}. It is clear that the holonomy along the $A$-cycle is still given by
\begin{align}
M_A = M_0 ^{-1} = \sum_{\alpha} ^N \left(\fm_0 ^{(\alpha)} \right)^{-1} \Pi_0 ^{(\alpha)}.
\end{align}
Hence we obtain
\begin{align} \label{eq:tran}
\text{Tr} \, M_A ^k = \sum_{\alpha=1} ^N \left( \fm_0 ^{(\alpha)} \right)^{-k}, \quad k =1, \cdots, N-1.
\end{align}
The holonomy along the $B$-cycle is written as
\begin{align} \label{eq:mbn}
\begin{split}
M_B &= g_1 ^{-1} g_0 ^{-1} \\
&=\fm_0 ^{-1} \fm_1 ^{-1} \left( \mathds{1}_N +(\fm_1 ^N -1) \Pi_1 \right) \left( \mathds{1}_N +(\fm_0 ^N -1) \Pi_0 \right).
\end{split}
\end{align}
Due to the properties of the projection operators, we have
\begin{align}
\text{Tr}\, \left(\Pi_0 \Pi_1\right)^k =  \left( \text{Tr}\, \Pi_0 \Pi_1 \right)^k, \quad k \in \mathbb{Z}^{> 0}.
\end{align}
Thus we can expand the traces of \eqref{eq:mbn} as a polynomial in $\text{Tr}\, \Pi_0 \Pi_1$,
\begin{align} \label{eq:trbn}
\begin{split}
\text{Tr}\, M_B ^k &= \fm_0 ^{-k} \fm_1 ^{-k} (N-2 +\fm_0 ^{Nk} +\fm_1 ^{Nk}) \\
& \quad+ \cdots + \fm_0 ^{-k} \fm_1 ^{-k} (\fm_0 ^N-1)^k (\fm_1 ^N-1)^k \left( \text{Tr}\, \Pi_0 \Pi_1 \right)^k,
\end{split}
\end{align}
for any $k=1, \cdots, N-1$. Since we can express $\text{Tr}\, \Pi_0 \Pi_1$ by the $\boldsymbol\beta$ coordinates,
\begin{align}
\begin{split}
\text{Tr}\, \Pi_0 \Pi_1 &= \sum_{\alpha=1} ^N \text{Tr}\, \Pi_0 \Pi_0 ^{(\alpha)} \Pi_1 \\
&=\sum_{\alpha=1} ^N e^{-\tilde{\boldsymbol\beta}_0 ^{(\alpha)} + \tilde{\boldsymbol\beta}}\, \text{Tr}\, \Pi_0 \Pi_0 ^{(\alpha)} \\
&=\sum_{\alpha=1} ^N \text{Tr} \, \Pi_0 \Pi_0 ^{(\alpha)}\, \text{Tr}\, \Pi_1 \Pi_0 ^{(\alpha)} + \sum_{\alpha\neq\beta} e^{\tilde{\boldsymbol\beta}_0 ^{(\alpha)} -\tilde{\boldsymbol\beta}_0 ^{(\beta)} } \text{Tr}\, \Pi_0 \Pi_0 ^{(\beta)} \, \text{Tr}\, \Pi_1 \Pi_0 ^{(\alpha)},
\end{split}
\end{align}
we obtain the representation of the traces \eqref{eq:trbn} in terms of the generalized NRS coordinates $\boldsymbol\alpha_\alpha$, $\boldsymbol\beta_\alpha \equiv \tilde{\boldsymbol\beta}_0 ^{(\alpha)} -\tilde{\boldsymbol\beta}_0 ^{(N)}$.

Next, we evaluate the monodromies of the oper $\widehat{\mathfrak{D}}_N$. By shifting $z \mapsto z \; e^{2 \pi i}$ for $\boldsymbol{\widetilde{\EuScript{Z}}}^{L \to M}$ we compute $\mathbf{M}_A (\widehat{\widehat{\mathfrak{D}}}_N)$. The $A$-monodromy matrix for the oper $\widehat{\mathfrak{D}}_N$ is then
\begin{align}
M_A (\widehat{\mathfrak{D}}_N) = \lim_{\varepsilon_2 \to 0} \mathbf{M}_A (\widehat{\widehat{\mathfrak{D}}}_N),
\end{align}
which can be expressed in terms of the $\boldsymbol\alpha$ coordinates by comparing its traces with \eqref{eq:tran}. We also compute the $B$-monodromy for $\widehat{\widehat{\mathfrak{D}}}_N$ by
\begin{align}
\mathbf{M}_B (\widehat{\widehat{\mathfrak{D}}}_N) = \mathbf{R}_\infty \; \mathbf{C}_\infty \; \mathbf{S} \; \mathbf{C}_0 ^{-1} \; \mathbf{R}_0  \; \mathbf{C}_0 \; \mathbf{S}^{-1} \; \mathbf{C}_\infty ^{-1} ,
\end{align}
from which we compute the $B$-monodromy matrix for the oper $\widehat{\mathfrak{D}}_N$ as
\begin{align}
\begin{split}
M_B (\widehat{\mathfrak{D}}_3 ) &= \lim_{\varepsilon_2 \to 0} \mathbf{M}_B (\widehat{\widehat{\mathfrak{D}}}_N)  \; e^{\frac{\widetilde{\EuScript{W}}}{\varepsilon_2}} \\
&= \lim_{\varepsilon_2 \to 0}  \mathbf{R}_\infty \; \mathbf{C}_\infty \; \mathbf{S} \; \mathbf{C}_0 ^{-1} \; \mathbf{R}_0  \; \mathbf{C}_0 \; \mathbf{S}^{-1} \; \mathbf{C}_\infty ^{-1} \; e^{\frac{\widetilde{\EuScript{W}}}{\varepsilon_2}}.
\end{split}
\end{align}
Then we find the expressions for the traces
\begin{align} \label{traces}
\text{Tr}\,M_B (\widehat{\mathfrak{D}}_N) ^k, \;\;\quad k=1, \cdots, N-1.
\end{align}
By comparing these expressions with \eqref{eq:trbn}, we find that
\begin{align} \label{betagen}
\boldsymbol{\beta}_\alpha = \frac{1}{\varepsilon_1} \frac{\partial \EuScript{\widetilde{W}}^{\text{full}}}{\partial \boldsymbol{\alpha}_\alpha}, \quad \alpha=1, \cdots, N-1.
\end{align}
This relation verifies that the generating function for the variety $\mathcal{O}_N [\mathbb{P}^1 \backslash \{ 0, \underline{\qe} , \underline{1}, \infty \}]$ of opers in the generalized NRS coordinate system $\{ \boldsymbol{\alpha}_\alpha, \boldsymbol{\beta}_\alpha \; \vert \; \alpha=1, \cdots, N-1 \}$ is identical to the effective twisted superpotential:
\begin{align} \label{equivn}
\mathcal{S}\left[ \mathcal{O}_N [\mathbb{P}^1 \backslash \{ 0, \underline{\mathfrak{q}} , \underline{1}, \infty \}] \right] = \frac{1}{\varepsilon_1} \widetilde{\EuScript{W}}^{\text{full}} \left[ \mathcal{T}[A_{N-1}, \mathbb{P}^1 \backslash \{ 0, \underline{\qe} , \underline{1}, \infty \} ] \right].
\end{align}

\section{Discussion} \label{dis}
We have shown that non-perturbative Dyson-Schwinger equations for the class-$\EuScript{S}$ theories with the insertion of a surface defect produce the operators $\widehat{\widehat{\mathfrak{D}}}$ annihilating their partition functions. These operators were reduced to the opers $\widehat{\mathfrak{D}}$ in the limit $\varepsilon_2 \to 0$, providing an explicit relation between the holomorphic coordinates on the variety of opers and the expectation values of the chiral observables in the limit $\varepsilon_2 \to 0$. The surface defect partition functions, i.e., the solutions to $\widehat{\widehat{\mathfrak{D}}}$, were analytically continued to different convergence domains and glued together in the intermediate domain. This procedure enabled the computation of the monodromies of the solutions to $\widehat{\widehat{\mathfrak{D}}}$, and therefore the monodromies of the opers $\widehat{\mathfrak{D}}$ by taking the limit $\varepsilon_2 \to 0$. We constructed a higher-rank generalization of the NRS coordinate system, and represented the monodromies of opers in terms of these coordinates. The effective twisted superpotential arose as the generating function for the variety of opers in the generalized NRS coordinate system by construction. 

We believe that the subject deserves more investigations in various aspects. Let us consider the example of $\mathfrak{g}=A_1$. We have constructed the Darboux coordinate system $(\boldsymbol{\alpha}, \boldsymbol{\beta})$ in which the generating function for the variety $\mathcal{O}_2 [ \EuScript{C}]$ oper $\widehat{\mathfrak{D}}_2$ is identified with the effective twisted superpotential. Meanwhile, $\mathcal{O}_2 [ \EuScript{C}]$ is a a Lagrangian submanifold of $\EuScript{M}_{\text{flat}} (SL(2), \EuScript{C})$, which is spanned by the off-shell spectra $u_2 = \lim_{\varepsilon_2 \to 0} \Big\langle \EuScript{O}_2 \Big\rangle$ for fixed gauge couplings $\mathfrak{q}$. The variation of the gauge couplings, i.e., the elements of the Teichm\"{u}ller space $\mathbb{T}[\EuScript{C}]$ of $\EuScript{C}$, gives the foliation of the moduli space $\EuScript{M}_{\text{flat}} (SL(2), \EuScript{C})$ by the leaves of the varieties of opers with varying gauge couplings. Thus there exists another Darboux coordinate system $(\tau_2 = \log \mathfrak{q}, u_2)$ on $\EuScript{M}_{\text{flat}} (SL(2), \EuScript{C})$ induced from the identification $\EuScript{M}_{\text{flat}} (SL(2), \EuScript{C}) \simeq T^* \mathbb{T}[\EuScript{C}]$. We observe that the relations
\begin{align}
\boldsymbol{\beta} = \frac{1}{\varepsilon_1} \frac{ \partial \widetilde{\EuScript{W}}}{\partial \boldsymbol{\alpha}}, \quad u_2 =  \frac{\partial \widetilde{\EuScript{W}}}{\partial \tau_2},
\end{align}
identify the effective twisted superpotential with the generating function for the canonical transformation of the Darboux coordinate systems,
\begin{align}
(\tau_2, u_2) \xlongleftrightarrow{\widetilde{\EuScript{W}}} (\boldsymbol{\alpha},\boldsymbol{\beta}).
\end{align}
Let us consider generalizing this relation to the higher rank $\mathfrak{g}=A_{2}$, for a fixed Riemann surface, say, $\mathbb{P} ^1 _{2, \underline{r+1}}$. We still have the generalized NRS coordinate system $\{ \boldsymbol{\alpha}_i ^{(\alpha)}, \boldsymbol{\beta}_i ^{(\alpha)} \: \vert \: i=0,1, \cdots, r-1, \; \alpha=1, \cdots, N-1 \}$ on one hand, but it is apparent that the variation on the Teichm\"{u}ller space $\mathbb{T} [ \mathbb{P}^1 _{2, \underline{r+1}} ]$ does not saturate the half of the dimension of the moduli space, since the dimension of the moduli space increases as the rank increases, $\dim \EuScript{M}_{\text{flat}} (SL(3), \mathbb{P}^1 _{2, \underline{r+1}}) = 2r(N-1)=4r$, while the dimension of the Teichm\"{u}ller space is independent of the rank, $\dim \mathbb{T} [ \mathbb{P}^1 _{2, \underline{r+1}} ] =r$. In other words, we need $r$ more parameters $\tau_{\mathbf{i},3}$ to form a Darboux coordinate system, 
\begin{align}
\{ \tau_{\mathbf{i},2}, \tau_{\mathbf{i},3}, u_{\mathbf{i},2}, u_{\mathbf{i},3} \; \vert \; \mathbf{i}=1, \cdots, r \},
\end{align}
in which the effective twisted superpotential produces the spectrum $u_{\mathbf{i},3} = \lim_{\varepsilon_2 \to 0} \Big\langle \EuScript{O}_{\mathbf{i},3} \Big\rangle$ of the higher Hamiltonian $\EuScript{O}_{\mathbf{i},3} = \text{Tr} \phi_{\mathbf{i}} ^3$ under the differentiation with respect to $\tau_{\mathbf{i},3}$. Then the effective twisted superpotential becomes the generating function for the canonical transformation between Darboux coordinate systems, through the relations
\begin{align}
&\boldsymbol{\beta}_i ^{(\alpha)} = \frac{1}{\varepsilon_1} \frac{ \partial \widetilde{\EuScript{W}}}{\partial \boldsymbol{\alpha}_i ^{(\alpha)}}, \:\quad\quad\quad\quad\quad\, i=0,1, \cdots, r-1,\; \alpha=1,\cdots, N-1, \\
&u_{\mathbf{i},2} =  \frac{\partial \widetilde{\EuScript{W}}}{\partial \tau_{\mathbf{i},2}},\:\:  u_{\mathbf{i},3}  \stackrel{?}{=} \frac{\partial \widetilde{\EuScript{W}}}{\partial \tau_{\mathbf{i},3}}, \quad\;\; \mathbf{i}=1, \cdots, r.
\end{align}
But what is the meaning of the parameters $\tau_{\mathbf{i},3}$?

In the gauge theory side, the meaning of $\tau_{\mathbf{i},3}$ is clear. As investigated in \cite{marnek}, we may extend the theory by manually adding \textit{the higher times} to the microscopic action,
\begin{align}
\EuScript{L} = \sum_{\mathbf{i}=1} ^r \tau_{\mathbf{i},2} \int d^4 \theta \: \text{Tr}\: \boldsymbol{\Phi}_{\mathbf{i}} ^2 + \tau_{\mathbf{i},3} \int d^4 \theta \: \text{Tr} \: \boldsymbol{\Phi}_{\mathbf{i}} ^3,
\end{align} 
whose partition function can still be computed by equivariant localization as, schematically,
\begin{align}
\EuScript{Z}^{\text{inst}} (\mathbf{a}, \mathbf{m}, \varepsilon_1, \varepsilon_2; \tau_2, \tau_3) = \sum_{\boldsymbol{\lambda}} \prod_{\mathbf{i}=1} ^r \mathfrak{q}_{\mathbf{i}} ^{\vert \boldsymbol{\lambda}^{({\mathbf{i}})} \vert} \exp \left[ \sum_{\mathbf{i}=1} ^r \tau_{\mathbf{i},3} \: \EuScript{O}_{\mathbf{i},3} [\boldsymbol{\lambda}] \right] \boldsymbol{\mu}_{\boldsymbol{\lambda}} (\mathbf{a},\mathbf{m}, \varepsilon_1, \varepsilon_2).
\end{align}
Under the limit $\varepsilon_2 \to 0$, the partition function shows the asymptotic behavior,
\begin{align}
\EuScript{Z}^{\text{inst}} (\mathbf{a}, \mathbf{m}, \varepsilon_1, \varepsilon_2; \tau_2, \tau_3)  = e^{\frac{\widetilde{\EuScript{W}} (\mathbf{a}, \mathbf{m}, \varepsilon_1; \tau_2, \tau_3)}{\varepsilon_2}} \left( 1+ \mathcal{O}(\varepsilon_2 )\right).
\end{align}
Then it is straightforward that we produce the relation
\begin{align}
u_{\mathbf{i},3} = \lim_{\varepsilon_2 \to 0} \Big\langle \EuScript{O}_{\mathbf{i},3} \Big\rangle = \frac{\partial \widetilde{\EuScript{W}}}{\partial \tau_{\mathbf{i},3}}.
\end{align}
Therefore, the extra parameters that foliate the remaining orthogonal directions to the varieties of opers are the higher times of the extended theory. The  varieties $\mathcal{O}_3 [\EuScript{C}] $ of opers such as \eqref{heun3} are located at $\tau_{\mathbf{i},3}=0$ and only probe the $\tau_{\mathbf{i},2}$-variations. 

The question is, then, what \textit{the extended opers} are, which rise under the flow along the directions of the higher times. When re-phrased in terms of the $qq$-characters, the problem is to derive proper extended operators $\widehat{\widehat{\mathfrak{D}}}$ from the non-perturbative Dyson-Schwinger equations of the extended theories with an insertion of a surface defect. The limit $\varepsilon_2 \to 0$ of these objects would yield the desired extended opers. Note that the expectation values of $\EuScript{O}_{\mathbf{i},3}$ would be compensated by the derivatives with respect to $\tau_{\mathbf{i},3}$, so that the issue of equating the analytically continued expectation values in the intermediate domain would also be resolved with this enhancement. It is not clear, however, how to derive meaningful expressions for these extended quantized opers $\widehat{\widehat{\mathfrak{D}}}$ as of yet, so we leave this to future work. The variation along the higher times has many different manifestations. It corresponds to varying the higher Teichm\"{u}ller structures studied in \cite{fg1,fg2}, flowing along the higher Hamiltonians in the isomonodromic deformation of Fuchsian systems, and properly extending the Hamilton-Jacobi formulation of the Painlev\'{e} equations discussed in \cite{llnz} to the higher order Painlev\'{e}-type equations. It would be interesting to see how the extended gauge theory ties up these different realms of mathematical physics.

In the context of the BPS/CFT correspondence, the subject reveals still another feature along the line of \cite{agt}. The well-established relation between the partition functions of $\mathcal{T}[A_{N-1}, \EuScript{C}]$ and the correlation functions of $A_{N-1}$-Toda CFTs has to be extended when we deal with the higher ranks $N \geq 3$. Namely, we start to face the expectation values of higher chiral observables in the gauge theory side, which cannot be compensated by the derivatives of gauge couplings, and they are supposed to correspond to the correlation functions with inclusion of $\mathcal{W}$-descendant fields in the CFT side. The precise dictionary between the two objects are yet to be accomplished. The realization of the higher times of the extended theories in the CFT side is even more unclear. The free field representation of the monodromies of degenerate fields studied in \cite{cpt} can be relevant for this study.

Another problem related to this work is the generalized NRS coordinate systems corresponding to the non-Lagrangian theories. It is well-known that the higher rank class-$\EuScript{S}$ theories do not always admit Lagrangian descriptions. Our computation of monodromy data of opers heavily utilized the availability of the exact computations of the partition functions and the expectation values of the chiral observables. For the non-Lagrangian theories, it is not even clear what \textit{the instanton counting} means. Nevertheless, the Fuchsian systems with the prescribed monodromies around the punctures are still well-defined, and we may wonder if it is possible to explicitly link the accessory parameters of the corresponding opers and the expectation values of the chiral observables in the non-Lagrangian theories. In the case when the non-Lagrangian theory is S-dual to a Lagrangian theory, it is desirable to explicitly construct the coordinate transformation \cite{nrs} between the relevant generalized NRS coordinate systems and investigate their field theoretical meaning.

\appendix
\section{Partition functions of $\EuScript{N}=2$ supersymmetric quiver gauge theories} \label{appA}
We give a brief review on the partition functions of the $\EuScript{N}=2$ quiver gauge theories. For more details on this subject, see \cite{nekpessha, nek2}.

 For an oriented graph $\gamma$, we denote the sets of its vertices and edges and $\text{Vert}_\gamma$ and $\text{Edge}_\gamma$, respectively. We define $s,t:\text{Edge}_\gamma \to \text{Vert}_\gamma$ as the maps which send an edge to its source and target, respectively. For each vertex we assign two integers,
\begin{align}
\mathbf{n} = (\mathpzc{n}_\mathbf{i})_{\mathbf{i} \in \text{Vert}_\gamma} \in \left( \mathbb{Z}^{>0} \right) ^{\text{Vert}_\gamma}, \quad \mathbf{m} = (\mathpzc{m}_\mathbf{i})_{\mathbf{i} \in \text{Vert}_\gamma} \in \left( \mathbb{Z}^{\geq 0} \right) ^{\text{Vert}_\gamma}.
\end{align}
The $\EuScript{N}=2$ quiver gauge theory associated to $\gamma$ is the four-dimensional $\EuScript{N}=2$ supersymmetric gauge theory, whose gauge group is 
\begin{align} \label{globalgauge}
G_{g} = \bigtimes_{\mathbf{i} \in \text{Vert}_\gamma} U(\mathpzc{n}_\mathbf{i}),
\end{align}
and whose flavor group is
\begin{align} \label{flavor}
G_f = \left( \bigtimes_{\mathbf{i} \in \text{Vert}_\gamma} U(\mathpzc{m}_\mathbf{i}) \times U(1) ^{\text{Edge}_\gamma} \right) \Bigg/ U(1) ^{\text{Vert}_\gamma}.
\end{align}
Here the overall $U(1)^{\text{Vert}_\gamma}$ transformation has been mod out due to the gauge symmetry,
\begin{align}
(u_\mathbf{i})_{\mathbf{i} \in \text{Vert}_\gamma} : \left( (g_\mathbf{i})_{\mathbf{i} \in \text{Vert}_\gamma} , ( u_\mathbf{e})_{\mathbf{e} \in \text{Edge}_\gamma} \right) \mapsto \left( (u_\mathbf{i} g_\mathbf{i})_{\mathbf{i} \in \text{Vert}_\gamma} , (u_{s(\mathbf{e})} u_\mathbf{e} u_{t(\mathbf{e})} ^{-1})_{\mathbf{e} \in \text{Edge}_\gamma} \right).
\end{align}
The field contents of the theory are the following: the vector multiplets $\boldsymbol{\Phi} = (\Phi_\mathbf{i})_{\mathbf{i} \in \text{Vert}_\gamma}$ in the adjoint representation of $G_g$, the fundamental hypermultiplets $\boldsymbol{Q}_{\text{fund}} = (Q_\mathbf{i})_{\mathbf{i} \in \text{Vert}_\gamma}$ in the fundamental representation of $G_g$ and the antifundamental representation of $G_f$, and finally the bifundamental hypermultiplets $\boldsymbol{Q}_{\text{bifund}} = (Q_\mathbf{e})_{\mathbf{e} \in \text{Edge}_\gamma}$ in the bifundamental representation $(\overline{n_{s(\mathbf{e})}} ,n_{t(\mathbf{e})})$ of $G_g$. The $\EuScript{N}=2$ supersymmetric action is then fixed up to the gauge couplings,
\begin{align}
\mathfrak{q}_\mathbf{i} = \text{exp}(2 \pi i \tau_\mathbf{i}) \quad \left( \tau_\mathbf{i} = \frac{\vartheta_\mathbf{i}}{2\pi} + \frac{4\pi i}{g_\mathbf{i} ^2} \right), \quad \mathbf{i} \in \text{Vert}_\gamma,
\end{align}
and the masses of the hypermultiplets,
\begin{align} \label{mass}
&\textbf{\textit{m}} = ((\mathbf{m}_\mathbf{i})_{\mathbf{i} \in \text{Vert}_\gamma}, (m_\mathbf{e})_{ \mathbf{e} \in  \text{Edge}_\gamma}), \nonumber \\
&\mathbf{m}_\mathbf{i} = \text{diag}(m_{\mathbf{i},1} , \cdots, m_{\mathbf{i},\mathpzc{m}_\mathbf{i}}) \in \text{End}(\mathbb{C}^{\mathpzc{m}_\mathbf{i}}), \quad m_\mathbf{e} \in \mathbb{C}.
\end{align}
The global symmetry group of the theory is 
\begin{align}
H = G_g \times G_f \times G_{\text{rot}},
\end{align}
where $G_g$ \eqref{globalgauge} is the group of global gauge symmetry, $G_f$ \eqref{flavor} is the group of flavor symmetry, and $G_{\text{rot}} = SO(4)$ is the group of the Lorentz symmetry. We turn on equivariant parameters for the maximal torus $T_H \subset H$. The equivariant parameters for $G_g$ is the vacuum expectation values of the complex scalars,
\begin{align}
\langle \Phi_\mathbf{i} \rangle = \mathbf{a}_\mathbf{i}, \quad  \mathbf{a}_\mathbf{i} = \text{diag}(a_{\mathbf{i},1}, \cdots, a_{\mathbf{i},\mathpzc{n}_\mathbf{i}}) \in \text{End}(\mathbb{C}^{\mathpzc{n}_\mathbf{i}}), \quad \mathbf{i} \in \text{Vert}_\gamma.
\end{align}
The equivariant parameters for $G_f$ is the masses of the hypermultiplets \eqref{mass}. Finally the equivariant parameters for $G_{\text{rot}}$ is the $\Omega$-deformation parameters $\varepsilon_1, \varepsilon_2$. The partition function of the theory is a function of these parameters $(\mathbf{a}, \textbf{\textit{m}}, \boldsymbol{\varepsilon}) \in \text{Lie}(T_H)$. In expressing the partition function, we abuse our notation and denote the vector spaces and their $T_H$-equivariant characters in the same letters. Hence we write
\begin{align}
N_{\mathbf{i}} = \sum_{\alpha =1} ^{\mathpzc{n}_{\mathbf{i}}} e^{\beta a_{\mathbf{i},\alpha}}, \quad M_{\mathbf{i}} = \sum_{f=1} ^{\mathpzc{m}_{\mathbf{i}}} e^{\beta m_{\mathbf{i}, f}}.
\end{align}
It is helpful to use the following notation for abbreviated expressions,
\begin{align}
\begin{split}
&q_{i} \equiv e^{\beta \varepsilon_{i}}, \quad P_{i} \equiv 1-q_{i} \quad i=1,2, \\
&q_{12} \equiv q_1 q_2,\quad P_{12} \equiv (1-q_1)(1-q_2).
\end{split}
\end{align}
The partition function factors into the classical, one-loop, and the instanton parts:
\begin{align}
\EuScript{Z}(\mathbf{a}, \boldsymbol{m}, \boldsymbol{\varepsilon}, \mathfrak{q}) = \EuScript{Z}^{\text{classical}}\; \EuScript{Z}^{\text{1-loop}} \; \EuScript{Z}^{\text{inst}}.
\end{align}
The classical part is given by
\begin{align}
 \EuScript{Z}^{\text{classical}}(\mathbf{a}, \boldsymbol{\varepsilon}, \mathfrak{q}) = \prod_{\mathbf{i} \in \text{Vert}_\gamma} \mathfrak{q}_{\mathbf{i} } ^{ -\frac{1}{2\varepsilon_1 \varepsilon_2} \sum_{\alpha=1} ^{\mathpzc{n}_{\mathbf{i}}} a_{\mathbf{i}, \alpha} ^2 }.
\end{align}
The one-loop part is given by
\begin{align} \label{1loop}
\begin{split}
&\EuScript{Z}^{\text{1-loop}}(\mathbf{a}, \boldsymbol{m}, \boldsymbol{\varepsilon}) \\
& = \epsilon \left[ \frac{1}{(1-e^{-\beta \varepsilon_1} )(1-e^{-\beta \varepsilon_2})} \left( \sum_{\mathbf{i} \in \text{Vert}_\gamma } (M_{\mathbf{i}} -N_{\mathbf{i}})N_{\mathbf{i}} ^* + \sum_{\mathbf{e} \in \text{Edge}_\gamma} e^{\beta m_{\mathbf{e}}} N_{t(\mathbf{e})} N_{s(\mathbf{e})} ^* \right) \right],
\end{split}
\end{align}
where the $\epsilon$-symbol is defined by
\begin{align} \label{epsilon}
\epsilon\left[ \cdots \right] \equiv \exp \left[ \frac{d}{ds} \Bigg\vert_{s=0} \frac{1}{\Gamma(s)} \int_0 ^\infty d\beta \beta^{s-1} [\cdots] \right],
\end{align}
which converts a character into a product of weights. In particular, the $\epsilon$-symbol regularizes an infinite product of weights such as \eqref{1loop} by the Barnes double gamma function,
\begin{align}
\Gamma_2 (x;\varepsilon_1,\varepsilon_2) \equiv \exp \left[ -\frac{d}{ds} \Bigg\vert_{s=0} \frac{1}{\Gamma(s)} \int_0 ^{\infty} d\beta \beta^{s-1} \frac{ e^{-\beta x}}{(1-e^{-\beta \varepsilon_1})(1-e^{-\beta \varepsilon_2})} \right].
\end{align}

The instanton part $\EuScript{Z}^{\text{inst}}$ is obtained by a $T_H$-equivariant integral over the instanton moduli space. Given the vector of the instanton charges $\mathbf{k} = (k_\mathbf{i})_{\mathbf{i} \in \text{Vert}_\gamma} \in \mathbb{Z}^{\geq 0}$, the total framed noncommutative instanton moduli space of the quiver gauge theory for $\gamma$ is
\begin{align}
\EuScript{M}_\gamma (\mathbf{n}, \mathbf{k}) \equiv \bigtimes_{\mathbf{i} \in \text{Vert}_\gamma} \EuScript{M}(\mathpzc{n}_\mathbf{i} , k_\mathbf{i}) \label{quivmodul},
\end{align}
where $\EuScript{M} (\mathpzc{n}_\mathbf{i} , k_\mathbf{i})$ is the ADHM moduli space
\begin{align}
&\EuScript{M} (n , k)  = \begin{cases} \quad \quad B_{1,2}  : K \rightarrow K, \\ I : N \rightarrow K, J : K \rightarrow N\end{cases} \Bigg\vert \begin{rcases} \left[ B_1 , B_2   \right] + I J=0, \quad \quad \quad \quad \quad   \\ [B_1 , {B_1 } ^\dagger ] + [B_2  , {B_2 } ^ \dagger] + I {I} ^\dagger - {J} ^\dagger J = \zeta \end{rcases} \Bigg/ U(k). \\
&(N = \mathbb{C}^n, K = \mathbb{C}^k) \nonumber
\end{align}
Solving the real moment map equation $[B_1 , {B_1 } ^\dagger ] + [B_2  , {B_2 } ^ \dagger] + I {I} ^\dagger - {J} ^\dagger J = \zeta$ and dividing by the compact $U(k)$ is equivalent to imposing the stability condition and dividing by the complex group $GL(k)$,
\begin{align} \label{adhmmod}
&\EuScript{M} (n , k)  = \begin{cases} \quad \quad B_{1,2}  : K \rightarrow K, \\ I : N \rightarrow K, J : K \rightarrow N\end{cases} \Bigg\vert \begin{rcases} \left[ B_1 , B_2   \right] + I J=0,  \quad\quad \\\quad  K = \mathbb{C} [B_1, B_2]\: I (N) \quad \end{rcases} \Bigg/ GL(k).
\end{align}
The $T_H$-equivariant integration over the instanton moduli space \eqref{quivmodul} localizes on the set of fixed points of $T_H$-action, $\EuScript{M}_\gamma (\mathbf{n}, \mathbf{k})^{T_H}$, which is the set of colored partitions $\boldsymbol{\lambda}=((\lambda^{(\mathbf{i}, \alpha)})_{\alpha=1} ^{\mathpzc{n}_\mathbf{i}})_{\mathbf{i} \in \text{Vert}_\gamma}$, where each $\lambda^{(\mathbf{i}, \alpha)}$ is a partition,
\begin{align} \lambda^{(\mathbf{i}, \alpha)} = \left( \lambda^{(\mathbf{i},\alpha)} _1 \geq \lambda^{(\mathbf{i}, \alpha)}_2 \geq \cdots \geq \lambda^{(\mathbf{i},\alpha)}_{l(\lambda^{(\mathbf{i},\alpha)})} > \lambda^{(\mathbf{i}, \alpha)}_{l(\lambda^{(\mathbf{i},\alpha)})+1} = \cdots = 0   \right),
\end{align}
with the size $\vert \lambda^{(\mathbf{i}, \alpha)} \vert = \sum_{i=1} ^{l(\lambda^{(\mathbf{i}, \alpha)})} \lambda^{(\mathbf{i},\alpha)} _i = k_{\mathbf{i}, \alpha}$ constrained by $k_\mathbf{i} = \sum_\alpha k_{\mathbf{i}, \alpha} = \vert \boldsymbol{\lambda} ^{(\mathbf{i})} \vert$ \cite{nek1, nekokoun}. At each fixed point $\boldsymbol{\lambda}$, the vector space $K_{\mathbf{i}}$ carries a representation of $T_H$ with the weights given by the formula
\begin{align}
K_{\mathbf{i}} [\boldsymbol{\lambda}] = \sum_{\alpha=1} ^{\mathpzc{n}_\mathbf{i}} \sum_{\Box \in \lambda^{(\mathbf{i}, \alpha)}} e^{\beta c_{\square}},
\end{align}
where we have defined the content of the box,
\begin{align} 
c_{\square} = a_{\mathbf{i}, \alpha} + \varepsilon_1 (i-1) + \varepsilon_2 (j-1) \quad \text{for} \quad \Box = (i,j) \in \lambda^{(\mathbf{i},\alpha)} \iff 1 \leq j \leq \lambda_i ^{(\mathbf{i},\alpha)}.
\end{align}
The tangent bundle and the matter bundle comprise the character
\begin{align}
\EuScript{T}[\boldsymbol{\lambda}] &= \sum_{\mathbf{i} \in \text{Vert}_\gamma} \left( N_\mathbf{i} K_\mathbf{i} ^*+ q_{12} N_\mathbf{i} ^* K_\mathbf{i} - P_{12} K_\mathbf{i} K_\mathbf{i} ^* -M_\mathbf{i} ^* K_\mathbf{i} \right) \nonumber \\
&  - \sum_{\mathbf{e} \in \text{Edge}_\gamma} e^{\beta m_\mathbf{e}} (N_{t(\mathbf{e})} K_{s(\mathbf{e})} ^* +q_{12} N_{s(\mathbf{e})} ^*  K_{t(\mathbf{e})} - P_{12} K_{t(\mathbf{e})} K_{s(\mathbf{e})} ^*)  \label{charac},
\end{align}
associated to each fixed point $\boldsymbol{\lambda} \in \EuScript{M}_\gamma (\mathbf{n}, \mathbf{k})^{T_H}$. At last the instanton part of the partition function is evaluated by
\begin{align}
\EuScript{Z}^{\text{inst}} ( \mathbf{a}; \textbf{\textit{m}}; \boldsymbol{\varepsilon} ; \mathfrak{q}) = \sum_{\boldsymbol{\lambda}} \prod_{\mathbf{i} \in \text{Vert}_\gamma} \mathfrak{q}_\mathbf{i} ^{\vert \boldsymbol{\lambda}^{(\mathbf{i})} \vert} \; \epsilon \left[ \EuScript{T}[\boldsymbol{\lambda}] \right], \label{instpart}
\end{align}
where we have used the $\epsilon$-symbol \eqref{epsilon}. Note that the one-loop part and the instanton part can be combined into
\begin{align}\label{1loopinst}
\begin{split}
&\EuScript{Z}^{\text{1-loop}}(\mathbf{a}, \boldsymbol{m}, \boldsymbol{\varepsilon})  \: \EuScript{Z}^{\text{inst}} ( \mathbf{a}; \textbf{\textit{m}}; \boldsymbol{\varepsilon} ; \mathfrak{q}) \\
&=\sum_{\boldsymbol{\lambda}} \prod_{\mathbf{i} \in \text{Vert}_\gamma} \mathfrak{q}_\mathbf{i} ^{\vert \boldsymbol{\lambda}^{(\mathbf{i})} \vert} \; \epsilon\left[ \frac{1}{(1-e^{-\beta \varepsilon_1})(1-e^{-\beta \varepsilon_2})} \left( \sum_{\mathbf{i} \in \text{Vert}_\gamma } (M_{\mathbf{i}} -S_{\mathbf{i}})S_{\mathbf{i}} ^* + \sum_{\mathbf{e} \in \text{Edge}_\gamma} e^{\beta m_{\mathbf{e}}} S_{t(\mathbf{e})} S_{s(\mathbf{e})} ^* \right) \right],
\end{split}
\end{align}
with the character $S_{\mathbf{i}} \equiv N_\mathbf{i} - P_{12} K_{\mathbf{i}}$.

The regularized characteristic polynomials of the adjoint scalars form important chiral observables, called the $\EuScript{Y}$-observables, are defined by
\begin{align}
\EuScript{Y}_{\mathbf{i}}(x) \equiv x^{\mathpzc{n}_\mathbf{i}} \exp \sum_{l=1} ^\infty -\frac{1}{lx ^l} \text{Tr}\: \Phi_\mathbf{i} ^l \vert_0,
\end{align}
Their expressions at the fixed point $\boldsymbol{\lambda}$ are written as
\begin{align} \label{yobs}
\EuScript{Y}_\mathbf{i} (x)[\boldsymbol{\lambda}] = \prod_{\alpha=1} ^{\mathpzc{n}_\mathbf{i}} \left( (x-a_{\mathbf{i},\alpha}) \prod_{\square \in \lambda^{(\mathbf{i},\alpha)}} \frac{(x- c_\square -\varepsilon_1)(x-c_\square-\varepsilon_2)}{(x-c_\square)(x-c_\square-\varepsilon)}  \right).
\end{align}
which shows that upon the regularization, the instanton contribution makes the polynomials into rational functions of the auxiliary variable $x$. 
The $\EuScript{Y}$-observable can be simply written as
\begin{align}
\EuScript{Y}_\mathbf{i} (x)[\boldsymbol{\lambda}] = \beta^{-\mathpzc{n}_\mathbf{i}}\: \epsilon [-e^{\beta x} S_\mathbf{i} ^*].
\end{align}
Note that the $\EuScript{Y}$-observables are the generating functions for the chiral observables
\begin{align}
\begin{split}
\EuScript{O}_{\mathbf{i},k} [\boldsymbol{\lambda}] &\equiv \text{Tr} \: \Phi_{\mathbf{i}} ^k \vert_0 [\boldsymbol{\lambda}] \\
& =  \sum_{\alpha=1} ^{\mathpzc{n}_\mathbf{i}} \left[ a_{\mathbf{i},\alpha} ^k + \sum_{\Box \in \lambda^{(\mathbf{i},\alpha)}} \left( (c_{\Box} + \varepsilon_1)^k + ( c_{\Box} + \varepsilon_2)^k - c_{\Box} ^k - (c_{\Box} + \varepsilon)^k \right) \right].
\end{split} 
\end{align}
The $qq$-characters for the quiver gauge theories are given as certain Laurent polynomials of the $\EuScript{Y}$-observables.

In section \ref{mono}, it is important to correctly identify the 1-loop contribution to the effective twisted superpotential in the $A_1$-theory. The formula \eqref{1loop} tells that
\begin{align} \label{1looppart}
\EuScript{Z}^{\text{1-loop}} _{A_1} = \frac{\prod_{\alpha, \beta =1} ^N \Gamma_2 (a_\alpha -a_\beta ; \varepsilon_1, \varepsilon_2)}{\prod_{\alpha,\beta=1} ^N \Gamma_2 (a_{\alpha} -a_{0,\beta}; \varepsilon_1, \varepsilon_2) \Gamma_2 (a_{3,\alpha} -a_\beta; \varepsilon_1, \varepsilon_2)}.
\end{align}
Note that we have the following identity,
\begin{align}
\frac{\partial}{\partial x} \left(  \lim_{\varepsilon_2 \to 0} \varepsilon_2 \log \Gamma_2 (x;\varepsilon_1, \varepsilon_2) \right) = - \log \Gamma_1(x;\varepsilon_1),
\end{align}
where we have defined
\begin{align}
\begin{split}
\Gamma_1 (x;\varepsilon_1) &\equiv \exp \left[ -\frac{d}{ds} \Bigg\vert_{s=0} \frac{1}{\Gamma(s)} \int_0 ^\infty d\beta \beta^{s-1} \frac{e^{-\beta x}}{1- e^{-\beta \varepsilon_1}} \right] =\frac{\sqrt{2\pi/\varepsilon_1}}{\varepsilon_1 ^{\frac{x}{\varepsilon_1}} \Gamma \left( \frac{x}{\varepsilon_1} \right)}.
\end{split}
\end{align}
Thus for the 1-loop part of the effective twisted superpotential,
\begin{align}
\widetilde{\EuScript{W}} ^{\text{1-loop}} \equiv  \lim_{\varepsilon_2 \to 0} \varepsilon_2 \log \EuScript{Z}^{\text{1-loop}} _{A_1},
\end{align}
we derive the identity,
\begin{align} \label{1loopformula}
\left( \frac{\partial}{\partial a_\alpha} - \frac{\partial}{\partial a_\beta} \right)\widetilde{\EuScript{W}} ^{\text{1-loop}} = \log \prod_{\alpha' \neq \alpha} \frac{\Gamma\left( \frac{a_{\alpha}-a_{\alpha'}}{\varepsilon_1} \right)}{\Gamma\left( \frac{a_{\alpha'}-a_\alpha}{\varepsilon_1} \right)}  \prod_{\beta' \neq \beta} \frac{\Gamma\left( \frac{a_{\beta'}-a_{\beta}}{\varepsilon_1} \right)}{\Gamma\left( \frac{a_{\beta}-a_\beta'}{\varepsilon_1} \right)} \prod_{\gamma=1} ^N \frac{\Gamma \left( \frac{a_{3,\gamma}-a_{\alpha}}{\varepsilon_1} \right) \Gamma\left( \frac{a_\beta -a_{0,\gamma}}{\varepsilon_1} \right)}{\Gamma\left( \frac{a_{\alpha}-a_{0,\gamma}}{\varepsilon_1} \right) \Gamma\left( \frac{a_{3,\gamma}-a_\beta}{\varepsilon_1} \right)},
\end{align}
which was used in section \ref{mono} to absorb the 1-loop contribution $\widetilde{\EuScript{W}}^{\text{1-loop}}$ into $\widetilde{\EuScript{W}}^{\text{full}}$.

\section{Computing $\EuScript{G}$(x;t)} \label{appB}
Even though we only consider the $A_2$-quiver in section \ref{npdsquiver}, it is possible to compute the generating function $\EuScript{G}_r (x;t)$ for general $A_r$-quiver gauge theory. The fundamental $qq$-characters for the $A_r$-theory are written as \cite{nek2}
\begin{equation}
\mathscr{X}_{\ell}(x)=\frac{\mathscr{Y}_{0}\left(x+\varepsilon\left(1-\ell\right)\right)}{z_{0}z_{1}\cdots z_{\ell-1}}\sum_{\substack{I\subset[0,r]\\
|I|=\ell
}
}\prod_{i\in I}\left[z_{i} \: \Xi_{i}\left(x+\varepsilon\left(h_{I}(i)+1-\ell\right)\right)\right],\quad\ell=0,1,\cdots,r+1, \label{eq:qq-ch}
\end{equation}
where we have defined $\mathfrak{q}_i \equiv \frac{z_i}{z_{i-1}}$, $h_I (i) \equiv \vert \{ j \in I \; \vert \; j < i \} \vert$, and
\begin{equation}
\Xi_{i}(x)\equiv\frac{\mathscr{Y}{}_{i+1}(x+\varepsilon)}{\mathscr{Y}{}_{i}(x)}=1+\sum_{n=1}^{\infty}\frac{\zeta_{i,n}}{x^{n}}.
\end{equation}
We form the generating function by
\begin{eqnarray}
\mathscr{G}_{r}(x;t) & = & \mathscr{Y}_{0}(x)^{-1}\Delta_{r}^{-1}\sum_{\ell=0}^{r+1}z_{0}z_{1}\cdots z_{\ell-1}t^{\ell}\mathscr{X}_{\ell}\left(x-\varepsilon(1-\ell)\right)\nonumber \\
 & = & \Delta_{r}^{-1}\sum_{I\subset[0,r]}\left[\left(\prod_{i\in I}tz_{i}\right)\prod_{i\in I}\Xi_{i}\left(x+\varepsilon h_{I}(i)\right)\right] \\
& = &\sum_{n=0}^{\infty}\frac{\mathscr{G}_{r}^{(-n)}(t)}{x^{n}},\label{eq:Gr}
\end{eqnarray}
where
\begin{equation}
\Delta_{r}=\sum_{I\subset[0,r]}\left(\prod_{i\in I}tz_{i}\right)=\prod_{i=0}^{r}\left(1+tz_{i}\right).
\end{equation}
For convenience, we define the parameter
\begin{equation}
u_{i}=\frac{tz_{i}}{1+tz_{i}}.
\end{equation}
Let us start from $r=0$. Straightforward computation shows that
\begin{equation}
\mathscr{G}_{0}^{(0)}(t)=1,\quad\mathscr{G}_{0}^{(-n)}(t)=u_{0}\zeta_{0,n},\quad n\in\mathbb{Z}^{> 0}.
\end{equation}
We proceed to higher $r$ by recursion. Divide the sum in \eqref{eq:Gr} over $I \subset [0,r]$ into two classes, where in the first $r \in I$ and in the second $r \notin I$.  From this decomposition follows the relation:
\begin{eqnarray}
 \mathscr{G}_{r}(x;t) & = & \mathscr{G}_{0}(x;t)+\sum_{j=1}^{r}u_{j}\Delta_{j-1}^{-1}\sum_{n=1}^{\infty}\frac{\zeta_{j,n}}{\left(x+\varepsilon t\frac{\partial}{\partial t}\right)^{n}}\left(\Delta_{j-1}\mathscr{G}_{j-1}(x;t)\right).
\end{eqnarray}
Explicitly pulling out the coefficients of each negative powers of $x$, we can write the recursive relations as
\begin{subequations} \label{recursive}
\begin{align} 
&\mathscr{G}_{r}^{(0)}(t)  =  \mathscr{G}_{0}^{(0)}(t)=1,\\
&\mathscr{G}_{r}^{(-1)}(t)  =  \sum_{j=0}^{r}u_{j}\zeta_{j,1},\\
&\mathscr{G}_{r}^{(-2)}(t)  = \sum_{j=0}^{r}u_{j}\zeta_{j,2}+\sum_{j=1}^{r}u_{j}\zeta_{j,1}\mathscr{G}_{j-1}^{(-1)}(t) -\varepsilon\sum_{j=1}^{r}u_{j}\zeta_{j,1}\Delta_{j-1}^{-1}t\frac{\partial}{\partial t}\Delta_{j-1},\\
&\mathscr{G}_{r}^{(-3)}(t)  =  \sum_{j=0}^{r}u_{j}\zeta_{j,3}+\sum_{j=1}^{r}u_{j}\left[\zeta_{j,1}\mathscr{G}_{j-1}^{(-2)}(t)+\zeta_{j,2} \mathscr{G}_{j-1}^{(-1)}(t)\right]\nonumber \\
& \quad \quad -\varepsilon\sum_{j=1}^{r}u_{j}\Delta_{j-1}^{-1}\left[\zeta_{j,1}t\frac{\partial}{\partial t}\left(\Delta_{j-1}\mathscr{G}_{j-1}^{(-1)}(t)\right)+2\zeta_{j,2}t\frac{\partial}{\partial t}\Delta_{j-1}\right]\nonumber \\
 &\quad \quad +\varepsilon^{2}\sum_{j=1}^{r}u_{j}\zeta_{j,1}\Delta_{j-1}^{-1}\left(t\frac{\partial}{\partial t}\right)^{2}\Delta_{j-1},\\
&\mathscr{G}_{r}^{(-4)}(t)  =  \sum_{j=0}^{r}u_{j}\zeta_{j,4}+\sum_{j=1}^{r}u_{j}\left[\zeta_{j,1}\mathscr{G}_{j-1}^{(-3)}(t)+\zeta_{j,2}\mathscr{G}_{j-1}^{(-2)}(t)+\zeta_{j,3}\mathscr{G}_{j-1}^{(-1)}(t)\right]\nonumber \\
&\quad \quad -\varepsilon\sum_{j=1}^{r}u_{j}\Delta_{j-1}^{-1}\left[\zeta_{j,1}t\frac{\partial}{\partial t}\left(\Delta_{j-1}\mathscr{G}_{j-1}^{(-2)}(t)\right)+2\zeta_{j,2}t\frac{\partial}{\partial t}\left(\Delta_{j-1}\mathscr{G}_{j-1}^{(-1)}(t)\right)+3\zeta_{j,3}t\frac{\partial}{\partial t}\Delta_{j-1}\right]\nonumber \\
 &\quad \quad +\varepsilon^{2}\sum_{j=1}^{r}u_{j}\Delta_{j-1}^{-1}\left[\zeta_{j,1}\left(t\frac{\partial}{\partial t}\right)^{2}\left(\Delta_{j-1}\mathscr{G}_{j-1}^{(-1)}(t)\right)+3\zeta_{j,2}\left(t\frac{\partial}{\partial t}\right)^{2}\Delta_{j-1}\right]\nonumber \\
& \quad \quad-\varepsilon^{3}\sum_{j=1}^{r}u_{j}\zeta_{j,1}\Delta_{j-1}^{-1}\left(t\frac{\partial}{\partial t}\right)^{3}\Delta_{j-1}.
\end{align}
\end{subequations}
Introduce the notation:
\begin{equation}
U_{r}\left[s_{1},s_{2},\cdots,s_{\ell}\right]\equiv\sum_{0\leq i_{1}<\cdots<i_{\ell}\leq r}\prod_{n=1}^{\ell}\left(u_{i_{n}}\zeta_{i_{n},s_{n}}\right),
\end{equation}
where $s_i \in \mathbb{Z} ^{\geq 0}$ and $\zeta_{i,0} \equiv 1$ by definition. It is straightforward to derive the useful identities:
\begin{eqnarray}
t\frac{\partial}{\partial t}\Delta_{r} & = & \Delta_{r}U_{r}[0],\\
t\frac{\partial}{\partial t}\left(\Delta_{r}U_{r}[s_{1},\cdots,s_{\ell}]\right) & = & \Delta_{r}\left(\ell U_{r}[s_{1},\cdots,s_{\ell}]+U_{r}^{\oplus}[s_{1},\cdots,s_{\ell}]\right),\\
\sum_{j=1}^{r}u_{j}\zeta_{j,m}U_{j-1}[s_{1},\cdots,s_{\ell}] & = & U_{r}\left[s_{1},s_{2},\cdots,s_{\ell},m\right],
\end{eqnarray}
where
\begin{equation}
U_{r}^{\oplus}[s_{1},\cdots,s_{\ell}]\equiv U_{r}[0,s_{1},\cdots,s_{\ell}]+U_{r}[s_{1},0,\cdots,s_{\ell}]+\cdots+U_{r}[s_{1},\cdots,s_{\ell},0].
\end{equation}
Using them we simplify \eqref{recursive} and finally arrive at the expressions\begin{subequations}
\begin{align}
&\mathscr{G}_{r}^{(0)}(t) = 1,\\
&\mathscr{G}_{r}^{(-1)}(t) =  U_{r}[1],\\
&\mathscr{G}_{r}^{(-2)}(t) =  U_{r}[2]+U_{r}[1,1]-\varepsilon U_{r}[0,1],\\
&\mathscr{G}_{r}^{(-3)}(t) =  U_{r}[3]+U_{r}[2,1]+U_{r}[1,2]-\varepsilon\left(U_{r}[1,1]+2U_{r}[0,2]\right)+\varepsilon^{2}U_{r}[0,1]\nonumber \\
&  \quad\quad+U_{r}[1,1,1]-\varepsilon\left(2U_{r}[0,1,1]+U_{r}[1,0,1]\right)+2\varepsilon^{2}U_{r}[0,0,1],\\
&\mathscr{G}_{r}^{(-4)}(t) = U_{r}[4]+U_{r}[1,3]+U_{r}[2,2]+U_{r}[3,1]\nonumber \\
 &\quad\quad -\varepsilon\left(U_{r}[2,1]+2U_{r}[1,2]+3U_{r}[0,3]\right)+\varepsilon^{2}\left(U_{r}[1,1]+U_{r}[0,2]\right)-\varepsilon^{3}U_{r}[0,1]\nonumber \\
 &\quad\quad +U_{r}[2,1,1]+U_{r}[1,2,1]+U_{r}[1,1,2]\nonumber \\
 & \quad\quad -\varepsilon\left(3U_{r}[1,1,1]+3U_{r}[0,2,1]+3U_{r}[0,1,2]+2U_{r}[1,0,2]+U_{r}[2,0,1]\right)\nonumber \\
 &  \quad\quad  +\varepsilon^{2}\left(6U_{r}[0,1,1]+3U_{r}[1,0,1]+6U_{r}[0,0,2]\right)-6\varepsilon^{3}U_{r}[0,0,1]\nonumber \\
 &  \quad\quad  +U_{r}[1,1,1,1]-\varepsilon\left(3U_{r}[0,1,1,1]+2U_{r}[1,0,1,1]+U_{r}[1,1,0,1]\right)\nonumber \\
 &  \quad\quad  +\varepsilon^{2}\left(6U_{r}[0,0,1,1]+3U_{r}[0,1,0,1]+2U_{r}[1,0,0,1]\right)-6\varepsilon^{3}U_{r}[0,0,0,1].
\end{align}
\end{subequations}
 for the coefficients of the generating function. 
For the results in section \ref{npdsquiver}, we simply set $r=2$.

\section{The accessory operator $\widehat{H}_2$} \label{H2}
We present the full expression for the accessory operator $\widehat{H}_2 (z,\mathfrak{q})$ in $\widehat{\widehat{\mathfrak{D}}}_3$ below.
\footnotesize
\begin{align}
\begin{split}
&\widehat{H}_2  (z, \mathfrak{q}) \\
&= -(1-\mathfrak{q}) \left[ \frac{1}{3} \Big\langle  \EuScript{O}_3 \Big\rangle_{A_2} + \frac{ \varepsilon_1 \varepsilon_2 (12\bar{a}_0 -4 \mathcal{A}_1 ^{(1)} +2 \mathcal{A}_2 ^{(1)} +15 \varepsilon_1 +8\varepsilon_2)}{6} \mathfrak{q} \frac{\partial}{\partial \mathfrak{q}} -\frac{1}{3} \sum_{\alpha=1} ^3 a_{2,\alpha} ^3 +\frac{2 \left( \mathcal{A}_1 ^{(1)} \right)^2 \mathcal{A}_2 ^{(1)}}{9} \right. \\
& \quad\quad\quad\quad - \left(\mathcal{A}_2 ^{(1)} \right)^2 \left( \frac{3\bar{a}_0 - \mathcal{A}_1 ^{(1)}}{3} +\frac{33 \varepsilon_1 + 22\varepsilon_2}{36} \right) + \frac{\mathcal{A}_1 ^{(2)} \mathcal{A}_2 ^{(1)}}{3} +\mathcal{A}_2 ^{(2)} \left(  -\bar{a}_0 +\frac{4\mathcal{A}_1 ^{(1)} +2 \mathcal{A}_2 ^{(1)} -9\varepsilon_1 -2\varepsilon_2}{12} \right) +\frac{\mathcal{A}_2 ^{(3)}}{3} \\
& \quad\quad\quad\quad -\frac{8\bar{a}_0 +7\varepsilon_1 +2\varepsilon_2}{6} \mathcal{A}_1 ^{(1)} \mathcal{A}_2 ^{(1)}  -\frac{2\varepsilon_1 \varepsilon_2}{3} (\Delta_0 -\Delta_a) \mathcal{A}_1 ^{(1)} -\frac{\left(\mathcal{A}_2 ^{(1)}\right)^3}{18}  + \frac{\varepsilon_1 \varepsilon_2 (\Delta_0 -\Delta_a)}{6} (12\bar{a}_0 + 15\varepsilon_1 + 8\varepsilon_2) \\
&  \quad\quad\quad\quad \left.  + \left( \frac{\varepsilon_1 \varepsilon_2 (\Delta_0+\Delta_{\infty} -\Delta_a)}{3} +\frac{7 \bar{a}_0 \varepsilon_1}{2} +\frac{\prod_{\alpha<\alpha'} a_{0,\alpha} a_{0,\alpha'} + \varepsilon_2 (2a_{0,\beta} +3\bar{a}_0)}{3} +\frac{51 \varepsilon_1 ^2 +48 \varepsilon_1 \varepsilon_2 +2 \varepsilon_2 ^2}{18} \right) \mathcal{A}_2 ^{(1)}  \right] \\
&- \mathfrak{q} \left[ \varepsilon_1 \varepsilon_2 \left(-\mathcal{A}_1 ^{(1)} +\mathcal{A}_2 ^{(1)} +2\varepsilon \right) \mathfrak{q} \frac{\partial}{\partial \mathfrak{q}} +\frac{\mathcal{A}_1 ^{(3)} +\mathcal{A}_2 ^{(2)} }{3} + \frac{\mathcal{A}_1 ^{(2)} \left( -12\bar{a}_0 +2\mathcal{A}_1 ^{(1)} +8 \mathcal{A}_2 ^{(1)} +3\varepsilon_1 +10\varepsilon_2 \right)}{12} \right. \\
& \quad\quad\quad+\frac{\mathcal{A}_2 ^{(2)} \left( -12\bar{a}_0 +8 \mathcal{A}_1 ^{(1)} +  2\mathcal{A}_2 ^{(1)} -21\varepsilon_1 -14\varepsilon_2 \right)}{12} + \frac{\left(\mathcal{A}_1 ^{(1)} \right)^3 -5 \left( \mathcal{A}_2 ^{(1)} \right)^3}{18} +\frac{(-12\bar{a}_0 +4 \mathcal{A}_2 ^{(1)} -3\varepsilon_1 -10\varepsilon_2)\left(\mathcal{A}_1 ^{(1)} \right)^2}{36} \\
& \quad\quad\quad+\frac{\left(-12\bar{a}_0 +28 \mathcal{A}_1 ^{(1)} -39 \varepsilon_1 -46 \varepsilon_2\right) \left(\mathcal{A}_2 ^{(1)}\right)^2}{36} + \mathcal{A}_1 ^{(1)} \mathcal{A}_2 ^{(1)} \left( -2\bar{a}_0 - \frac{\varepsilon_1}{6} + \frac{8\varepsilon_2}{9} \right) \\
& \quad\quad\quad+\mathcal{A}_1 ^{(1)} \left( \varepsilon_1 \varepsilon_2 (\Delta_a -\Delta_0) -\frac{\bar{a}_0 \varepsilon_1}{2} -2 \bar{a}_0 \varepsilon_2 + \frac{a_{0,\beta} \varepsilon_2}{3} +\frac{3\varepsilon_1 ^2 +9\varepsilon_1 \varepsilon_2 +38 \varepsilon_2 ^2}{18} +\frac{\varepsilon_1 \varepsilon_2 \Delta_\infty -\varepsilon^2 +\prod_{\alpha<\alpha'} a_{0,\alpha}a_{0,\alpha'}}{3} \right) \\
&\quad\quad\quad+\mathcal{A}_2 ^{(1)} \left( \varepsilon_1 \varepsilon_2 (\Delta_0 - \Delta_a) +\bar{a}_0 \left( \frac{3\varepsilon_1}{2} -\varepsilon_2 \right) +\frac{4a_{0,\beta} \varepsilon_2}{3} +\frac{12\varepsilon_1 ^2 -15 \varepsilon_1 \varepsilon_2 -16\varepsilon_2 ^2}{18} +\frac{\varepsilon_1 \varepsilon_2 \Delta_\infty -\varepsilon^2 +\prod_{\alpha<\alpha'} a_{0,\alpha}a_{0,\alpha'}}{3} \right) \\
&\quad\quad\quad\left.+2\varepsilon_1  \varepsilon_2 \varepsilon (\Delta_0-\Delta_a) +\frac{\varepsilon_2 (a_{0,\beta}-\bar{a}_0 )(2a_{0,\beta}-2\bar{a}_0-3\varepsilon_1)}{2} +\frac{\varepsilon_2 ^2 (4a_{0,\beta} -4\bar{a}_0 +3\varepsilon_1)}{6} - \frac{35 \varepsilon_2 ^3}{9}  \right] \\
&+ \frac{\mathfrak{q}^2}{3(1-\mathfrak{q})} \mathcal{A}_2 ^{(1)} \left( \mathcal{A}_1 ^{(1)} -3\varepsilon \right) \left( \mathcal{A}_1 ^{(1)} - \mathcal{A}_2 ^{(1)} -2\varepsilon \right),
\end{split}
\end{align}
\normalsize
where we have defined
\begin{align}
\Delta_a \equiv \frac{1}{\varepsilon_1 \varepsilon_2} \left( \varepsilon^2 -\frac{(a_{2,1}-a_{2,2})^2 +(a_{2,1}-a_{2,3})^2-(a_{2,1}-a_{2,2})(a_{2,1}-a_{2,3})}{3} \right).
\end{align}
The accessory parameter $H_2$ can be obtained simply by taking the limit $\varepsilon_2 \to 0$. 

\section{Computing the non-regular parts of $\EuScript{X}_\omega$} \label{appC}
We present the explicit expressions for the non-regular parts of the fundamental refined $qq$-character $\EuScript{X}_\omega$ for the $N=3$ case, i.e., the $(2,1)$-type $\mathbb{Z}_2$-orbifold surface defect. The computation for the $N=2$ case is easier and can be done in a similar way.
\begin{align}
\begin{split}
&[x^{-1}] \; \EuScript{X}_0 (x)  \\
&= \frac{\varepsilon_1 ^2}{2} (k_0 -k_1)^2 -\frac{\varepsilon_1 ^2}{2}(k_0 -k_1) +\varepsilon_1 \varepsilon_2 k_1 + (\varepsilon-a_\beta) \varepsilon_1 (k_0 -k_1) +\varepsilon_1 \left( \sum_{K_0} c_{\square} - \sum_{K_1} c_{\square} \right) \\
& + \mathfrak{q}_0 \left[ \frac{1}{2} \left( \sum_{\bar{\beta} \neq \beta} a_{\bar{\beta}} -\sum_{\alpha} m_{+,\alpha} +\varepsilon_1 (k_0-k_1) \right)^2 +\frac{1}{2}\sum_{\bar{\beta}\neq\beta}  a_{\bar{\beta}}^2 - \frac{1}{2}\sum_{\alpha} m_{+,\alpha}^2 +\frac{\varepsilon_1 ^2}{2}(k_0-k_1) -\varepsilon_1 \varepsilon_2 k_1  \right. \\
&\left.  \quad\quad +\varepsilon_1 \left( \sum_{K_0} c_{\square} -\sum_{K_1} c_{\square}  \right) \right]
\end{split}
\end{align}

\begin{align}
\begin{split}
&[x^{-2}]\; \EuScript{X}_0 (x) \\
&= \frac{\varepsilon_1 ^3}{6} (k_0 -k_1)^3 -\frac{\varepsilon_1 ^3}{2} (k_0-k_1)^2 +\varepsilon_1 ^2 \varepsilon_2 k_1 (k_0-k_1) +\frac{\varepsilon_1 ^3}{3} (k_0-k_1) -\varepsilon_1 \varepsilon_2 \varepsilon k_1 + 2\varepsilon_1 \varepsilon_2 \sum_{K_1} c_{\square} \\
&+\varepsilon_1 ^2 (k_0-k_1) \left(\sum_{K_0} c_{\square} -\sum_{K_1} c_{\square} \right) -\varepsilon_1 ^2 \left( \sum_{K_0} c_{\square} -\sum_{K_1} c_{\square} \right) + \varepsilon_1 \left( \sum_{K_0} c_{\square} ^2 -\sum_{K_1} c_{\square} ^2 \right) \\
&+(\varepsilon-a_\beta) \left( \frac{\varepsilon_1 ^2}{2} (k_0-k_1)^2 -\frac{\varepsilon_1 ^2}{2} (k_0-k_1) +\varepsilon_1 \varepsilon_2 k_1 +\varepsilon_1 \left( \sum_{K_0} c_{\square} -\sum_{K_1} c_{\square}\right) \right) \\
&+\mathfrak{q}_0 \left[ \frac{1}{6} \left( \sum_{\bar{\beta}\neq\beta} a_{\bar{\beta}} -\sum_{\alpha} m_{+,\alpha} +\varepsilon_1 (k_0-k_1) \right)^2 +\frac{\varepsilon_1}{2} \left( \sum_{\bar{\beta} \neq\beta} a_{\bar{\beta}}^2 -\sum_{\alpha} m_{+,\alpha}^2  \right)(k_0-k_1) \right. \\
&+\frac{\varepsilon_1 ^3}{2} (k_0-k_1)^2 -\varepsilon_1 ^2 \varepsilon_2 k_1 (k_0-k_1) +\varepsilon_1 ^2 (k_0-k_1)\left( \sum_{K_0} c_{\square} -\sum_{K_1} c_{\square} \right) +\frac{\varepsilon_1 ^3}{3}(k_0-k_1) - 2\varepsilon_1 \varepsilon_2 \sum_{K_1} c_{\square} \\
&+\frac{1}{3} \left( \sum_{\bar{\beta}\neq\beta} a_{\bar{\beta}} ^3 -\sum_{\alpha} m_{+,\alpha} ^3 \right) +\varepsilon_1 ^2 \left( \sum_{K_0} c_{\square} -\sum_{K_1} c_{\square}\right) +\varepsilon_1 \left( \sum_{K_0} c_{\square}^2 -\sum_{K_1} c_{\square}^2\right) -\varepsilon_1 \varepsilon_2 \varepsilon k_1 \\
&\left.+ \left( \sum_{\bar{\beta}\neq\beta} a_{\bar{\beta}} -\sum_{\alpha} m_{+,\alpha} \right) \left( \frac{1}{2} \sum_{\bar{\beta} \neq \beta} a_{\bar{\beta}}^2 -\frac{1}{2} \sum_{\alpha} m_{+,\alpha} ^2 +\frac{\varepsilon_1 ^2}{2} (k_0-k_1) -\varepsilon_1 \varepsilon_2 k_1 +\varepsilon_1 \left( \sum_{K_0} c_{\square} -\sum_{K_1} c_{\square}\right) \right)\right]
\end{split}
\end{align}

\begin{align}
\begin{split}
&[x^{-1}] \; \EuScript{X}_1 (x) \\
& = -\frac{\varepsilon_1 ^3}{6} (k_0-k_1)^3 -\frac{\varepsilon_1 ^3}{2} (k_0-k_1)^2 - \frac{\varepsilon_1 ^3}{3}(k_0-k_1) -\varepsilon_1 ^2 \varepsilon_2 k_0 (k_0-k_1) +\varepsilon_1 ^2 (k_0-k_1)\left( \sum_{K_0} c_{\square} -\sum_{K_1} c_{\square}\right) \\
& +\varepsilon_1 ^2 \left( \sum_{K_0} c_{\square} -\sum_{K_1} c_{\square}\right) - \varepsilon_1 \left( \sum_{K_0} c_{\square}^2 -\sum_{K_1} c_{\square}^2\right) +2\varepsilon_1 \varepsilon_2 \sum_{K_0} c_{\square} -\varepsilon_1 \varepsilon_2 \varepsilon k_0 -\varepsilon_1 (k_0-k_1) \prod_{\bar{\beta} \neq \beta} (\varepsilon - a_{\bar{\beta}} )  \\
&+\left(2\varepsilon -\sum_{\bar{\beta} \neq \beta} a_{\bar{\beta}} \right) \left( \frac{\varepsilon_1 ^2}{2} (k_0-k_1)^2 +\frac{\varepsilon_1 ^2}{2} (k_0-k_1) +\varepsilon_1 \varepsilon_2 k_0 -\varepsilon_1 \left( \sum_{K_0} c_{\square} -\sum_{K_1} c_{\square}\right) \right) \\
&+\mathfrak{q}_1 \left[ \frac{1}{6} \left( a_\beta -\sum_{\alpha} m_{-,\alpha} -\varepsilon_1 (k_0-k_1) \right)^3  +\frac{\varepsilon_1 ^3}{2}(k_0-k_1)^2 +\varepsilon_1 ^2\varepsilon_2 k_0 (k_0-k_1) -\frac{\varepsilon_1 ^3}{3} (k_0-k_1)\right. \\
&-\frac{\varepsilon_1}{2} \left( a_\beta ^2 -\sum_{\alpha} m_{-,\alpha} ^2 \right)(k_0-k_1)  -\varepsilon_1 \varepsilon_2 \varepsilon k_0 -2\varepsilon_1 \varepsilon_2 \sum_{K_0} c_{\square} +\frac{1}{3} a_{\beta} ^3 -\frac{1}{3} \sum_{\alpha} m_{-,\alpha} ^3 \\
&+ \varepsilon_1 ^2 (k_0-k_1)\left( \sum_{K_0} c_{\square} -\sum_{K_1} c_{\square}\right) -\varepsilon_1 ^2 \left( \sum_{K_0} c_{\square} -\sum_{K_1} c_{\square}\right)-\varepsilon_1 \left( \sum_{K_0} c_{\square}^2 -\sum_{K_1} c_{\square}^2\right) \\
&\left. +\left( a_\beta -\sum_{\alpha} m_{-,\alpha} \right) \left( -\frac{\varepsilon_1 ^2 }{2} (k_0-k_1) -\varepsilon_1 \varepsilon_2 k_0 -\varepsilon_1 \left( \sum_{K_0} c_{\square} -\sum_{K_1} c_{\square}\right) +\frac{a_\beta ^2}{2}-\frac{\sum_{\alpha} m_{-,\alpha} ^2}{2} \right) \right]
\end{split}
\end{align}

\section{Computing the Poisson brackets} \label{appD}
Let us first recap some definitions. Let $N \approx {\BC}^{N}$ be a vector space with a volume form. 
Let 
\beq
g_{i}, M_{i} \in {\rm End}(N), , \  \, i = -1, 0, 1, \ldots , r+1
\eeq
be $SL(N)$ matrices, such that
\begin{align}
\begin{split}
&M_{-1} = g_{-1} \\
& g_{i} = {\fm}_{i} \left( 1  + \left( {\fm}_{i}^{-N}-1 \right) {\Pi}_{i} \right) \, , \qquad i = 0, 1, \ldots, r \\
&M_i = g_{-1} g_0 g_1 \ldots g_{i} = \sum_{\alpha=1} ^N \fm_i ^{(\alpha)} \Pi_i ^{(\alpha)} \, , \qquad i = 0, 1, \ldots, r \\
&M_{r+1}  = \mathds{1}_N,
\end{split}
\end{align}
where the projection operators $\Pi$ are written in terms of
\begin{align}
\begin{split}
&E_i, E_i ^{(\alpha)} \in N, \quad \tilde{E}_i, \tilde{E}_i ^{(\alpha)} \in N^*, \\ & \tilde{E}_i (E_i) =1, \quad \tilde{E}_i ^{(\alpha)} (E_i ^{(\beta)}) = \delta_{\alpha,\beta}
\end{split}
\end{align}
as
\begin{align}
\begin{split}
&\Pi_i = E_i \otimes \tilde{E}_i, \quad i=0,1,\cdots, r-1, \\
&\Pi_i ^{(\alpha)} = E_i ^{(\alpha)} \otimes \tilde{E}_i ^{(\alpha)}, \quad i=-1,0,1,\cdots, r, \,\; \alpha=1,\cdots, N.
\end{split}
\end{align}
The following formulas are useful throughout the computation: for any $\alpha, \beta  = 1, \ldots, N$,  
\begin{align}
\begin{split}
&{\tilde E}_{i+1}^{(\alpha)}(E_{i}^{(\beta)}) = \frac{\fm_{i+1} ^{(\alpha)}({\fm}_{i+1}^{N}-1)}{{\fm}_{i+1}{\fm}_{i}^{(\beta)} - {\fm}_{i+1}^{(\alpha)}} \, {\tilde E}_{i+1}^{(\alpha)}(E_{i+1}) {\tilde E}_{i+1}(E_{i}^{(\beta)}) \\
&\tilde{E} _{i} ^{(\alpha)} (E_{i+1} ^{(\beta)} ) = \frac{ \fm_i ^{(\alpha)} (\fm_{i+1} ^{-N} -1)}{\fm _{i+1} ^{-1} \fm_{i+1} ^{(\beta)} -\fm_i ^{(\alpha)} } \tilde{E} _i ^{(\alpha)} (E_{i+1}) \tilde{E}_{i+1} (E_{i+1} ^{(\beta)}).
\end{split}
\end{align}
We packaged the Darboux coordinates into
\begin{align}
\begin{split}
&{\BA}_{i}(x) \equiv {\Tr}_{N} \, \left( x - M_{i}\right)^{-1} = \sum_{l=0} ^\infty \frac{1}{x^{l+1}} \Tr_N \, M_i ^l \\
&{\BB}_{i}(x) \equiv {\Tr}_{N}\, {\Pi}_{i} \left( x- M_{i}\right)^{-1} {\Pi}_{i+1} = e^{\tilde{\boldsymbol\beta}_i} \sum_{\alpha=1}^{N} e^{-{\tilde{\boldsymbol\beta}}_{i}^{(\alpha)}} \, \frac{{\Tr}_{N} \, {\Pi}_{i}{\Pi}_{i}^{(\alpha)}}{x-m_{i}^{(\alpha)}},
\end{split}
\end{align}
where we express ${\BB}_{i}(x)$ via
\begin{align}
\begin{split}
{\BD}_{i}(x) &\equiv {\Tr}_{N}\, g_{i} \left( x- M_{i}\right)^{-1} g_{i+1} \\
&= {\fm}_{i}{\fm}_{i+1} ({\fm}_{i}^{-N}-1) ({\fm}_{i+1}^{-N}-1) 
{\BB}_{i}(x) + \fm_i \fm_{i+1} x^{-1} \left( \frac{P_{i-1} (\fm_i ^{-1} x)}{P_i (x)} -1 \right) \\
& \quad - \fm_i \fm_{i+1} ^{1-N} x^{-1} \left( \frac{P_{i+1} (\fm_{i+1} x)}{P_i (x)} -1 \right) + \fm_i \fm_{i+1} \BA_i (x).
\end{split}
\end{align}

\begin{figure}
\centering
\begin{tikzpicture}
\node[inner sep=0pt] (figure) at (0,0) {\includegraphics[width=0.8\textwidth]{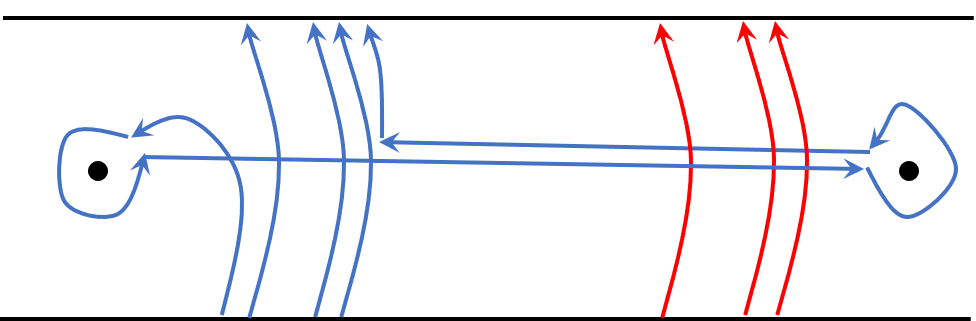}};
\node at (-5,-0.5) {\scalebox{1.0}{$i$}};
\node at (5.6,-0.5) {\scalebox{1.0}{$i+1$}};
\node at (-2.4,1) {\scalebox{1.0}{$....$}};
\node at (2.9,1) {\scalebox{1.0}{$....$}};
\node at (2.75,0.35) {\scalebox{1.0}{$-$}};
\node at (3.75,0.35) {\scalebox{1.0}{$-$}};
\node at (4.2,0.35) {\scalebox{1.0}{$-$}};
\node at (2.75,-0.3) {\scalebox{1.0}{$+$}};
\node at (3.75,-0.3) {\scalebox{1.0}{$+$}};
\node at (4.2,-0.3) {\scalebox{1.0}{$+$}};
\end{tikzpicture} \caption{The geometric picture for $\left\{ {\BD}_{i}(x) , {\BA}_{i}(y) \right\}$.} \label{fig:dabrac}
\end{figure}

The brackets remained to be computed are
\beq
\Biggl\{ {\BD}_{i}(x) , {\BA}_{i}(y) \Biggr\}, \quad \quad \Biggl\{ {\BD}_{i}(x), {\BD}_{i+1}(y) \Biggr\}, \quad \text{and}\quad \Biggl\{ {\BD}_{i}(x), {\BD}_{i}(y) \Biggr\}.
\eeq
Using the geometric representation \eqref{eq:poigeo}, the first Poisson bracket is computed as (see Figure \ref{fig:dabrac})
\begin{align}
\begin{split}
\Biggl\{ {\BD}_{i}(x) , {\BA}_{i}(y) \Biggr\}  &= {\Tr}_{N} \left( \frac{M_{i}}{(y-M_{i})^{2}} g_{i} \frac{1}{x-M_{i}} g_{i+1} \right) - {\Tr}_{N} \left( g_{i} \frac{M_{i}}{(y-M_{i})^{2}}  \frac{1}{x-M_{i}} g_{i+1} \right)   \\
&= {\fm}_{i}{\fm}_{i+1} ( {\fm}_{i}^{-N}-1 ) ({\fm}_{i+1}^{-N}-1) \, {\Tr}_{N} \left[ \frac{M_{i}}{(y-M_{i})^{2}}, \Pi _i \right] \frac{1}{x-M_{i}} {\Pi}_{i+1}.
\end{split}
\end{align}
On the other hand, a direct computation gives (we omit the $2\pi i $ in front of the $\boldsymbol\alpha$ coordinates)
\begin{align}
\begin{split}
\Biggl\{ {\BD}_{i}(x) , {\BA}_{i}(y) \Biggr\} &=\fm_i \fm_{i+1} (\fm_i ^{-N}-1)(\fm_{i+1} ^{-N}-1) \Biggl\{ e^{\tilde{\boldsymbol\beta}_i} \sum_{\alpha=1} ^N \frac{e^{-\tilde{\boldsymbol\beta}_i ^{(\alpha)}} \text{Tr}\, \Pi_i \Pi_i ^{(\alpha)}}{x-\fm_i ^{(\alpha)}}, \sum_{\beta=1} ^N \frac{1}{x-\fm_i ^{(\beta)}} \Biggr\} \\
&=\fm_i \fm_{i+1} (\fm_i ^{-N}-1)(\fm_{i+1} ^{-N}-1) \sum_{\alpha,\beta} \frac{\fm_i ^{(\alpha)} \, \text{Tr}\, \Pi_i \Pi_i ^{(\beta)} \Pi_{i+1}}{(x-\fm_i ^{(\beta)}) (y-\fm_i ^{(\alpha)})^2} \{ \boldsymbol\alpha_i ^{(\alpha)} , \tilde{\boldsymbol\beta}_i ^{(\beta)} \} \\
& \quad -\fm_i \fm_{i+1}(\fm_i ^{-N}-1)(\fm_{i+1} ^{-N}-1) \sum_{\alpha, \beta,\gamma}  \frac{\fm_i ^{(\alpha)} \text{Tr}\, \Pi_i \Pi_i ^{(\beta)} \Pi_{i+1} \Pi_i ^{(\gamma)}}{(x-\fm_i ^{(\beta)})(y-\fm_i ^{(\alpha)})^2} \{ \boldsymbol\alpha_i ^{(\alpha)}, \tilde{\boldsymbol\beta}_i ^{(\gamma)} \} 
\end{split}
\end{align}
By comparing the two expressions, we derive:
\beq
\Biggl\{  \tilde{\boldsymbol\beta}_i ^{(\alpha)}  , \,  \boldsymbol\alpha_i ^{(\beta)} \Biggr\} = {\delta}_{\alpha,\beta}, \quad i=0,1, \cdots, r-1, \; \alpha,\beta=1, \cdots, N.
\eeq

\begin{figure}
\centering
\begin{tikzpicture}
\node[inner sep=0pt] (figure) at (0,0) {\includegraphics[width=0.8\textwidth]{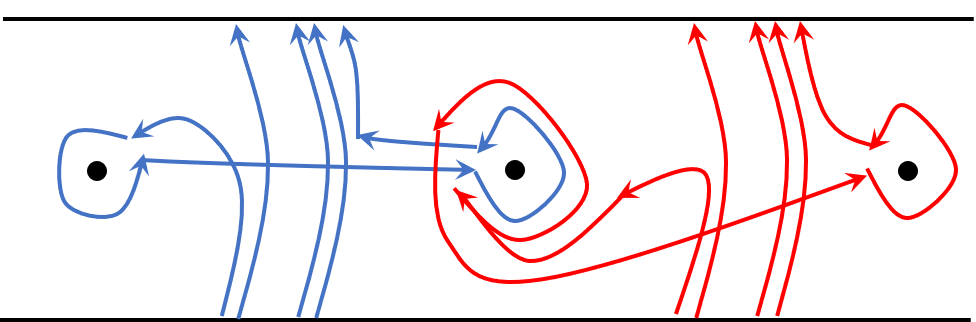}};
\node at (0.5,-0.5) {\scalebox{1.0}{$i+1$}};
\node at (-5,-0.5) {\scalebox{1.0}{$i$}};
\node at (5.5,-0.5) {\scalebox{1.0}{$i+2$}};
\node at (-2.6,1) {\scalebox{1.0}{$....$}};
\node at (3.25,1) {\scalebox{1.0}{$....$}};
\node at (-0.85,0.45) {\scalebox{1.0}{$+$}};
\node at (-0.85,-0.35) {\scalebox{1.0}{$-$}};
\end{tikzpicture} \caption{The geometric picture for $\left\{\BD_i (x), \BD_{i+1} (y) \right\}$.} \label{fig:dd1}
\end{figure}

Next, we compute from the geometric representation (see Figure \ref{fig:dd1})
\begin{align}
\Biggl\{ {\BD}_{i}(x), {\BD}_{i+1}(y) \Biggr\} = {\Tr}_{N}\, \left(  \Biggl[ g_{i+1}, g_{i} (x-M_{i})^{-1} \Biggr] g_{i+1} (y- M_{i+1})^{-1} g_{i+2} \right).
\end{align}
On the other hand, a direct computation gives
\begin{align}
\begin{split}
\Biggl\{ {\BD}_{i}(x), {\BD}_{i+1}(y) \Biggr\} &= \fm_i \fm_{i+1} ^2 \fm_{i+2} (\fm_i ^{-N}-1)(\fm_{i+1} ^{-N}-1)(\fm_{i+2}^{-N}-1) \{ \BB_i (x), \BB_{i+1} (y) \} \\
& + \fm_i \fm_{i+1} ^2 \fm_{i+2} (\fm_i ^{-N}-1)(\fm_{i+1} ^{-N}-1) y^{-1} \left\{ \BB_i (x), \frac{P_{i} (\fm_{i+1} ^{-1} y)}{P_{i+1} (y)} \right\} \\
& - \fm_i \fm_{i+1} ^{2-N} \fm_{i+2} (\fm_{i+1} ^{-N}-1)(\fm_{i+2} ^{-N}-1) x^{-1} \left\{ \frac{P_{i+1} (\fm_{i+1} x)}{ P_i (x)}  , \BB_{i+1} (y) \right\}.
\end{split}
\end{align}
Each term can be explicitly computed. By comparing the results we derive
\begin{align}
\Biggl\{ \tilde{\boldsymbol\beta}_i ^{(\alpha)} , \tilde{\boldsymbol\beta}_{i+1} ^{(\beta)} \Biggr\} = 0, \quad i =0, 1, \cdots, r-1, \; \alpha, \beta=1, \cdots, N.
\end{align}

\begin{figure}
\centering
\begin{tikzpicture}
\node[inner sep=0pt] (figure) at (0,0) {\includegraphics[width=0.8\textwidth]{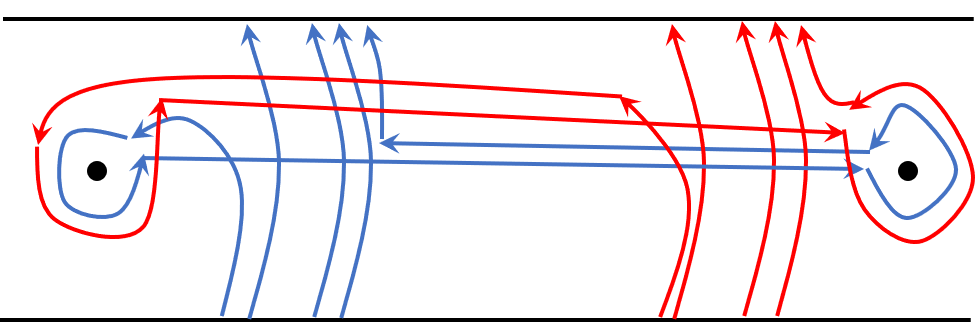}};
\node at (-5,-0.5) {\scalebox{1.0}{$i$}};
\node at (5.6,-0.5) {\scalebox{1.0}{$i+1$}};
\node at (-2.4,-1) {\scalebox{1.0}{$....$}};
\node at (3.1,-1) {\scalebox{1.0}{$....$}};
\node at (-4,0.3) {\scalebox{1.0}{$-$}};
\node at (-4,-0.25) {\scalebox{1.0}{$+$}};
\node at (-2.75,1.3) {\scalebox{1.0}{$+$}};
\node at (-2,1.3) {\scalebox{1.0}{$+$}};
\node at (-1.6,1.3) {\scalebox{1.0}{$+$}};
\node at (-1.2,1.3) {\scalebox{1.0}{$+$}};
\node at (-2.55,0.5) {\scalebox{1.0}{$-$}};
\node at (-1.75,0.5) {\scalebox{1.0}{$-$}};
\node at (-1.45,0.5) {\scalebox{1.0}{$-$}};
\node at (-1.15,0.5) {\scalebox{1.0}{$-$}};
\node at (4.7,0.3) {\scalebox{1.0}{$+$}};
\node at (4.75,-0.3) {\scalebox{1.0}{$-$}};
\node at (3.9,-0.3) {\scalebox{1.0}{$+$}};
\node at (3.45,-0.3) {\scalebox{1.0}{$+$}};
\node at (2.3,-0.3) {\scalebox{1.0}{$+$}};
\node at (2.9,-0.3) {\scalebox{1.0}{$+$}};
\node at (3.9,0.3) {\scalebox{1.0}{$-$}};
\node at (3.45,0.3) {\scalebox{1.0}{$-$}};
\node at (2.3,0.3) {\scalebox{1.0}{$-$}};
\node at (2.9,0.3) {\scalebox{1.0}{$-$}};
\end{tikzpicture} \caption{The geometric picture for $\left\{ {\BD}_{i}(x) , {\BD}_{i}(y) \right\}$.} \label{fig:dd2}
\end{figure}

Finally, we compute from the geometric representation (see Figure \ref{fig:dd2})
\begin{align}
\begin{split}
\Biggl\{ {\BD}_{i}(x), {\BD}_{i}(y) \Biggr\}&= \Tr_N \left( \left[ \frac{1}{y-M_i}  g_{i+1}, g_{i+1} \right] \frac{1}{x-M_i} g_{i+1} g_i \frac{M_i}{x-M_i}   \right) \\
&+ \Tr_N \left( \left[ g_i \frac{1}{x-M_i} , g_{i+1} \right] \frac{M_i}{y-M_i} g_{i+1}g_i \frac{1}{y-M_i} \right) \\
& + \Tr_N \left( g_i \frac{1}{y-M_i} g_{i+1} \left[ g_i , \frac{1}{x-M_i} g_{i+1} \right] \right) \\
& + \Tr_N \left( g_i \frac{1}{y-M_i} g_{i+1} \left[ g_{i+1} , g_i \frac{1}{x-M_i} \right] \right).
\end{split}
\end{align}
On the other hand, a direct computation gives
\begin{align}
\begin{split}
&\Biggl\{ {\BD}_{i}(x), {\BD}_{i}(y) \Biggr\} \\ &= \fm_i ^2 \fm_{i+1} ^2 (\fm_i ^{-N}-1)(\fm_{i+1} ^{-N}-1)\left( \{\BB_i (x), \BA_i (y) \} + \{ \BA_i (x), \BB_i (y) \} \right) \\
&+\fm_i ^2 \fm_{i+1} ^2 (\fm_i ^{-N}-1)(\fm_{i+1} ^{-N}-1) \left( y^{-1} \left\{ \BB_i (x), \frac{P_{i-1} (\fm_i ^{-1} y)}{P_i (y)} \right\}+ x^{-1} \left\{ \frac{P_{i-1} (\fm_i ^{-1} x)}{P_i (x)} , \BB_i (y) \right\}  \right) \\
&- \fm_i ^2 \fm_{i+1} ^{2-N} (\fm_i ^{-N}-1)(\fm_{i+1} ^{-N}-1) \left( y^{-1} \left\{ \BB_i (x), \frac{P_{i+1} (\fm_{i+1} y)}{P_i (y)} \right\} +x^{-1} \left\{  \frac{P_{i+1} (\fm_{i+1} x)}{P_i (x)} , \BB_i (y) \right\}  \right) \\
&+ \fm_i ^2 \fm_{i+1} ^2 (\fm_i ^{-N}-1)^2 (\fm_{i+1} ^{-N}-1)^2 \{ \BB_i (x), \BB_i (y) \},
\end{split}
\end{align}
in which all the brackets are explicitly computable. By comparing the results we derive
\begin{align}
\Biggl\{ \widetilde{\boldsymbol\beta}_i ^{(\alpha)} , \widetilde{\boldsymbol\beta}_i ^{(\beta)} \Biggr\} = 0, \quad i=0, 1, \cdots, r=1, \; \alpha, \beta=1, \cdots, N.
\end{align}
Therefore, we confirm that the Poisson brackets for the coordinates $\boldsymbol\alpha_i ^{(\alpha)}, \tilde{\boldsymbol\beta}_i ^{(\alpha)}$ are canonical.



\end{document}